\documentclass[twocolumn]{aa}
\usepackage{graphicx}  
\usepackage{wrapfig}
\usepackage[pdftex]{color}
\usepackage{mathrsfs}
\usepackage{mathtools}
\DeclarePairedDelimiter\abs{\lvert}{\rvert}
\usepackage{dblfloatfix}
\usepackage[right]{eurosym}
\usepackage{natbib}   
\usepackage{enumitem}
\usepackage[colorinlistoftodos,prependcaption,textsize=tiny]{todonotes}
 \usepackage{flushend} 
\usepackage{verbatimbox}
\usepackage{aalongtable}

\def\degs{\ifmmode ^{\circ}\else$^{\circ}$\fi}
\def\amin{\ifmmode ^{\prime}\else$^{\prime}$\fi}
\def\asec{\ifmmode ^{\prime\prime}\else$^{\prime\prime}$\fi}
\def\fdg{\hbox{$.\!\!^\circ$}}          
\def\farcm{\hbox{$.\!\!^{\prime}$}}  
\newbox\grsign \setbox\grsign=\hbox{$>$}
\newdimen\grdimen \grdimen=\ht\grsign
\newbox\laxbox \newbox\gaxbox
\setbox\gaxbox=\hbox{\raise.5ex\hbox{$>$}\llap
     {\lower.5ex\hbox{$\sim$}}}\ht1=\grdimen\dp1=0pt
\setbox\laxbox=\hbox{\raise.5ex\hbox{$<$}\llap
     {\lower.5ex\hbox{$\sim$}}}\ht2=\grdimen\dp2=0pt
\def\gax{$\mathrel{\copy\gaxbox}$}
\def\lax{$\mathrel{\copy\laxbox}$}

\begin{document}

\title{A proposed network of Gamma-ray Burst detectors \\ on the  Global Navigation
  Satellite System Galileo G2\thanks{The work described in this document was done 
    under ESA contract with funding from the EU Horizon-2020 program
    (H2020-038.09). Responsibility for the content
  resides in the author or organization that prepared it.}}

\author{J. Greiner\inst{1}, U. Hugentobler\inst{2}, J.M. Burgess\inst{1},
  F. Berlato\inst{1}, M. Rott\inst{2}, A. Tsvetkova\inst{1, 3}}

\authorrunning{Greiner et al.}

\titlerunning{Network of GRB detectors on Galileo G2}

\offprints{J. Greiner, jcg@mpe.mpg.de}

\institute{
  Max-Planck Institute for extraterrestrial Physics, 
    Giessenbachstr. 1,  85748 Garching, Germany
  \and
  Technical Univ. Munich, Institute for Astronomical and
  Physical Geodesy, Arcisstr. 21, 80333 Munich, Germany
  \and   
  Ioffe Institute, Polytechnikheskaya 26, St. Petersburg 194021, Russia}

\date{Received 12 Dec 2021  / Accepted 17 May 2022 }

\abstract{The accurate localization of gamma-ray bursts remains a
  crucial task. While historically, improved localization have led
  to the discovery of afterglow emission and the realization of
  their cosmological distribution via redshift measurements, a more recent
  requirement comes with the potential of studying the kilonovae of
  neutron star mergers. Gravitational wave detectors are expected to
  provide locations to not better than 10 square degrees over the next
  decade. With their increasing horizon for merger detections also the
  intensity of the gamma-ray and kilonova emission drops, making their
  identification in large error boxes a challenge. Thus, a
  localization via the gamma-ray emission seems to be the best chance to
  mitigate this problem.
  Here we propose to equip some of the second generation Galileo satellites
  with dedicated GRB detectors.
  This saves costs for launches and satellites for a dedicated GRB
   network, the large orbital radius is beneficial for triangulation,
   and perfect positional and timing accuracy come for free.
  We present simulations of the
  triangulation accuracy, demonstrating that short GRBs as faint as
  GRB 170817A can be localized to 1 degree radius (1$\sigma$).}


\maketitle

\section{Introduction}

The coincident detection of gravitational waves (GW) from a binary
neutron star merger with aLIGO/Virgo and short-lived gamma-ray (GR) emission
with Fermi-GBM (called GW 170817) in August 2017 is a milestone for the
establishment of multi-messenger astronomy \citep{Abbott2017b},
i.e. the measurement of electromagnetic radiation, gravitational waves
and/or particles or neutrinos from the same astrophysical source.
Merging neutron stars (NS) represent the standard scenario \citep{Eichler1989}
for short-duration ($<2$ s) gamma-ray bursts (sGRBs) which are produced in a
collimated, relativistically expanding jet with an opening angle of a
few degrees and a bulk Lorentz factor $\Gamma$ of 300--1000.
While the aLIGO detection
is consistent with predictions, the measured faint gamma-ray emission from
GW 170817A is about 1000x less luminous than known short-duration GRBs.
Hence, the presence of this sGRB in the local Universe is either a very rare
event, or points to a dramatic mis-understanding of the emission properties
of sGRBs outside their narrow jets. By now we know that the jet in this
GRB had an opening angle of $<$5\degs, but we observed it from
$\sim$20--30\degs\ offaxis \citep{Mooley+2018}. In all previous models,
no emission was predicted to occur outside the opening angle.

Thus, also the previous estimates of the volume density of NS-NS mergers
was wrong, and needs to be corrected \citep{Burgess+2020}.
This has important implications on our understanding of the chemical
evolution of our Galaxy and the Universe, as NS-NS mergers are believed to be
the main source of heavy elements \citep{Kasen+2017},
so-called r-process elements (like gold and platinum). This material is
expelled both during the tidal disruption of the NSs and through winds
during the subsequent disk accretion onto the compact core.
Further progress in our understanding of NS-NS mergers will depend
on measurements in the electromagnetic regime, and these in turn
will only be possible if the localizations of these events can
be reduced to of order a few square degrees on the sky. While there exist
several large-field-of-view optical sky surveys, covering
up to several thousands square degrees, the challenge is to find the kilonova
among the many other transient sources. Future NS-NS mergers will
likely all(!) be at larger distance than GW 170817, and thus their kilonova
much fainter. Already for the only 3x more distant
four NS-NS merger events from 2019--2020, none detected in gamma-rays,
the optical emission would peak at 23rd mag
(if at identical luminosity as GW 170817).
Except for one particularly poor localisation, the error regions of the
other three events encompass 2300--14700 square degrees each \citep{Wiki2020}.
At the expected optical brightness, there will be about 3--60 transient alerts
per square degree down to 21 mag \citep{Masci+2019}, or estimated
5x more at 23rd mag, against which the kilonova will have to be identified.
Thus, a pre-requisite to identify the kilonova is the fast and precise
localization of the GW/GR event.

Expectations for the fourth observing run O4 are 10$^{+52}_{-10}$ BNS
mergers, with a median 33$^{+5}_{-5}$ square degrees localization.
Likely not before 2026 \citep{Abbott+2020},
the GW detector network of LIGO, Virgo and KAGRA is
looking forward to include LIGO-India, which promises a reduction of
the GW error regions to of order $<$10 square degrees. 
Further reduction of the localization error is foreseen with the
Einstein Telescope in Europe, or the Cosmic Explorer in the USA,
both not earlier than the mid 2030s.

Thus, accurate localization of the GW events should be sought elsewhere.
Gamma-rays provide an interesting
alternative, at least for those NS-NS mergers for which the jet would
be broadly pointed towards us. With $\gamma$-ray emission at large
off-axis angles as in GRB 170817A, up to 30\% of mergers will be
simultaneously detectable in $\gamma$-rays \cite[][]{Howell+2019, Burgess+2020}.
Obviously, accurate measurements of many GRBs will be beneficial
for other science questions beyond kilonova physics, such as
(1) the structure of jets in GRBs \citep[e.g.,][]{Janka+2006} and the
origin of the off-axis emission which is distinctly different to on-axis
emission \citep{Begue+2017}, or
(2) the potential emission of high-energy neutrinos as measured by
IceCube \citep{Aartsen+2013}, promising a potential 
'triple'-messenger, i.e. electromagnetic radiation, gravitational waves,
and particles: 
\cite{Kimura+2017} estimated that GRB 170817A could
have been detectable by IceCube if the jet had been viewed on-axis
instead of the $\sim$30\degs\ off-axis.

Here, we propose adding a GRB detector on some of the next
two dozen 2nd generation Galileo satellites (G2) in order to improve
the localization capability for short GRBs to the 1-degree level, 
reducing the error region by a factor of 100--1000.

\section{Prospects of accurate GRB localisation}

\subsection{Challenges of short GRBs}

Short-duration GRBs (sGRB) have three properties which make their localisation
in large numbers more difficult than that of long-duration GRBs:\\
(1) Their short duration, of order 0.01--2 s, implies that their observable
fluence is of order 5--50x smaller than in long-duration GRBs.\\
(2) Their peak fluxes during their maximum spike is typically a factor 10
smaller than long-duration GRBs, making the discrepancy of (1) even larger.\\
(3) sGRBs are also harder, with their spectral peak at
higher energies. This implies that the flux at soft gamma-rays (20--100 keV)
is smaller than that in long-duration GRBs even if the energy-integrated
flux is equal.

These factors together imply detection and localisation disadvantages
in various detector types:
(i) in coded-mask imagers like Swift/BAT or INTEGRAL/IBIS, the mask
elements are getting increasingly more transparent at higher energy,
leading to less ``sharp'' shadows, and thus detection sensitivity.
Thus, the ratio of long-to-short GRBs in Swift/BAT
is about 10:1, while it is 10:2.5 in Fermi/GBM.
(ii) in counting experiments like Fermi/GBM, short spikes can more easily
be mistaken for noise spikes. Moreover, at the higher photon energies,
the cosine dependence of the
effective area is much less pronounced in detectors with slab-like
scintillators, due to the larger absorption probability 
at inclined incidence angles.

\begin{table*}[th]
  \smallskip
  \caption{Comparison of different $\gamma$-localization methods in the
    200--2000 keV band. The sensitivity column reports the peak flux threshold
    over the 1--1000 keV band (for a Band function with $\alpha$=--1,
    E$_{\rm peak}$=300 keV, $\beta$=--2) of the listed detectors
    \citep{Band2003, Bosnjak+2014}. \label{methods}}
  \vspace{-0.3cm}
   \begin{tabular}{llllclc}
    \hline
     \noalign{\smallskip}
     Method & Accuracy & Comments & E-range & GRBs & Example & Sensitivity \\
            &          &          & (keV)   & (1/yr) &       & (ph/cm$^2$/s) \\
      \noalign{\smallskip}
      \hline
      \noalign{\smallskip}
      Triangulation    & arcsec  & cheap, all-sky & 10--1000  & 20-50 & IPN & 2.0 \\
      Relative rates   & degrees & cheap, half sky & 8--500  &300 & BATSE, GBM & 1.0, 3.0 \\
      Coded-mask       & arcmin  & small FOV & 10--200  & 10--100 & Swift/BAT, ISGRI & 1.2, 0.6 \\
      Photon-by-Photon$\!\!$ & degrees & heavy, big & 100--2000 & 10--30 & COMPTEL & 180 \\
       \noalign{\smallskip}
       \hline
    \end{tabular}
\end{table*}

\subsection{Localisation methods}

Rather independent of the different $\gamma$-ray detection technology
(gas detectors, scintillation detectors or solid-state detectors)
are the methods with which gamma-rays can be localized. 
The four main methods with their advantages and disadvantages
(Tab. \ref{methods}) are described below. The summarizing statement
is: large field-of-view (FOV) instruments with high GRB detection rates
are operating at softer energies, not appropriate for short-duration GRBs,
while detectors at higher energies are suffering from either bad
localization capabilities or low detection rates. Over the last 20 years,
all techniques except triangulation have been used in space applications
with the maximum possible capability.

\noindent{\bf Triangulation:} 
Among the first methods of localizing sources in gamma-ray astronomy
was triangulation, i.e. measuring the time difference
of a signal arriving at different detectors. This was the method used
by the Vela satellites in the 1960s to verify the Nuclear Arm Treaty
between USA-Russia, which then led to the discovery of GRBs.
This method requires at least 3 detectors/satellites, and accurate
knowledge of time and the relative position of the detectors; it allows
to cover all-sky, and provides localizations in the arcsec--arcmin range for
widely spaced satellites \citep{Hurley+2017}.
But since GRB detectors on interplanetary
spacecraft are auxiliary instruments, and thus small, triangulation
offers substantial improvements.

\noindent{\bf Orientation-dependent rate measurements:}
Measuring relative rates of orientation dependent $\gamma$-ray
detectors, typically scintillation crystals, was used for
GRB localizations with the BATSE instrument on the Compton
Observatory, and is used presently with the GRB Monitor (GBM)
on Fermi. This method requires $>$4--6 detectors with different
orientations on the sky, and the localization accuracy is on the
degree-scale at best \citep{Berlato+2019}.

\noindent{\bf Coded-mask imaging:}
Coded-mask imaging also allows a 2D reconstruction on the sky,
and was frequently used over the last 30 years, such as Granat, Swift
and INTEGRAL. It also requires large (m$^3$ scale) detector sizes, has
a restricted field-of-view, but allows localizations in the arcmin range.

\noindent{\bf Photon-by-photon imaging:}
 Proper imaging (2D reconstruction) of individual photons on the
sky. This method was used by the COMPTEL telescope in the
1990s. Improved versions require electron tracking, and thus will
be large and heavy telescopes. However, localizations (degrees) and
field of view (up to 70\degs\ radius) are advantageous.

\subsection{Missions with GRB capabilities}

Except for planned missions beyond 2027, 
the near future can be summarized by the following three strategies:
(i) new large(r) missions just represent a replication of existing missions,
  such as GECAM (replicating Fermi/GBM) and SVOM (Swift).
(ii) new small(er) missions are mainly driven by enhancing the
  sky coverage, not improving localizations
(iii) an euphoric engagement in CubeSat swarms using triangulation
  which due to their size and LEO will not provide accurate (degree)
  localizations.
All of these strategies will not change
 the lack of well-localized short-duration GRBs.
The operational and planned (to our knowledge) missions 
are shortly sketched below.

The dedicated GRB mission {\bf Swift} (USA) uses a coded-mask imager
(BAT = Burst
Alert Telescope) in the 15--150 keV range for GRB localization, to an
accuracy of 3 arcmin radius \citep{Barthelmy+2005}.
It has a 1.4 steradian field-of-view (half-coded), and detects about
100 GRBs/yr, predominantly as rate triggers (excess counts in the total rate
of a detector module). Due to the soft energy band
and the combined noise of the 32768 CdZnTe detector cells, the detection
rate of short-duration GRBs is only $\sim$10\% (10 sGRBs/yr).

The gamma-ray observatory Fermi (USA) features a Gamma-ray Burst Monitor
({\bf Fermi/GBM})
aimed at localizing GRBs outside of the zenith-looking field-of-view of the
prime instrument LAT (Large Area Telescope, 100 MeV -- 10 GeV). GBM consists
of two sub-systems: (i) a collection of 12 NaI scintillation detectors
for the energy range 8--500 keV, and (ii) two thick BGO scintillation detectors
for the high-energy range up to 40 MeV \citep{Meegan+2009}. It is presently
the most prolific GRB detector, with the detection and localization of
about 240 GRBs/yr, among those about 40 short-duration GRBs
\citep{Kienlin+2020}. 
The usual, 30-yr-long used localization method (based on orientation-dependent
rates in different detectors) comes with large
systematic errors \citep{Connaughton+2015}. The cause of these systematics
have recently been understood \citep{Burgess+2018}, but even after
correction the typical error regions have 5\degs--10\degs\ radius
\citep{Berlato+2019}, with the 17\degs\ error radius for GRB 170817A
completely dominated by the statistical error.

The Interplanetary Network ({\bf IPN}) is the logistic combination of different
spacecrafts equipped with GRB detectors. The locations of GRBs are determined
by the comparison of the arrival times of the event at the locations of the
GRB detectors. The precision is proportional to the distance of spacecraft
separations, so that the localisation accuracy of a network with baselines
of thousands of light-seconds can be equal or superior to that of any other
technique \citep{Hurley+2017}. 
A major disadvantage of the IPN method is the 1--2 day delay in the
downlink of the GRB data from the spacecraft.
At present, the main IPN contributors are Konus-WIND, Mars Odyssey,
INTEGRAL, RHESSI, Swift, AGILE, BepiColombo, and Fermi/GBM. 

The European gamma-ray satellite {\bf INTEGRAL} can detect GRBs with
three of its instruments, i.e. in the field-of-view of ISGRI (a 15-300 keV
coded-mask imager with few arcmin localization accuracy) or SPI (a
200--8000 keV coded mask imager with degree localization accuracy but
very high energy resolution), and the SPI anti-coincidence system ACS
(working at $>$80 keV). Due to the small field-of-view of ISGRI and SPI,
their combined GRB detection rate is only $\sim$10 GRBs/yr
\citep{Bosnjak+2014}. The ACS detects about 150 GRBs/yr, but has no
localization capability \citep{Savchenko+2012}.

\noindent
{\bf CALET} (Japan),  {\bf Insight-HXMT} (China) and {\bf AstroSat/CZTI} (India)
are operational satellite experiments with the capability of detecting GRBs
in their particle detectors or shields, without localizations. Due to their
low-Earth orbit, they do not provide constraints via
triangulation, and thus are not (or very rarely) used in the IPN.

\noindent{\bf GECAM} (China): The Gravitational Wave Electromagnetic
Counterpart All-sky Monitor GECAM is a twin spacecraft mission to monitor GRBs
coincident with GW events \citep{ZhengXiong2019}. With a dome-shaped
distribution of multiple scintillators it reaches an effective area
(and energy range) similar to that of Fermi/GBM.
The planned main advantage was the $\approx$100\% sky coverage due to the
180 deg phasing of the two spacecrafts in their orbit. 
Launched on Dec. 9th, 2020, only one of the spacecrafts
returns data.

\noindent{\bf GRBAlpha} (Hungary/Czech/Slovakia/Japan):
GRBAlpha, launched on 2021 March 22, is a 1U CubeSat demonstration
mission \citep{Pal+2020} for a future CubeSat constellation
\citep{Werner+2018}. The detector consists of a 75 x 75 x 5 mm$^3$ CsI
scintillator read out by a SiPM array, covering the energy
range 50--1000 keV.

\noindent{\bf BurstCube} (USA) is a planned 6U CubeSat to be released into
low-Earth orbit from the ISS to detect GRBs. The instrument is
composed of 4 CsI scintillator plates, each 9 cm diameter, read out by
arrays of silicon photo-multipliers \citep{Smith2019}.
It reaches an effective area of 70\% of Fermi/GBM at 15\degs\ incidence,
but the localisation accuracy is substantially worse, with 7\degs\
radius at best for the brightest GRBs
(launch 2022).

\noindent{\bf SVOM} (China/France): The Space-based multi-band astronomical
Variable Objects Monitor (SVOM) is a Swift-like mission with a wide field
of view
$\gamma$-ray detector for GRB localization, and an X-ray and an optical
telescope for rapid follow-up of the GRB afterglow \citep{Yu+2020}.
60 GRBs/yr will be localized to 10\amin\ accuracy with a coded-mask telescope
with a 89\degs x89\degs\ field of view, working in the 4--250 keV band.
Due to this soft energy coverage the focus is on high-redshift GRBs
(launch early 2023).

\noindent{\bf HERMES} (Italy) is an Italian-led project to launch
100 CubeSats with X-/$\gamma$-ray detectors to localize GRBs,
and to derive limits
on Quantum Gravity \citep{Fuschino+2019}. Presently, six 3U CubeSats
are funded for a 2-year
pathfinder mission, with $\sim$56 cm$^2$ effective area per CubeSat
in the 3--1000 keV band (launch 2022). The anticipated
  localisation accuracy for transients with ms variability is 3\degs\ for
 the pathfinder and 10 arcsec for the full fleet in LEO
\citep{Fuschino+2019, Burderi+2020, Burderi+2021},
though this seems very optimistic given the detection of 0.05 ph/ms
even for the brightest bursts per CubeSat.

\noindent{\bf Glowbug} (NASA) is a funded small (30x30x40 cm$^3$)
satellite to detect GRBs and  other transients in the 30 keV to 2 MeV
band \citep{Grove+2020}. With an effective area about 2.5x that of
Fermi/GBM, about 70 short GRBs are expected per year. The localization
accuracy is expected to be slightly better than GBM, 
in the 5\degs\ (1$\sigma$ radius) range. The nominal lifetime is 1 yr
(launch 2023).

\noindent{\bf POLAR-2} (China/Switzerland) is a dedicated GRB polarimeter
to be flown onboard China's space station. With a field of view
of half the sky, the position determination will be a few degrees only.
Detailed polarization measurements are expected for 50 GRBs/yr, though more
GRBs are expected to be detected \citep{Kole2019} (launch 2024).

\noindent{\bf COSI} (USA): The Compton Spectrometer and Imager is an
approved NASA/SMEX mission,
working in the 0.2--5 MeV band, and scheduled for launch in 2025.
Its wide field of view of 3 sr will allow to detect 7--10 short GRBs per year, 
at sub-degree localisation \citep{Tomsick+2021}.

\noindent{\bf eXTP} (China/ESA): The enhanced X-ray Timing and Polarimetry
mission (eXTP) will study the X-ray
sky with 4 different instruments, covering the 0.5--50 keV band.
It will likely be the first to simultaneously measure the
spectral-timing-polarimetry properties of cosmic sources (launch 2027).
Relevant for GRB detection is the Wide-field monitor (WFM): with a field of view
of 1 sr (fully-coded) the detection of 16 GRBs/day(!) is predicted
\citep{Zhang+2017}, the brighter ones with 1 arcmin localization accuracy.
While this GRB rate is far ($\sim$5x) above the predicted total number
of GRBs in the Universe, the soft energy response of the WFM implies
a small fraction of short  GRBs (5\%-10\%).

\noindent{\bf HSP} (USA):
The proposed High Resolution Energetic X-ray Imager SmallSat Pathfinder
(HSP) is a wide-field hard X-ray (3-–200 keV) coded aperture telescope
with 1024 cm$^2$ CdZnTe detectors and a Tungsten mask \citep{Grindlay+2020a}.
With 4\farcm7
resolution covering 36\degs x 36\degs\ (FWHM), HSP localizes transients
and GRBs within $<$30\asec\ in less than 10 min.
(launch $>$2025).

Summarizing, there is a need to better localize short-duration GRBs.
We propose that GRB triangulation with the Galileo satellite network
provides such an opportunity.

\section{The Galileo system as a perfect host for triangulation}

The Global Navigation Satellite System (GNSS) Galileo has five parameters
which makes it a nearly ideal satellite constellation for triangulation:
\begin{itemize}
 \vspace{-0.2cm}\itemsep-.4ex
\item all satellites are synchronized with a very accurate atomic clock,
  ensuring time-stamps for the GRB signal at the 10$^{-9}$\,s level;
\item the satellites are distributed over three orbital planes, perpendicular
  to each other, making triangulation positions close to round;
\item the position knowledge of all satellites is accurate to sub-meter
  accuracy, and thus do not contribute to the error budget in
  realistic GRB measurements, similar to the timing;
\item the knowledge of the orientation of each satellite is known to
  better than 1\degs,
  removing any ambiguity in the relative rate measurements for GRBs;
\item the orbital radius is large enough that one can realistically expect
  sub-degree localizations, but small enough that light travel time
  distances are short and communication (data transfer) is quick.
  This differentiates it from the canonical Interplanetary Network (IPN),
  where the baseline is much longer and thus allows arcmin-scale localizations,
  but the triangulation can only be done 1-2 days after the GRB.
\vspace{-0.2cm}
\end{itemize}

\begin{figure}[th]
  \centering
  \vspace{-0.2cm}
  \includegraphics[width=0.5\textwidth]{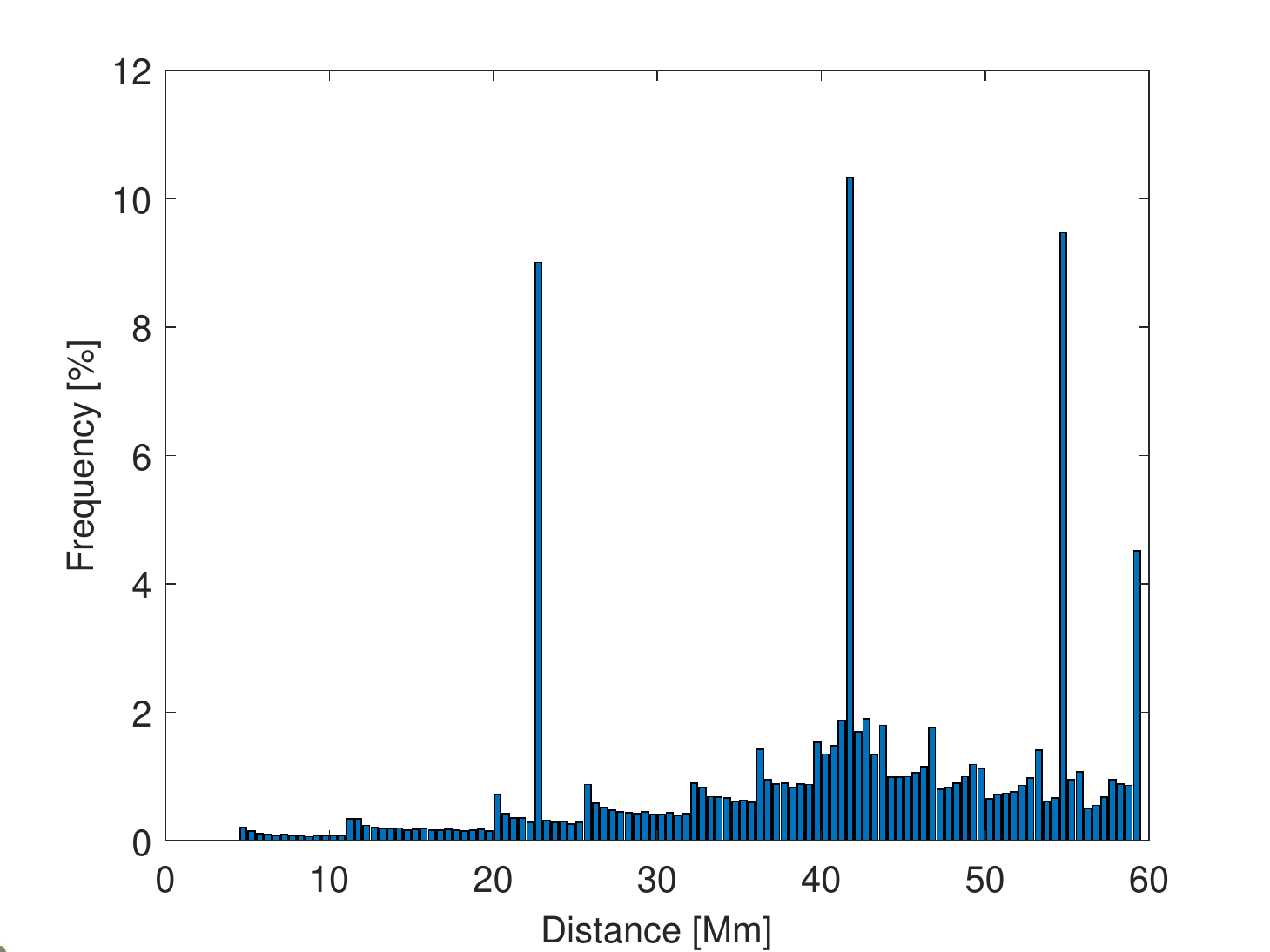}
  \vspace{-0.4cm}
  \caption[]{Histogram of pair-wise distances between Galileo satellites.
    The peaks indicate constant distances between satellites in the same
    orbital plane.
  \label{Fig_45histdist}}
\end{figure}

\noindent
The first satellites of the
present Galileo constellation were launched on October 21, 2011.
Today, 26 satellites are in orbit, among them
two unusable -– one with technical problems, one declared as spare due
to issues with clocks –- and two on non-nominal orbits due to launch
failure of the third rocket stage but otherwise fully operational and
usable. In December 2016, Initial Service Declaration was announced.
Currently 12 satellites, the so-called Batch 3 satellites are under
production, deployment shall start in 2021.
As the design lifetime of Galileo satellites is 12 years,
the constellation has to be replenished in the coming years.

The Galileo constellation is a Walker constellation \citep{Walker1984}.
This constellation type is characterized by the three numbers 24/3/1:
\begin{itemize}
 \vspace{-0.2cm}\itemsep-.4ex
\item[~$n_1$:] Total number of satellites (i.e. eventually 24 satellites), equally distributed over the orbital planes.
\item[~$n_2$:] Number of equally spaced orbital planes (i.e. 3), with 8 satellites each.
\item[~$n_3$:] Relative spacing between satellites in adjacent planes.
The difference in argument of latitude (in degrees) for equivalent
satellites in neighboring planes is equal to $n_3*360/n_1$.
\vspace{-0.2cm}
\end{itemize}

\noindent
The revolution period corresponds to 17 revolutions in 10
sidereal days, i.e., 14h04m. With a semi-major axis of 29.600~km,
the orbital inclination is 56\degs.
For current Galileo satellites the eccentricity is below 0.0007 except
for the two satellites on non-nominal orbits that have an eccentricity
of 0.166.
Fig.~\ref{Fig_45histdist} shows the histogram of pair-wise distances between
Galileo satellites in a Walker 24/3/1-constellation. Maximum distance is
59.000 km, i.e., the orbit diameter, while the mean distance is 42.000 km.
The peaks in the figure indicate the constant distances between satellites
in the same orbital plane.

Orbits and clock corrections for Galileo satellites are available with
high precision in real time. While precise orbits at a few cm level and
clock corrections at below the ns level are available in post
processing, sub-meter orbits and few ns clock corrections are available
through the broadcast messages that are updated every 10 minutes.
While the eccentric satellites show slightly larger broadcast
orbit errors, mainly in along-track, the rms of broadcast orbit errors
for the other satellites is at a level of 32~cm (12.5~cm radial,
25~cm along-track, 15~cm cross-track).

In contrast, the broadcast clock quality of eccentric Galileo satellites 
is comparable to the other satellites. The rms of the difference is
0.50~ns. For GRB triangulation it can thus be assumed that perfect
positions and time tagging are known in real time at any given time.

The attitude of the Galileo satellites is a nominal yaw steering in order
to point the navigation antenna (body-fixed z-axis) to the center of the
Earth and the solar panel axis (y-axis) in perpendicular direction to the
direction of the Sun. The Sun is thus always located in the body-fixed
x-z plane. While the positive x-surface is always illuminated by the Sun,
the negative x-surface constantly points to dark sky.
The nominal attitude is controlled by Earth- and Sun-sensors to below 0\fdg1,
except for non-nominal noon and midnight yaw maneuvers, if the Sun is
closer to the orbital plane than about 2\degs\ for IOV and about 4\degs\
for FOC, and the nominal yaw rate would exceed the maximum hardware
rate of 0\fdg2/s. The negative z-axis is always pointing in zenith
direction with a precision below 0\fdg1.

\section{Triangulation:  methods \& prospects \label{triangulation}}

The triangulation method uses measurements of differences of arrival times
of the same signal (GRB) at different clocks (each on a different satellite).
In general, time differences between three independent satellite pairs 
are needed to derive a unique position on the sky, and not all these satellites
shall be in the same plane.

The relation
cos\,$\theta$ = $c{\cdot}t / d$ holds for the time difference of a signal
between two satellites. Under the simplifying
assumption of perfect satellite clocks and perfectly known satellite
positions, the width of the
annulus $\Delta\theta$ is obtained as the derivative of the above equation,
i.e. is just determined by the error $\Delta t$ with which
the time delay $t$ between the two signals (light curves) can be measured:

\begin{equation}
\Delta\theta = \abs*{ \frac{-1}{\sqrt{1-(c\cdot t / d)^2}}} \Delta t \cdot \frac{c}{d} \label{eq:1}
\end{equation}

\noindent
There are two possible approaches to triangulation: Firstly, one can
compare pairs of two light curves to each other to find the time lag, i.e.
simple cross-correlation
of background subtracted time series \citep{Hurley+2013, Palshin+2013}.
Each satellite pair results in one time lag, and a
 corresponding triangulation ring. Combining multiple ($>$3) pairs then provides
 a unique sky position as the overlap of these triangulation rings.
 Cross-correlation is computationally fast, but
suffers from several draw-backs \citep{Burgess+2021}:
(i) they only work for binned light curves, at fixed binning;
(ii) no mathematical method exists to estimate the proper error of the
cross-correlation;
(iii) the approximation of $\chi^2$ rarely holds, in particular when
small bin sizes are choosen in order to ``increase'' the temporal accuracy;
(iv) the subtraction of two Poisson rates results in Skellam rather than
Poisson distributed data, often leading to over-confidence;
(v) it cannot take into account lightcurves taken at different energies.
As a second approach, one can forward-fold an
  identical model  through the (different) response of each detector and fit
  each observed light curve \citep{Burgess+2021}.
  This technique is computationally  expensive, but
  offers the major advantage that
it produces a complete posterior probability distribution allowing for a
very precise estimate of the uncertainty formally in $\Delta\theta$,
but due to the forward-folding directly in sky coordinates
$\Delta$RA, $\Delta$Decl.
This nazgul code \citep{Burgess+2021} has been made publicly
available\footnote{\url{https://github.com/grburgess/nazgul}}.

\section{Simulation set-up}

In order to test each of the localization methods and verify their
performance for different satellite configurations, we have developed
a simulation framework utilizing the Python package
PyIPN\footnote{\url{https://github.com/grburgess/pyipn}}.
This package allows for the
generation of synthetic GRB light curves as seen by detectors
distributed within the solar systems. We have added on top of this
frame work a procedure to generate realistic light curve shapes and
detector configurations that mimic the orbit of the Galileo
constellation. Below we detail the setup and procedure for the
generation of mock data sets which allow us to test our methods.

\subsection{Simulating GRB light curves}

The simulation of the triangulation capability of a network of
GRB detectors requires to create mock GRB light curves which then
hit differently-oriented detectors. These mock
light curves shall cover a peak flux range as bright as has
been seen with previous experiments (CGRO/BATSE, Swift/BAT, Fermi/GBM),
and down to our proposed sensitivity limit of
1x10$^{-7}$ erg/cm$^2$/s in the 25--150 keV band.
We pick the 256\,ms timescale for the peak flux
as a compromise between being short enough to cover spikes in short-duration
GRBs, and being general enough also for long-duration GRBs.

GRB light curves are generally very complex, and  unique  for  each  GRB.
In many cases the variability time-scale is significantly shorter than the
overall burst duration. Only in a minority of GRB lightcurves there is only one
peak, with no substructure. 
The most straightforward way is to ``assemble'' GRB light curves by the
superposition of different pulses.
We assume that candidate pulses can be modeled with the empirical
functional pulse form of \cite{Norris+1996, Norris+2005}: 
\begin{equation}
  I(t) = A \lambda {\rm e}^{-\tau_1/(t-t_s) - (t-t_s)/\tau_2} {\rm cts/s}
  \label{eq:lc}
\end{equation}
where $t$ is time since trigger, $A$ is the pulse amplitude, $t_s$ is the
pulse start time, $\tau_1$ and $\tau_2$ are characteristics of the pulse
rise and pulse decay, and the constant $\lambda = {\rm e}^{2(\tau_1/\tau_2)^{1/2}}$.
The pulse peak time occurs at time $\tau_{peak} = t_s + \sqrt(\tau_1\tau_2)$.
Typically, the rise times in individual GRBs are very short (steep rise),
and decay times substantially longer in most times. Thus, for single-pulse
GRBs, the decay time $\tau_2$ scales with the T90 duration of a GRB.

In order to implement the stochastic nature of the light emission
process, and to incorporate unavoidable background at the measurement
process, individual photon events are sampled according to an
inhomogeneous-Poisson distribution following the intrinsic pulse shape
specified. The photon arrival times are sampled via an inverse
cumulative distribution function rejection sampling scheme
\citep{Rubinstein+2016}. As the rate for the signal evolves with time,
a further rejection sampling step is implemented that thins the
arrival times according to the evolving light curve. This is done by
first sampling a waiting time $t$ and computing the light curve
intensity $I(t)$.  Another draw from $p \in \{0,1\}$ is made and the
sample is accepted if $p$ \lax\ $I(t)$ \citep{Burgess+2021}.

The cross-correlation of two light curves depends crucially on the
intensity of the GRB above background, and the structure of the light curves.
We therefore need a sample of different light curves.
In order to create a realistic sample, we need to make sure that
we reproduce
\begin{itemize}[leftmargin=12pt]
  \vspace{-0.2cm}\itemsep-.3ex
\item a rough --3/2 logN-logS intensity distribution
between the brightest GRBs
seen so far (2$\times$10$^{-4}$ erg/cm$^2$/s)
 and our aimed-at limit of 1x10$^{-7}$ erg/cm$^2$/s:
For a canonical GRB spectrum below E$_{\rm peak}$, i.e. a power law spectrum with
a slope in the range of -0.9...-1.1 (long) and 0.0...-0.2 (hard),
the following conversion holds for the 25–-150 keV band
with an detector size of  3600 cm$^2$ (see below):
1$\times$10$^{-7}$ erg/cm$^2$/s = 0.65$\pm$0.10 ph/cm$^2$/s.
Thus, we substitute the sampling over the
2$\times$10$^{-4}$ --- 1x10$^{-7}$ erg/cm$^2$/s range by that over
1300 --- 0.65 ph/cm$^2$/s.
\item the observed T90 duration distribution of GRBs \citep{Kouveliotou+1993}; and
\item some realistic distribution between single- and
  multi-pulse light curve structure:
  We assemble multi-pulse light curves by
  overlapping multiple single pulses, each
  with a shape a la \cite{Norris+1996}, but with different parameters
  and properly delayed to each other.
  \vspace{-0.2cm}
\end{itemize}

\begin{figure}[ht]
  \hspace{-0.2cm}\includegraphics[width=0.50\textwidth]{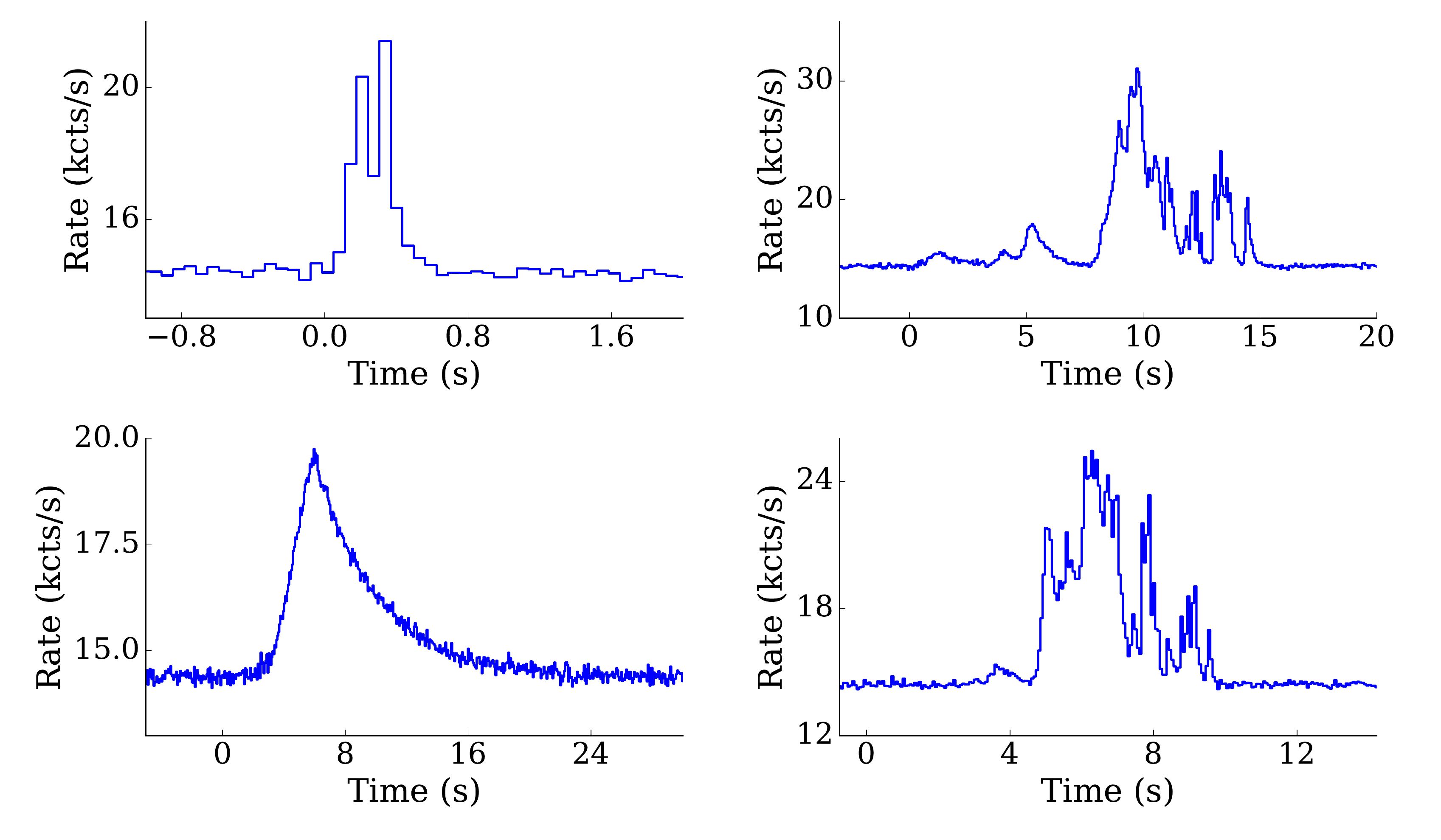}
  \vspace{-0.4cm}
  \caption[]{Example mock light curves of the long-duration class
    created with the pulse avalanche description of Norris-like pulses. 
    \label{example-lc}}
    \vspace{-0.2cm}
\end{figure}

\noindent
For the latter, we implement a pulse avalanche, a linear Markov
process, as proposed by \cite{SternSvensson1996},
and described in detail in appendix \ref{avalanche}.
Example light curves with this simulation set-up are shown in
Fig. \ref{example-lc}.

Energy dependent effects in GRB light curves are ignored.
Conceptually,
we treat our proposed energy band of 25--150 keV as a mono-energetic band.

\subsubsection{Lightcurve detectability by different detectors}

The previous steps provide theoretical light curves of GRBs (as emitted)
which are representative in intensity distribution, duration distribution
and pulse structure to the GRBs as measured over the last 30 years.
These light curves are now being measured by identical detectors
on a number of Galileo satellites. While the details of the Galileo
satellite network is described later, three effects combine together
to establish the measured light curves: (1) since the detectors
are oriented into different directions, each will detect photons
according to the cosine between the scintillator normal (we adopt
thin, but large-area scintillator plates as baseline) and the GRB,
and (2) the detector will measure quasi-isotropic $\gamma$-ray background
which has the effect of washing out
low-intensity features; (3) in the case of multiple detector plates
per satellite, the sensitivity can be improved by co-adding the data.
However, this
helps only for a certain incidence angle range, since the GRB signal
varies with the incidence angle, but the isotropic background does not.

\begin{figure}[th]
  \hspace{-0.2cm}\includegraphics[width=0.50\textwidth]{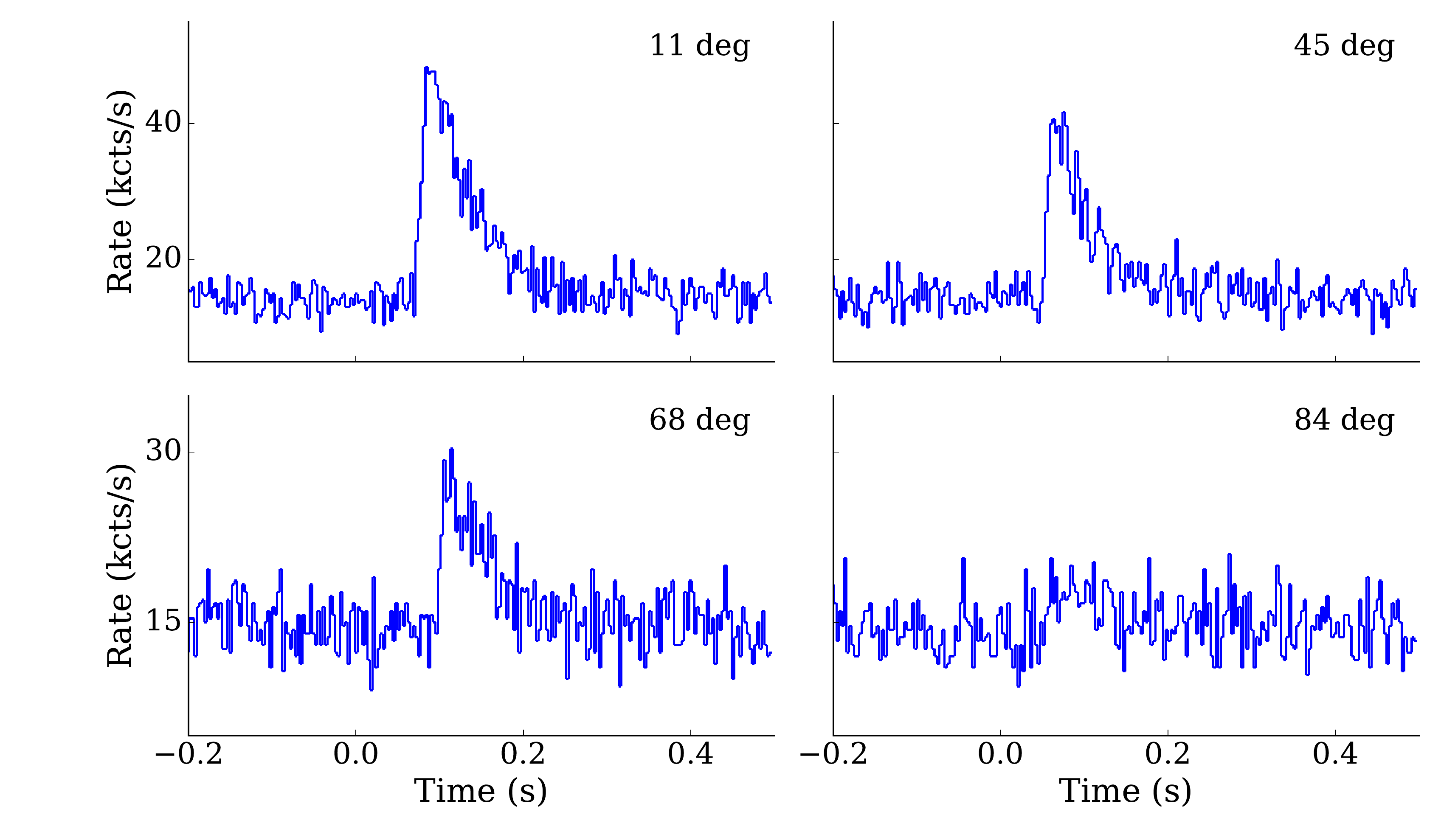}
  \caption[]{Light curves of the same GRB as seen with
    a flat detector plate from different incidence angles.
    The labels in each plot
    denote the angle under which the GRB impinges on the detector plane.
    With increasing angle, the effective area shrinks with the
    cosine of the angle, while the background remains the same.
    The ``mother'' lightcurve
    of this GRB has been generated with the above described
    pulse avalanche scheme.
  \label{cosine-cts}}
\end{figure}

An example of such a set of 'measured' light curves for a given GRB
and differently-oriented detectors is given in Fig. \ref{cosine-cts}.
These are the final 'measured' light curves
(in counts/s) over a certain duration in the 25--150 keV band,
which are then used for triangulation.

\subsubsection{Implementation of the discrete correlation function}

A modified version of the discrete correlation function method
\citep{EdelsonKrolik1988} has been implemented in PyIPN
with the following three parts: First, the model is initialized by
setting a GRB position and define all detectors. The actual
simulation creates the GRB signal (light curve) and computes the
arrival times at the detectors as detailed above.

The final step performs the cross-correlation and computes the
center point and opening angle of the circle for a 
specified pair of detectors. 
Rather than relying on mathematical covariance matrix estimation of
uncertainty, cross-correlation methods heuristically derive uncertainties
in one of two ways. First, the discretized time bins of the light curve
yield discrete estimates of the time lag between each pair of light curves.
For each value, a pseudo-$\chi^2$ statistics\footnote{Note that the
pseudo-$\chi^2$ values are incorrect in the first place due to the lack of
fidelity in the low-count regime and the fact that count data are fundamentally
Poisson distributed.} is derived yielding a grid of
statistics values hopefully in a parabolic shape. The minimum of these values
is taken as the true time lag (best fit). The 1, 2, and 3 $\sigma$ uncertainty
regions are recovered by moving up the grid of statistics at the appropriate
levels and reading off the implied time lags. This has several drawbacks:
first, the best-fit time lag can never be below the timing resolution of
the light curve. Additionally, the uncertainties are locked to the resolution
of the grid and can thus easily be over- or under-estimated. To get around
this, another heuristic can be introduced. One can fit this grid of
uncertainties with a parabolic shape to effectively interpolate to finer
timing resolution. While this alleviates the issues with discretization
in the previous method, it introduces the problem that the interpolation
has an associated uncertainty which is not accounted for. Moreover, the
chosen shape of this parabolic fit cannot incorporate asymmetries in the
grid and thus can easily over- or underestimate the true uncertainty. 
However, given the lack of a mathematically strict method, we use this
procedure, but keep the problems in mind.

\subsection{Gamma-ray background in the Galileo orbit
\label{sect8p1p1}}

The background which a $\gamma$-ray detector (whether scintillation
detector or other type) experiences in space is composed of several
different components
\citep[e.g.][]{Weidenspointner+2003, Weidenspointner+2005, Wunderer+2006, Cumani+2019}.
In the 10--250 keV band, the most important
components are the extragalactic diffuse $\gamma$-ray background, Earth albedo
photons (for LEO), Galactic cosmic-ray protons, and radioactive
decay of activated detector and spacecraft material due to cosmic-ray
bombardment. For a satellite in MEO, the diffuse $\gamma$-ray background
dominates below 100 keV,
while at $\sim$200 keV the rising proton contribution has reached the level
of the diffuse $\gamma$-ray background.
We therefore just incorporate the extragalactic diffuse $\gamma$-ray background
in our simulations.

We adopt the following smoothly broken powerlaw for the diffuse
background spectrum \citep{Ajello+2008}:
\begin{equation}
  E^2 {dN \over dE} = E^2 \times {C \over (E/E_{\rm B})^{\Gamma_1} + E/E_{\rm B})^{\Gamma_2}}
\end{equation}
\noindent with the following constants:
C = 0.102$\pm$0.008 ${\rm ph\ cm^{-2}\ s^{-1}\ sr^{-1}\ keV^{-1}}$,
$\Gamma_1 = -1.32 \pm 0.02$, $\Gamma_2 = -2.88 \pm 0.02$ and a break at
$E_{\rm B} = 30.0 \pm 1.1$ keV.
Integrating over the 25--150 keV energy range and the 2$\pi$ sky coverage
is consistent with both, the Konus-Wind \citep{Aptekar+1995}
as well as Fermi/GBM \citep{Burgess+2018} measurements,
and leads to $\approx$4 cts cm$^{-2}$ s$^{-1}$.
This background rate is then added to the scaled
light curve generated with
the pulse avalanche method (see previous subsection).

\subsection{Required detector timing}

\begin{figure}[!bh]
  \vspace{-0.2cm}
  \includegraphics[width=0.50\textwidth]{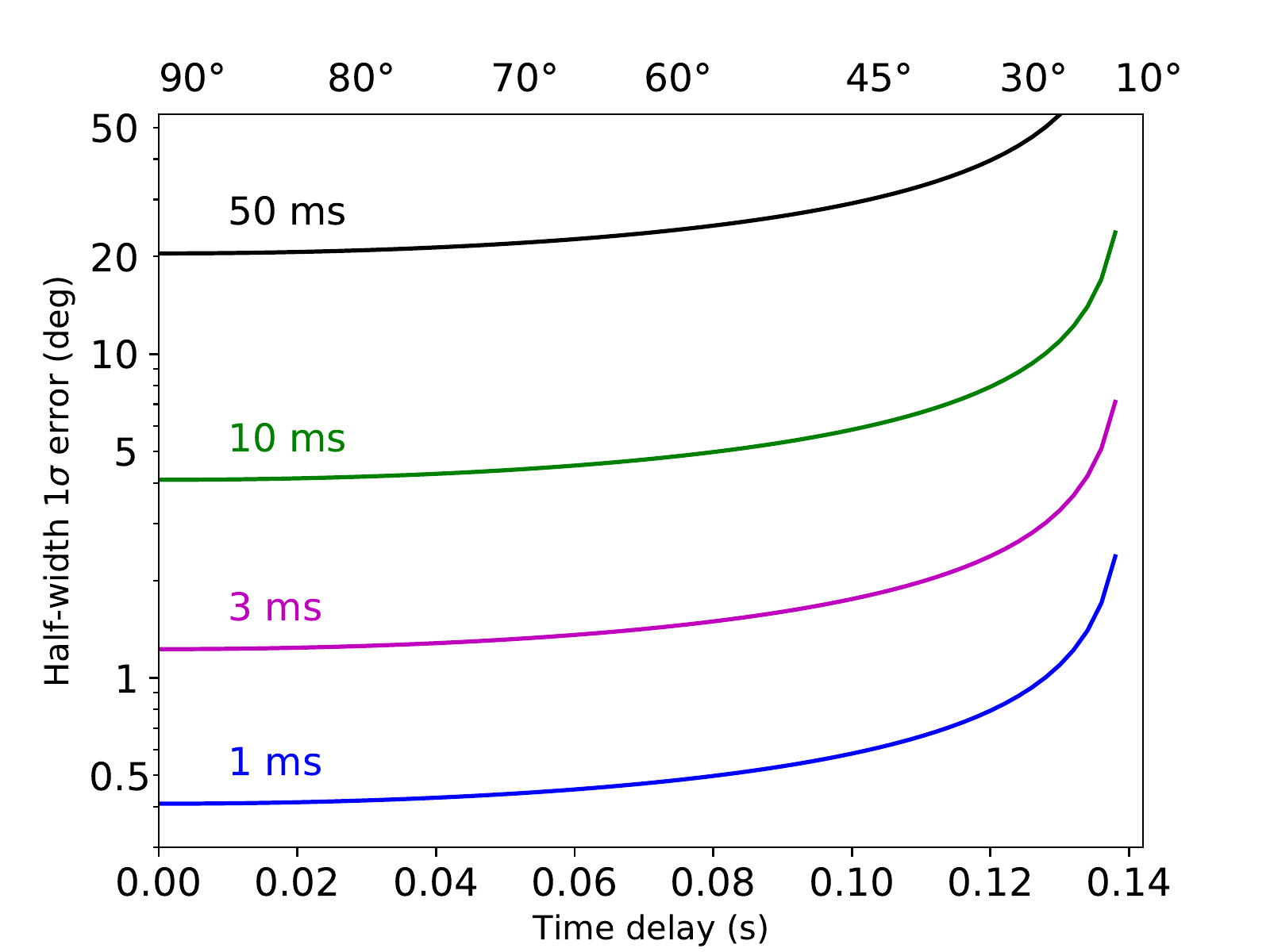}
  \vspace{-0.5cm}
  \caption[]{The 1$\sigma$ half-width error $\Delta\theta$ of the
    triangulation annulus is shown for a signal (GRB) arriving at a pair of
    satellites 42000 km apart (corresponding to the mean of
    Fig. \ref{Fig_45histdist}),
    with different delay times (bottom x-axis label) or angles relative
    to the line connecting the two satellites (top x-axis) for different
    accuracies $\Delta t$ of 1\,ms, 3\,ms, 10\,ms and 50\,ms
    with which the time delay can be measured. For comparison, the
    Anti-Coincidence system of the INTEGRAL spectrometer (SPI-ACS) has
    a time resolution of 50\,ms, and the shortest binning of Fermi-GBM is
    64\,ms, with individual events time-tagged at 2\,$\mu$s.
    \label{accuracydelay}}
    \vspace{-0.1cm}
\end{figure}

With a dedicated GRB detector on the Galileo satellites, we can
dramatically improve the localization accuracy.
Using the formal triangulation error (see  eq. \eqref{eq:1}),
it is easy
to compute the required temporal resolution, usually the bin size in 
the classical scheme, for a perfect system with satellites at known
distances. Fig. \ref{accuracydelay} shows
that sub-ms accuracy  in the determination of the time delay is
required to reach sub-degree localisation accuracy with two satellites at
a distance of 42000 km (which corresponds to the mean for
Galileo's Walker constellation; see Fig. \ref{Fig_45histdist}).

Fig. \ref{accuracydelay} only applies for a perfect system.
As discussed below, the classical triangulation method with its
use of a cross-correlation of binned data sets does not
provide a mathematically self-consistent error handling. In contrast, the
alternative method by \cite{Burgess+2020} does, but lacks the beauty
of a simple equation. We therefore will show with simulations below
how close this new method gets to the estimate of eq. \eqref{eq:1}.

\subsection{Required detector sensitivity}

Good timing resolution provides a necessary but not yet sufficient condition.
The detector needs to
(i) be large enough to detect a significant signal at these short
time scales, 
(ii) measure a significant signal independent on the arrival direction, and
(iii) provide this high time-resolution data for analysis, either on-board
or on the ground, rather than binning it up to save telemetry band width.

A simple estimate of the minimum detector size can be made
by recognizing that short-duration GRBs have structure, and do have
durations substantially longer than the 3\,ms time scale which
Fig. \ref{accuracydelay} implies as a requirement for sub-degree
localization accuracy. 
Assuming a canonical shape of a short GRB prompt emission lightcurve,
and knowing that for GRB 170817A a single Fermi/GBM detector measured
20--30 cts/0.1\,s in the 20--500 keV band at peak against $\sim$30 cts rms
from background fluctuations,
we estimate to need 2000 cts per short-duration (2\,s) GRB
or 10 cts/1\,ms at peak, so $\sim$30x the effective area of a single
GBM detector of 125 cm$^2$, that is 3500--4000 cm$^2$.
Incorporating the correspondingly higher background rate
at the Galileo orbit wrt. the LEO of Fermi will
modify this estimate, but for the simulations presented here,
we consider a detector of 60\,cm x 60\,cm geometrical area.

\begin{figure}[bh]
  \includegraphics[width=0.155\textwidth]{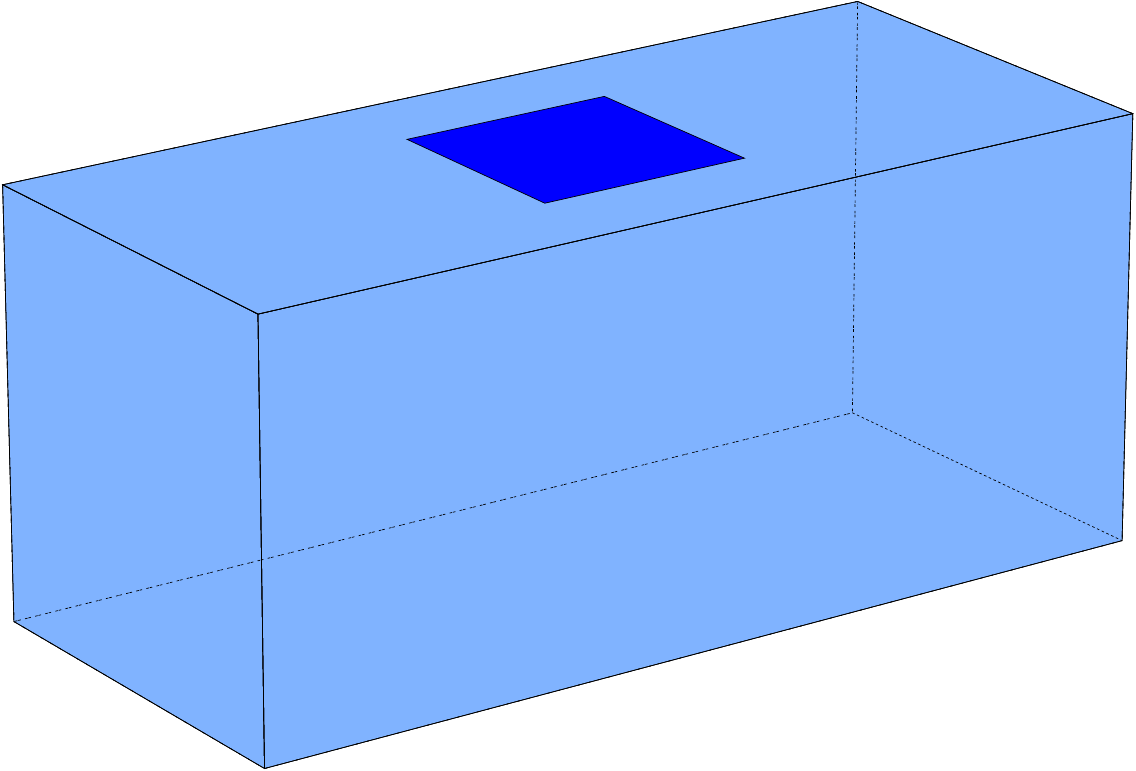}
  \includegraphics[width=0.155\textwidth]{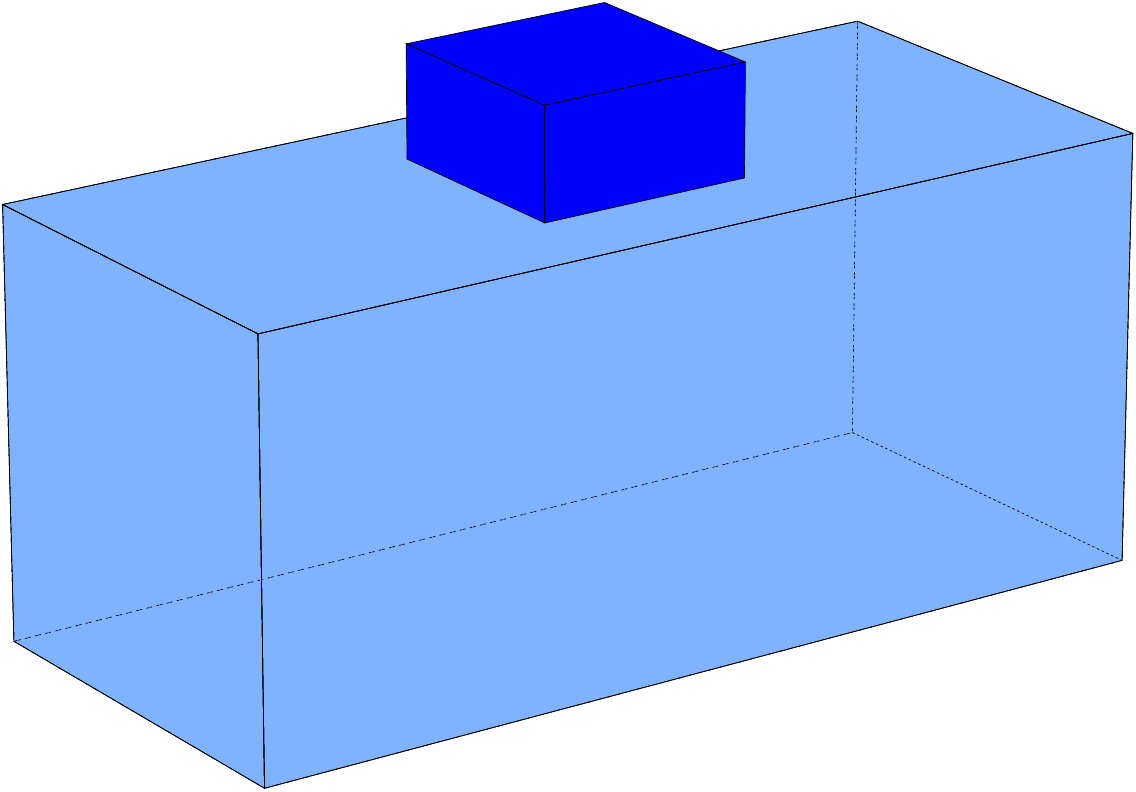}
  \includegraphics[width=0.155\textwidth]{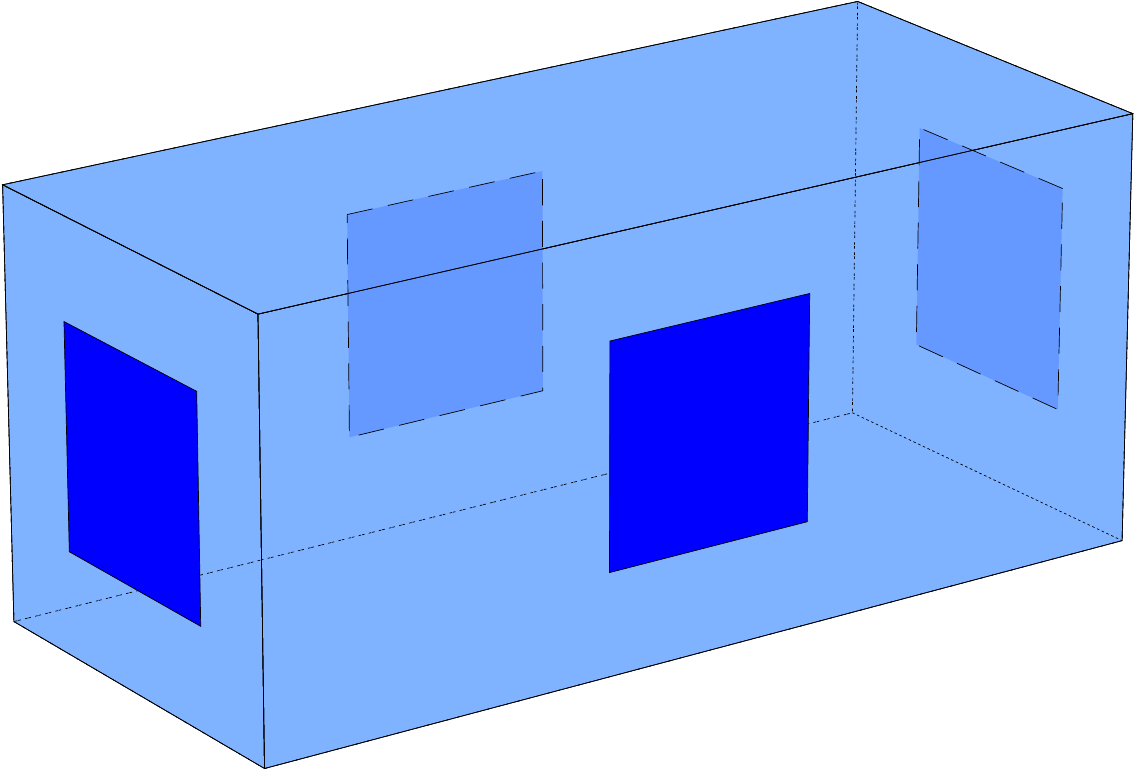}\\
  \includegraphics[width=0.155\textwidth]{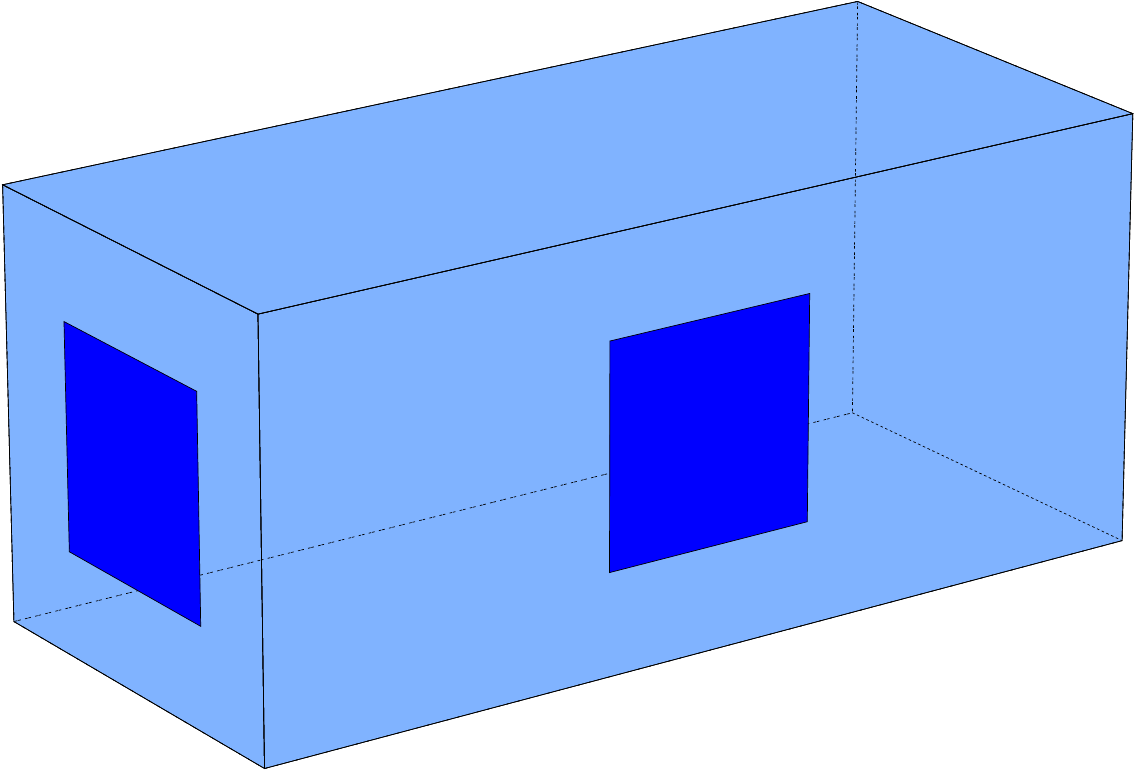}
  \includegraphics[width=0.155\textwidth]{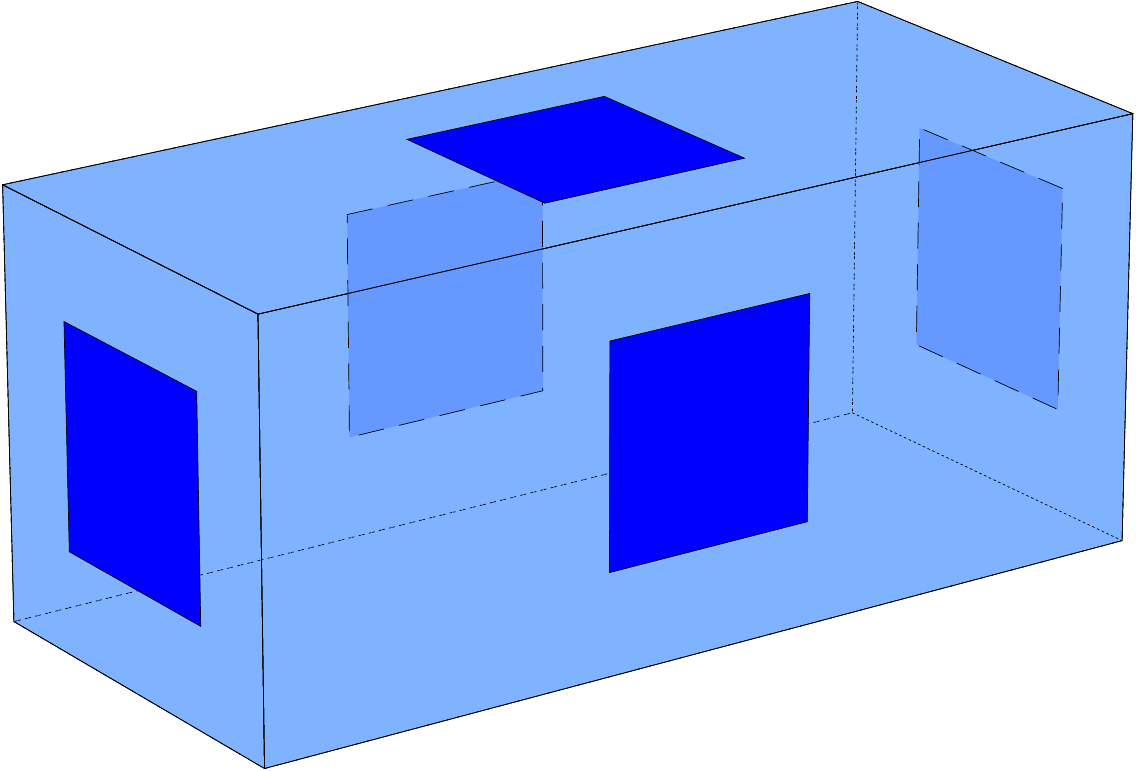}
  \includegraphics[width=0.155\textwidth]{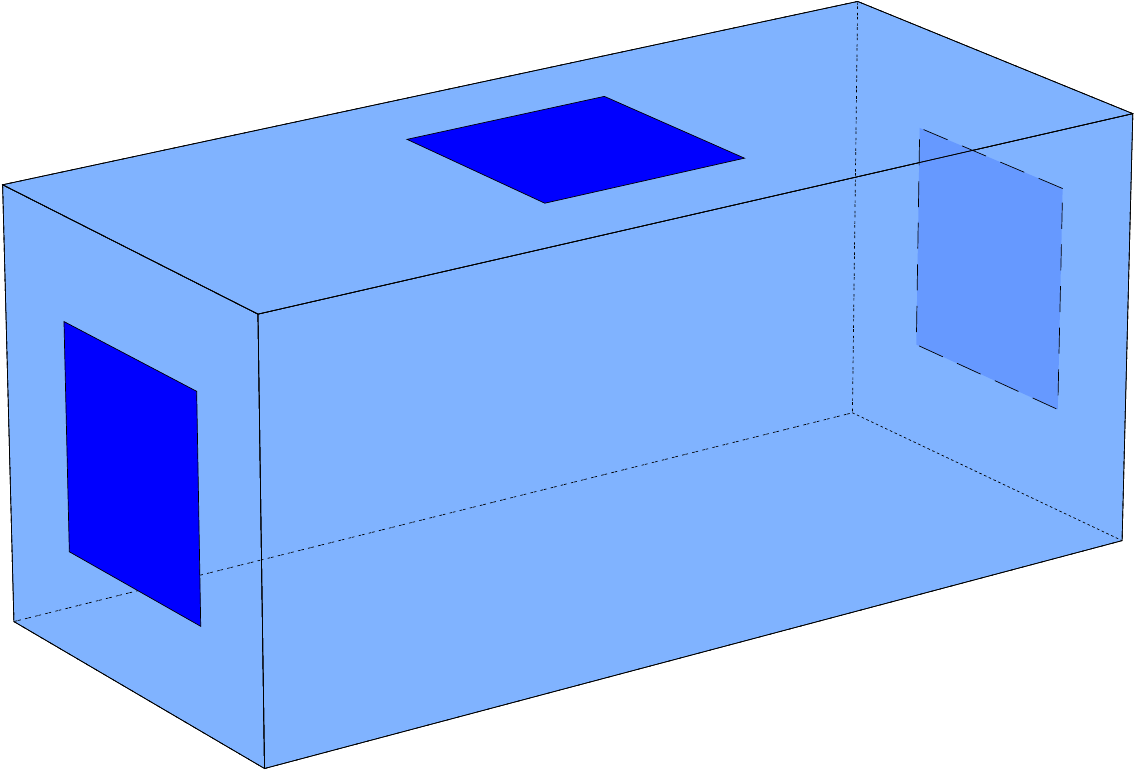}\\
  \includegraphics[width=0.155\textwidth]{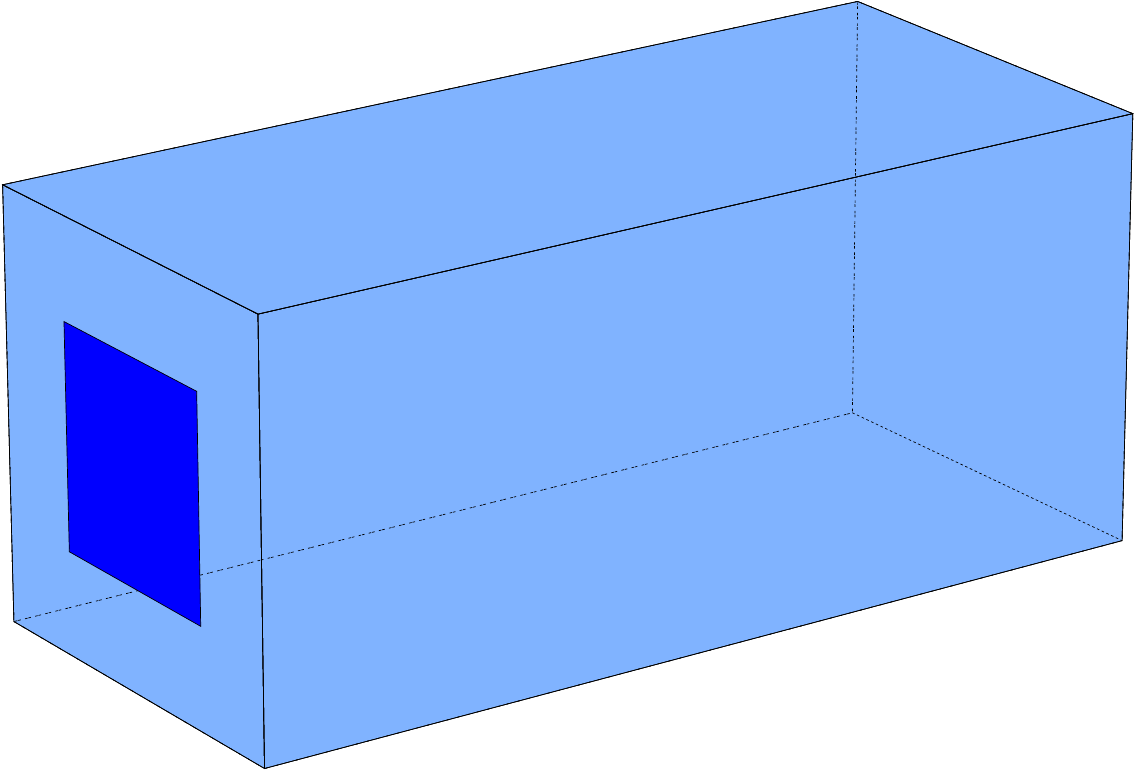}
  \includegraphics[width=0.155\textwidth]{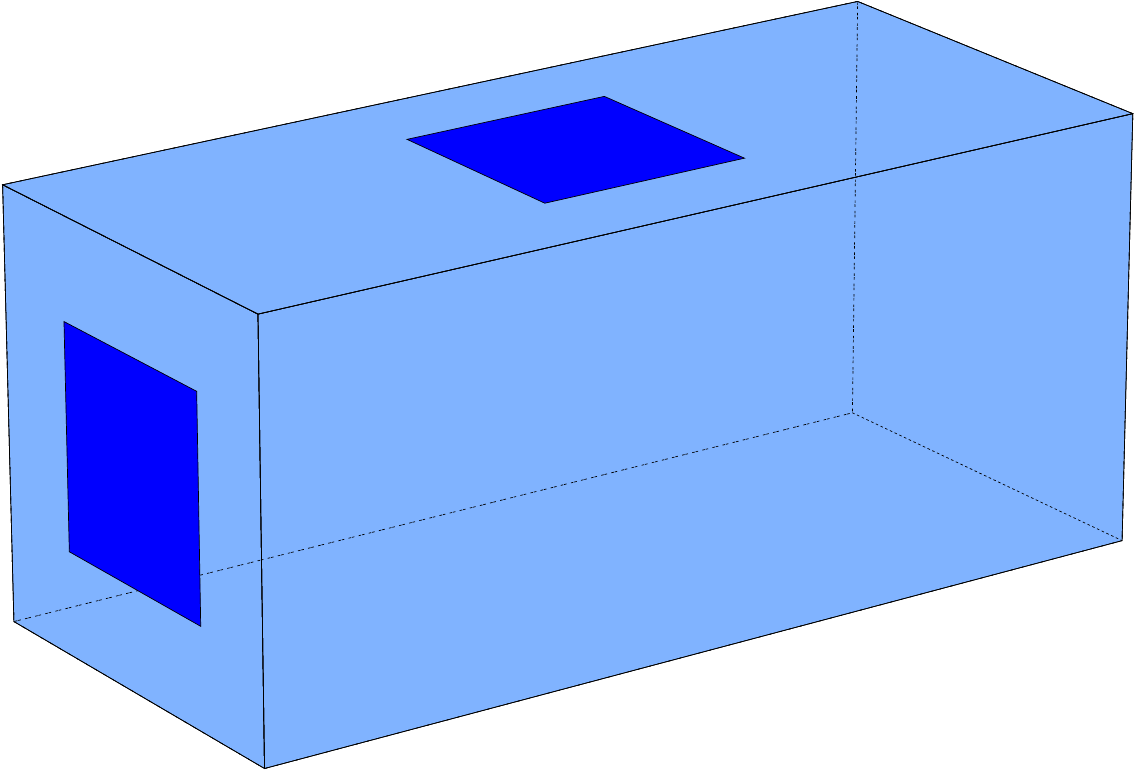}
  \includegraphics[width=0.155\textwidth]{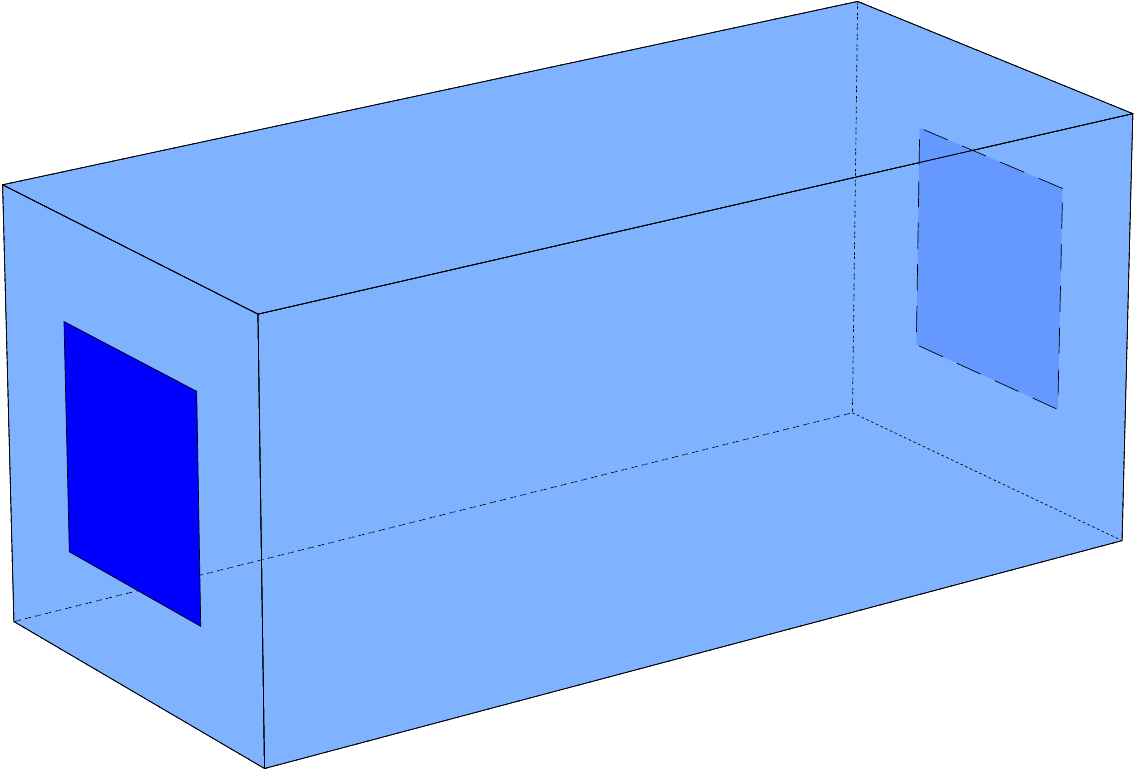}
  \caption[]{The different detector geometries simulated here,
    detector numbers 1--3 (top row) to 7--9 (bottom row).
    \label{geometry}}
\end{figure}

\subsection{Detector geometry}
\label{sect:Detector-geometry}

\begin{figure}[ht]
  \includegraphics[width=0.48\textwidth, clip]{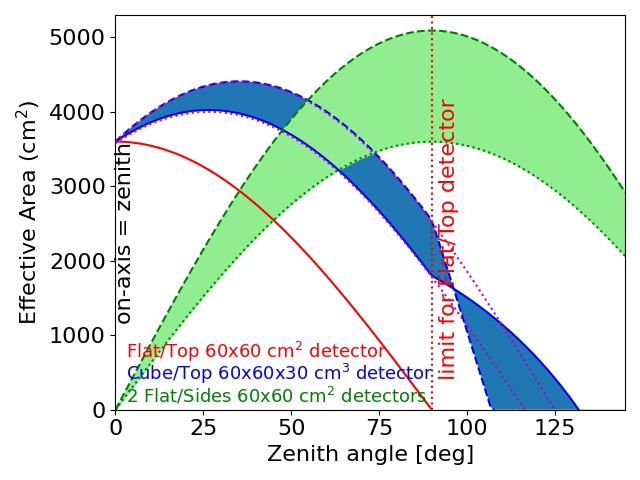}
  \vspace{-0.3cm}
  \caption[]{Off-axis dependence of the different detector geometries.
    The blue shaded range corresponds to the min-max range according
    to azimuth angle for the cube detector.
    Beyond 90\degs, the cube detector offers continued effective area,
    but shadowing by the satellite bus leads to a rapid drop.
    The green shaded area corresponds to equal-size detectors
    on two neighboring sides, with the top boundary corresponding
    to 45\degs\ view onto both, and the lower to 90\degs\
    on only one of the two detectors.
    \label{Eff_area}}
  \vspace{-0.3cm}
\end{figure}

As we will show, the generally preferred and assumed zenith-looking
detector is not a good choice. Since the best localisation accuracy is
reached at largest satellite separation and looking perpendicular to the
satellite-connecting line (see eq. \eqref{eq:1} and Fig. \ref{accuracydelay}),
detectors sensitive sidewards, i.e. 90\degs\ of zenith, are preferred.

Since the anticipated detector
size has $\approx$60 cm sidelength, adding a cube of that size to the
zenith-facing side of the Galileo satellite might be challenging in terms
of satellite momentum balance or station keeping, thus
we also consider configurations, where one-dimensional
detector plates are mounted on different sides of the Galileo satellite
(Fig. \ref{geometry}).
Any such 3D detector has several advantages:
(i) multiple detector units provide independent measurements
  to be used in the cross-correlation;
(ii) for the same reason, a
  coincidence veto against particle hits can be implemented, reducing
  the rate of false triggers;
(iii) since 3D detectors cover a field-of-view
  (FOV) of more than
  2$\pi$ of the sky, detectors on some 'behind-the-Earth' Galileo satellites
  will be able to detect the GRB, thus increasing not only the number
  of measuring detectors, but more crucially extending the baseline
  (maximum distance between detectors) for the time delay measurement.

We consider nine different detector geometries (Fig. \ref{geometry}):
a single detector facing zenith (called detector 01),
a hollow cube detector with 30\,cm height on the zenith-facing side (detector 02),
4 detector plates looking sideways (03),
2 neighboring sideways (+X, +Y) looking plates (04),
4 sideways plus a zenith-looking detector (05),
2 oppositely sideways and 1 zenith-looking (+X, -X; 06),
1 sideway only (+X; 07),
1 sideway plus 1 zenith-looking (+X; 08),
and 2 oppositely sideways looking detectors (+X, -X; 09).

All the side-looking plates have also 60\,cm $\times$ 60\,cm dimension
and 1 cm thickness.
These configurations obviously change the zenith-angle dependent
variation of the effective area; see Fig. \ref{Eff_area}:
the green area corresponds to two equal-size detectors on two neighboring
sides (Fig. \ref{geometry}). 
Two-dimensional versions of theses dependencies (including azimuthal variation)
are shown in Fig. \ref{3Deffarea} further below.

\subsection{Set-up of GNSS configuration}
\label{sect:GNSS-setup}

Since we use an existing satellite network, only one further
configuration choice needs to be considered in the simulations,
namely the number of satellites per orbital plane that shall be
equipped with GRB detectors to allow a $4\pi$-coverage of the sky.
We use the notation of [1] or [0] if a GRB detector is
placed on a given satellite or not. With 8 satellites per orbital plane,
and dealing with these planes consecutively, a configuration
of every second satellite equipped with a GRB detector would read
[10101010 10101010 10101010]. The set of simulated configurations is given
in  Tab. \ref{detconfigs}.

\begin{table}[th]
    \centering
  \caption{Overview on the simulated detector configurations 
    \label{detconfigs}}
  \vspace{-0.3cm}
  \begin{tabular}{ccc}
    \hline
    \noalign{\smallskip}
    Sat   & Configuration & Detectors \\
    \noalign{\smallskip}
    \hline
    \noalign{\smallskip}
    24 & 11111111 11111111 11111111 & 1,2,3,4,5,6,7,8,9 \\
    12 & 10101010 10101010 10101010 & 1,2,3,4,5,6 \\
     9 & 10010010 10010010 10010010 & 3,5 \\
       & 10010010 01001001 10100100 & 3,5 \\
     6 & 10001000 00100010 10001000 & 3,5 \\
       & 10001000 10001000 10001000 & 3,5 \\
       & 10010000 00010010 10010000 & 3,5 \\
    \noalign{\smallskip}
    \hline
   \end{tabular}
\end{table}

We compute two maps:
one 'instantaneous' snapshot map, and one averaged over one
orbital period.
Simulations are done in steps of 5\degs, which provides 72
subsequent snapshots for a full 14h04m orbital revolution of the Galileo
satellite network, i.e. the averaged map is the average of
such 72 snapshots. GRBs are distributed on the sky on a 2\degs\ grid,
thus providing a full sky map for each snapshot.

\begin{figure}[bh]
  \includegraphics[width=0.47\textwidth]{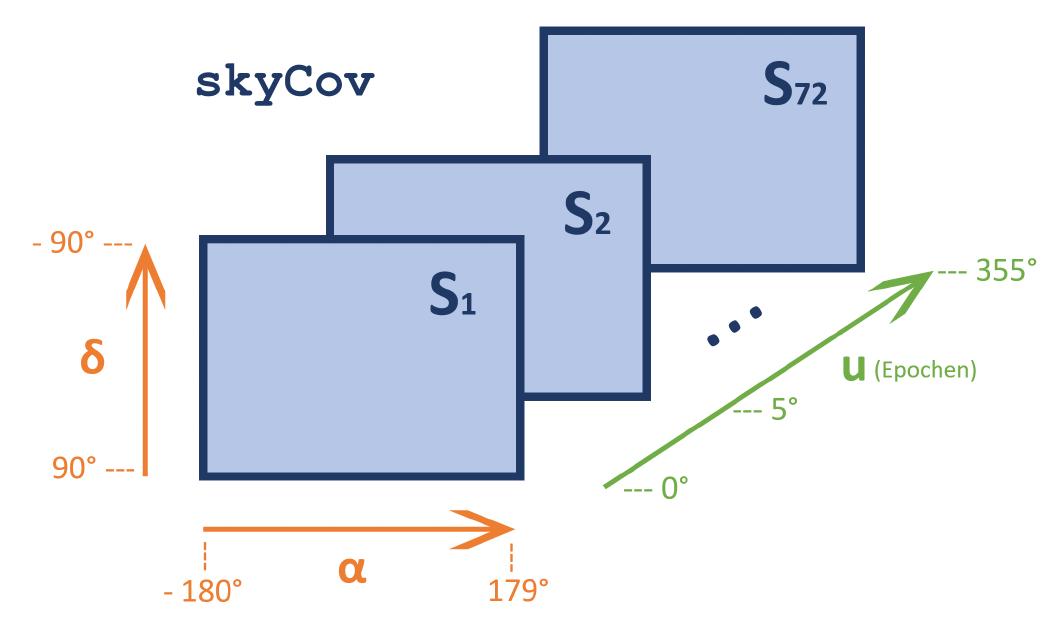}
  \vspace{-0.2cm}
  \caption[]{Structure of the skycov tensor, i.e. sky maps for the
    combinations of constellations and off-axis angle response.
    For this example, a step size (time step) of 5\degs\ is assumed,
    resulting in 360/5=72 sky maps.
    \label{skycov}}
\end{figure}

In order to allow any arbitrary combination of Galileo satellites
to be picked, a simulation tool has been set-up \citep{Rott2020}
which allows to switch on/off single Galileo satellites/detectors.
This is implemented as a
MATLAB function {\em galileo\_skyCoverage.m},
which computes the sky coverage of any Galileo constellation
and a (or several) given off-axis detector response(s)
(Fig. \ref{skycov}).

\begin{figure}[h]
  \includegraphics[width=0.48\textwidth]{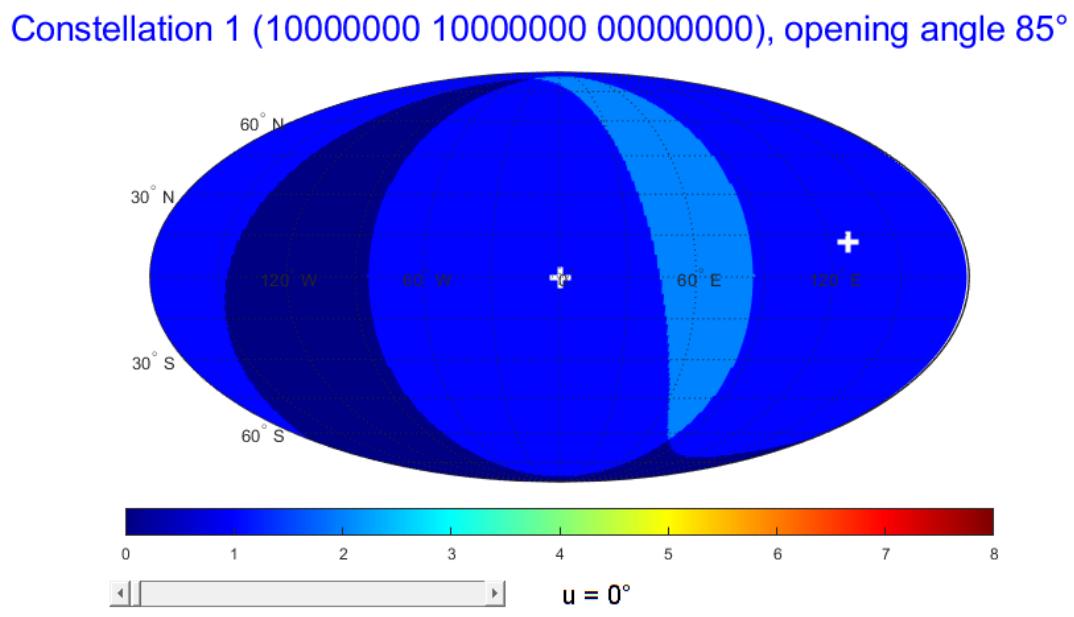}
  \vspace{-0.2cm}
  \caption[]{Example of an interactive plot for the sky coverage
    achieved with two Galileo satellites in separate planes,
    equipped with a flat GRB detector. The slide bar 
    moves through the 72 epochs at 5\degs\ steps.
    \label{cov2sat85d}}
\end{figure}

\begin{figure}[h]
  \vspace{-0.2cm}
  \includegraphics[width=0.49\textwidth, viewport=45 60 380 270, clip]{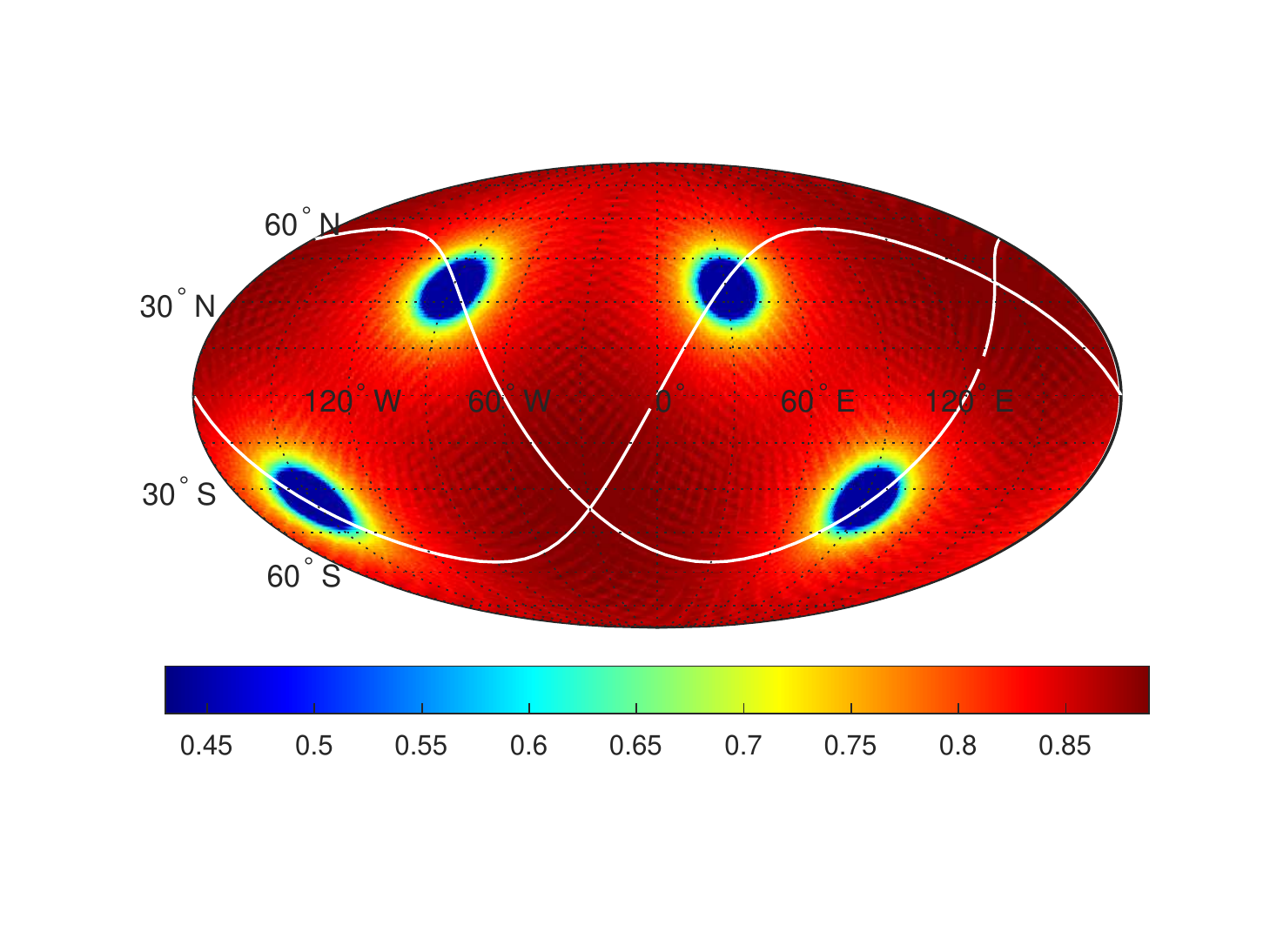}
  \vspace{-0.4cm}
  \caption[]{Mean sky coverage, i.e. the sum of 72 epochs similar to
    Fig. \ref{cov2sat85d}.
    Only 50\% coverage is achieved at the two poles of each of the two
    orbital planes in which the detectors move.
    \label{cov2sat85d_72}}
\end{figure}

\begin{figure}[ht]
  \centering
  \vspace{-0.2cm}
  \includegraphics[width=0.39\textwidth, viewport=45 60 385 270, clip]{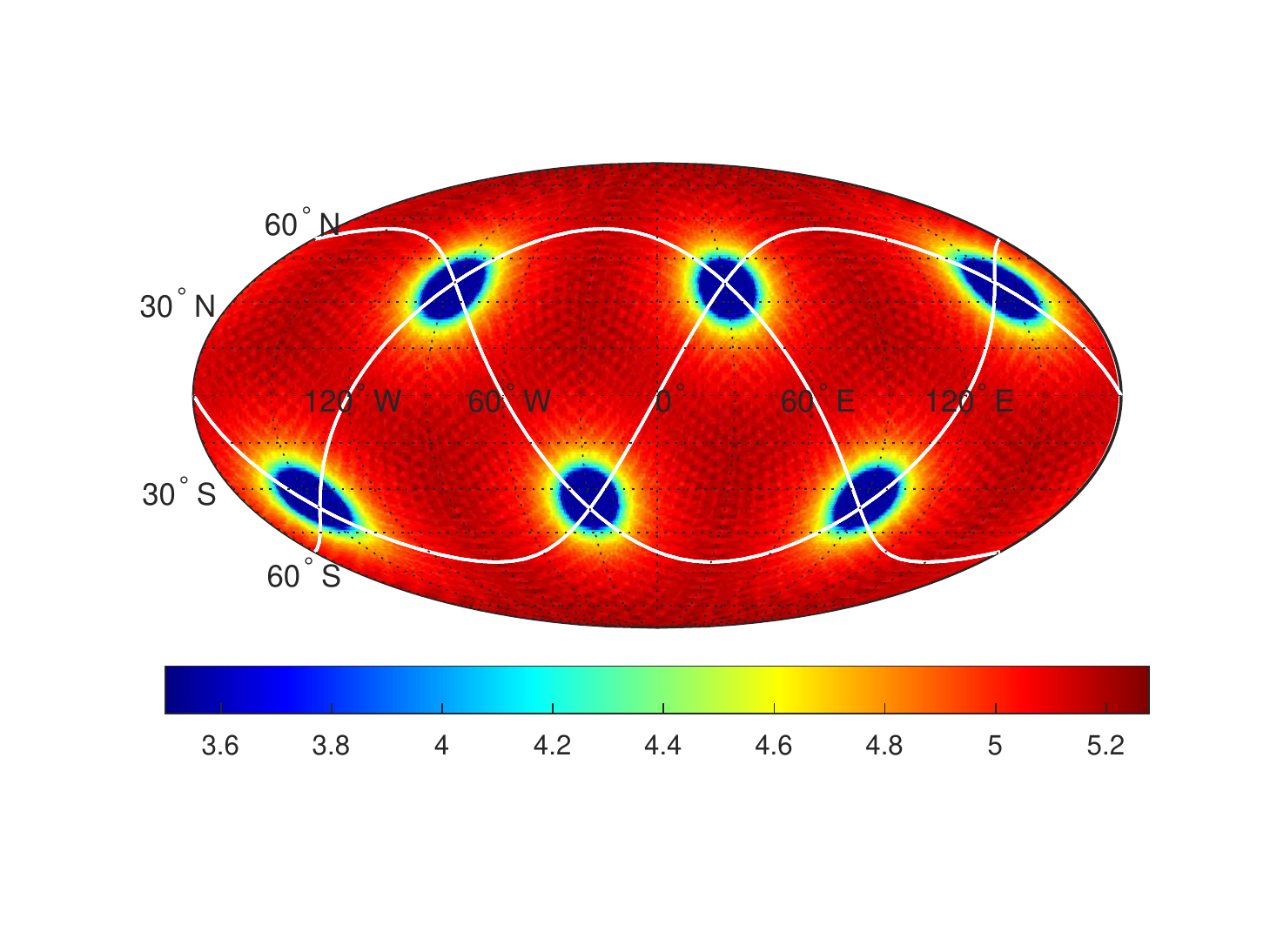}
  \includegraphics[width=0.39\textwidth, viewport=45 60 385 270, clip]{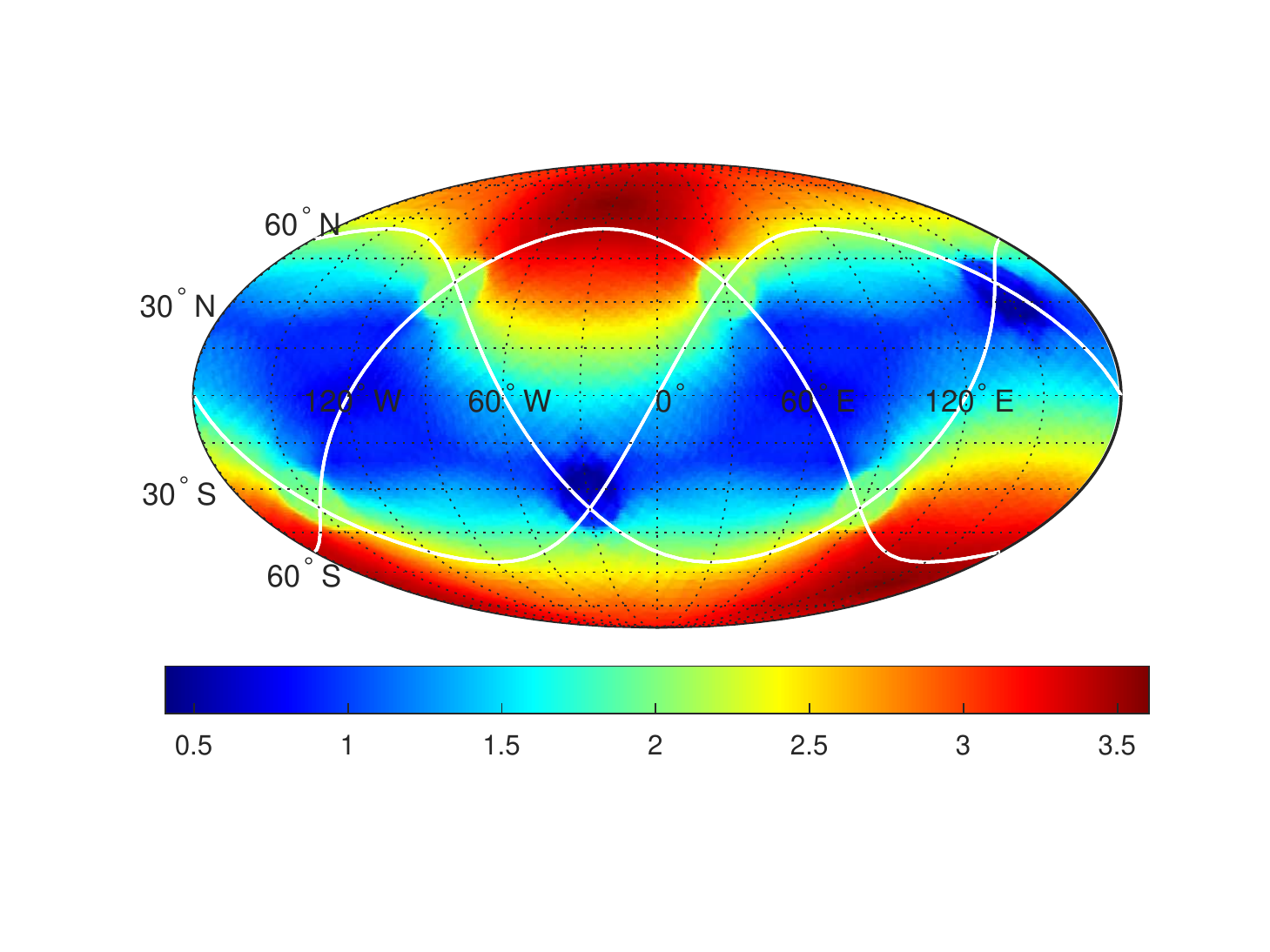}
  \includegraphics[width=0.39\textwidth, viewport=45 60 385 270, clip]{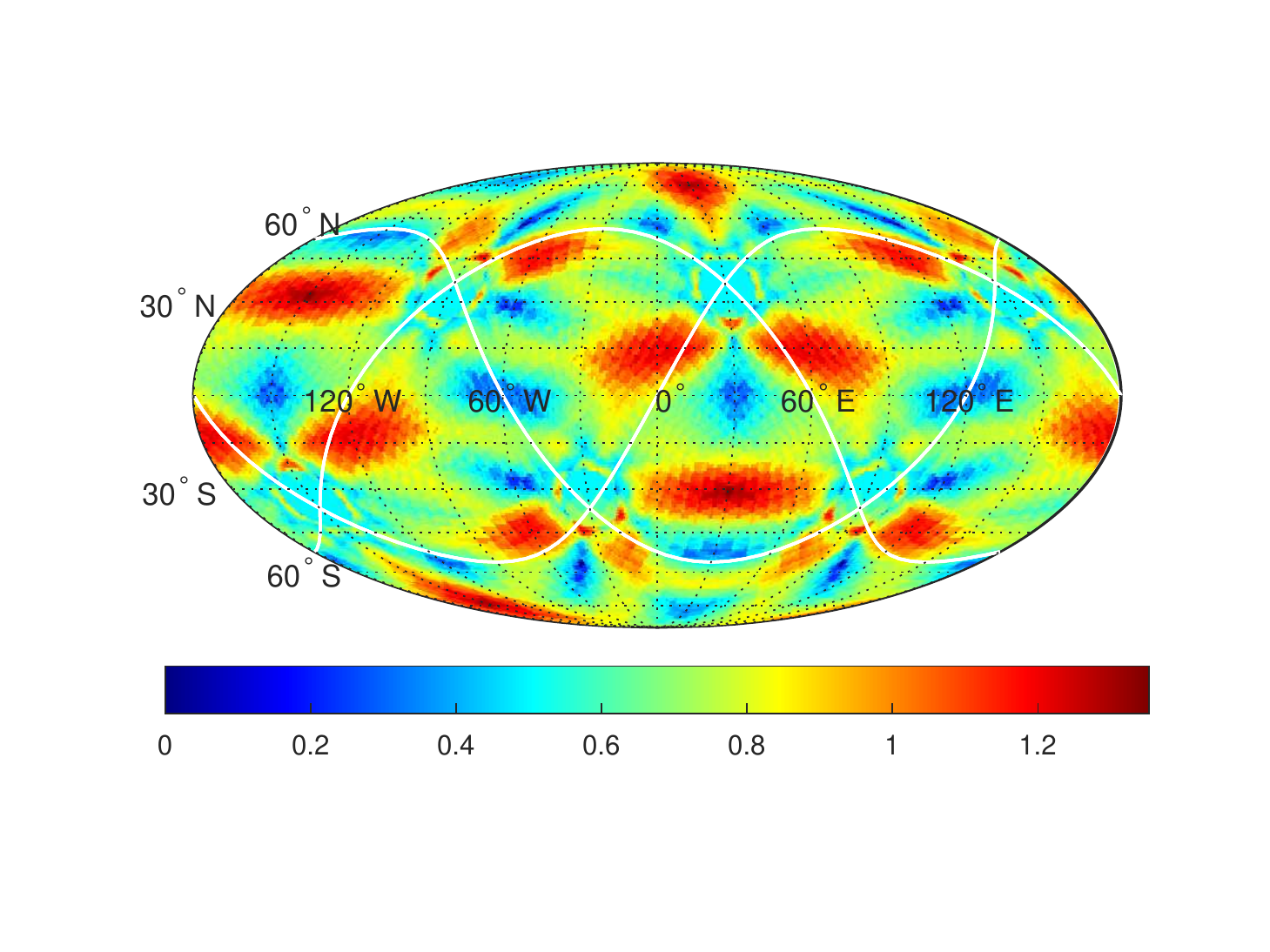}
  \caption[]{An extreme example of the influence of the distribution of
     GRB detectors over the orbital plane,
     with {\bf Top:} the identical mean sky coverage, ranging between 3.5 and
    $\sim$5.3;
     {\bf Middle:} all 12 detectors distributed over one pole,
     and {\bf Bottom:}
    all 12 equally distributed. 
    The lower two panels show the standard deviation for each point on
    the sky.
    \label{cov12satasym}}
    \vspace{-0.1cm}
\end{figure}

An example is shown in Fig. \ref{cov2sat85d} for a flat
GRB detector on two Galileo satellites in separate orbital planes,
for one single snapshot of the constellation.
Dark blue shows sky area not covered, light blue the sky covered
by both detectors. Summing over one 14h04m period, i.e.
72 snapshots with 5\degs\ rotation steps, results in Fig. \ref{cov2sat85d_72},
showing the mean sky coverage by two satellites.

We note that the distribution of detectors over the
satellites in a given orbital plane, or also between orbital planes,
has a substantial impact on the results. 
As a demonstration of the effect
Fig. \ref{cov12satasym} shows two different constellations with the
same number of satellites (with a GRB detector) per orbital plane,
but grossly different distribution: (1) one with equal distribution, i.e.
every second satellite has a GRB detector, and (2) all 4 satellites
(with GRB detector) per orbital plane are centered over one pole
of the Earth at the start configuration (epoch one out of the 72 epochs).
The top panel shows the total coverage  after 14h04m,
which obviously is equal for both options, as it depends
just on the number of satellites. The other two panels show the mean
variance of the coverage at any sky position.
In the equal distribution, there is no point on the sky which is covered
by less than 5 and more than 6 satellites,
while in the one-sphere constellation large regions of the
sky are covered only with 1--2 satellites at a given time,
with the consequence that triangulation would not be possible
with this total number of satellites.

Since the Earth is not infinitely small, there is a 5\% chance that
one Galileo satellite is Earth-occulted at any given time. This
is included in our computations.

\section{Simulations}

The simulations involve multiple steps:

\begin{enumerate}[leftmargin=14pt]
  \vspace{-0.28cm}\itemsep-.4ex
\item define different detector geometries
  (sect. \ref{sect:Detector-geometry})
  \item define the number of satellites  and which satellites per
    orbital plane are equipped (sect. \ref{sect:GNSS-setup})
  \item compute the effective area per detector configuration
      (sect. \ref{sect:effarea})
  \item compute the accuracy of the localisation
    via classical cross-correlation, depending on the effective
    areas of the detectors on separate satellites
    (sect. \ref{sect:accuracymatrix})
  \item simulate the sky coverage and the GRB localisation
    accuracy (sect. \ref{sect:covloc})
\vspace{-0.22cm}
\end{enumerate}

\noindent
In order to cover the range of GRB peak intensities (amplitude $A$
in eq. \ref{eq:lc})
four intensity intervals are created such that 
faint intensity levels can be differentiated,
see Tab. \ref{tab:12Det_det1_acc}.

\begin{table}[th]
  \caption{Four GRB intensity intervals
    \label{tab:12Det_det1_acc}}
  \vspace{-0.3cm}
  \begin{tabular}{ccc}
    \hline
    \noalign{\smallskip}
    Intensity ID & Peak count rate & Peak flux bin  \\
           & (ph/cm$^2$/s) & (10$^{-7}$ erg/cm$^2$/s)  \\
    \noalign{\smallskip}
    \hline
    \noalign{\smallskip}
    1 & 1.5--2 & 2.3--3.0  \\
    2 &  2--3  & 3.0--4.6  \\
    3 &  3--6  & 4.6--9.2  \\
    4 & 6--100 & 9.2--154  \\
    \noalign{\smallskip}
    \hline
   \end{tabular}
\end{table}

\noindent
We fix the detector temporal resolution at 3\,ms. Finally, we use the
present 24/3/1 walker configuration of the GNSS system, and
assume that satellites in all three orbital planes are indeed equipped
with a detector.

\noindent These simulations return sky maps which are used to
\begin{itemize}[leftmargin=14pt]
 \vspace{-0.2cm}\itemsep-.2ex
  \item verify the extent to which 4$\pi$ coverage is possible with a
   homogeneous localisation accuracy over the sky
  \item verify the extent to which 4$\pi$ coverage is possible with a
   homogeneous flux sensitivity level
 \item show the differences in sky coverage and localisation accuracy
   as a function of different number of satellites to be equipped
   with a detector and the detector geometry
 \item provide absolute values of the GRB localisation accuracy
   (distribution)  for both, single snapshots as well as time-averaged
   over the GNSS orbital period of 14h04m.
\vspace{-0.22cm}
\end{itemize}

\noindent
Given the CPU-intensive forward-folding triangulation technique, the
full range of parameter testing in the simulation is done by using
the classical cross-correlation. Only one individual set-up is
computed with the forward-folding triangulation technique
in order to obtain proper error estimates and compare the absolute
values of GRB location accuracy (distribution). Note that in this
case the above steps (3) and (4) are not necessary, since this
is part of the model forward-folding.

\begin{figure}[!ht]
  \vspace{-0.2cm}
  \includegraphics[width=0.23\textwidth]{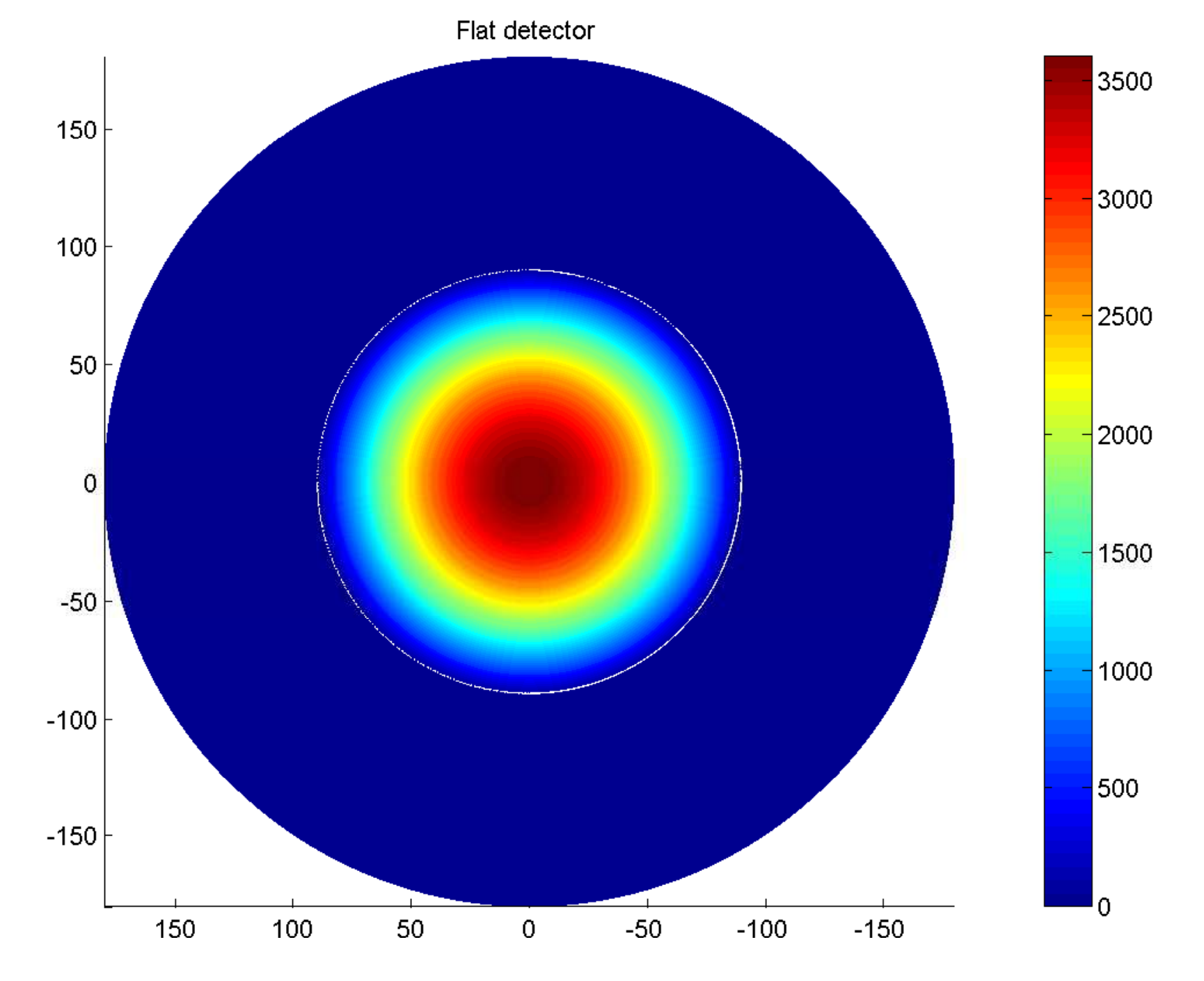}
  \includegraphics[width=0.23\textwidth]{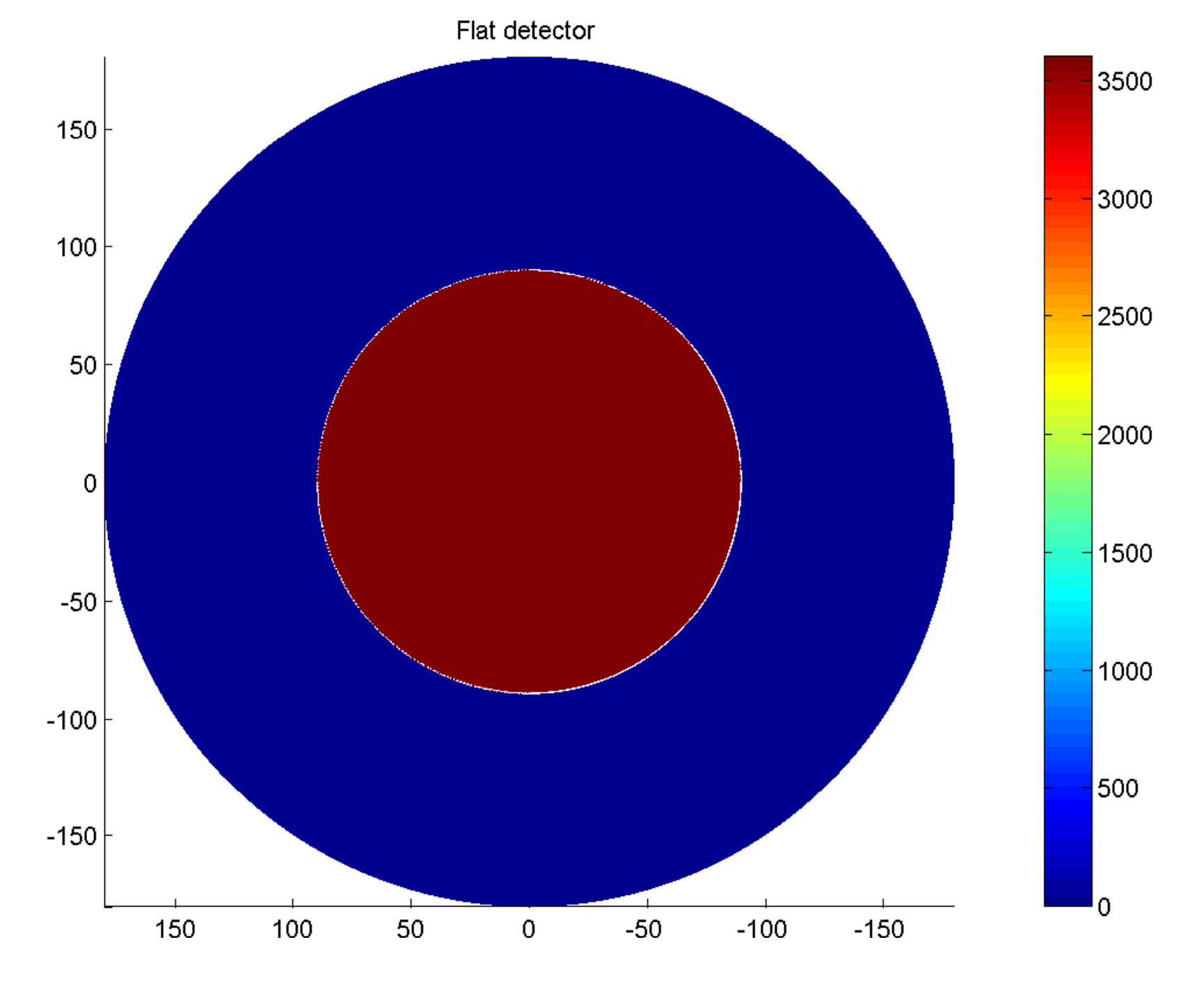}
  \includegraphics[width=0.23\textwidth]{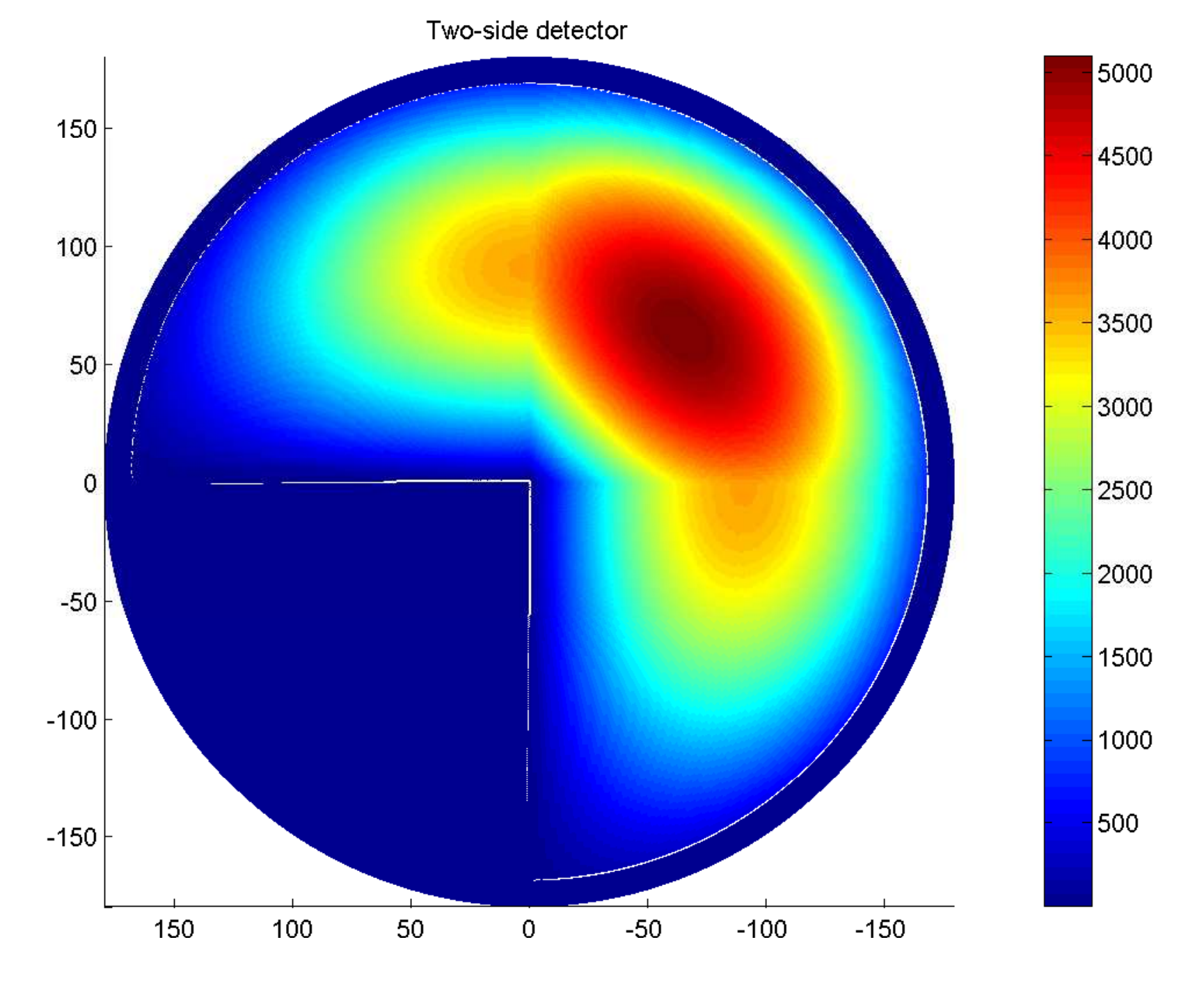}
  \includegraphics[width=0.23\textwidth]{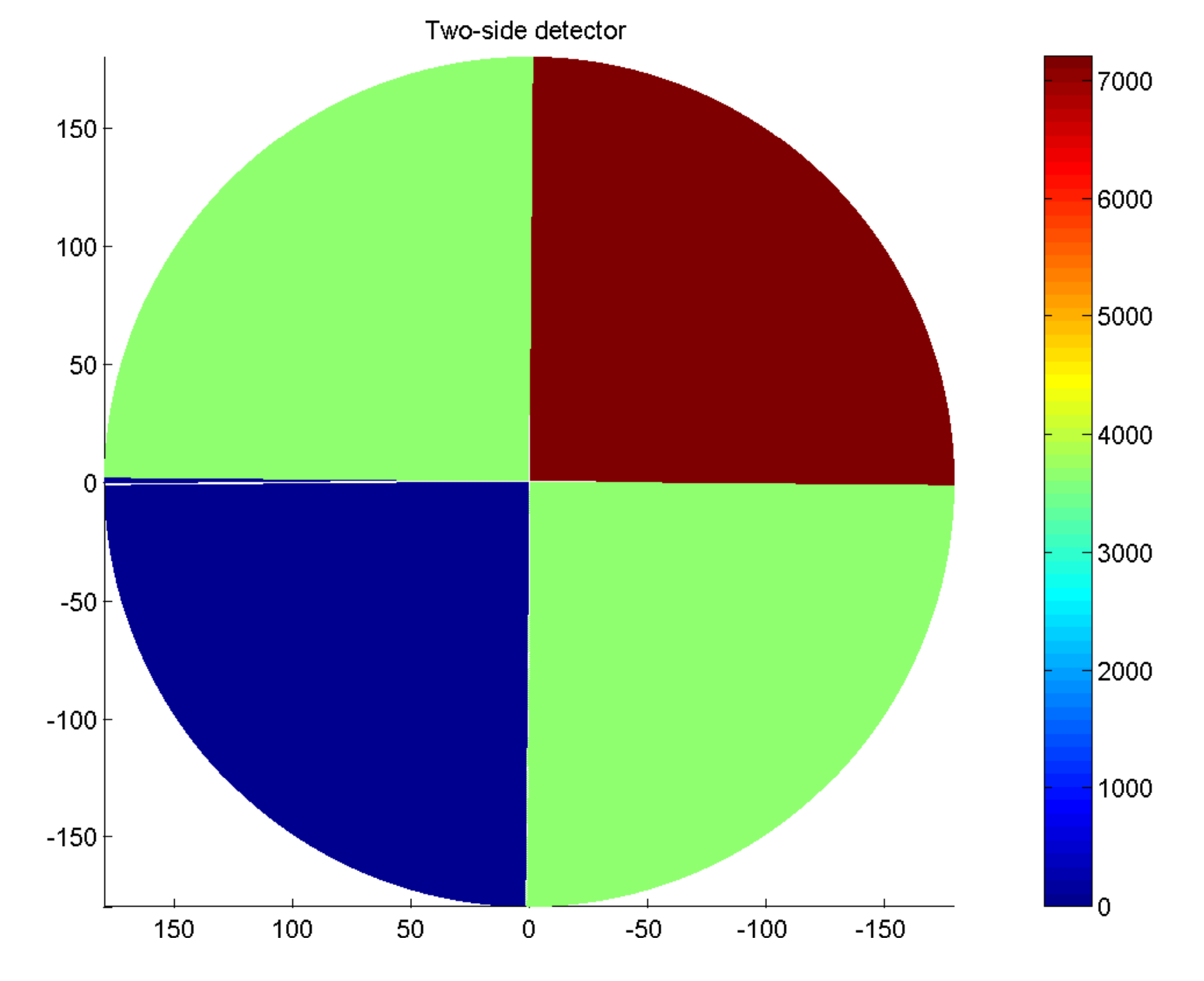}
  \includegraphics[width=0.23\textwidth]{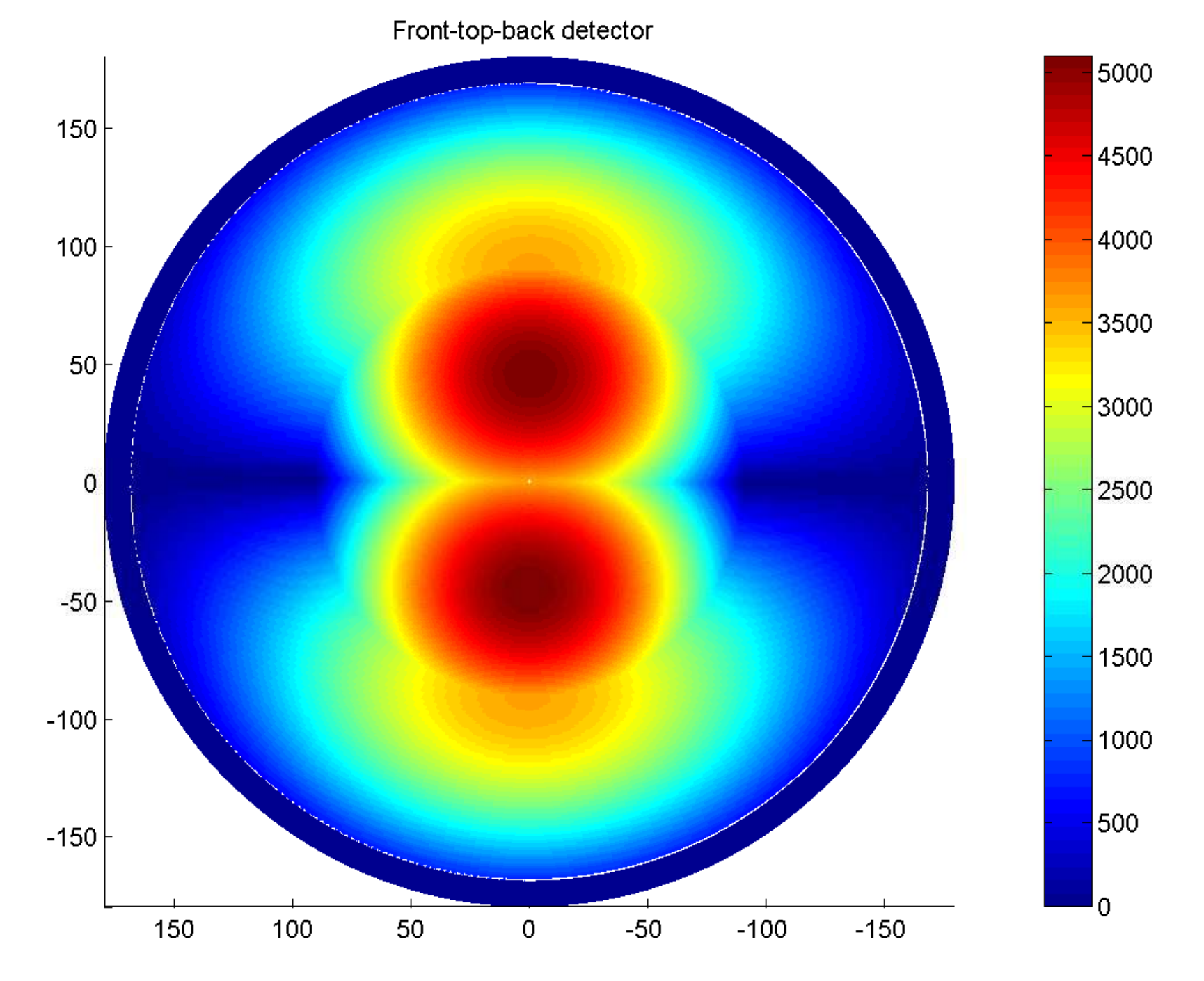}
  \includegraphics[width=0.23\textwidth]{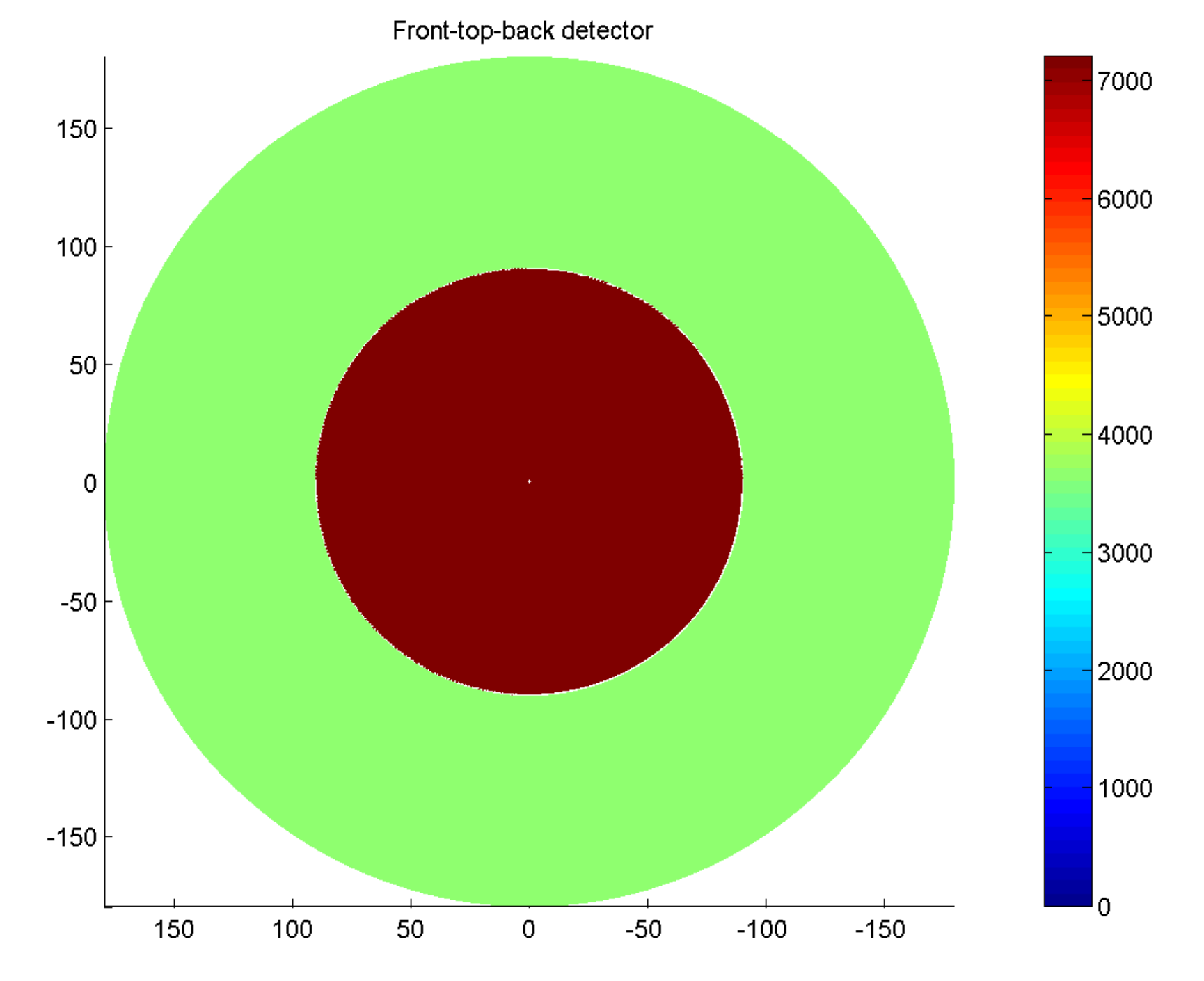}
  \includegraphics[width=0.23\textwidth]{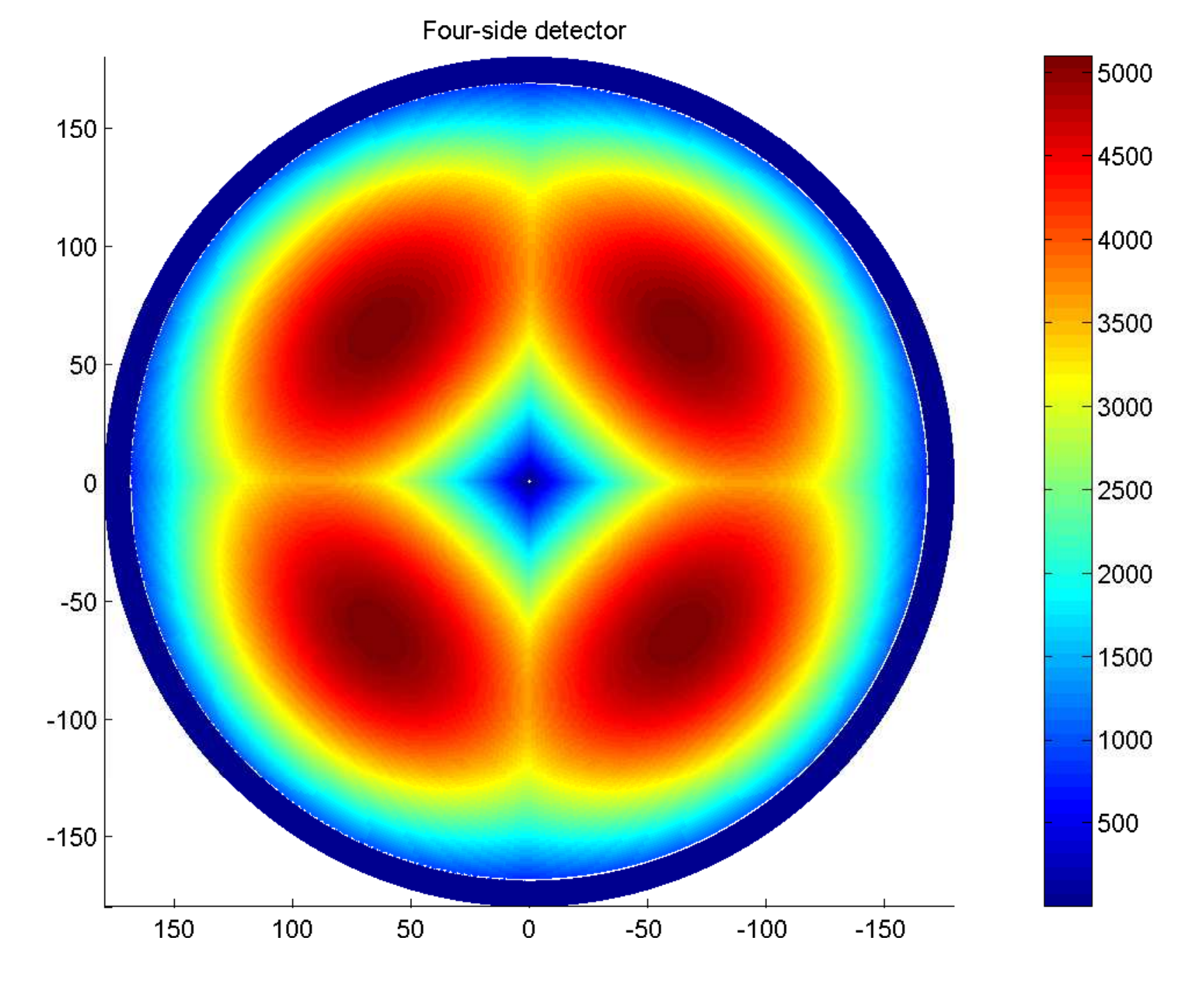}
  \includegraphics[width=0.23\textwidth]{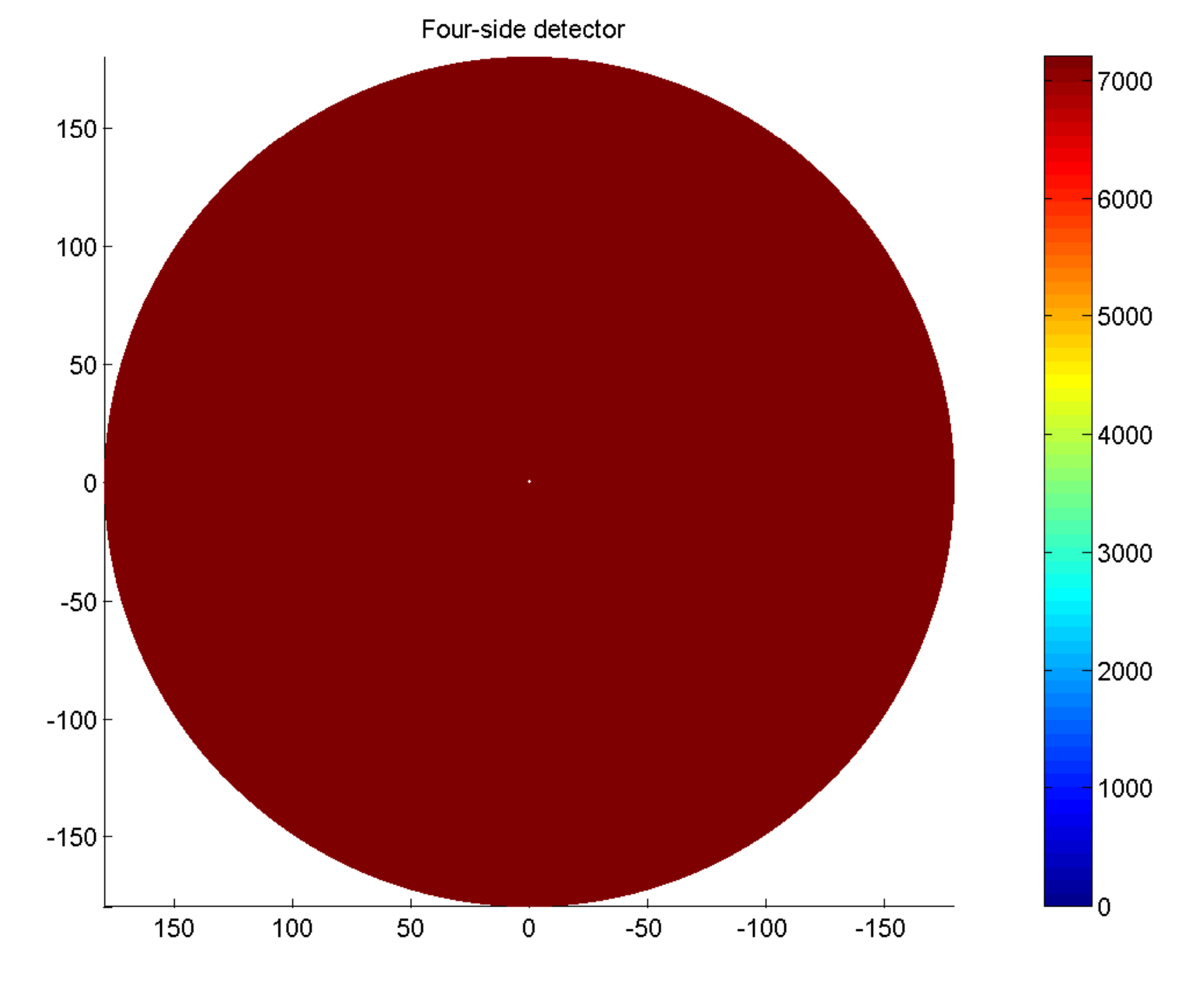}
  \includegraphics[width=0.23\textwidth]{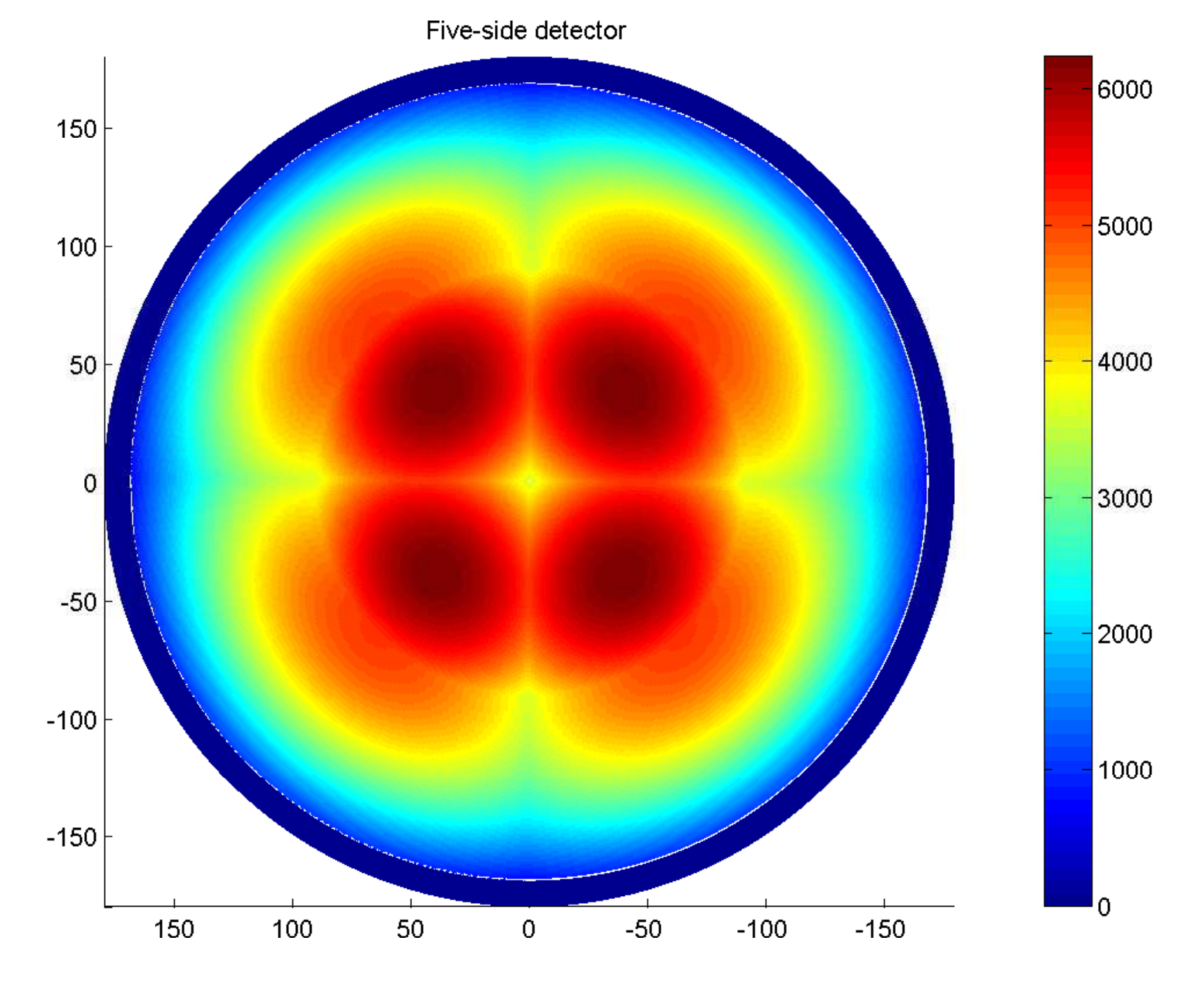}
  \includegraphics[width=0.23\textwidth]{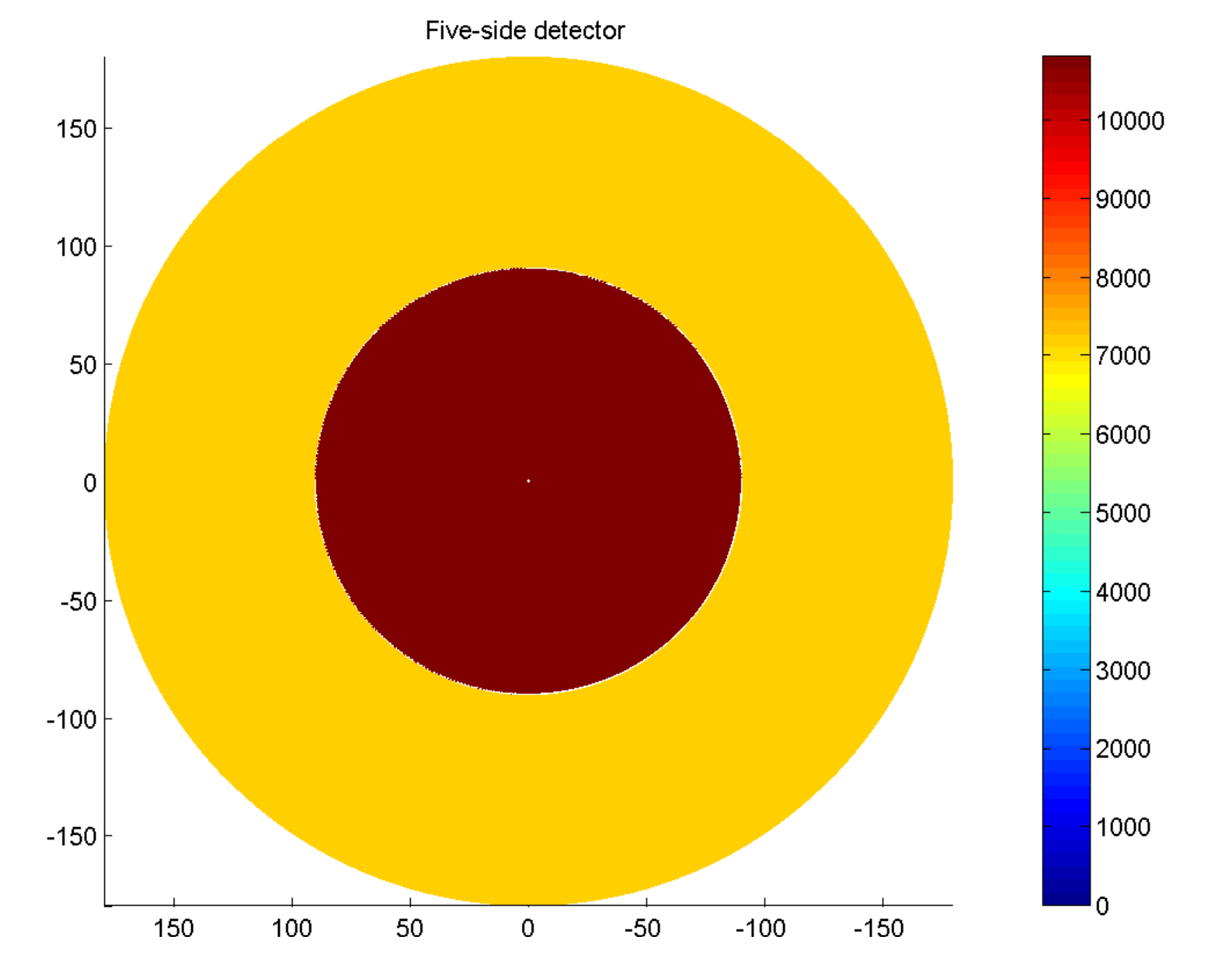}
  \includegraphics[width=0.23\textwidth]{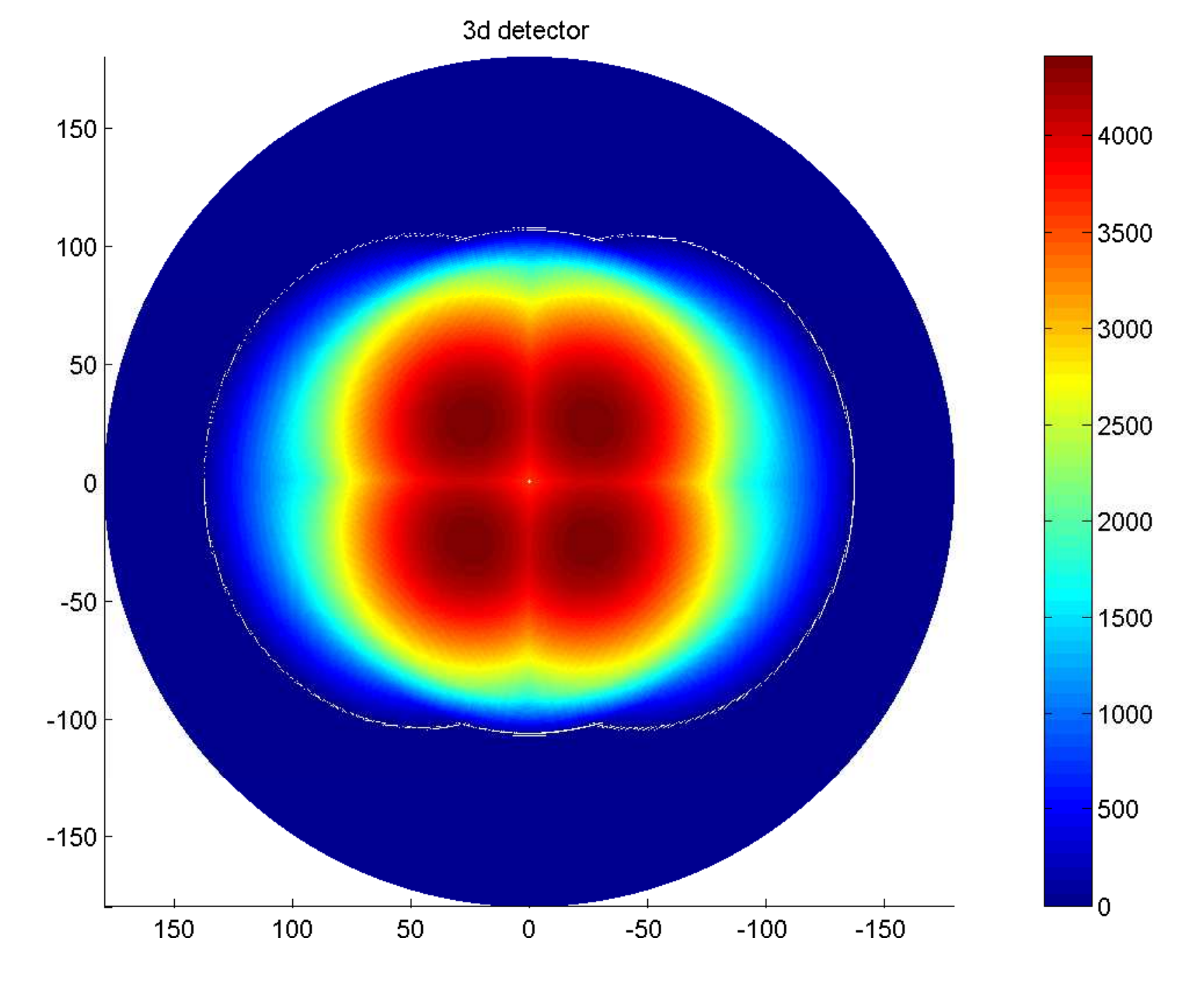}
  \hfill\includegraphics[width=0.235\textwidth]{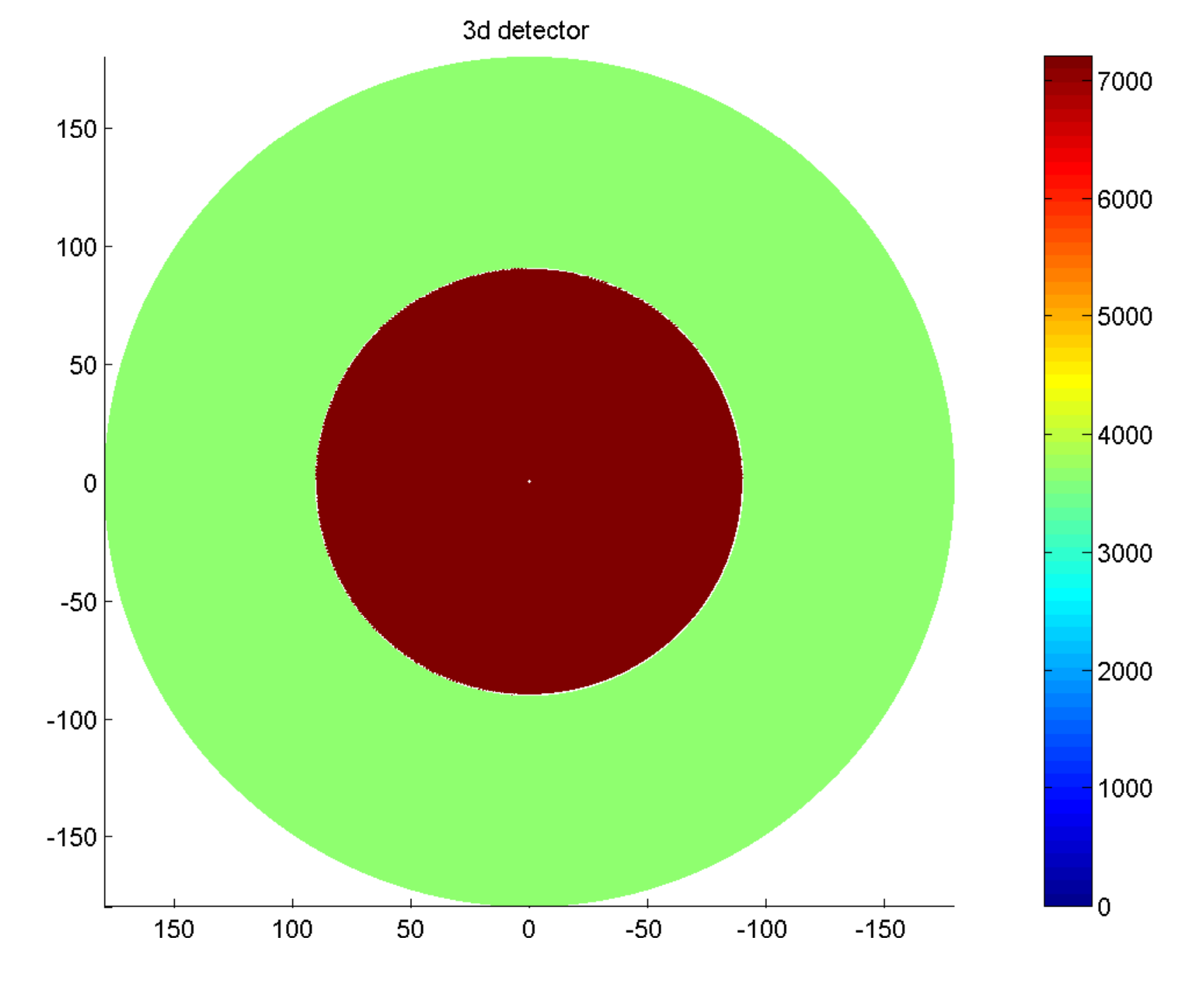}
  \vspace{-0.2cm}
  \caption[]{Two-dimensional variation of the effective area (left column)
    and the illuminated area (right) for our
    proposed geometries (each detector is 60cm x 60cm).
    {\bf Top row:} Flat detector, zenith looking.
    {\bf 2nd:} Two detectors on two neighboring satellite surfaces.
    {\bf 3rd:} Two detectors on opposite sides and zenith.
    {\bf 4th:}  Four detectors on each side of the satellite, none towards
    zenith. The Earth shadow is included.
    {\bf 5th:} Five detectors, one on each side and one zenith pointing.
    {\bf Last row:} A zenith-looking cube
     with 30\,cm height; shadowing included.
    \label{3Deffarea}}
  \vspace{-0.8cm}
\end{figure}

\subsection{Classical scheme using cross-correlation}

\subsubsection{Look-up table for direction-dependent effective area
\label{sect:effarea}}

Depending on the placement of single-plate  detectors on
different sides of a Galileo satellite, the three-dimensional distribution
of the effective area is grossly different. For illustration, we show
these in Fig. \ref{3Deffarea}: the left panels show the effective area
as a function of azimuth and zenith angle between 0\degs\ (figure center) to
180\degs\ (border of figure) of an illuminating source (GRB).
The right panel shows the corresponding area of the detectors that are
illuminated, i.e., the area relevant for the noise. This is different
from the left column figures, since the measured GRB counts per
detector scale with the cosine of the incidence angle, while the
background (noise) is isotropic.

The corresponding 360x180 degree matrices are used as detector-lookup tables
to identify the effective area for a given illumination direction. The
effective areas obtained in this way from the two detectors of a baseline
are used to access the lookup-table of the 
accuracy matrix, see next sub-section. With both together,
the different Galileo detector-equipment constellations are computed.

\begin{figure}[th]
  \centering
  \includegraphics[width=0.37\textwidth, viewport=10 80 710 632, clip]{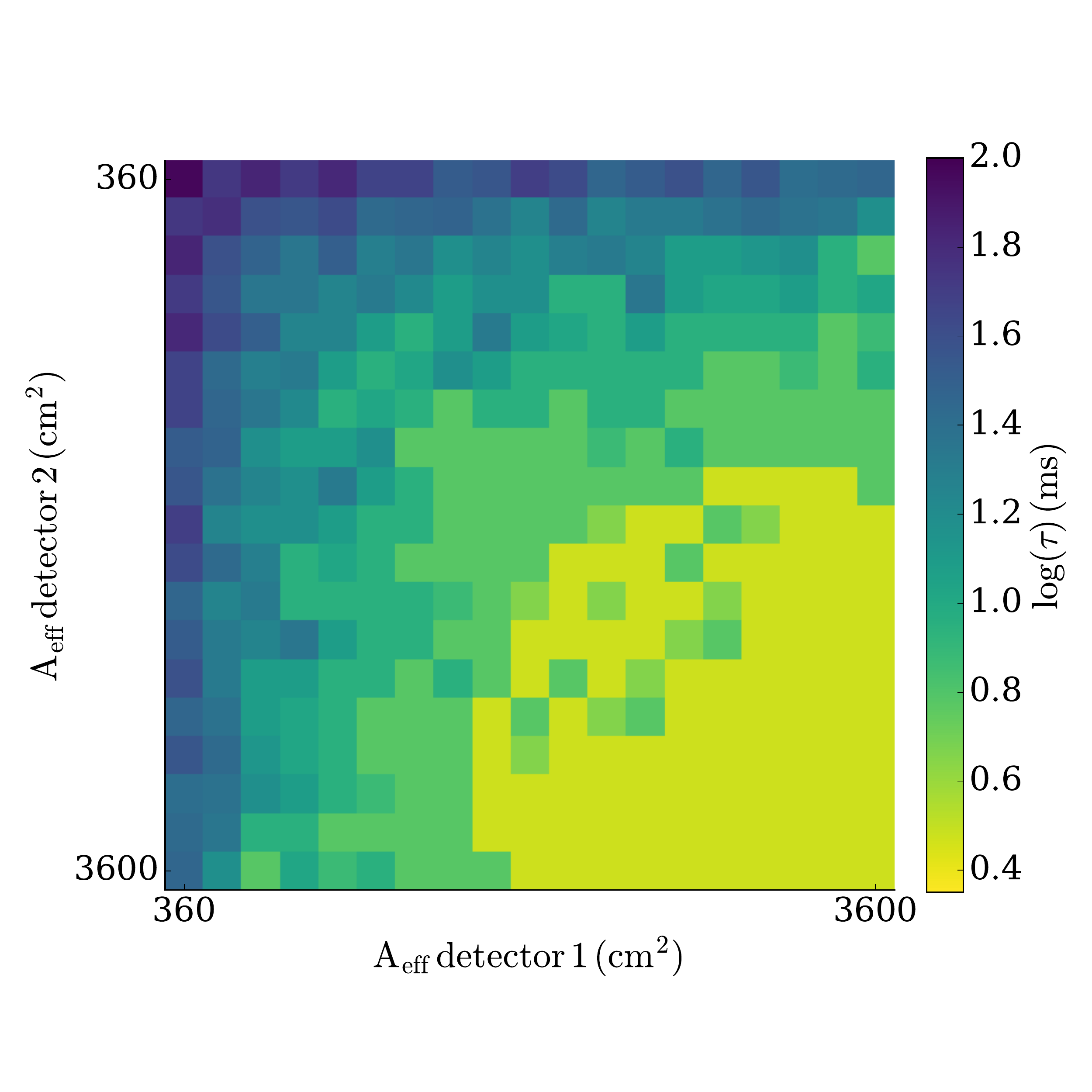}
  \includegraphics[width=0.37\textwidth, viewport=10 80 710 632, clip]{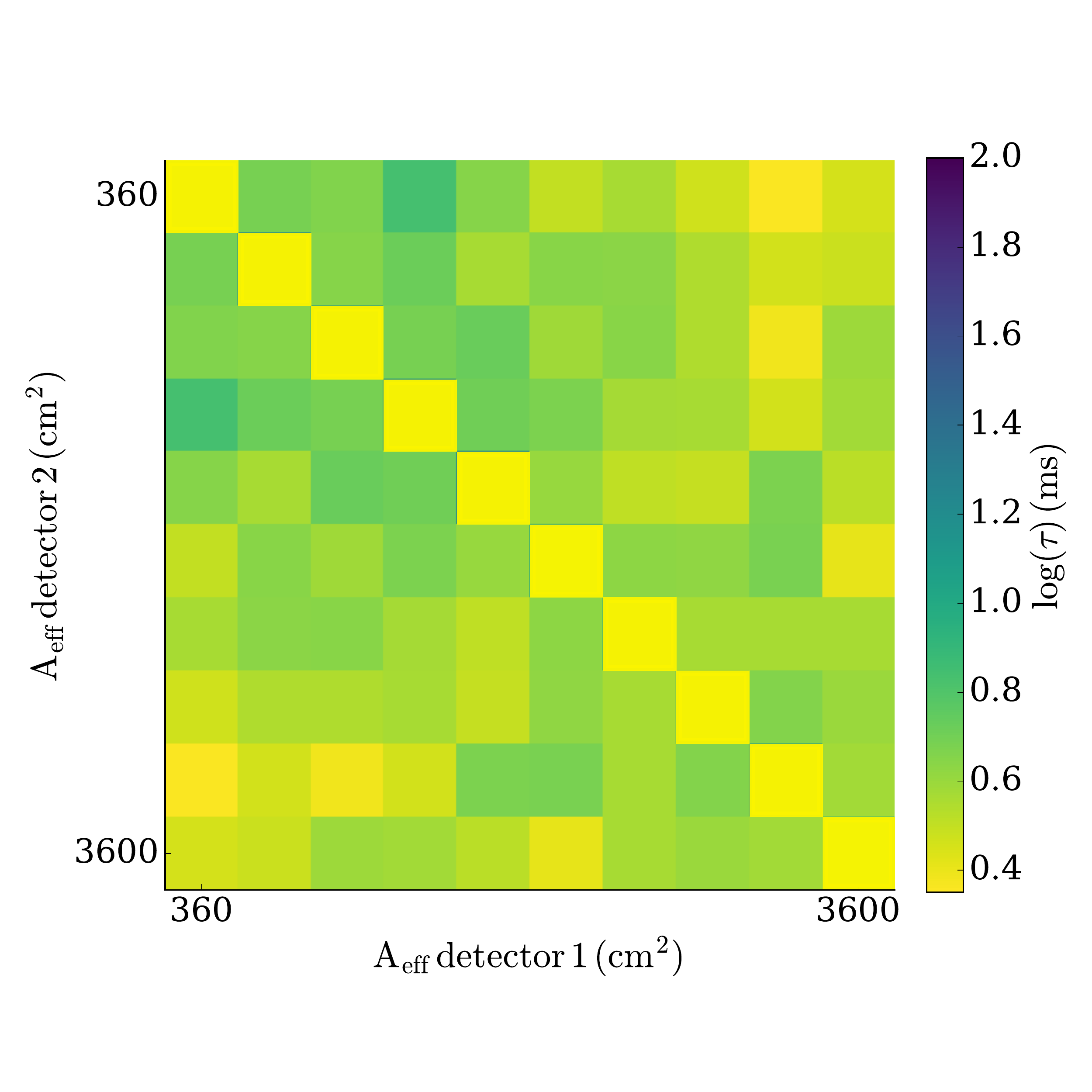}
  \vspace{-0.1cm}
  \caption[]{Distribution of the error of the time-delay (color coded)
    for different angles of two detectors (x- and y-axis) at the same
    position (=satellite), 
    (so the nominal time delay should be zero).
    The matrices shown are the median of
     all simulated GRBs in the bright fluence bin
      (\#4 in Tab. \ref{tab:12Det_det1_acc}), separately computed
    for the classical cross-correlation
    method (top) and the forward-folding nazgul method (bottom),
    and we have separate matrices for the other intensity bins.
    For identical effective areas, i.e.
    the diagonal, nazgul recovers the nominal time delay of zero, so it
    was set to 0.3 to avoid division by zero in the follow-up steps.
    The placement on the same satellite mimics (along the diagonal)
      also the net effect on the accuracy of two identical detectors on
      different satellites looking exactly towards the same sky location.
    Each pixel is the median of the time-delay of many different GRB light
    curves.
    The time resolution of the detector is assumed to be 3\,ms.
    The effective area distribution
    mimics a 1D detector with 3600 cm$^2$ seen at different off-axis angles.
  \label{delay3ms}}
\end{figure}

\subsubsection{Accuracy matrix \label{sect:accuracymatrix}}

For the computation of the effects of the relative orientations
of different detectors on different Galileo satellites according
to the given satellite equipment scheme,
we need to map the effect of detector-related parameters on the
localisation accuracy in  a way that they can be efficiently used.
Since this localisation quality depends at least on
two angles (the relative orientation of the detector normals
of 2 detectors relative to the GRB direction) and the total
intensity, this is a matrix rather than a factor.
It is straightforward to realize that
the cosine off-axis dependence of the detector sensitivity
is a similar geometrical effect as different detector geometries.
Thus, instead of computing effective area matrices per angle pair,
we can incorporate the detector geometry (in terms of total
effective area per direction) and compute the error of a
delay measurement per angle pair.
Such an ``accuracy matrix'' has been computed via
both methods
  (Fig. \ref{delay3ms}), and then  serves as input to the 
  Galileo satellite mapping simulation.
  
\subsubsection{Results with different detectors
  \label{sect:covloc} }

\begin{figure*}[h]
  \includegraphics[width=0.49\textwidth, viewport=0 90 825 515, clip]{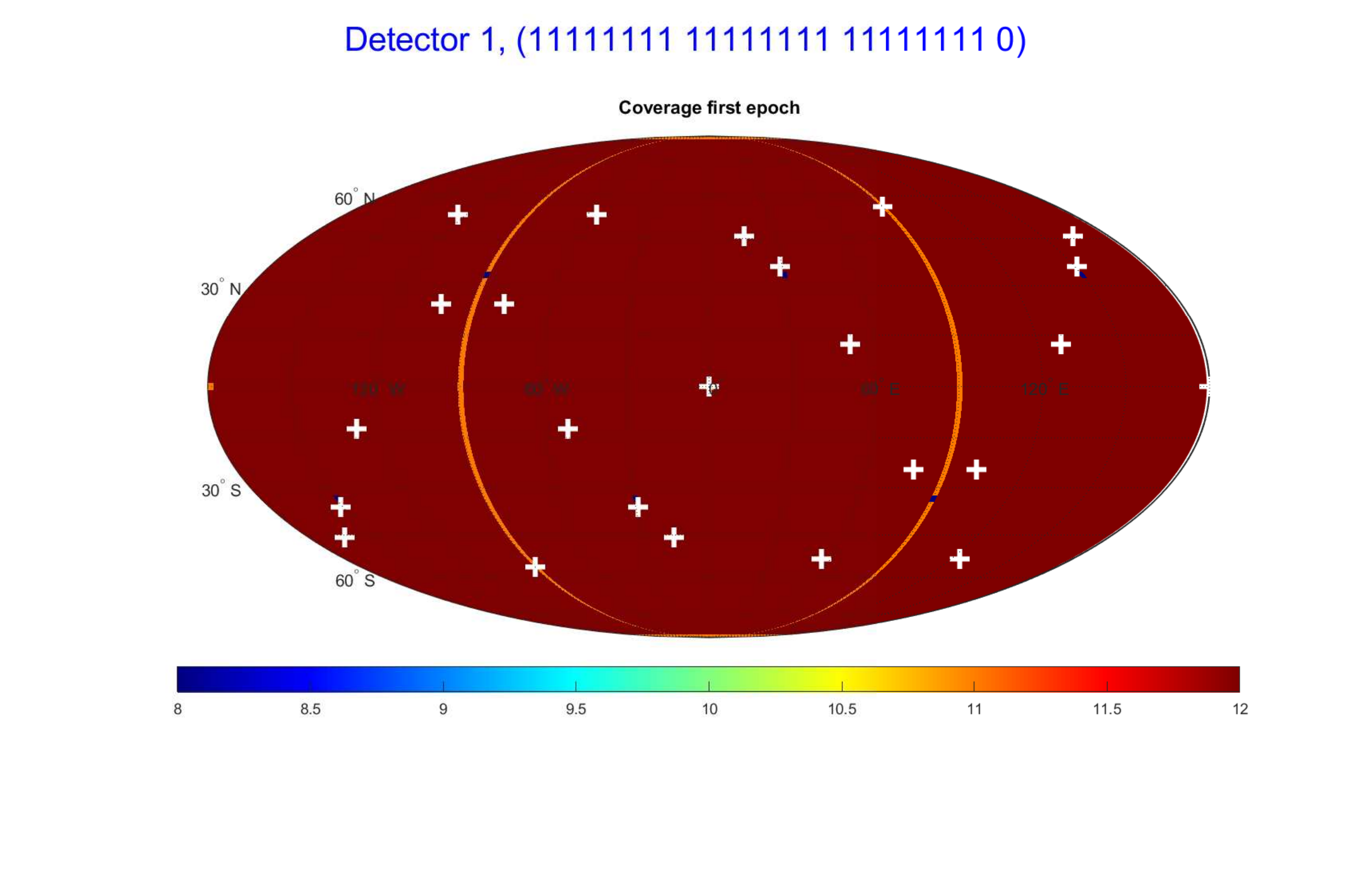}
  \includegraphics[width=0.49\textwidth, viewport=0 90 825 515, clip]{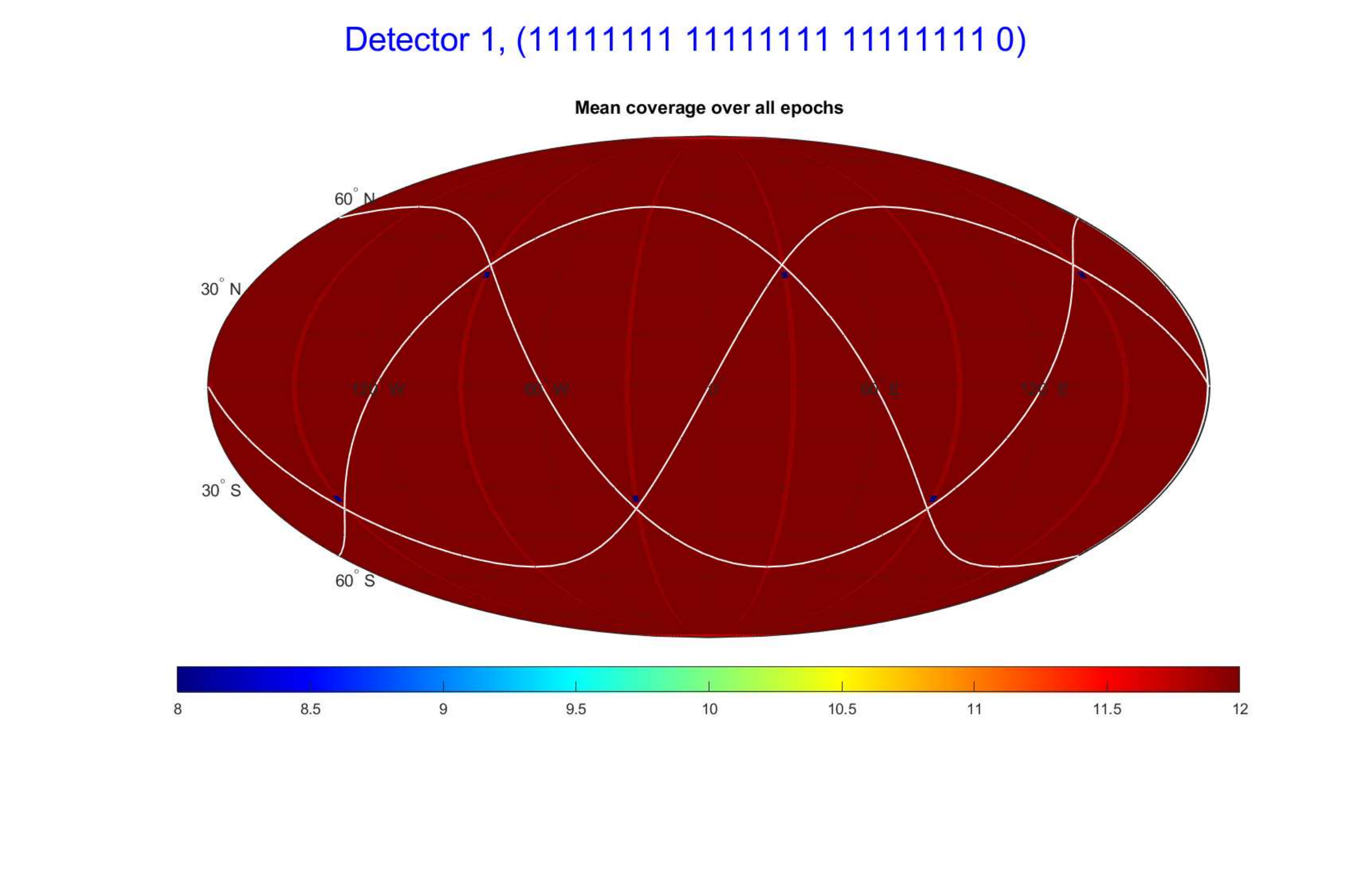}
  \includegraphics[width=0.49\textwidth, viewport=0 90 825 515, clip]{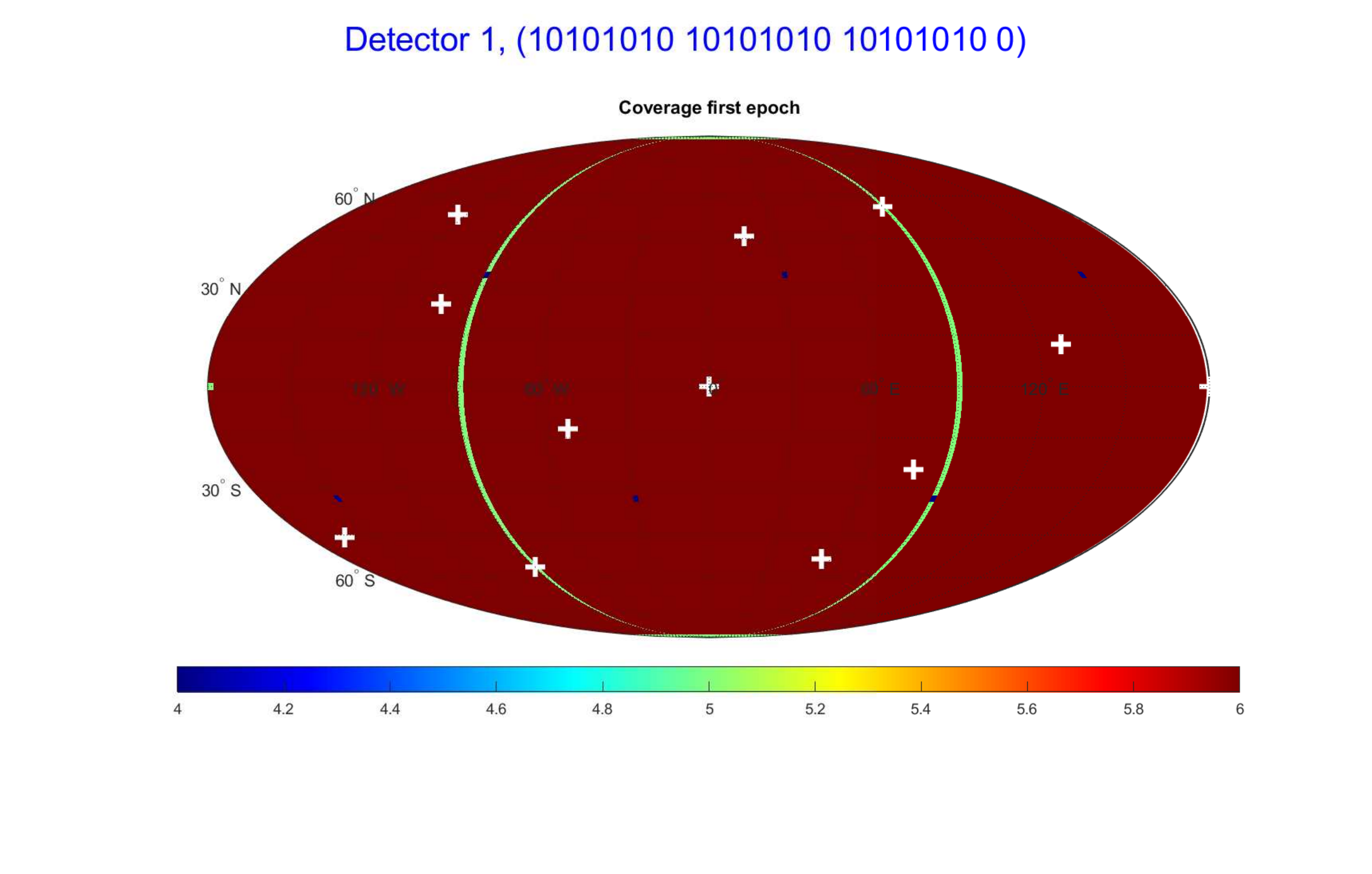}
  \hfill\includegraphics[width=0.49\textwidth, viewport=0 90 825 515, clip]{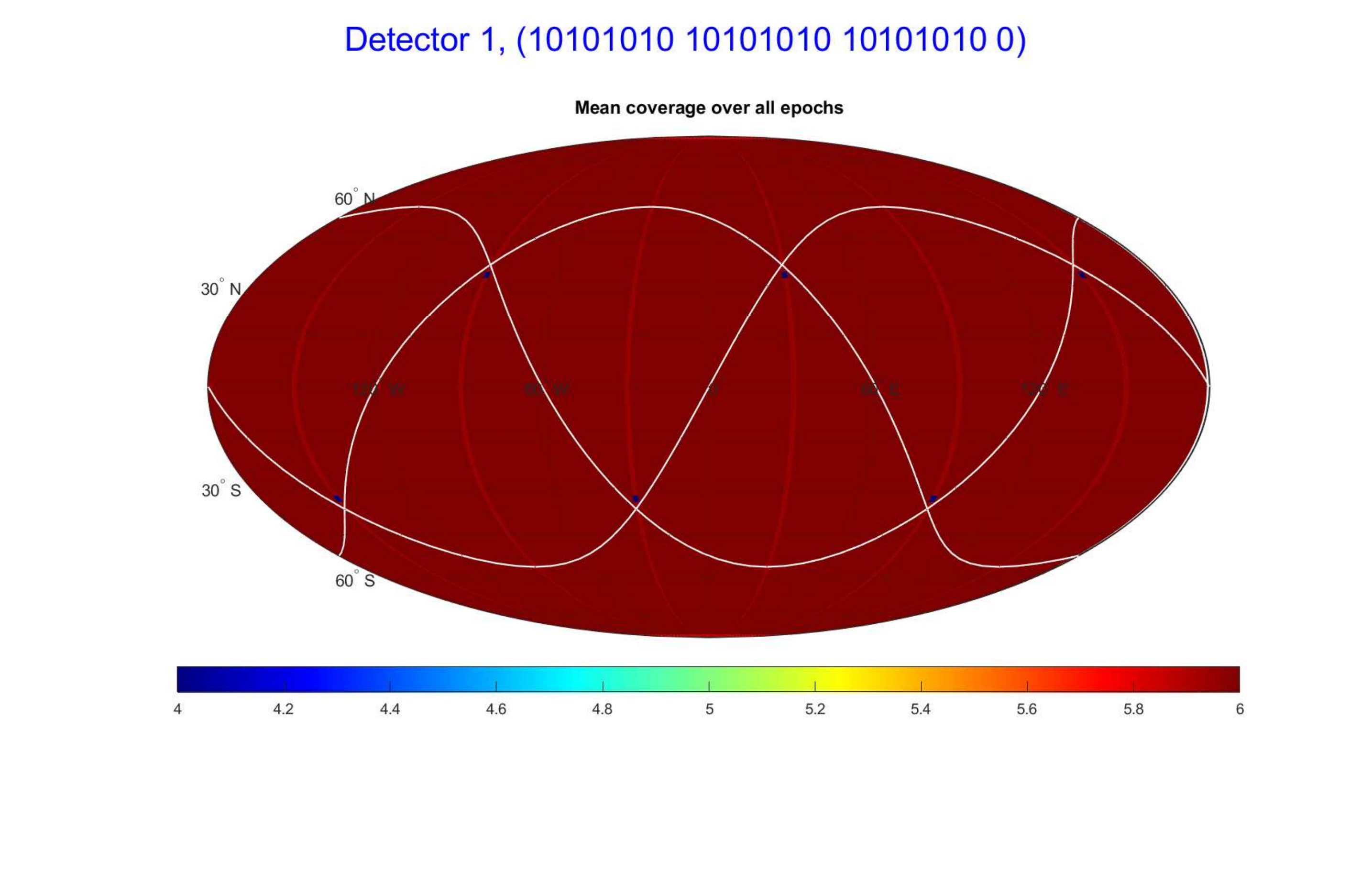}
  \caption[]{Sky coverage for a zenith-looking detector each on 24 (top) and
    12 satellites (bottom; every second along each orbital plane)
    for an instantaneous moment (left) and averaged over one orbit (right).
    The color-coding (note the different scales) gives the number of satellites
    which see a GRB depending on where the GRB happens on the sky.
    \label{12sat_Det1_skycov}}
\end{figure*}

\paragraph{Detector 1: zenith-looking}

We start with a single detector plate, looking at zenith,
with every as well as every second
of the 24 Galileo satellites equipped with one such detector.
We will use this constellation to show the different aspects of the
simulated data -- for the other detector geometries we will primarily
show example distributions and summarize the results in a table.

\begin{figure*}[h]
  \includegraphics[width=0.5\textwidth, viewport=0 90 825 515, clip]{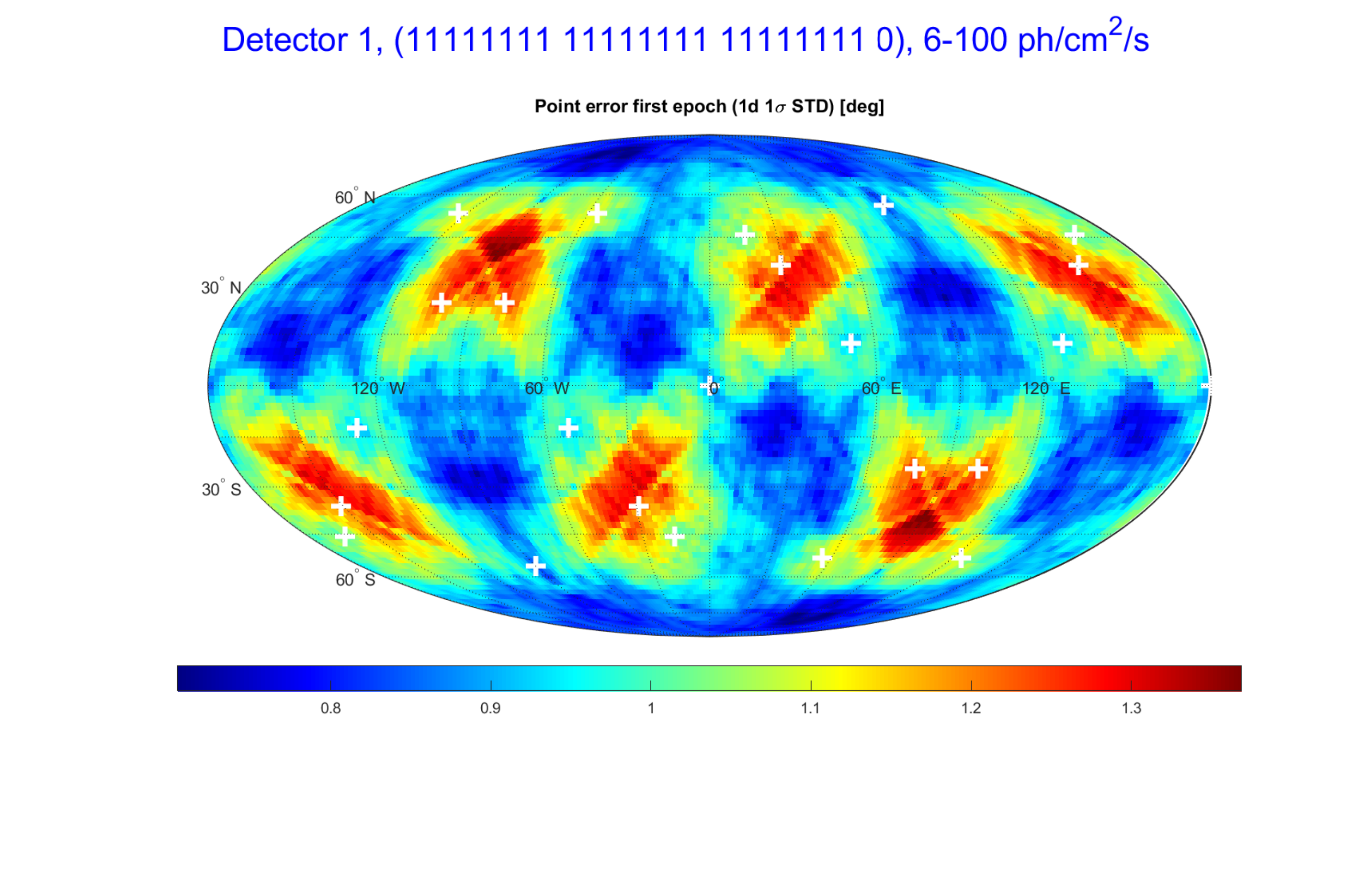}
  \includegraphics[width=0.5\textwidth, viewport=0 90 825 515, clip]{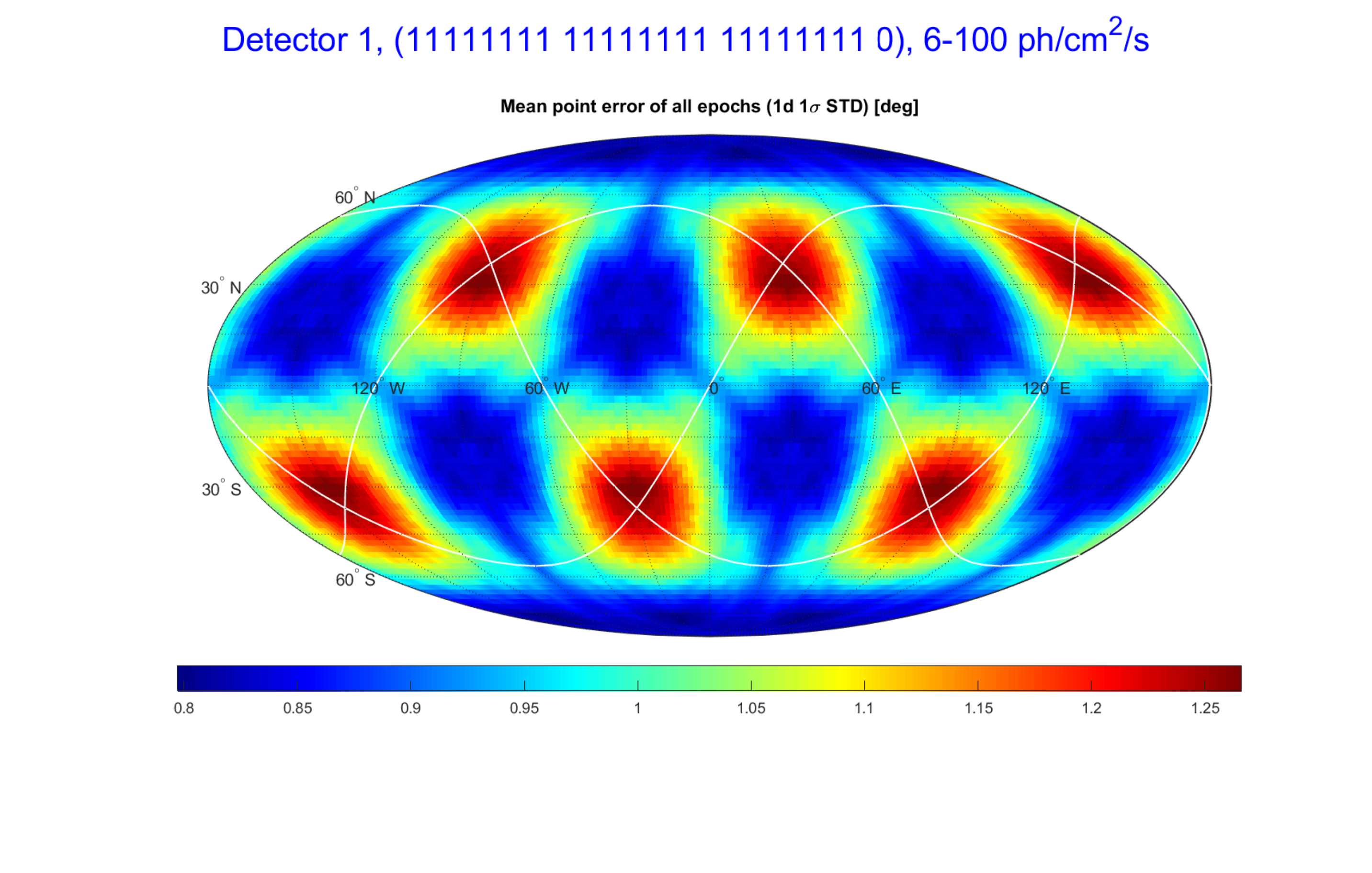}
  \includegraphics[width=0.5\textwidth, viewport=0 90 825 515, clip]{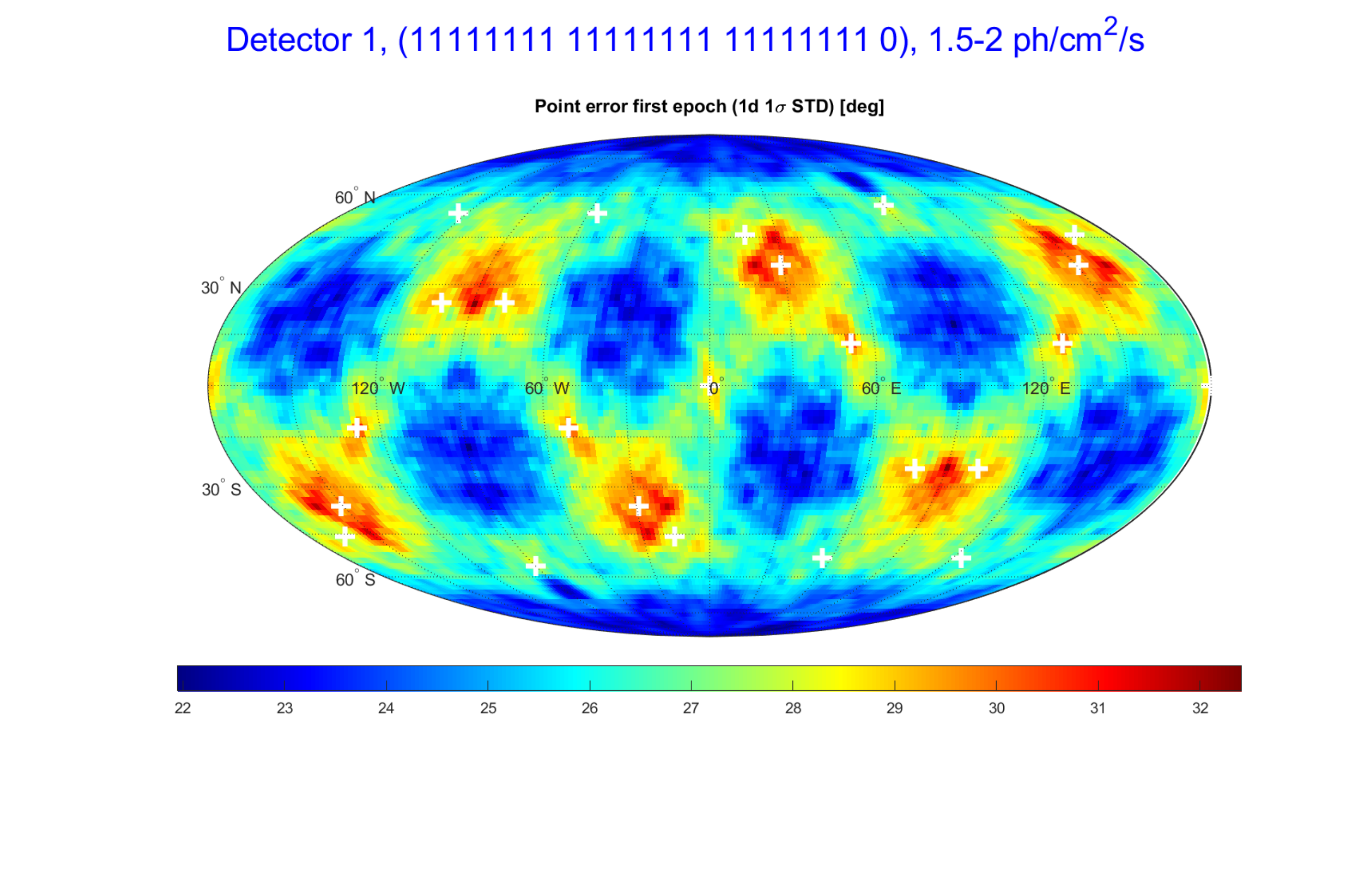}
  \includegraphics[width=0.5\textwidth, viewport=0 90 825 515, clip]{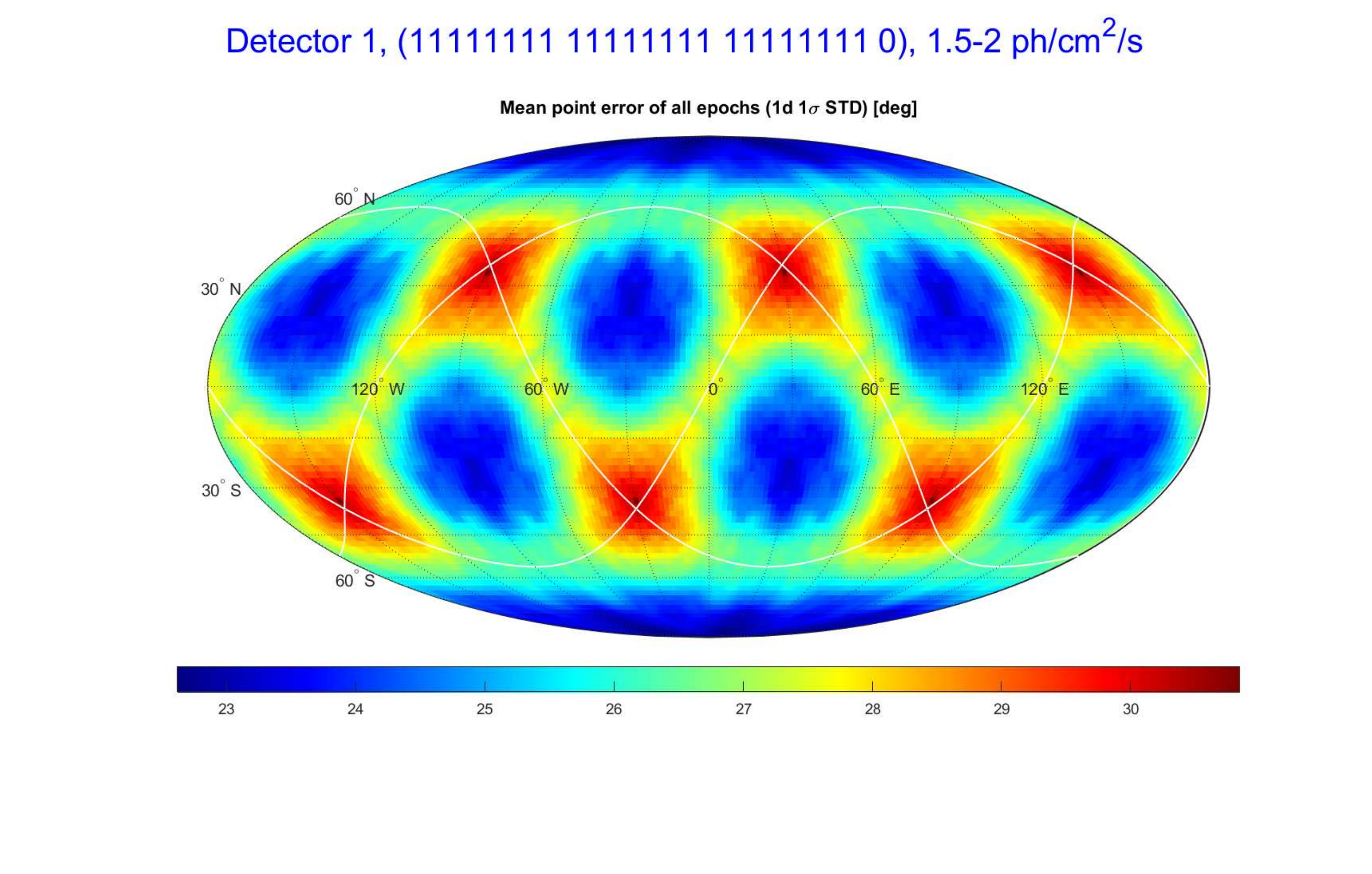}
  \caption[]{Localisation accuracy for a zenith-looking detector each on
    24 satellites 
    for an instantaneous moment (left) and averaged over one orbit (right),
    for GRBs in the brightest peak flux bin of 6--100 ph/cm$^2$/s (top row)
    and the faintest peak flux bin of 1.5--2 ph/cm$^2$/s (bottom row).
    \label{24sat_Det1}}
\end{figure*}

\begin{figure*}[th]
   \includegraphics[width=0.49\textwidth, viewport=0 90 825 515, clip]{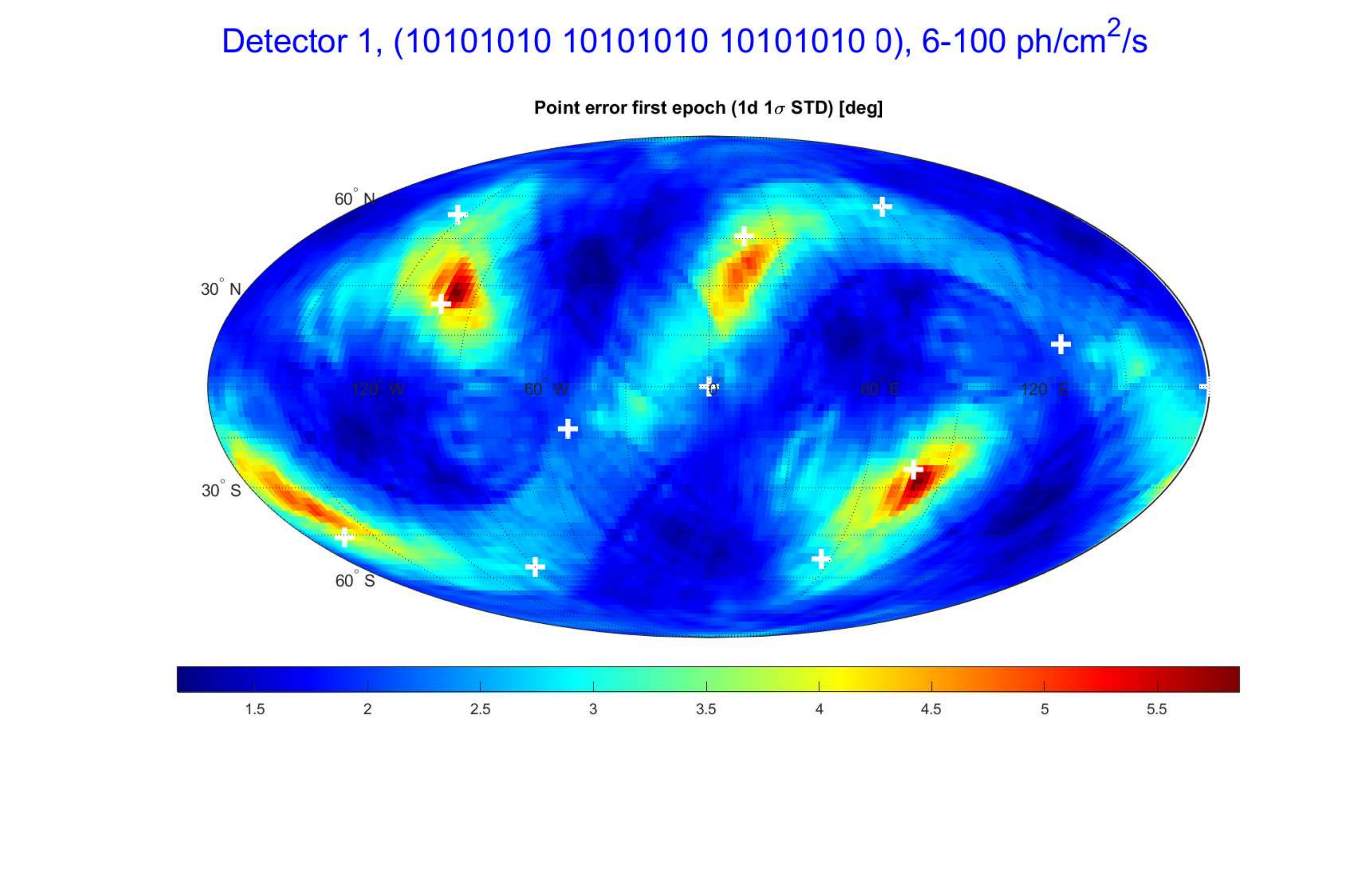}
   \includegraphics[width=0.49\textwidth, viewport=0 90 825 515, clip]{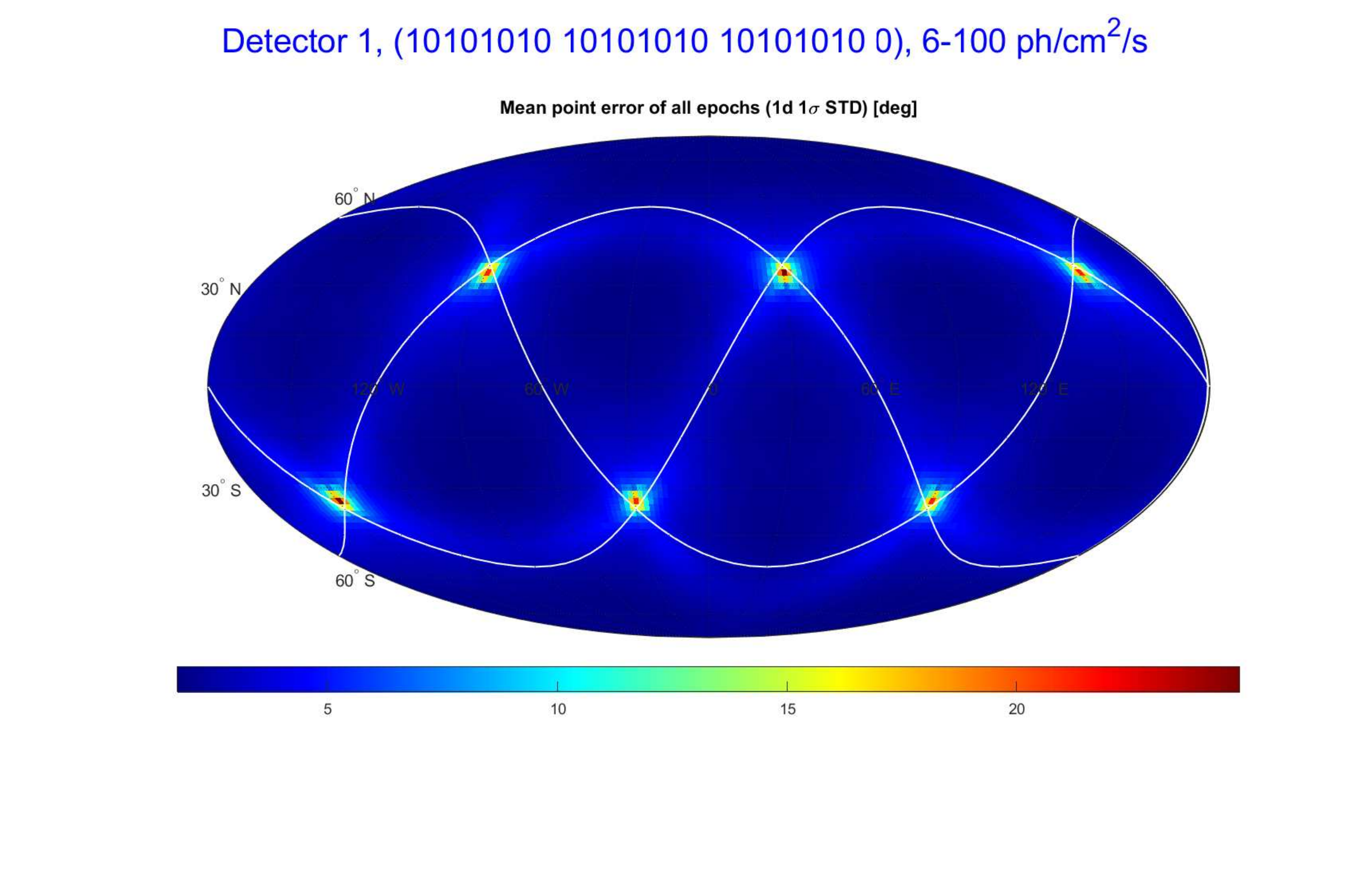}
   \includegraphics[width=0.49\textwidth, viewport=0 90 825 515, clip]{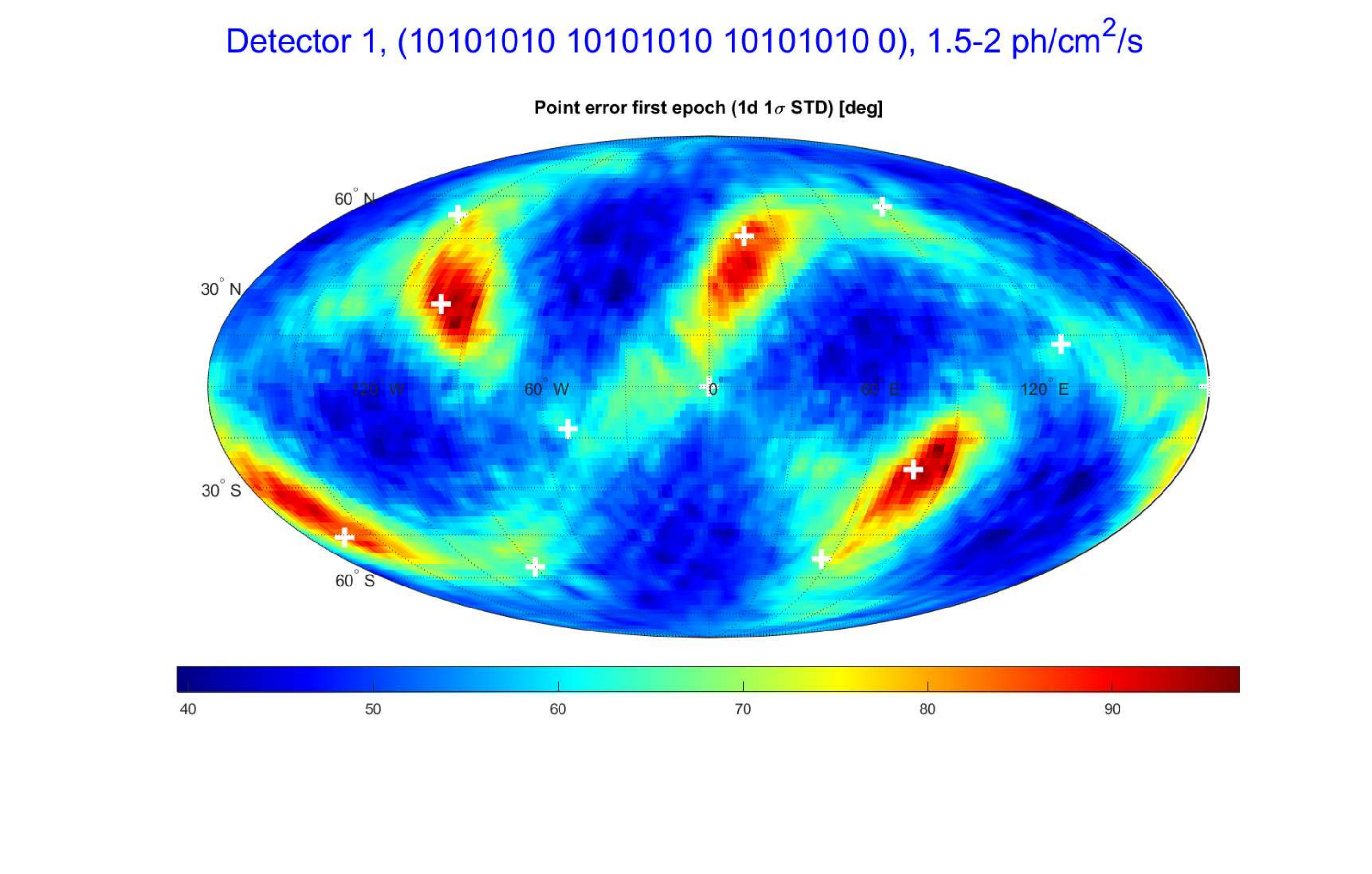}
   \hfill\includegraphics[width=0.49\textwidth, viewport=0 90 825 515, clip]{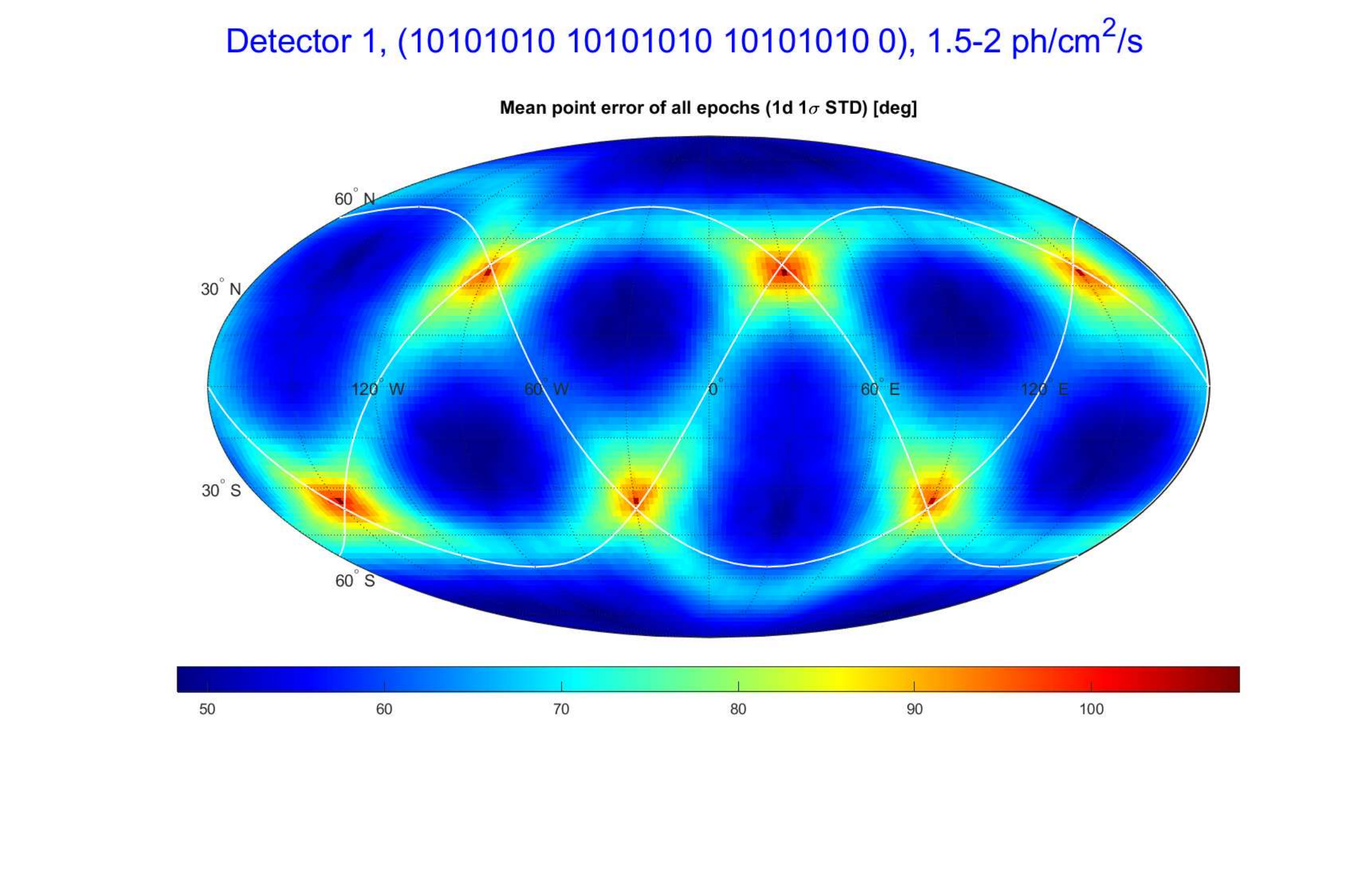}
  \caption[]{Same as Fig. \ref{24sat_Det1} but for 12 satellites (every second along each orbital plane).
    \label{12sat_Det1}}
\end{figure*}

Fig. \ref{12sat_Det1_skycov} shows the sky coverage for an instantaneous
moment (left) and averaged over the orbit (right) for the 24- and 12-satellite
options, and Figs. \ref{24sat_Det1}, \ref{12sat_Det1}
show the localisation accuracy for two different GRB flux intervals.

\begin{figure*}[h]
  \centering
  \hspace{-0.1cm}
  \includegraphics[width=0.28\textwidth,height=0.265\textwidth]{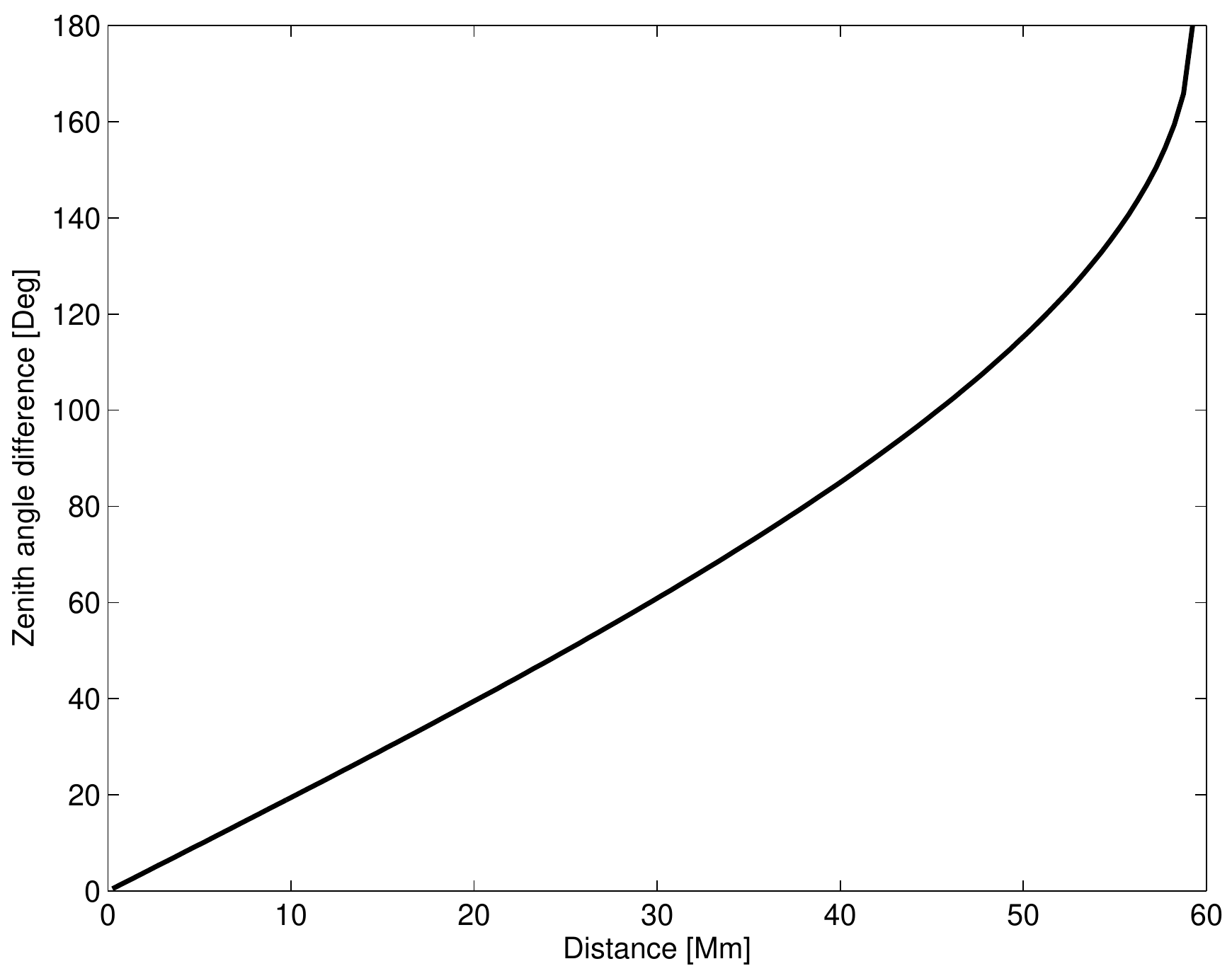}\hspace{0.1cm}
  \includegraphics[width=0.34\textwidth, viewport=0 0 390 310, clip]{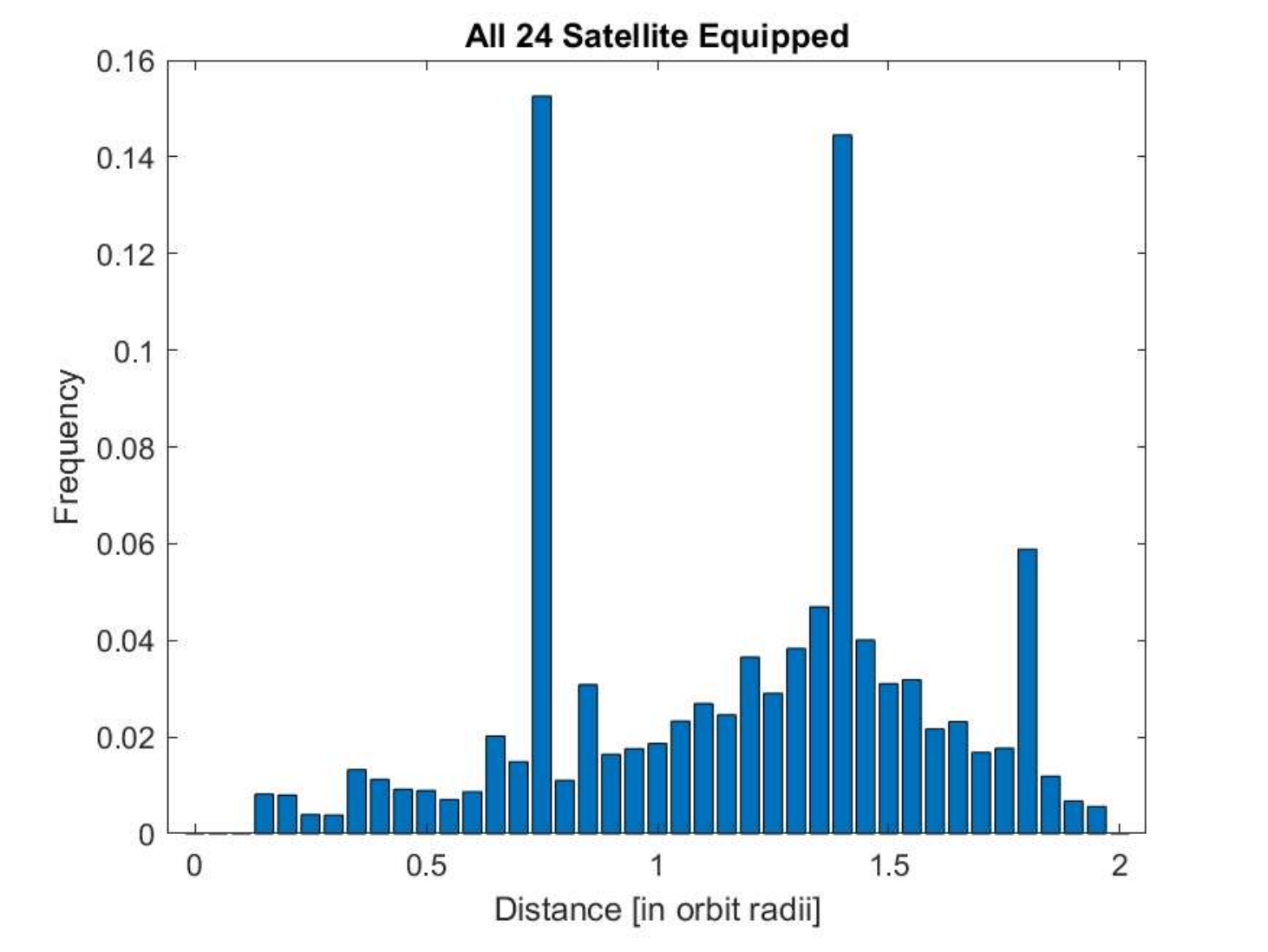}
  \includegraphics[width=0.34\textwidth, viewport=0 0 390 310, clip]{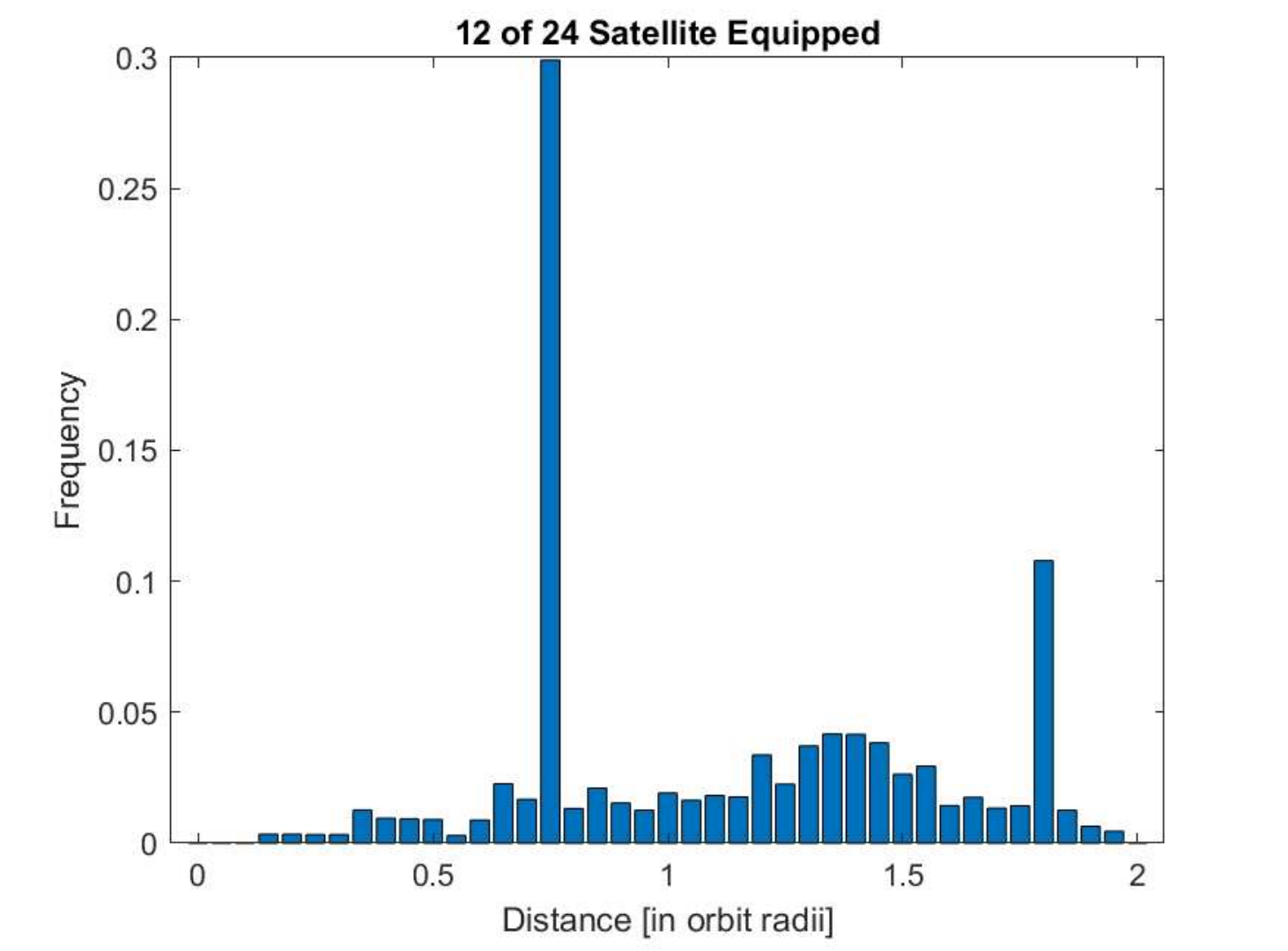}
   \caption[]{{\bf Left:} Relation between distance between Galileo satellites
     and zenith pointing difference.
    {\bf Middle:} Frequency of occurrence of satellite distances, averaged
    over one orbital cycle for an isotropic GRB distribution, if all
    Galileo satellites are equipped with a GRB detector.
    {\bf Right:} Same, but for every second Galileo satellite
    equipped with a GRB detector.
    \label{1D-statistics}}
\end{figure*}

For the simple 1D detector plate facing to zenith,
the geometry of the satellite kinematics leads to a
relation between the distance of Galileo satellites and the difference
of their zenith pointing direction, as shown in the left panel of
Fig. \ref{1D-statistics}: 
The larger the angle, the lower the area in
the sky that the GRB detectors on the two satellites jointly observe.
More importantly, since the sensitivity of triangulation is best
for GRBs occurring perpendicular to the connecting line of two
satellites, zenith-looking flat detectors
will not make use of the maximum baseline of the Galileo satellite system,
but use at most 2/3 of it ($<$1.3 orbital radii).
Our simulations over a full orbital cycle now return the frequency
of occurrence of these distances between pairs of Galileo satellites
for an isotropic distribution of GRBs. This shows that for the maximum
GRB-detector equipment rate on the GNSS, i.e.
a GRB detector on each of the 24 Galileo satellites, about half of the
detector pairs occur at satellite separations of $<$1.2 orbital radii
(middle panel of Fig. \ref{1D-statistics}). When reducing
the satellite equipment rate, this rate gets even worse (right panel of
Fig. \ref{1D-statistics}). Thus, a single zenith-facing detector per
satellite is far from optimal.

\paragraph{Overview of all detector geometries}

\begin{figure*}[!th]
  \centering
  \vspace{-0.1cm}
  \includegraphics[width=0.234\textwidth, viewport=80 90 780 515, clip]{24det_Detector01_11111111_11111111_11111111_SkyCov_first.pdf}
  \includegraphics[width=0.234\textwidth, viewport=80 90 780 515, clip]{24det_Detector01_11111111_11111111_11111111_SkyCov_mean.pdf}
  \includegraphics[width=0.234\textwidth, viewport=80 90 780 515, clip]{24det_Detector01_11111111_11111111_11111111_Bin1_mean_sqrt2.pdf} \\
  \includegraphics[width=0.234\textwidth, viewport=80 90 780 515, clip]{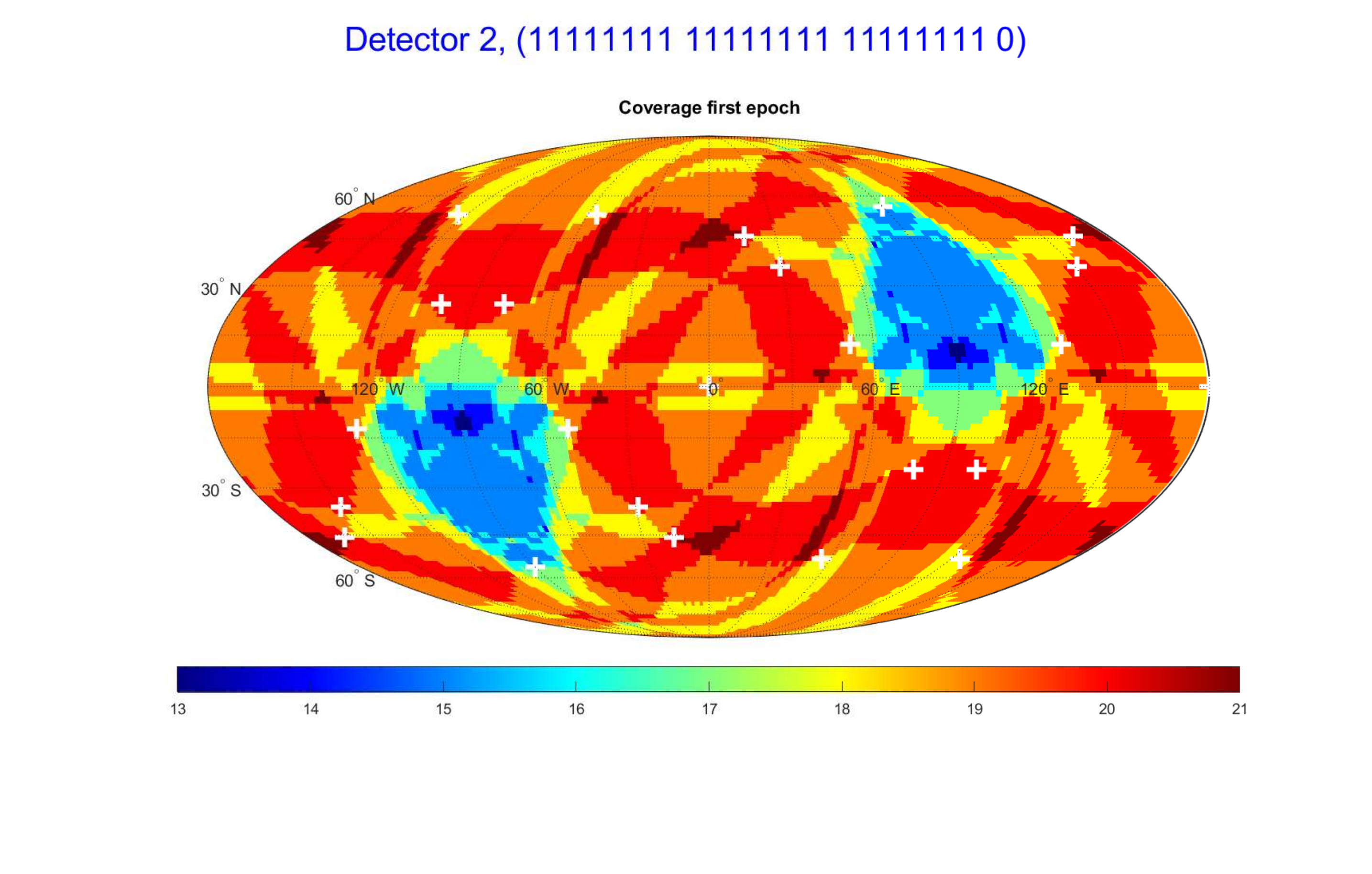}
  \includegraphics[width=0.234\textwidth, viewport=80 90 780 515, clip]{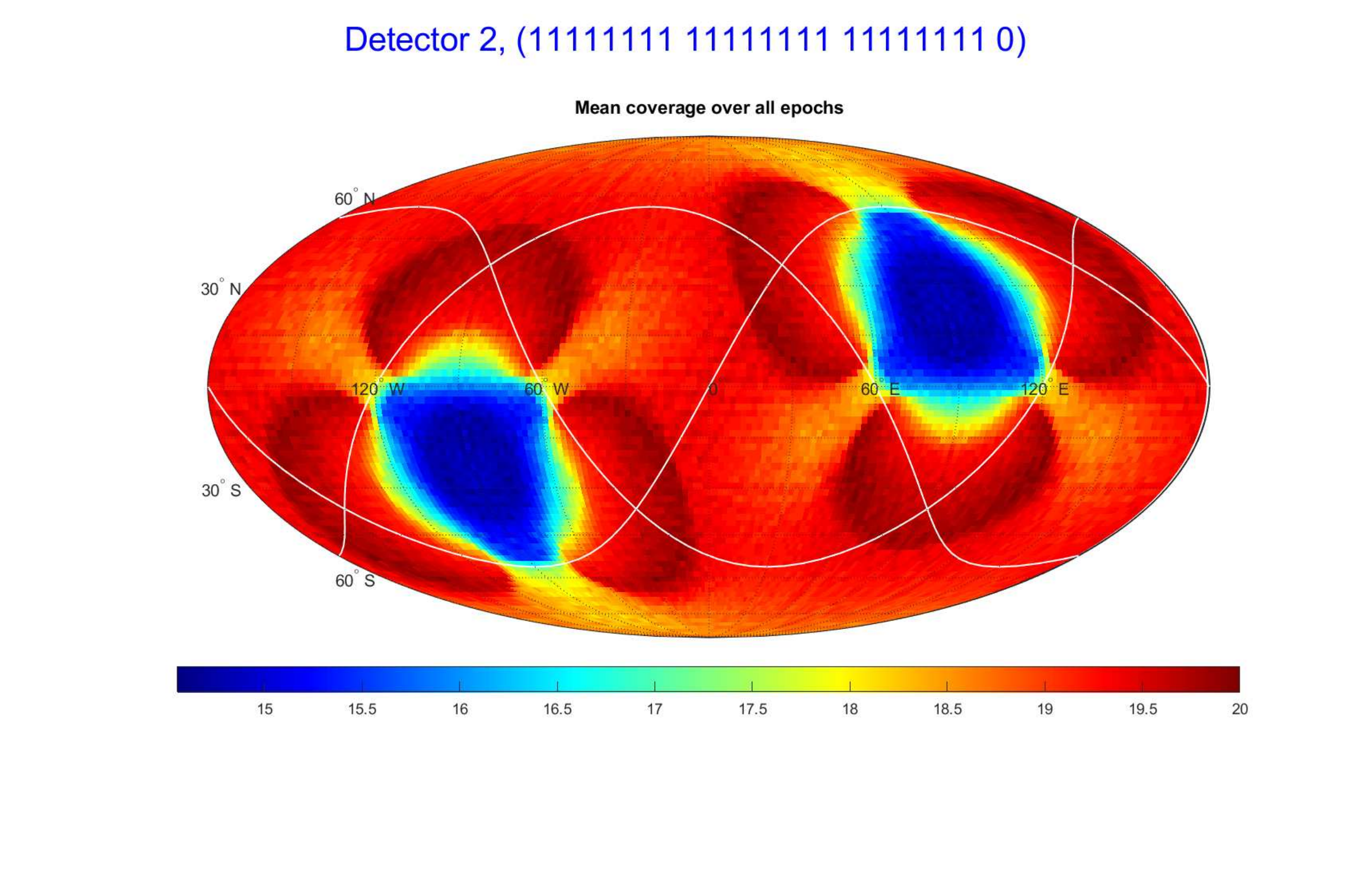}
  \includegraphics[width=0.234\textwidth, viewport=80 90 780 515, clip]{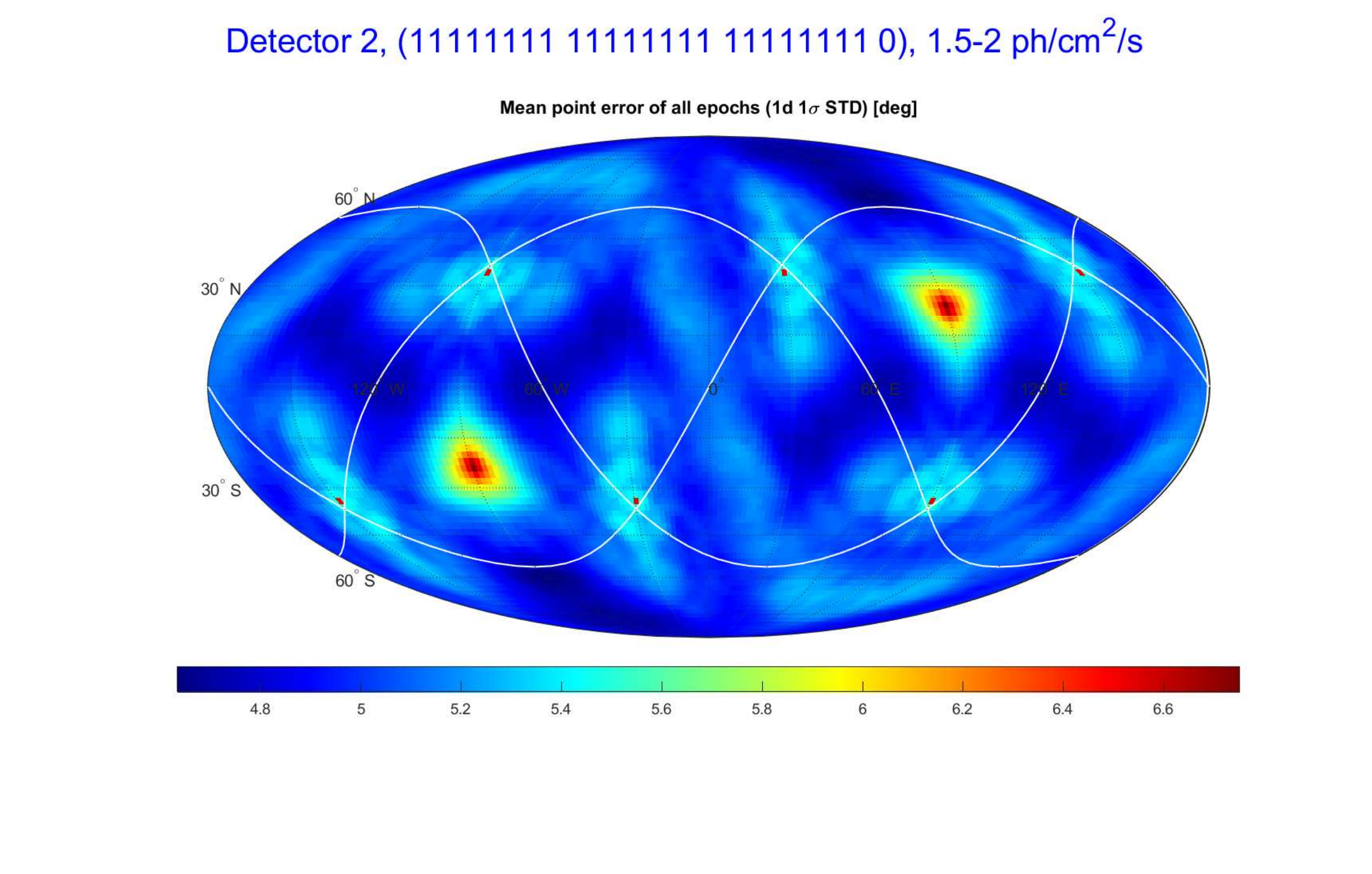} \\
  \includegraphics[width=0.234\textwidth, viewport=80 90 780 515, clip]{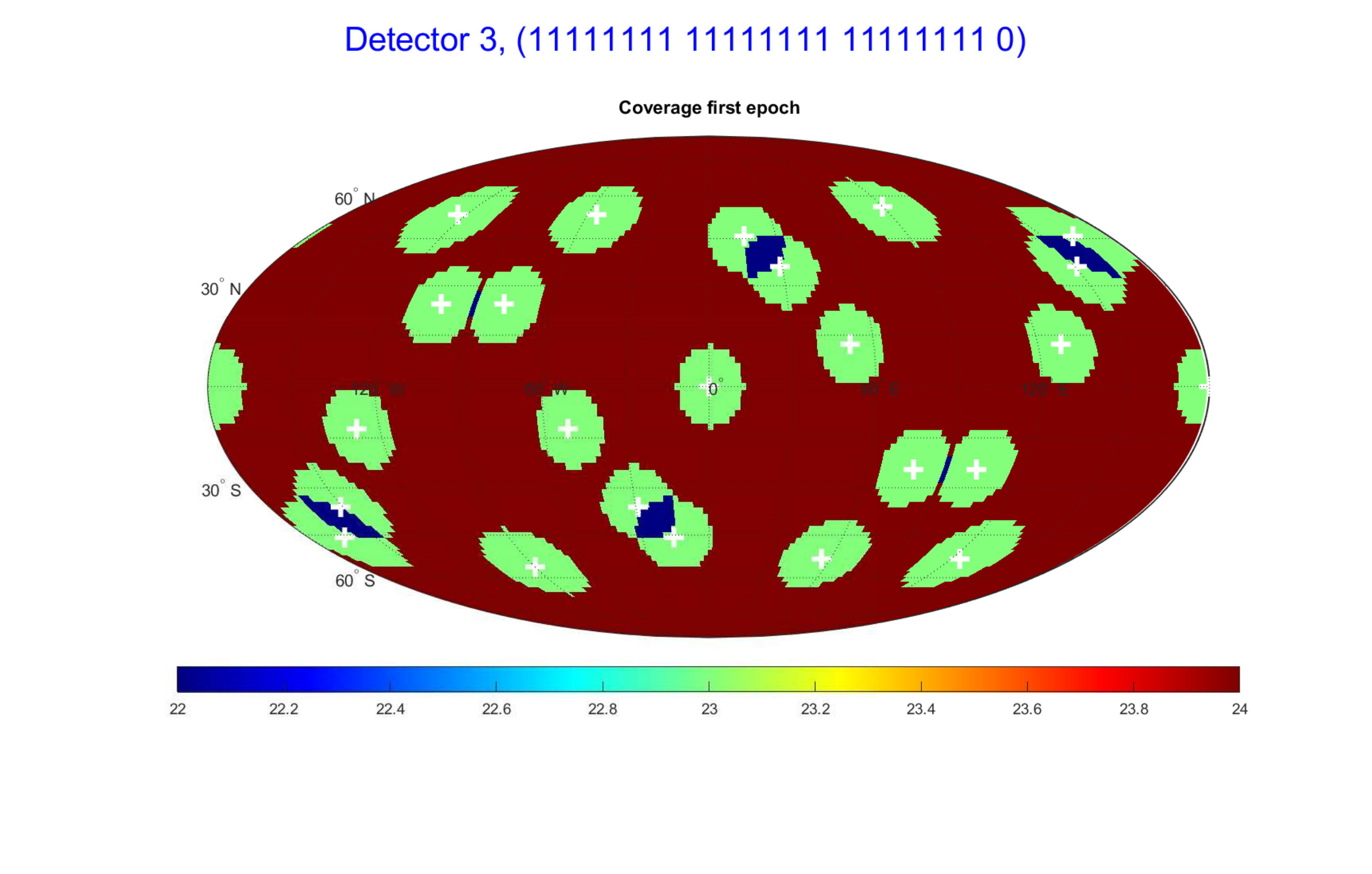}
  \includegraphics[width=0.234\textwidth, viewport=80 90 780 515, clip]{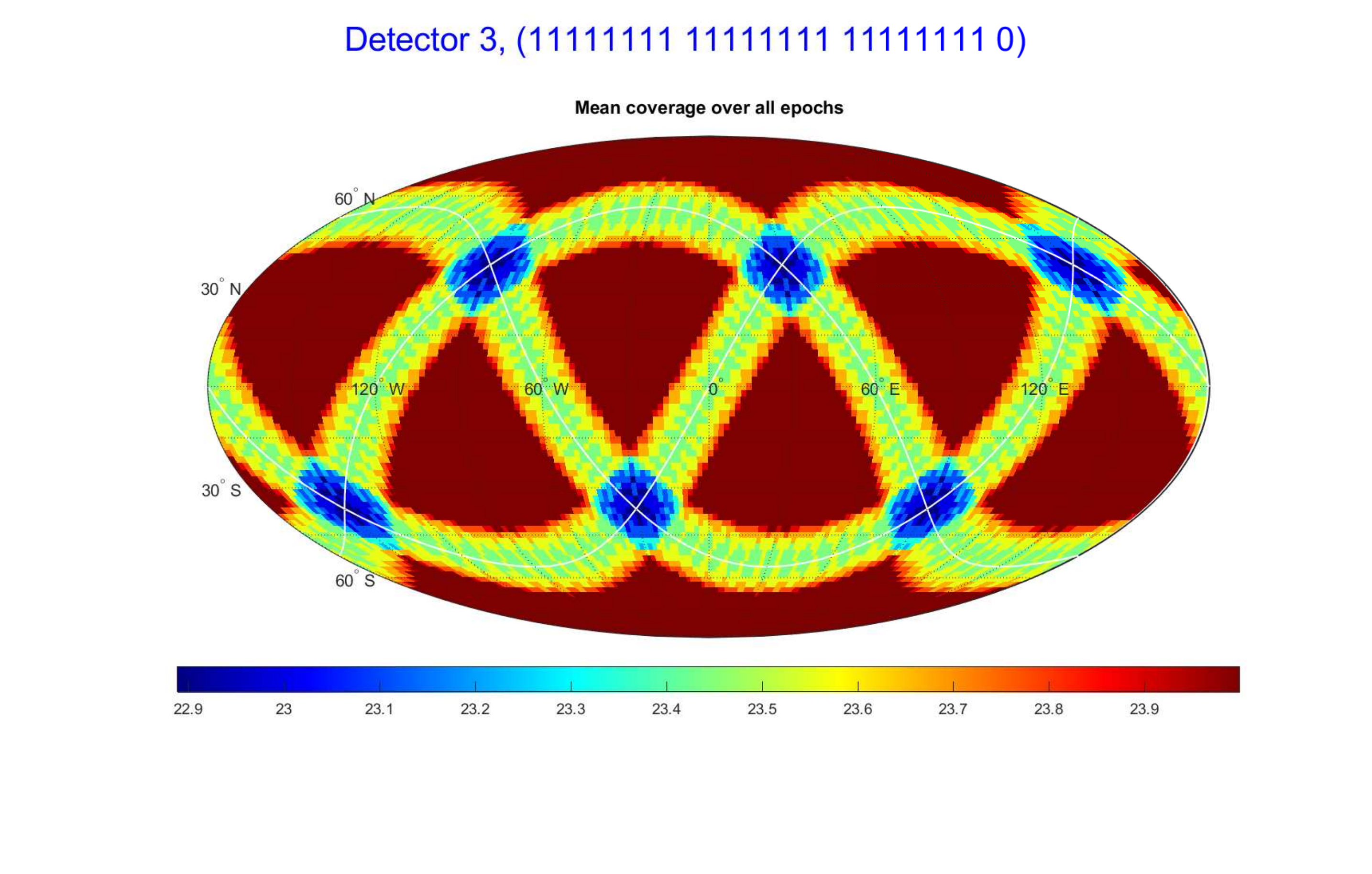}
  \includegraphics[width=0.234\textwidth, viewport=80 90 780 515, clip]{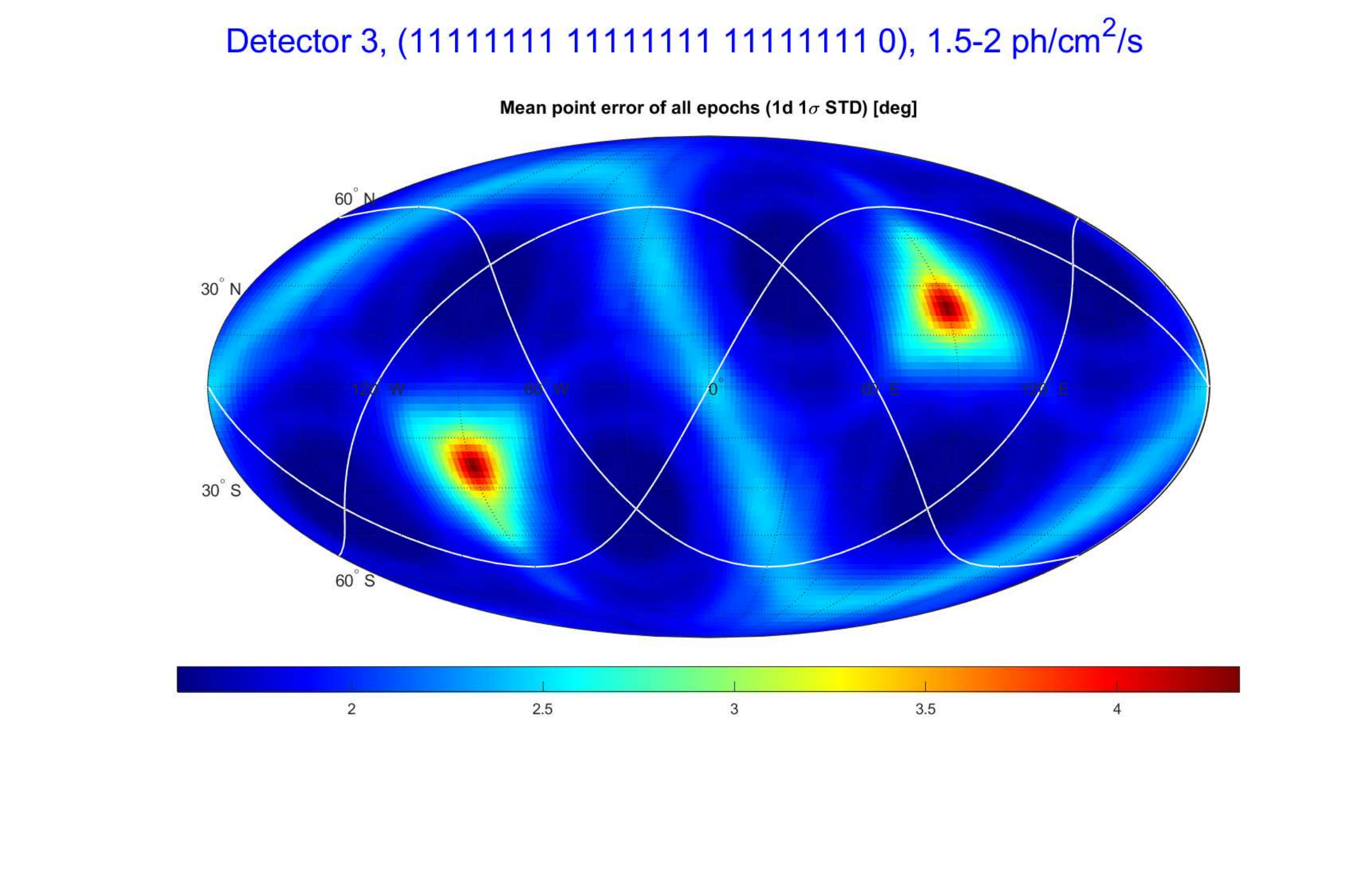} \\
  \includegraphics[width=0.234\textwidth, viewport=80 90 780 515, clip]{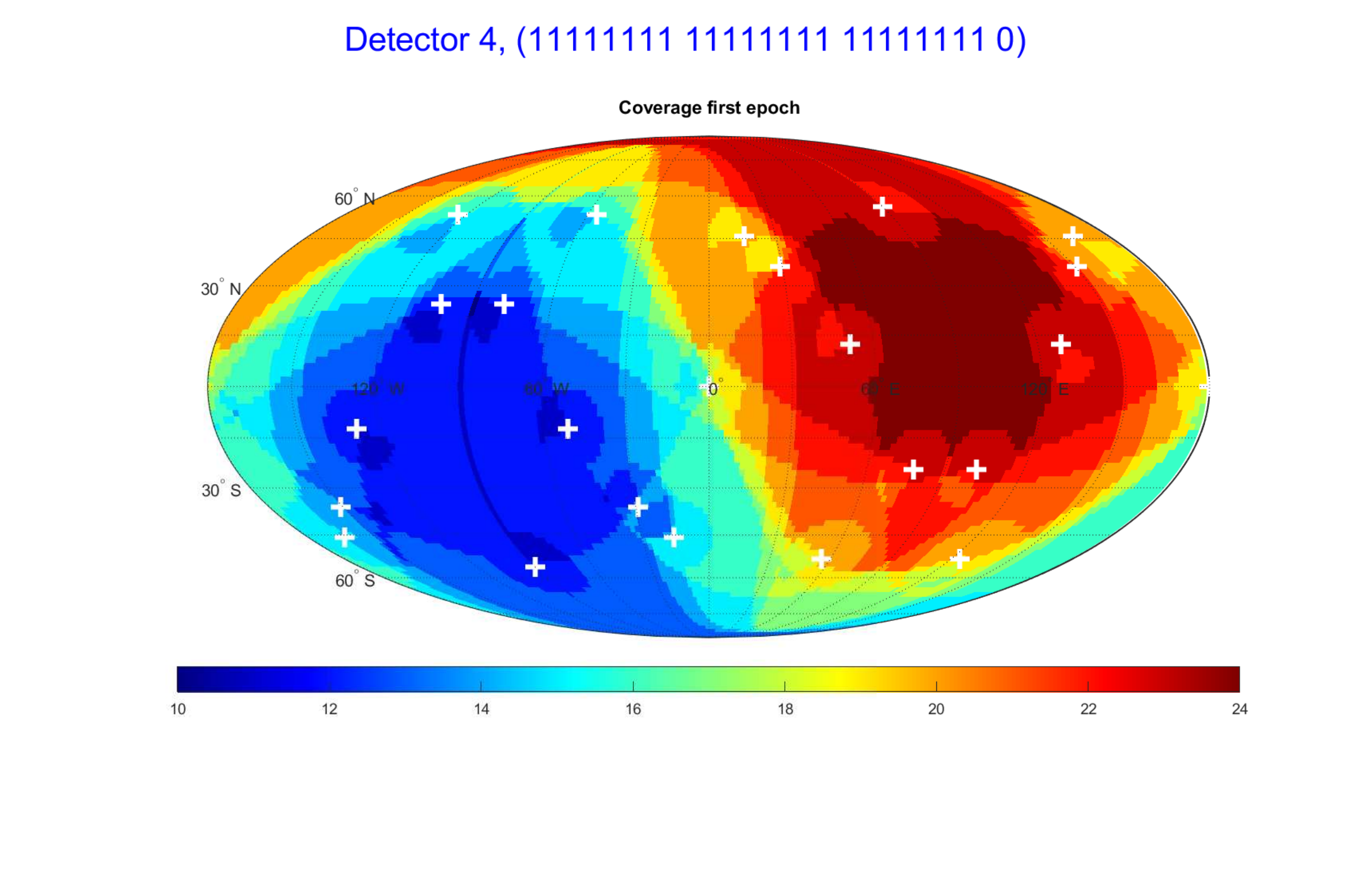}
  \includegraphics[width=0.234\textwidth, viewport=80 90 780 515, clip]{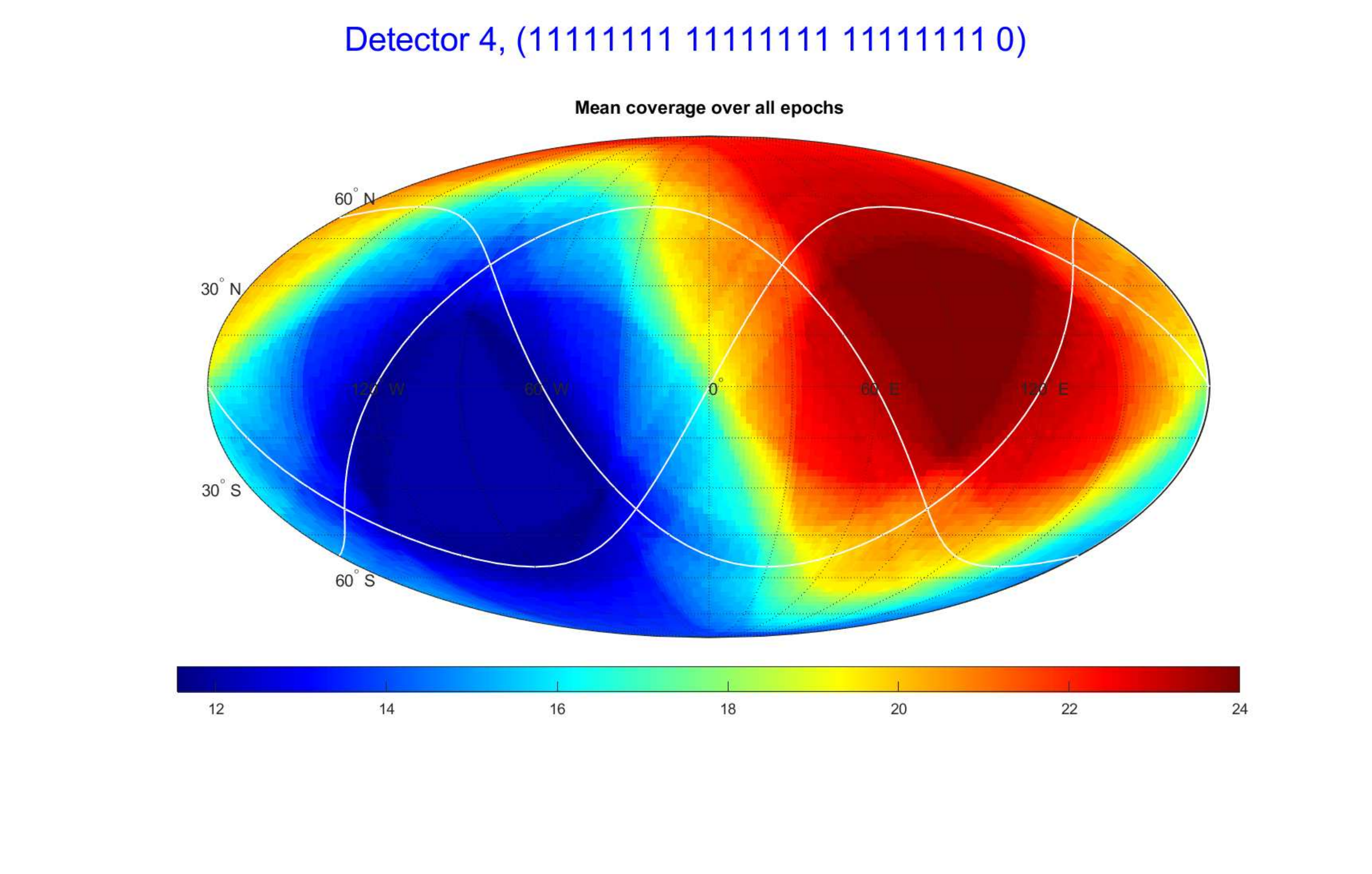}
  \includegraphics[width=0.234\textwidth, viewport=80 90 780 515, clip]{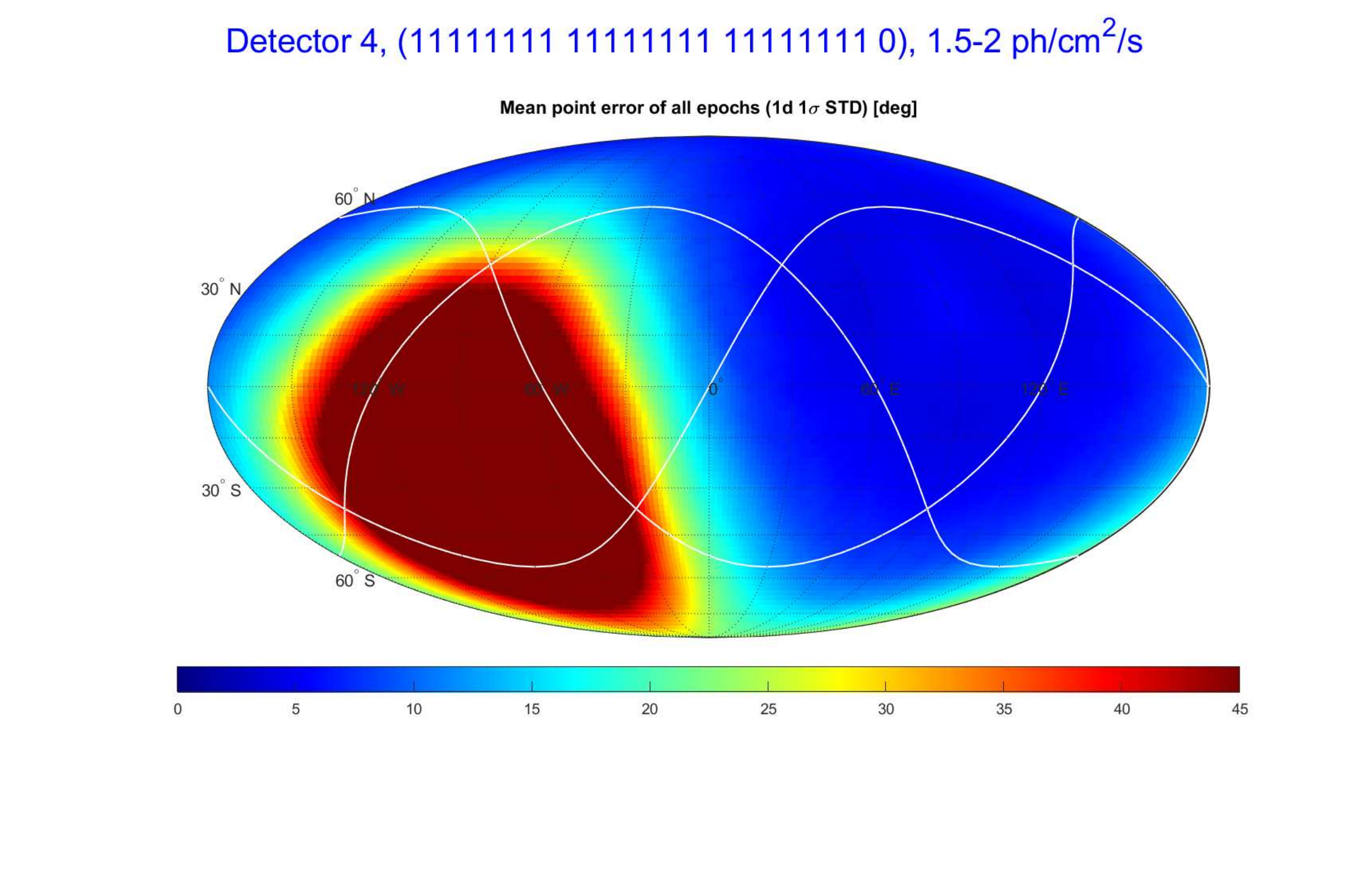} \\
  \includegraphics[width=0.234\textwidth, viewport=80 90 780 515, clip]{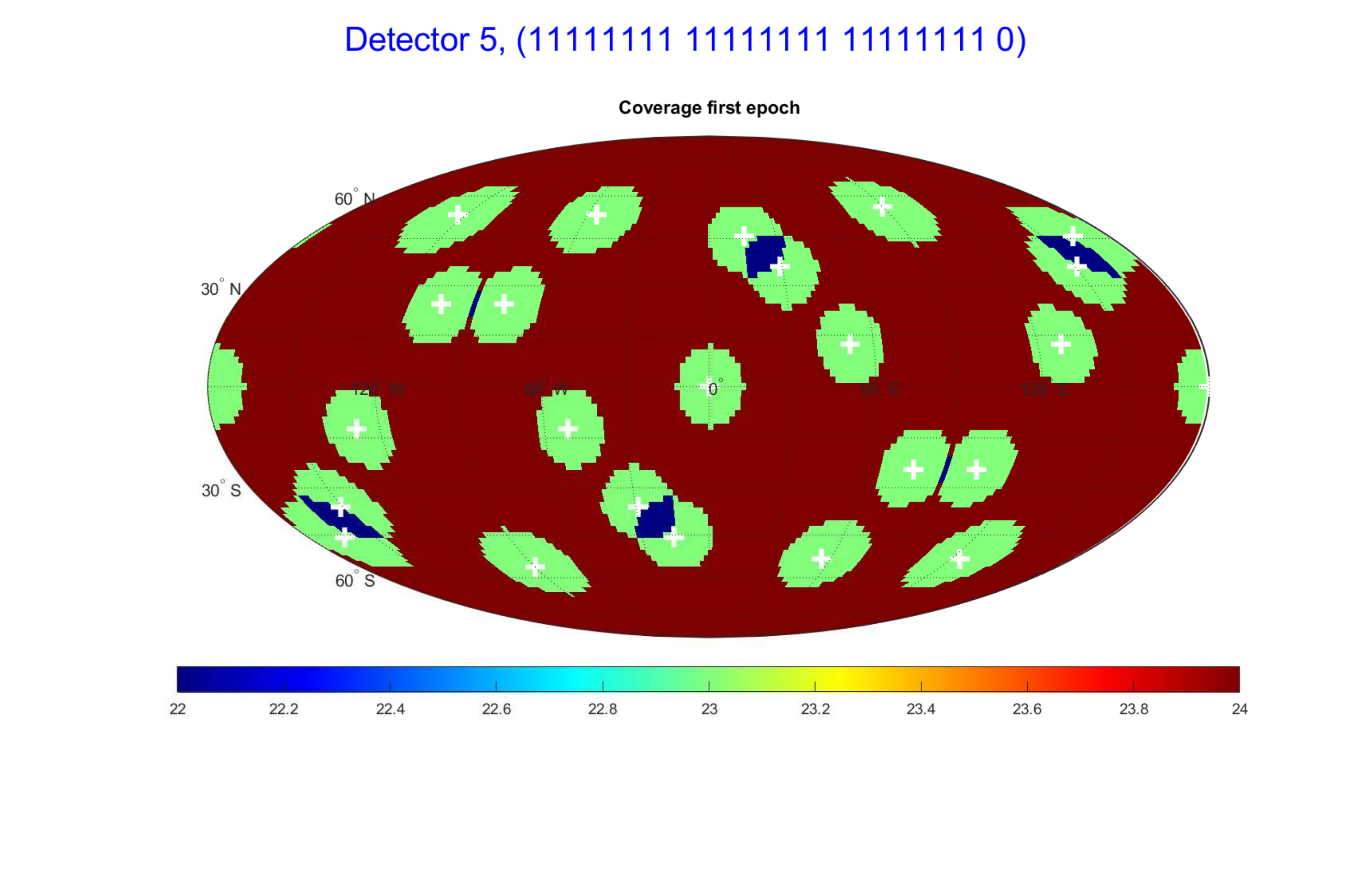}
  \includegraphics[width=0.234\textwidth, viewport=80 90 780 515, clip]{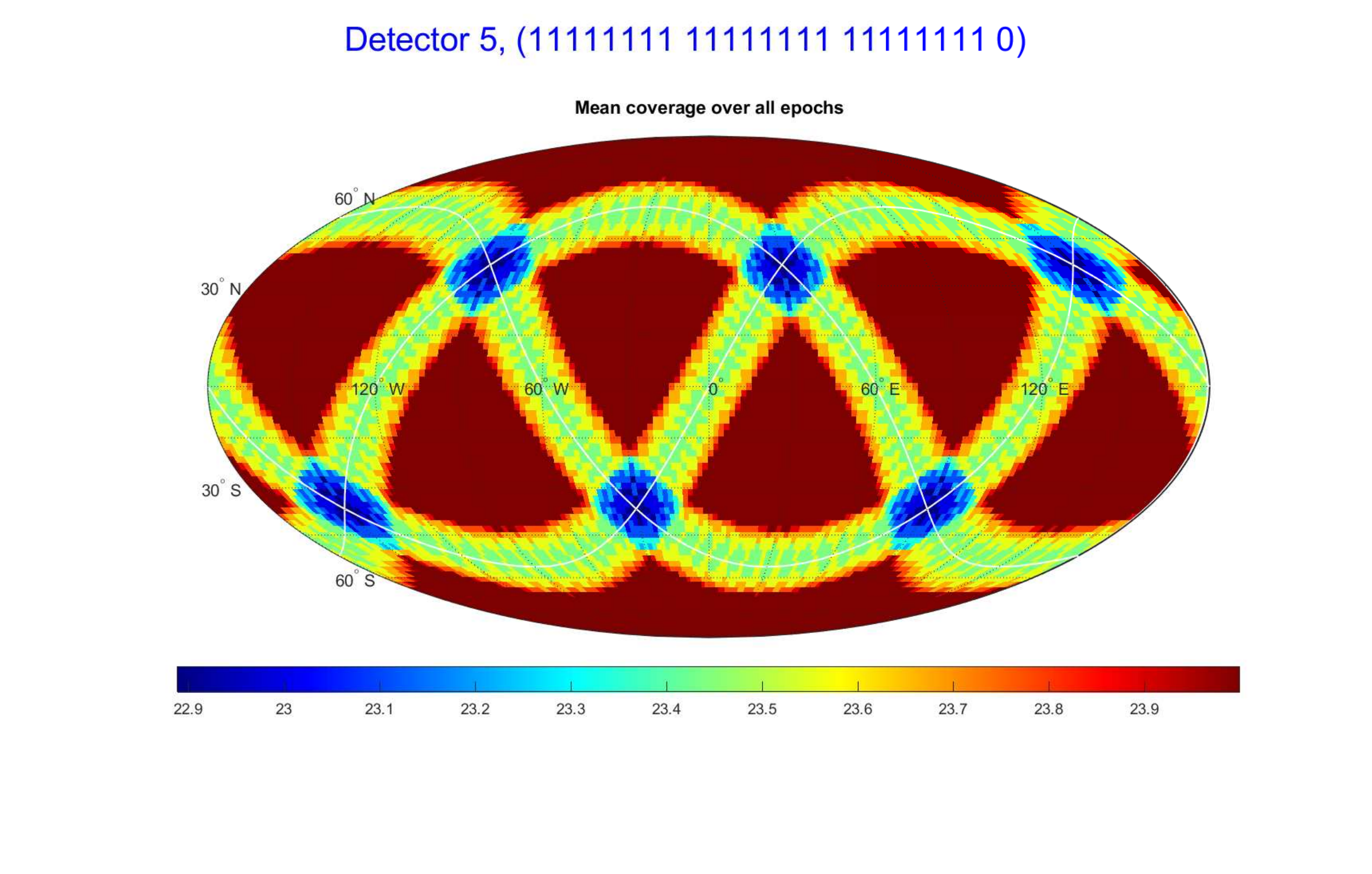}
  \includegraphics[width=0.234\textwidth, viewport=80 90 780 515, clip]{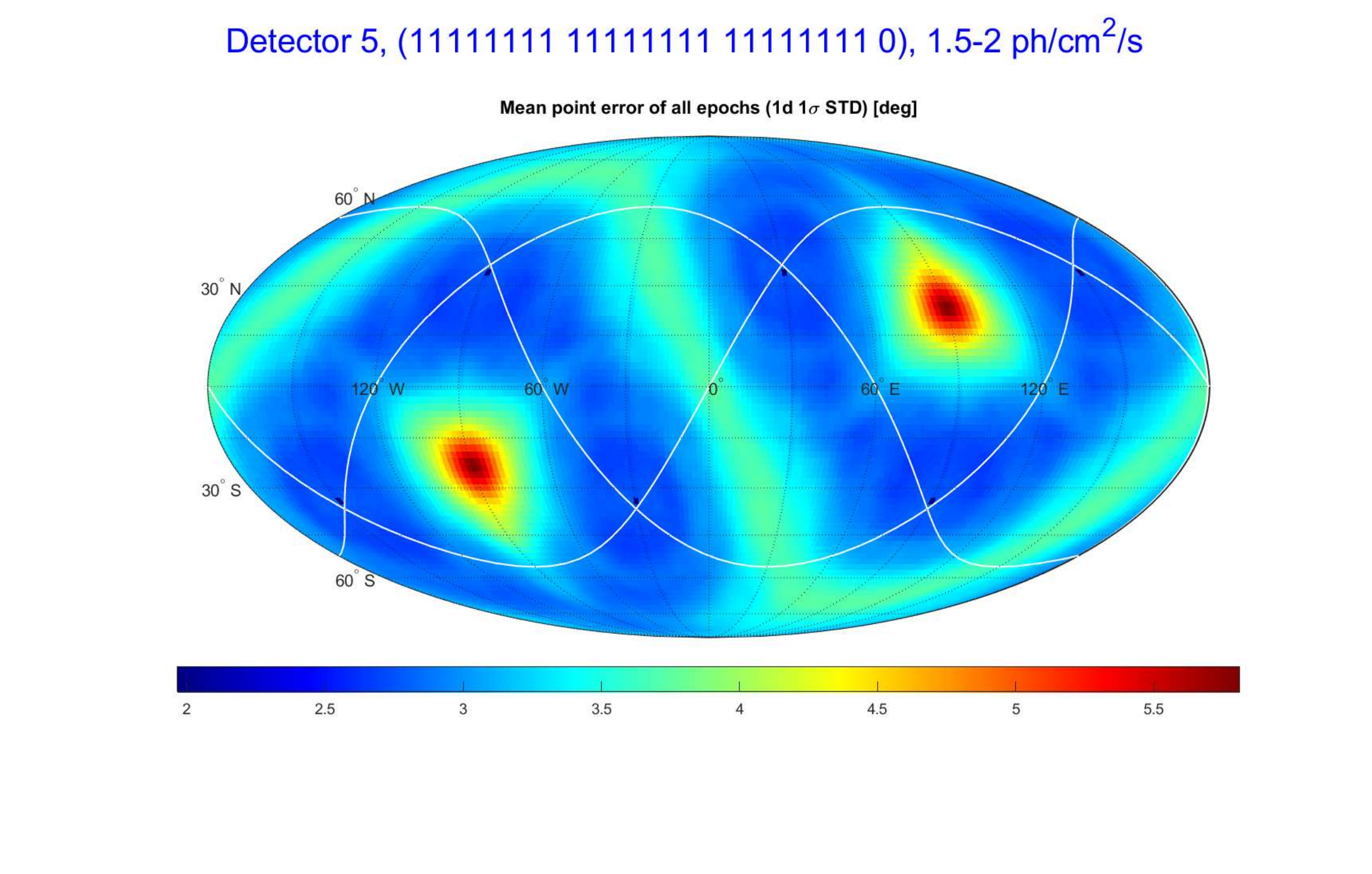} \\
  \includegraphics[width=0.234\textwidth, viewport=80 90 780 515, clip]{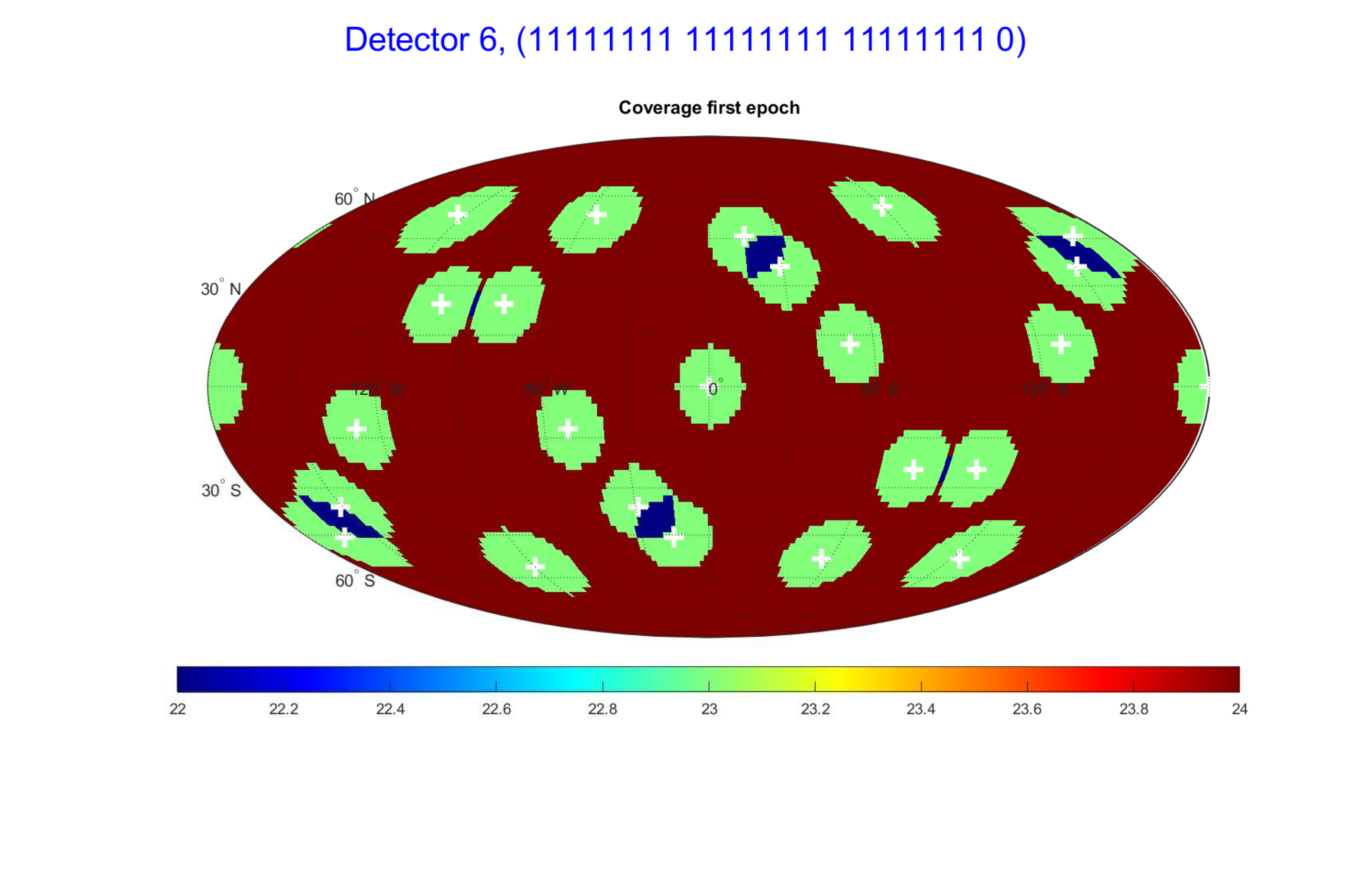}
  \includegraphics[width=0.234\textwidth, viewport=80 90 780 515, clip]{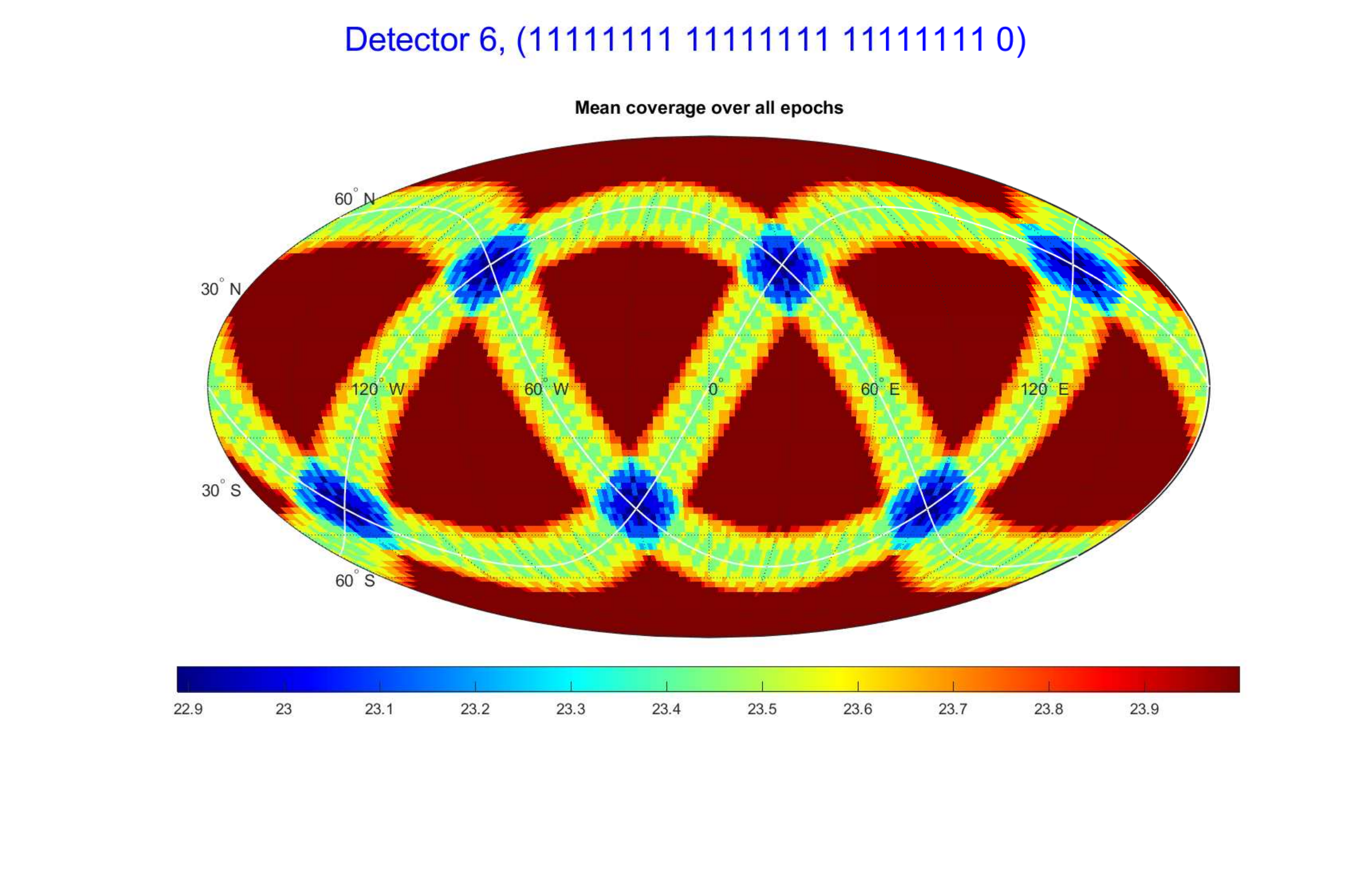}
  \includegraphics[width=0.234\textwidth, viewport=80 90 780 515, clip]{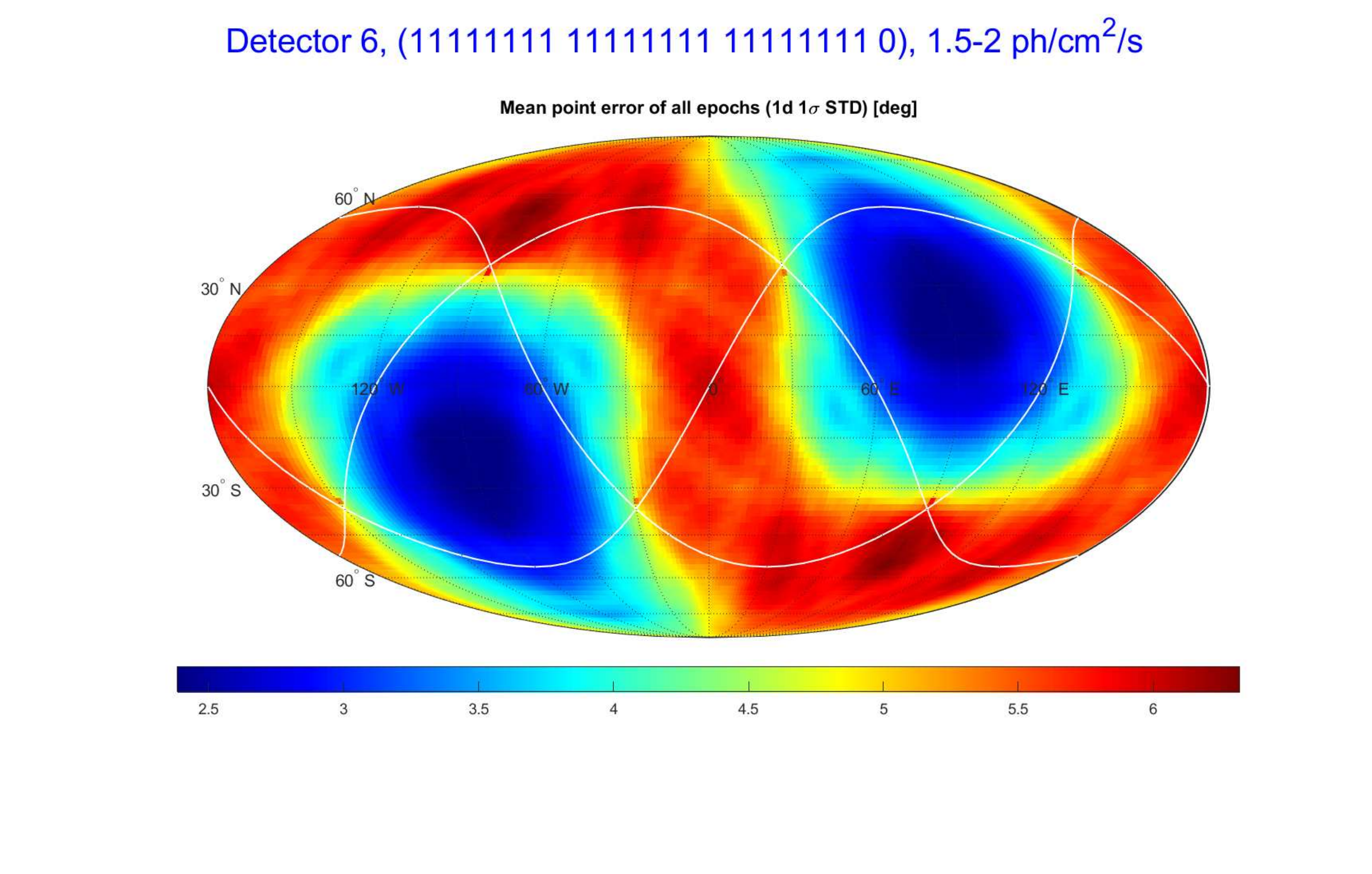} \\
  \includegraphics[width=0.234\textwidth, viewport=80 90 780 515, clip]{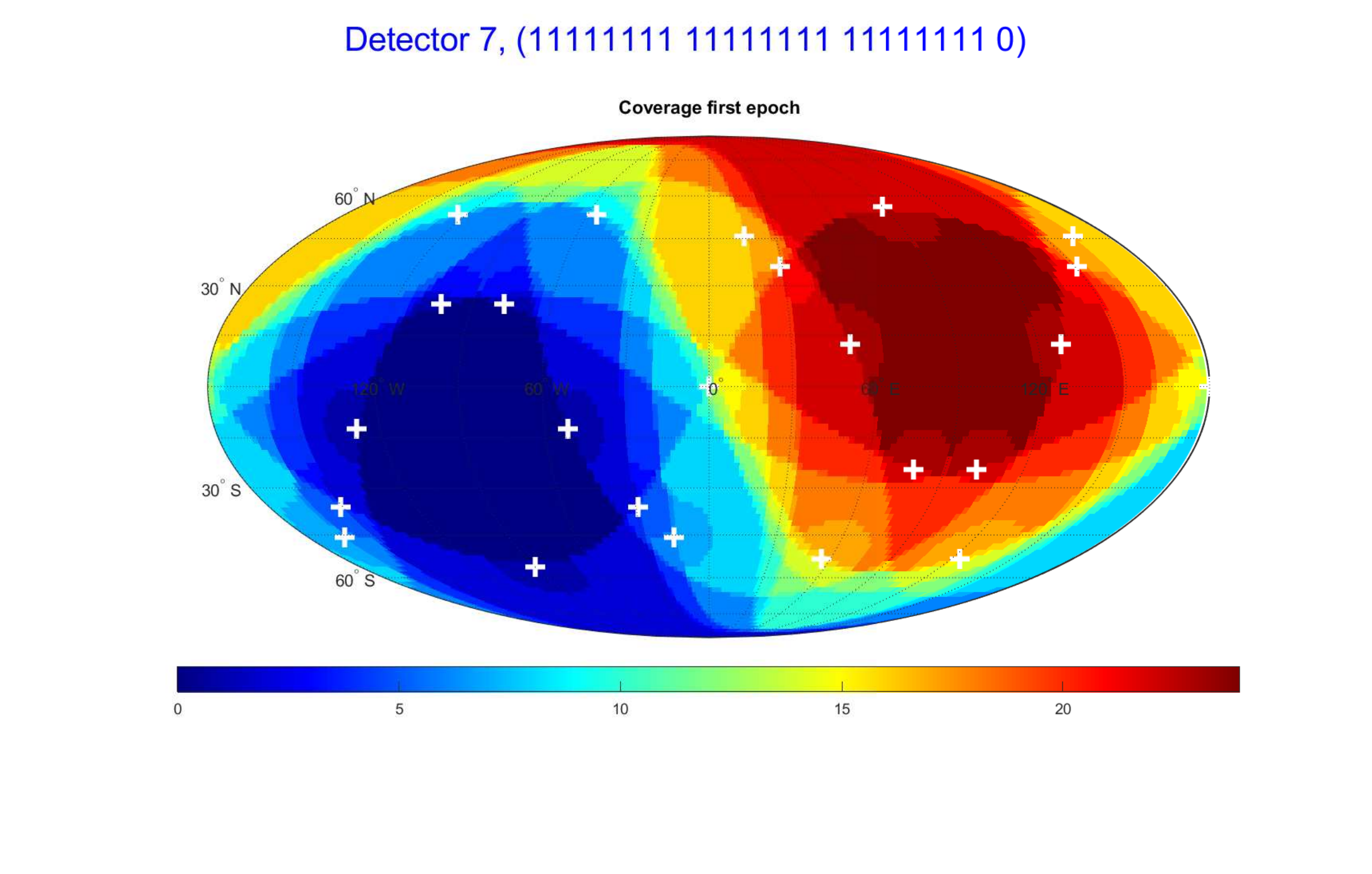}
  \includegraphics[width=0.234\textwidth, viewport=80 90 780 515, clip]{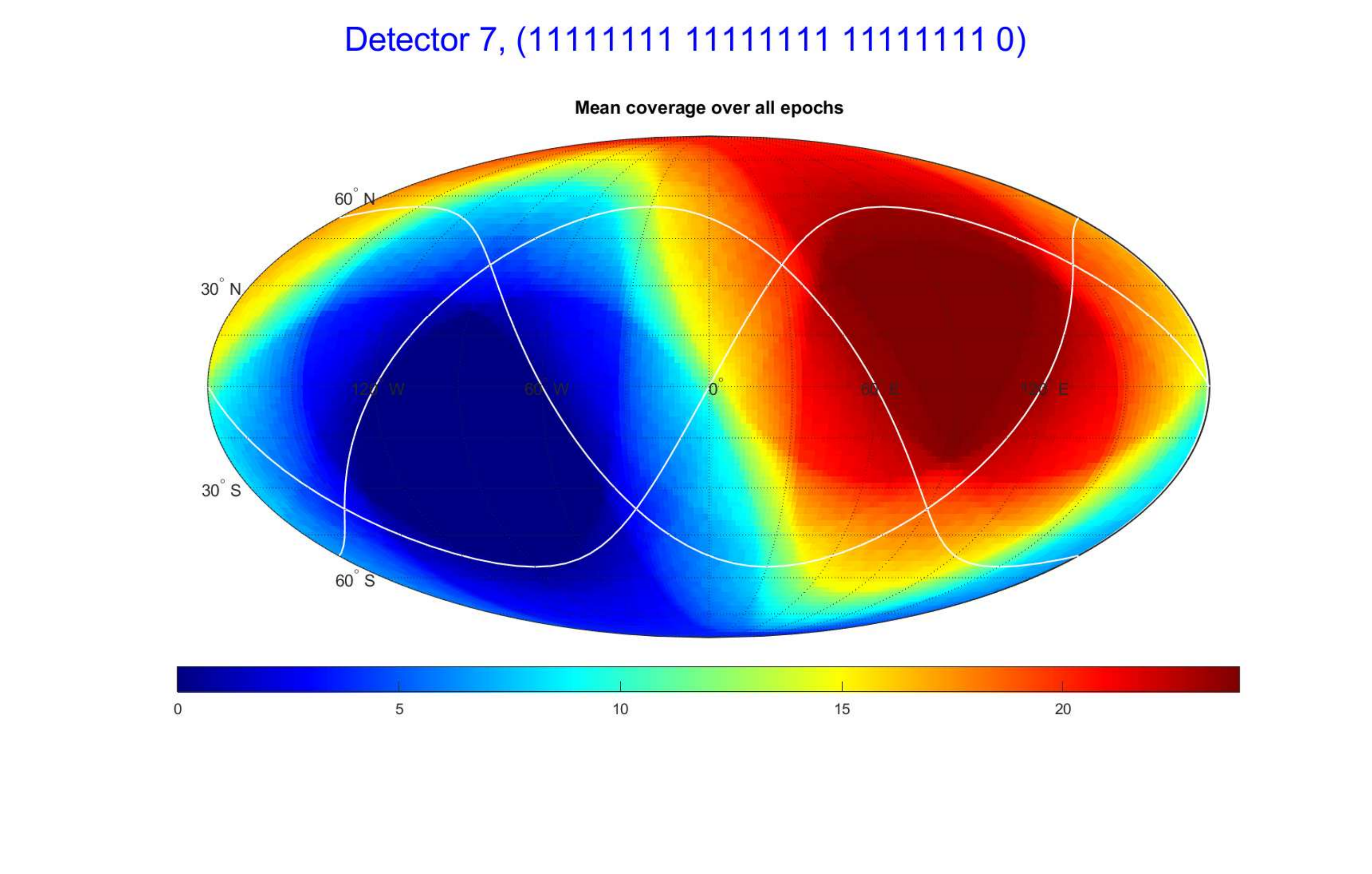}
  \includegraphics[width=0.234\textwidth, viewport=80 90 780 515, clip]{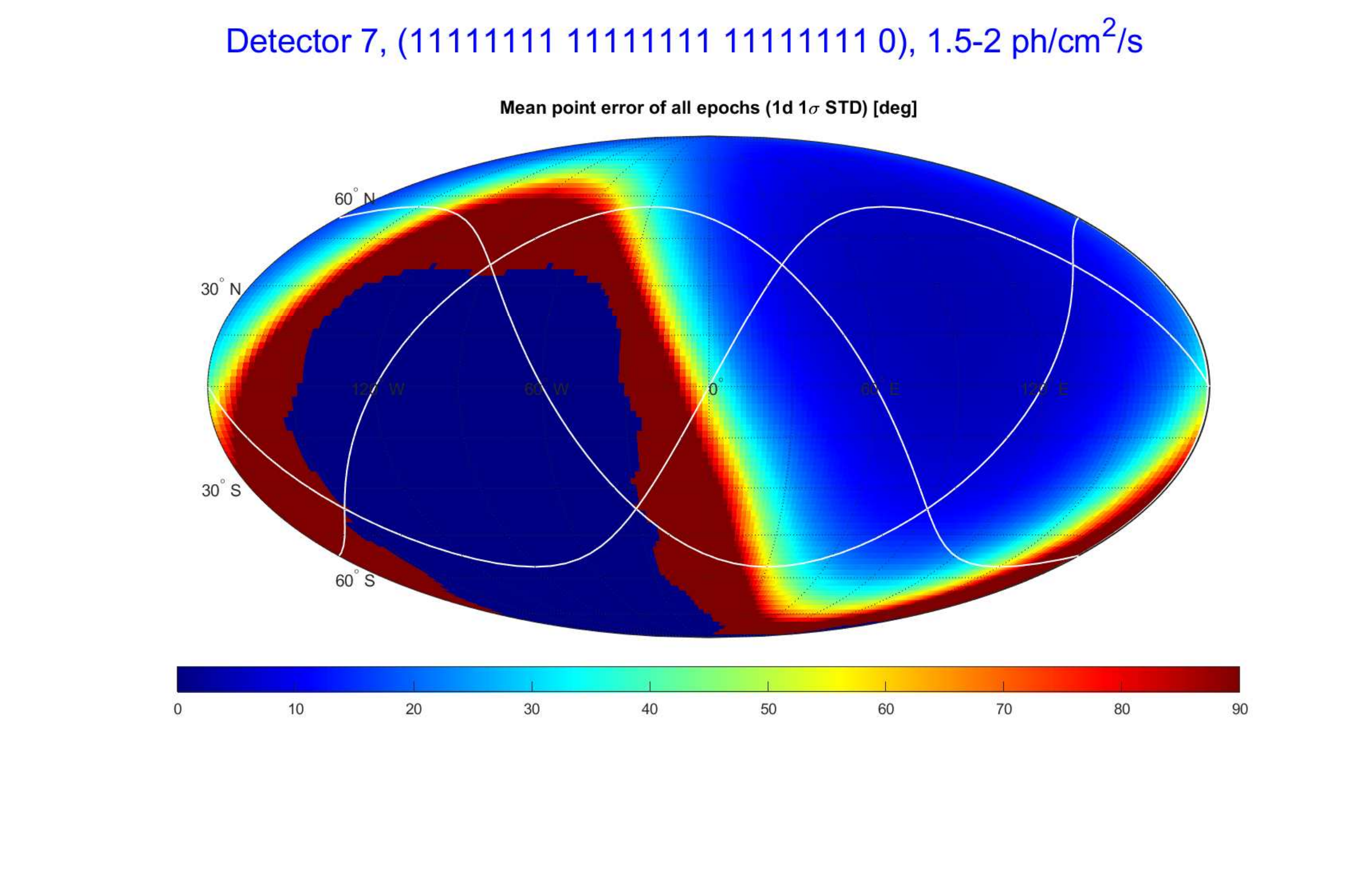} \\
  \includegraphics[width=0.234\textwidth, viewport=80 90 780 515, clip]{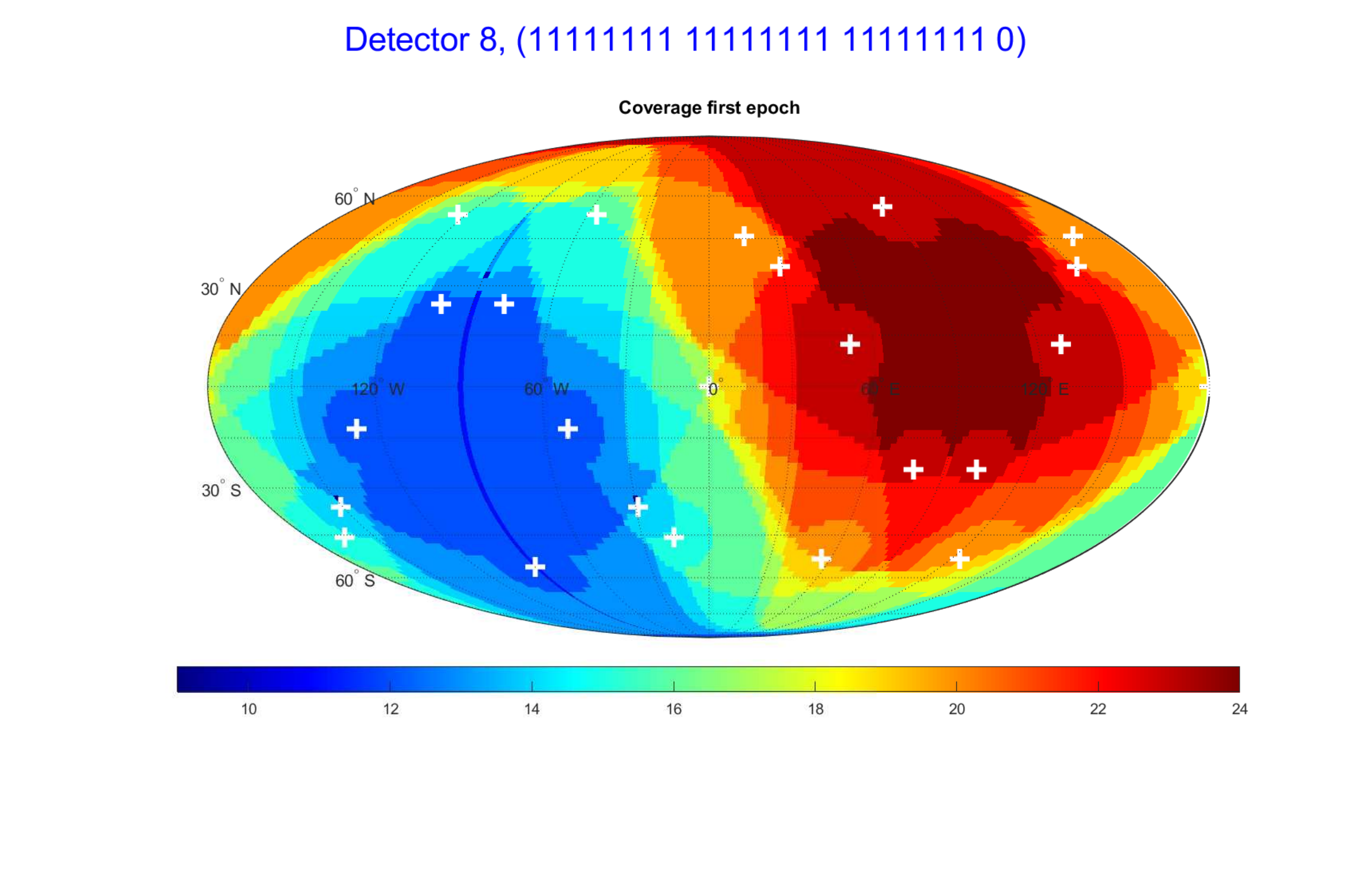}
  \includegraphics[width=0.234\textwidth, viewport=80 90 780 515, clip]{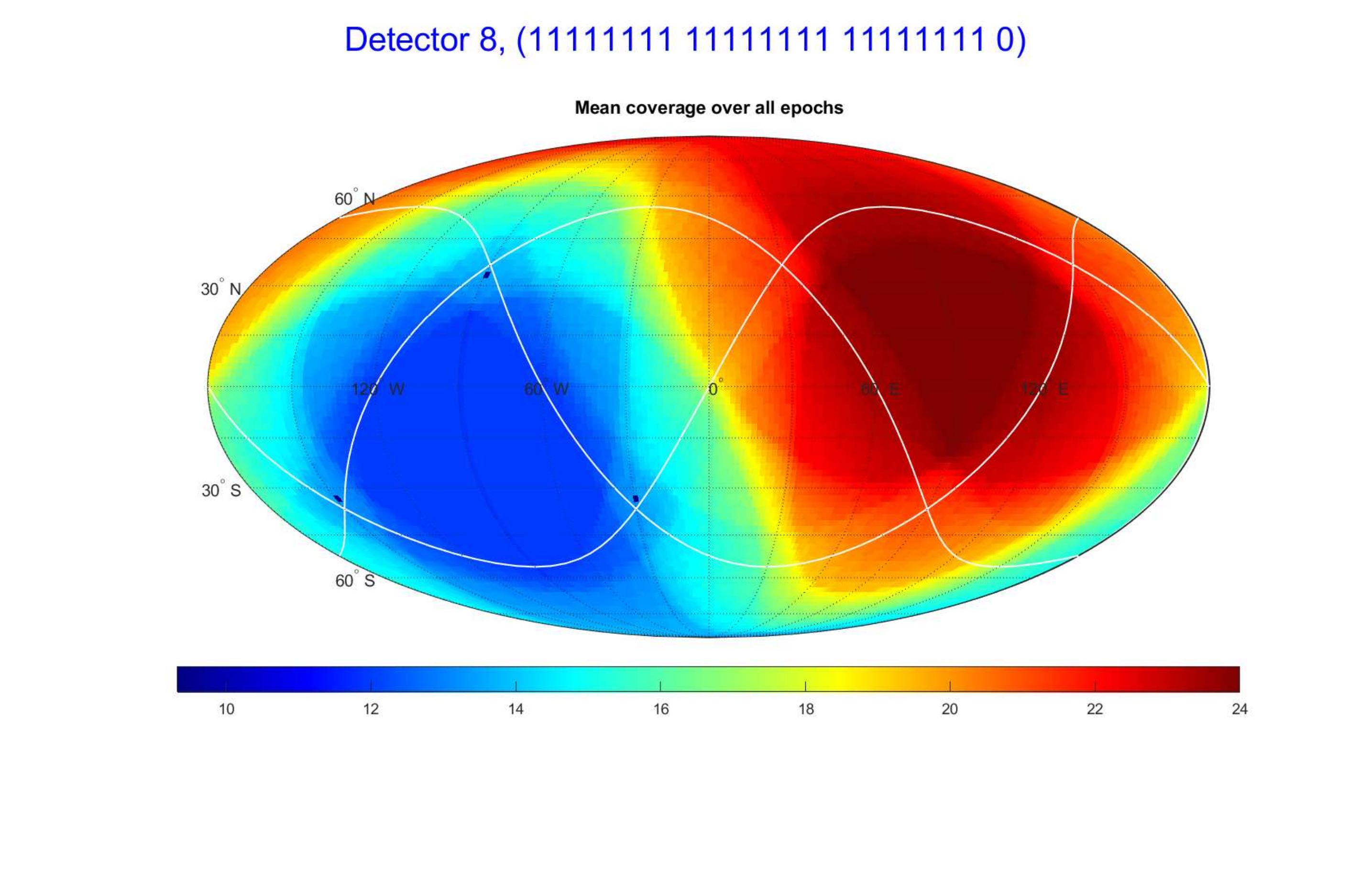}
  \includegraphics[width=0.234\textwidth, viewport=80 90 780 515, clip]{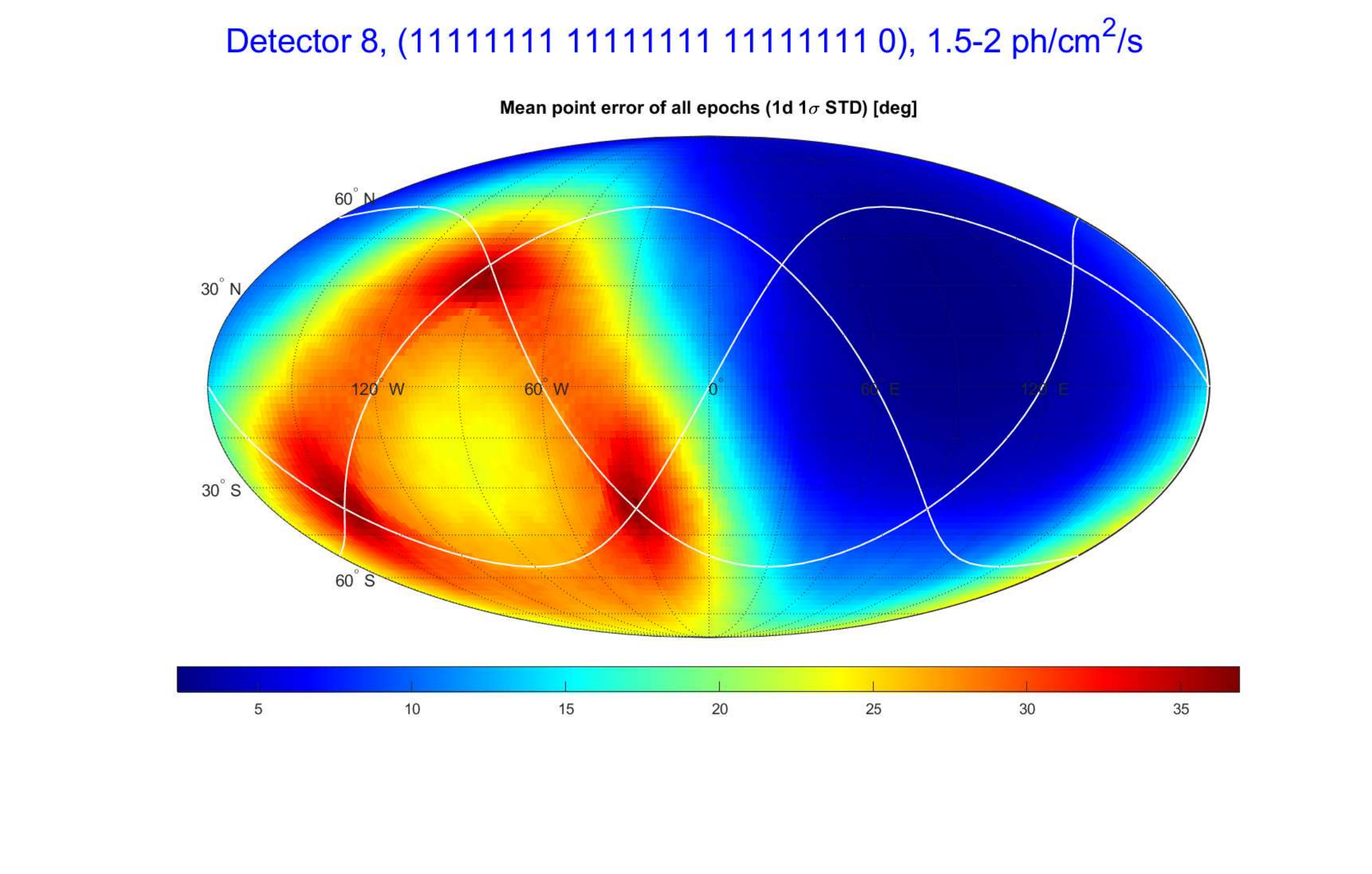} \\
  \includegraphics[width=0.234\textwidth, viewport=80 90 780 515, clip]{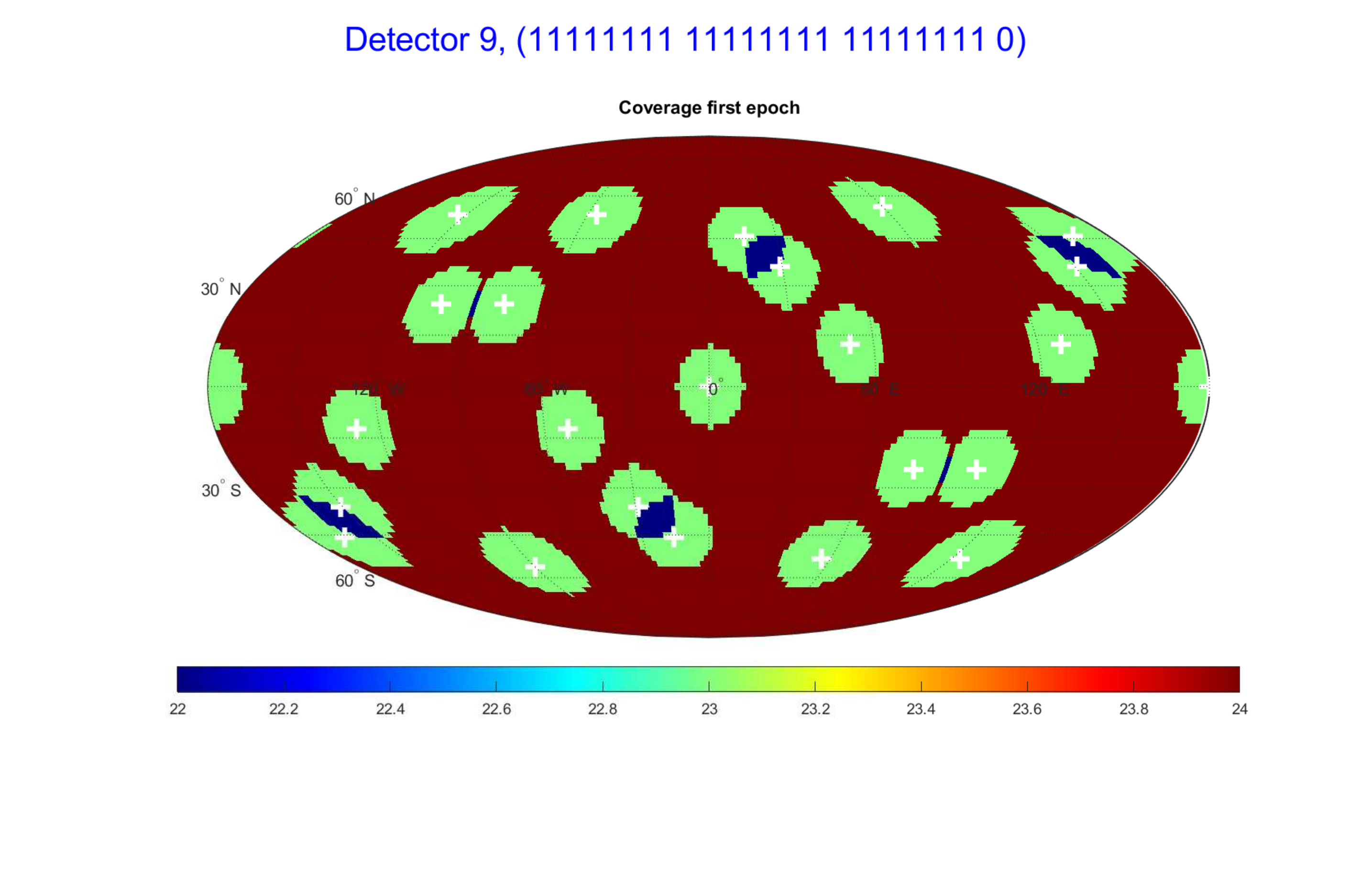}
  \includegraphics[width=0.234\textwidth, viewport=80 90 780 515, clip]{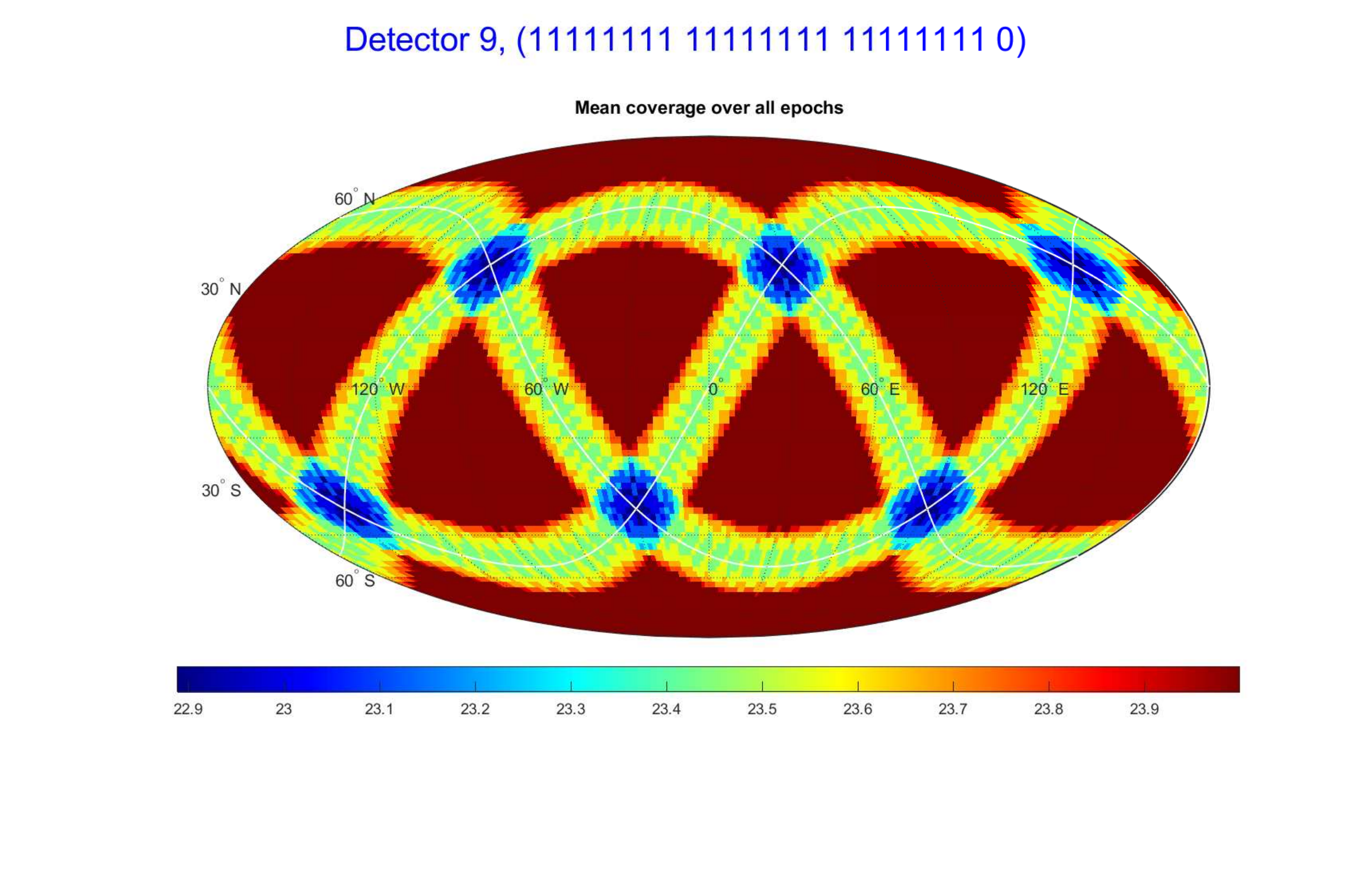}
  \includegraphics[width=0.234\textwidth, viewport=80 90 780 515, clip]{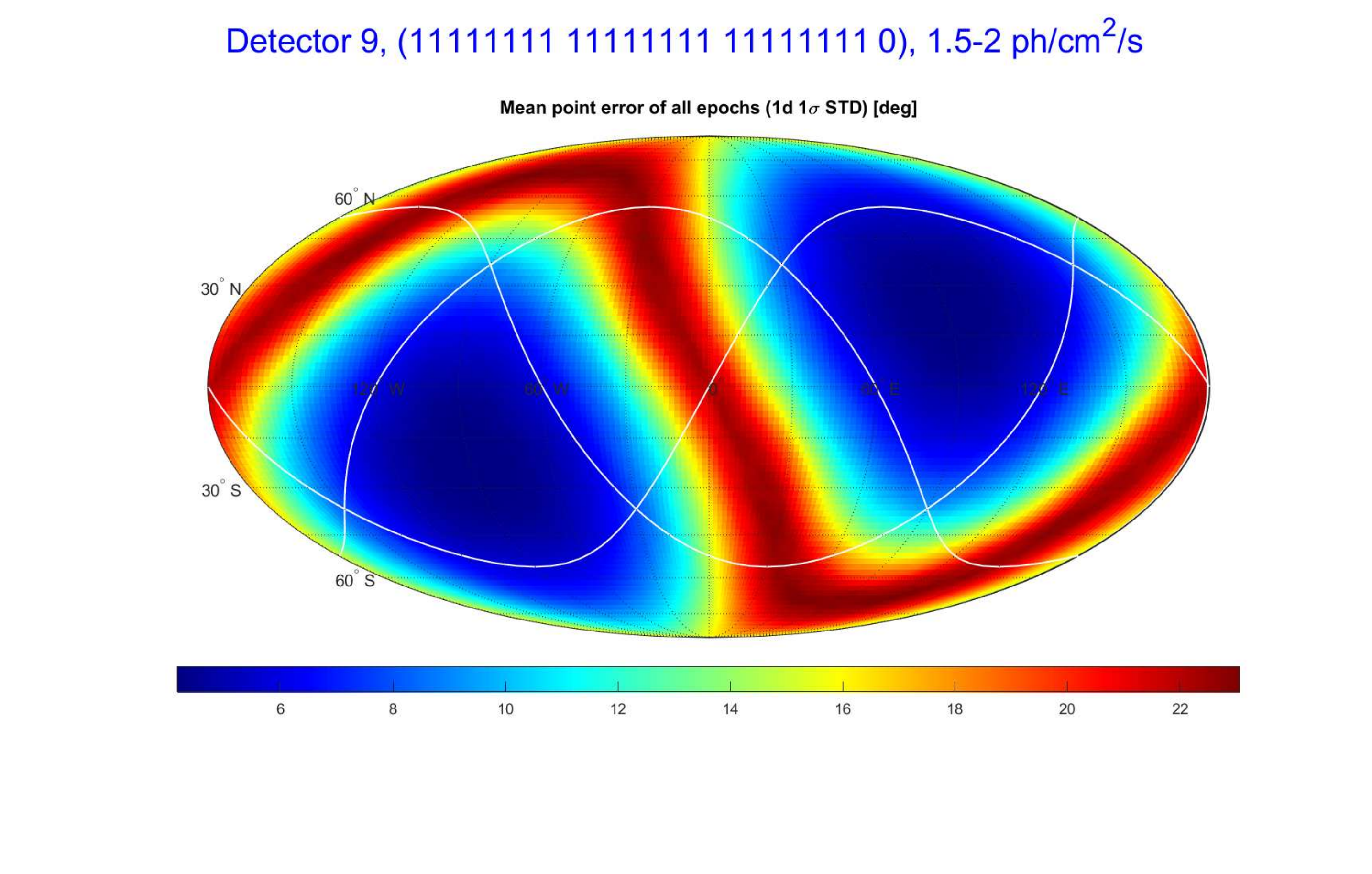}
  \caption[]{Sky coverage for a single time slice (left), averaged over one
    Galileo orbit (middle) and the averaged localisation accuracy 
     for the faintest
    GRB intensity bin (right) for detector geometries 1--9
    (from top to bottom) -- see also blue labels for each map.
    \label{24sat_allDet}}
\end{figure*}

Before elucidating the details of the other detector configurations,
we start with comparing the nine different GRB detector geometries
by using the maximum equipment rate, i.e. all 24 Galileo satellites
carry a GRB detector, in Fig. \ref{24sat_allDet}: 
the sky coverage for any given moment (left column),
the average of the sky coverage over one Galileo orbital period (middle),
and the mean localisation accuracy of the faintest GRB intensity bin (which
is the one where our goal is to obtain a sub-degree localisation).

One interesting pattern (Fig. \ref{24sat_allDet})
are the green filled circles on brown sky background in the left
column for detectors \#03, \#05, \#06, and \#09. These detectors 
all cover the whole sky (ignoring the cosine
dependence of the effective area), see right column of Fig. \ref{3Deffarea}.
The green circles reduce the coverage by one, namely due to the
shadowing of the Earth in nadir-direction, with a 12\degs\ radius.
Due to the three orbital planes being perpendicular to each other,
there are six positions in the sky where two satellites from two
orbital planes get close to each other, and their 12\degs\ radii overlap
to form a small region where the coverage is reduced by two satellites
(blue regions).

Another aspect is symmetry:
While a single detector looking towards zenith on all 24 satellites
produces a homogeneous sky coverage, this is not true anymore
for a single sidewards looking detector (\#07, \#08) or an
asymmetrical detector (\#04): given the Sun-pointing constraints
of the Galileo system, their sky coverage is very asymmetrical.

In the following sub-sections, we describe most of these configurations
in more detail.

\paragraph{Detector 2}

\begin{figure*}[th]
  \centering
  \includegraphics[width=0.99\columnwidth, viewport=0 90 825 515, clip]{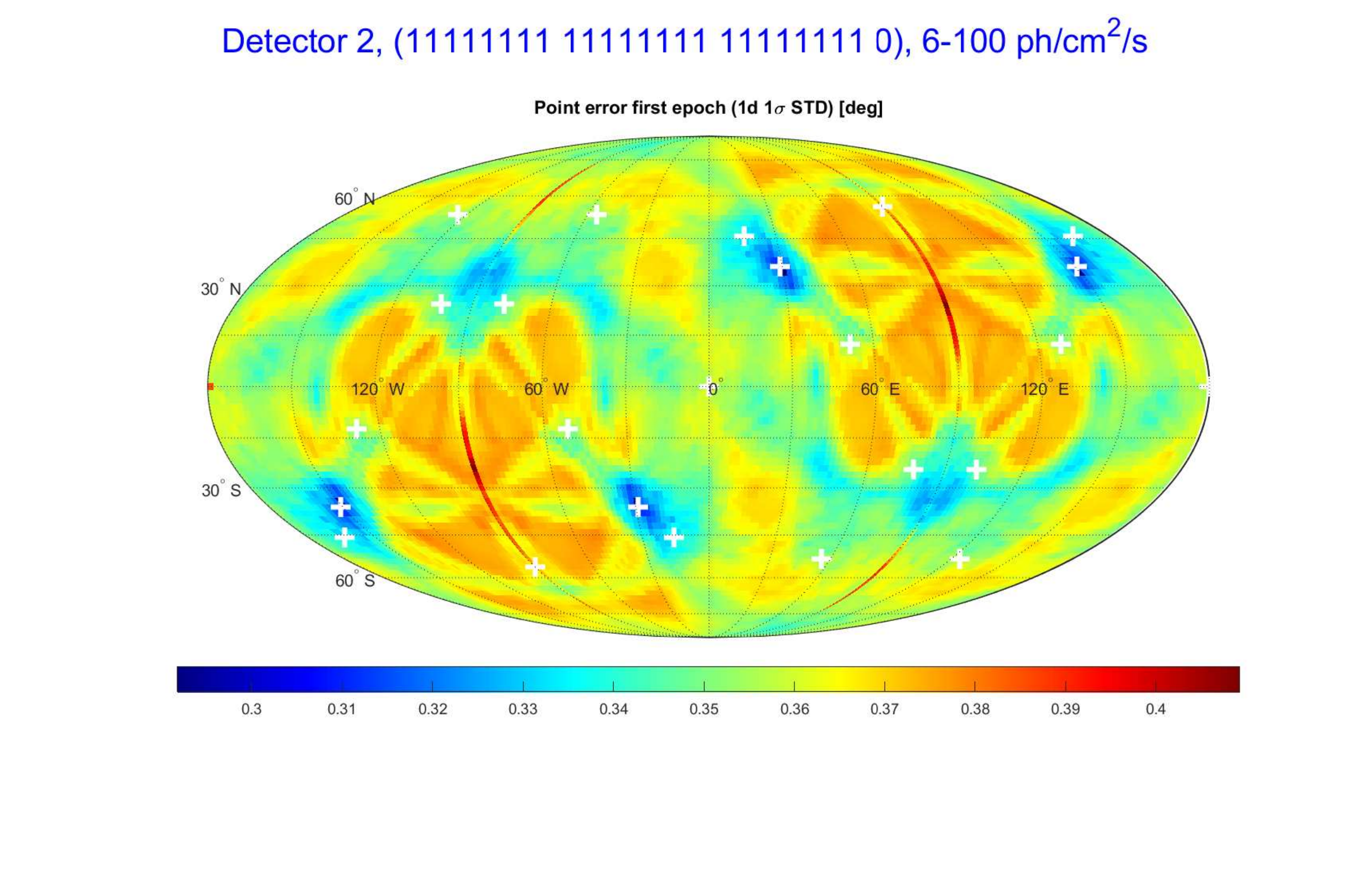}
  \includegraphics[width=0.99\columnwidth, viewport=0 90 825 515, clip]{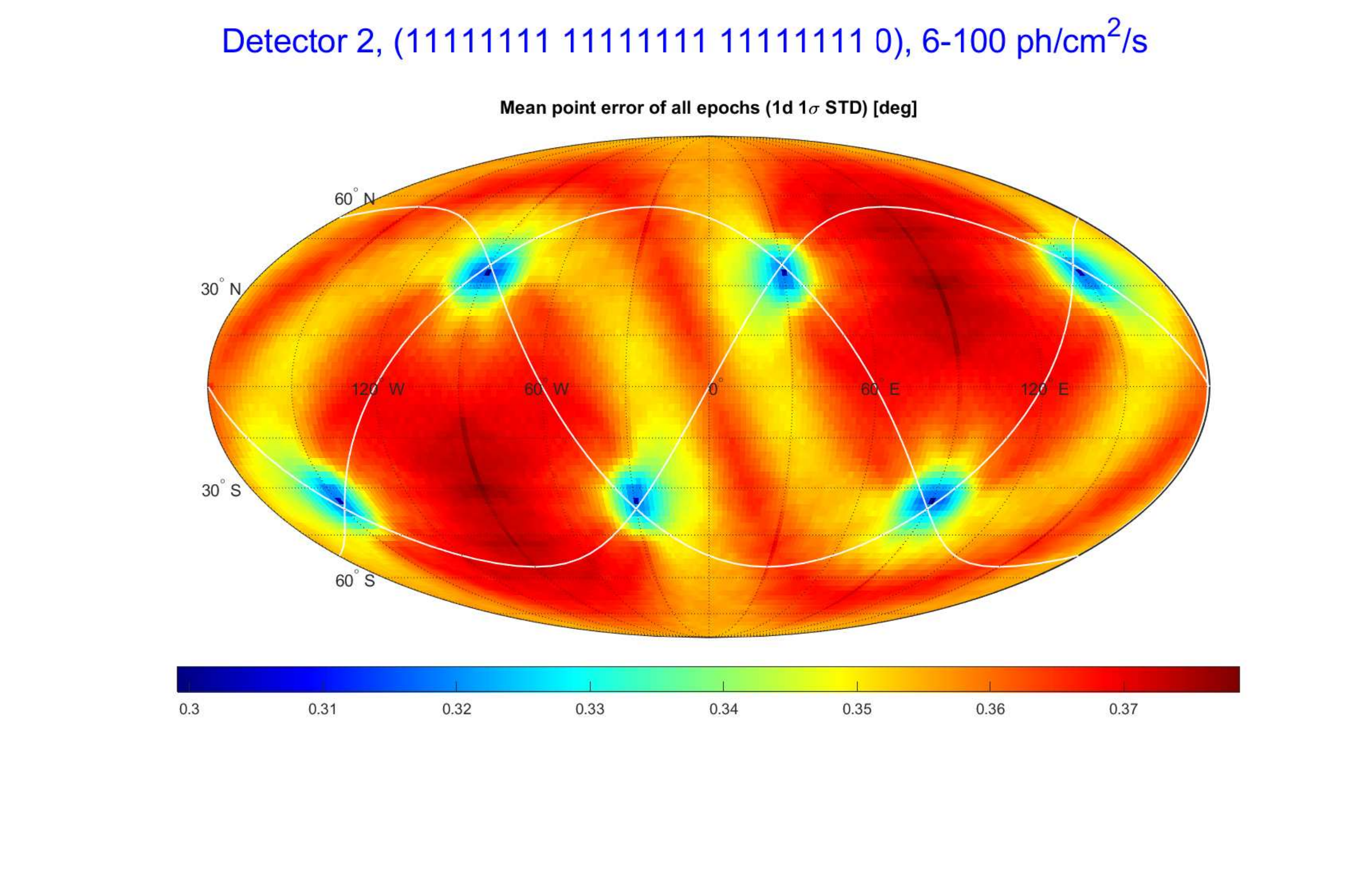}
  \includegraphics[width=0.99\columnwidth, viewport=0 90 825 515, clip]{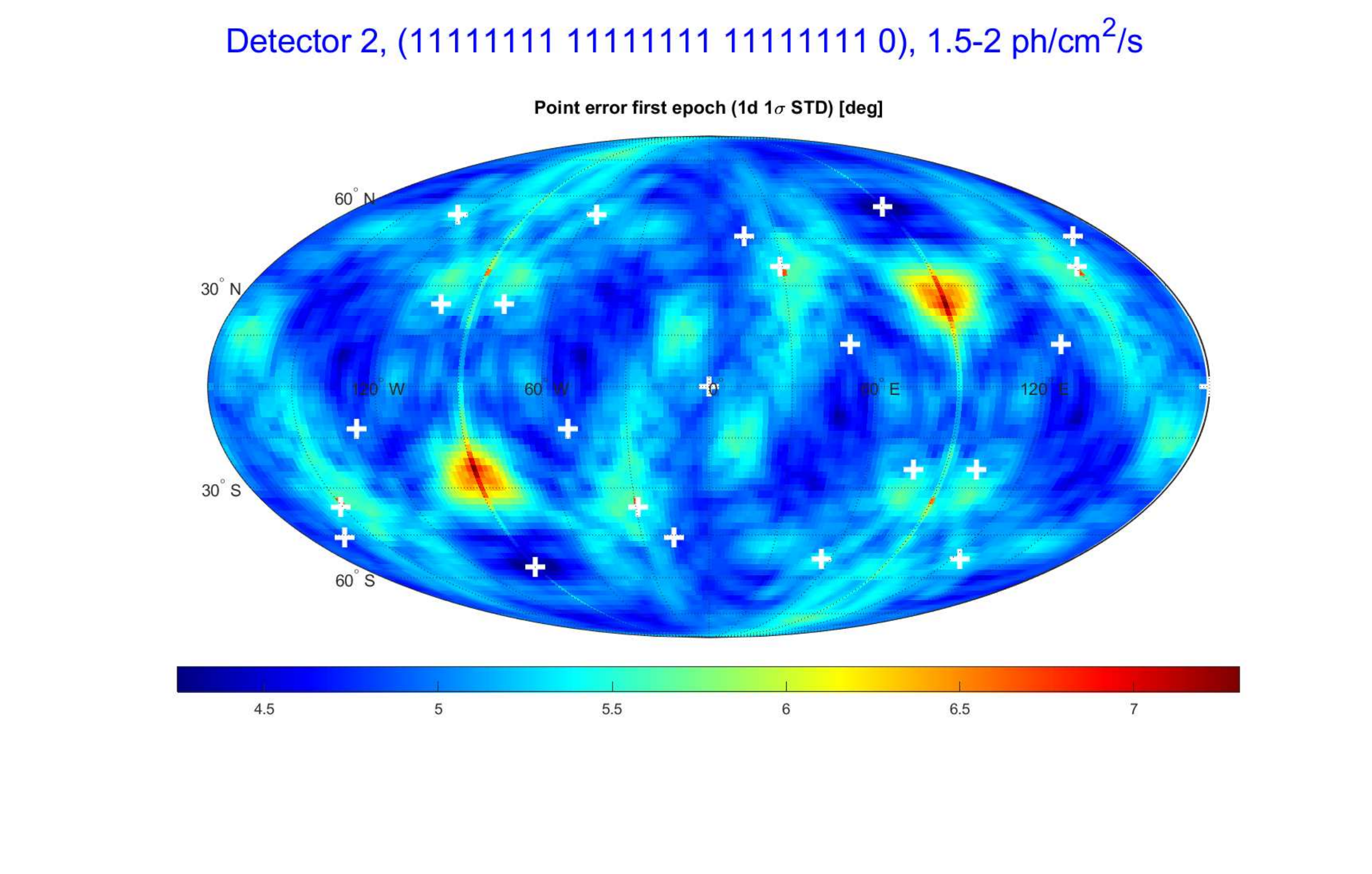}
  \includegraphics[width=0.99\columnwidth, viewport=0 90 825 515, clip]{24det_Detector02_11111111_11111111_11111111_Bin1_mean_sqrt2.pdf}
  \caption[]{Localisation accuracy for a zenith-looking cube detector each on
    24 satellites 
    for an instantaneous moment (left) and averaged over one orbit (right),
    for GRBs in the brightest peak flux bin of 6--100 ph/cm$^2$/s (top row)
    and the faintest peak flux bin of 1.5--2 ph/cm$^2$/s (bottom row).
    \label{24sat_Det2}}
\end{figure*}

\begin{figure*}[h]
   \centering
   \includegraphics[width=0.99\columnwidth, viewport=0 90 825 515, clip]{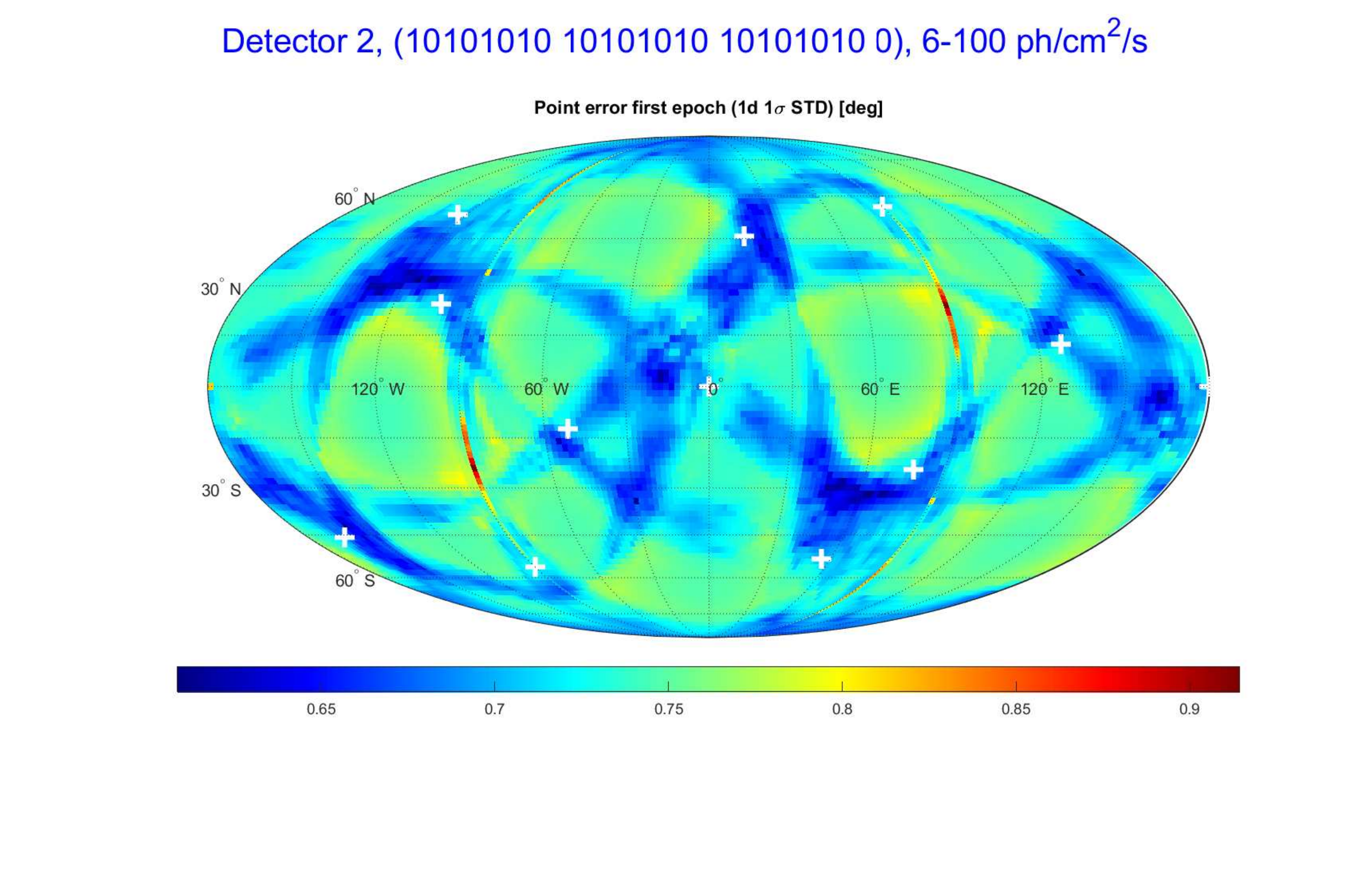}
   \includegraphics[width=0.99\columnwidth, viewport=0 90 825 515, clip]{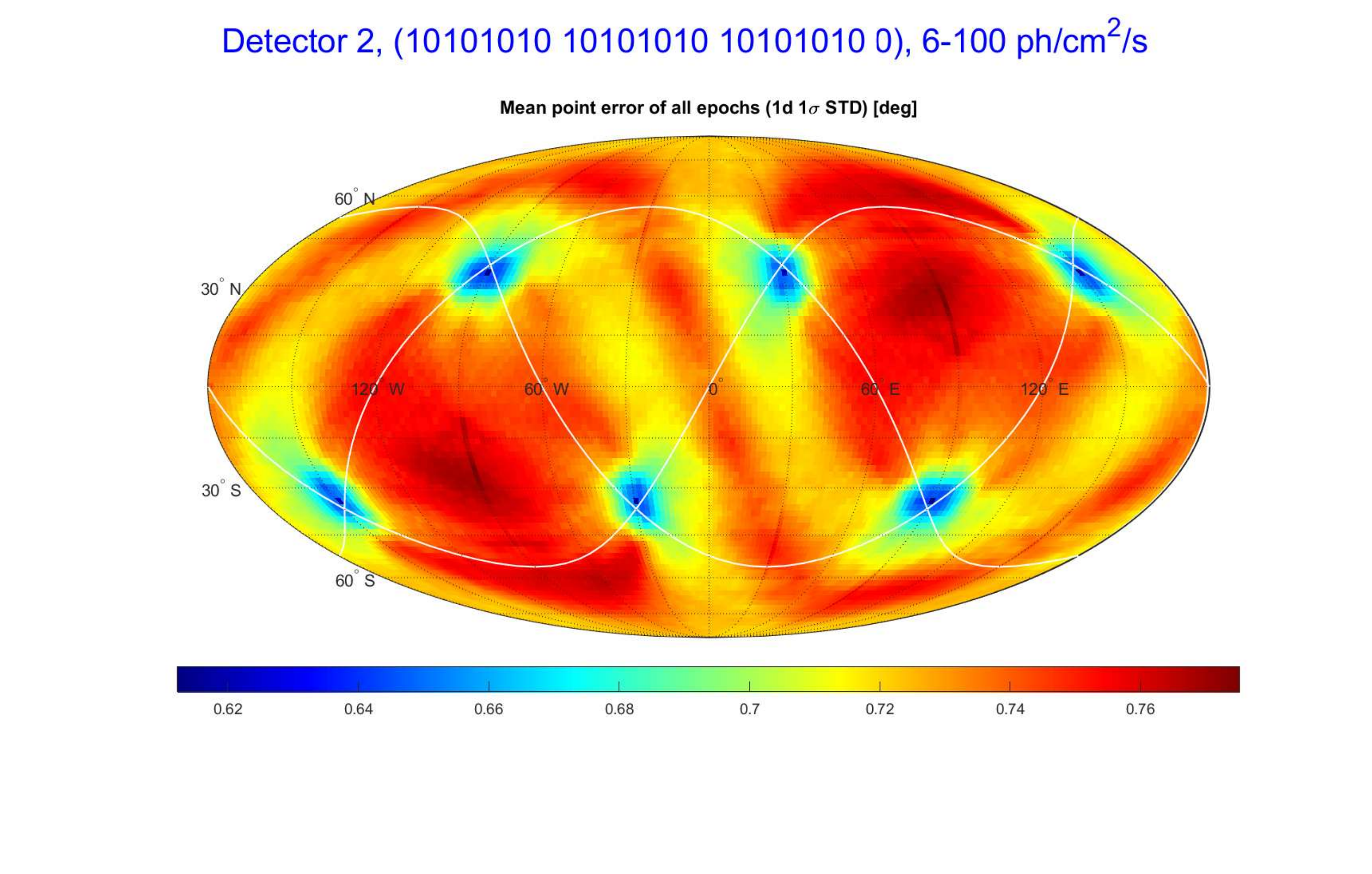}
   \includegraphics[width=0.99\columnwidth, viewport=0 90 825 515, clip]{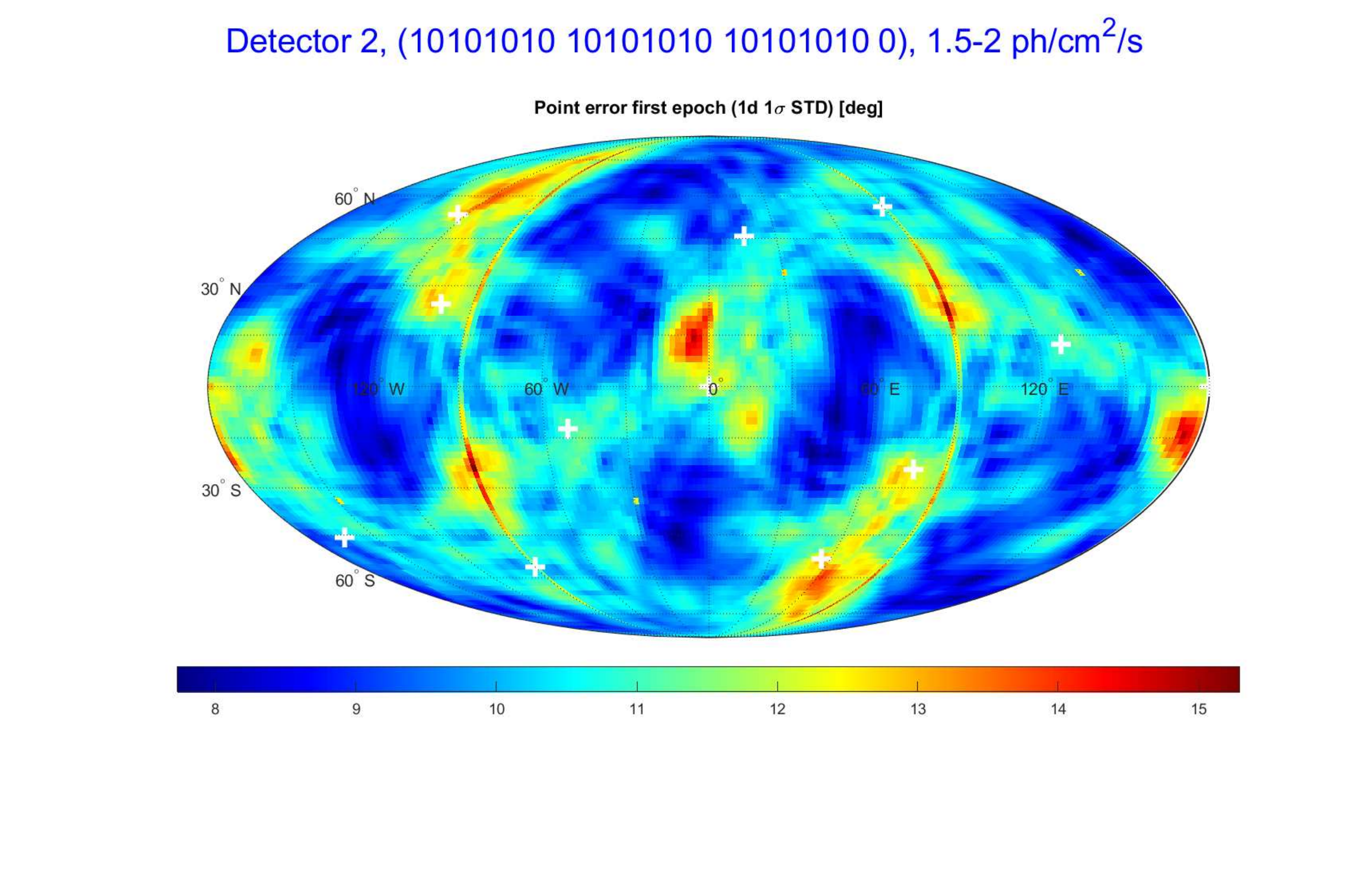}
   \includegraphics[width=0.99\columnwidth, viewport=0 90 825 515, clip]{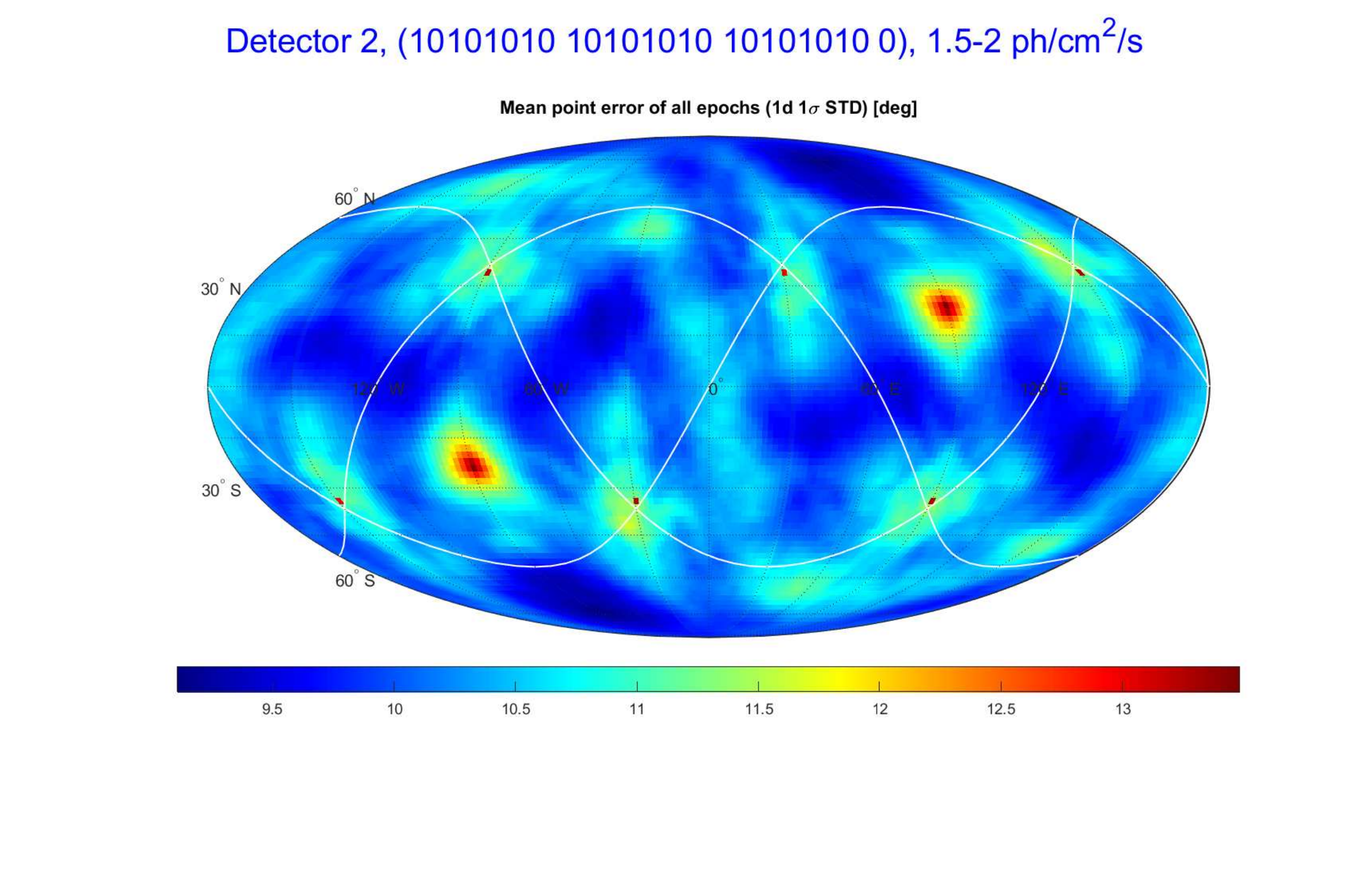}
  \caption[]{Same as Fig. \ref{24sat_Det2} but for 12 satellites (every second along each orbital plane).
    \label{12sat_Det2}}
\end{figure*}

As a consequence of the average short baselines  for a flat, zenith-facing
detector (Fig. \ref{1D-statistics}), we next test a cube detector on the
zenith-facing side of the
Galileo satellites: these have the same area as Detector 01 towards zenith,
and half of this (due to the height of only 30\,cm) towards all four sides.
With more satellites at large baselines and large effective area available
for large parts of the sky, this substantially improves the localisation accuracy
of the zenith-looking flat detector (see Figs. \ref{24sat_Det2}, \ref{12sat_Det2}).

\begin{figure*}[th]
  \centering
 \includegraphics[width=0.65\columnwidth]{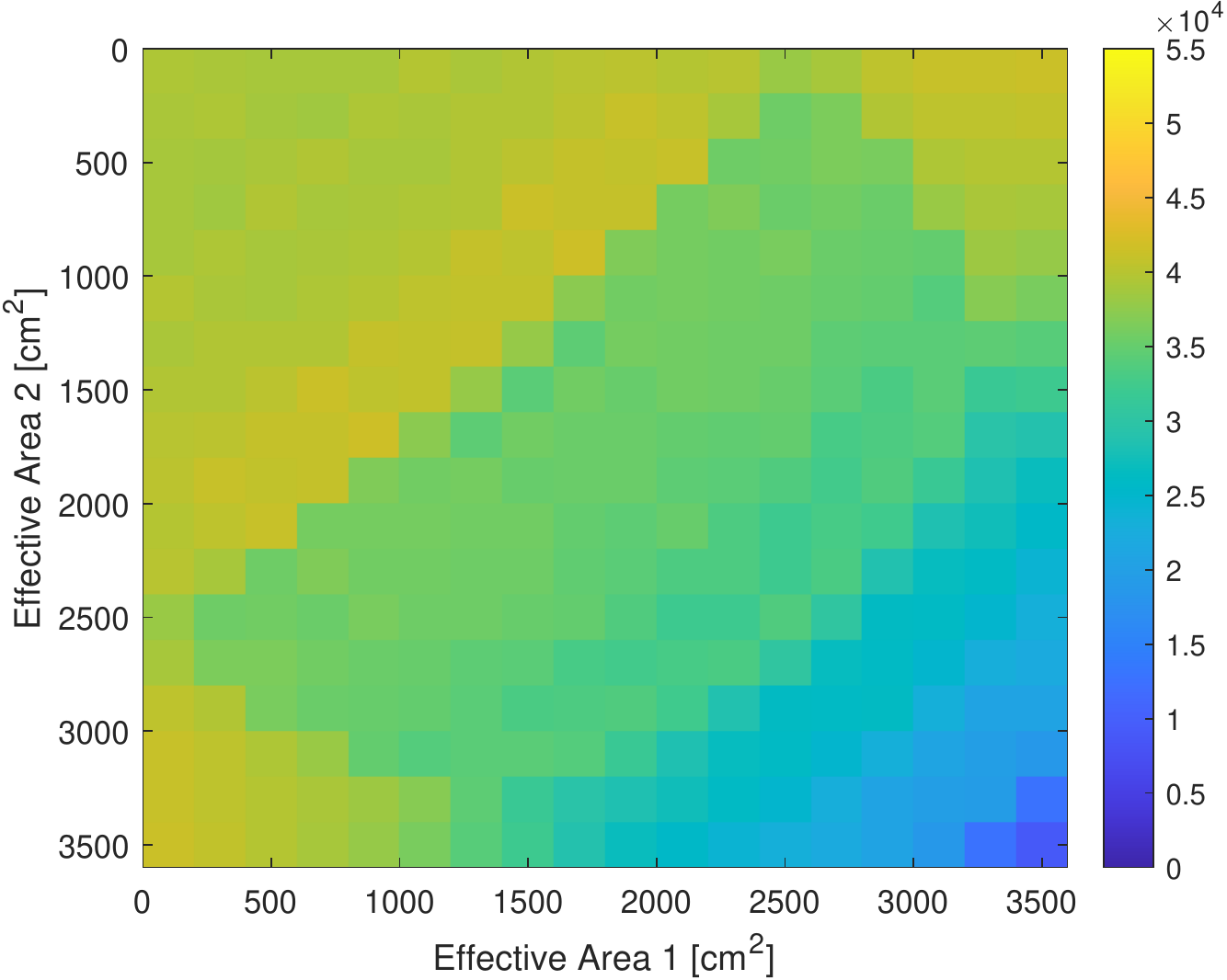}
 \includegraphics[width=0.65\columnwidth]{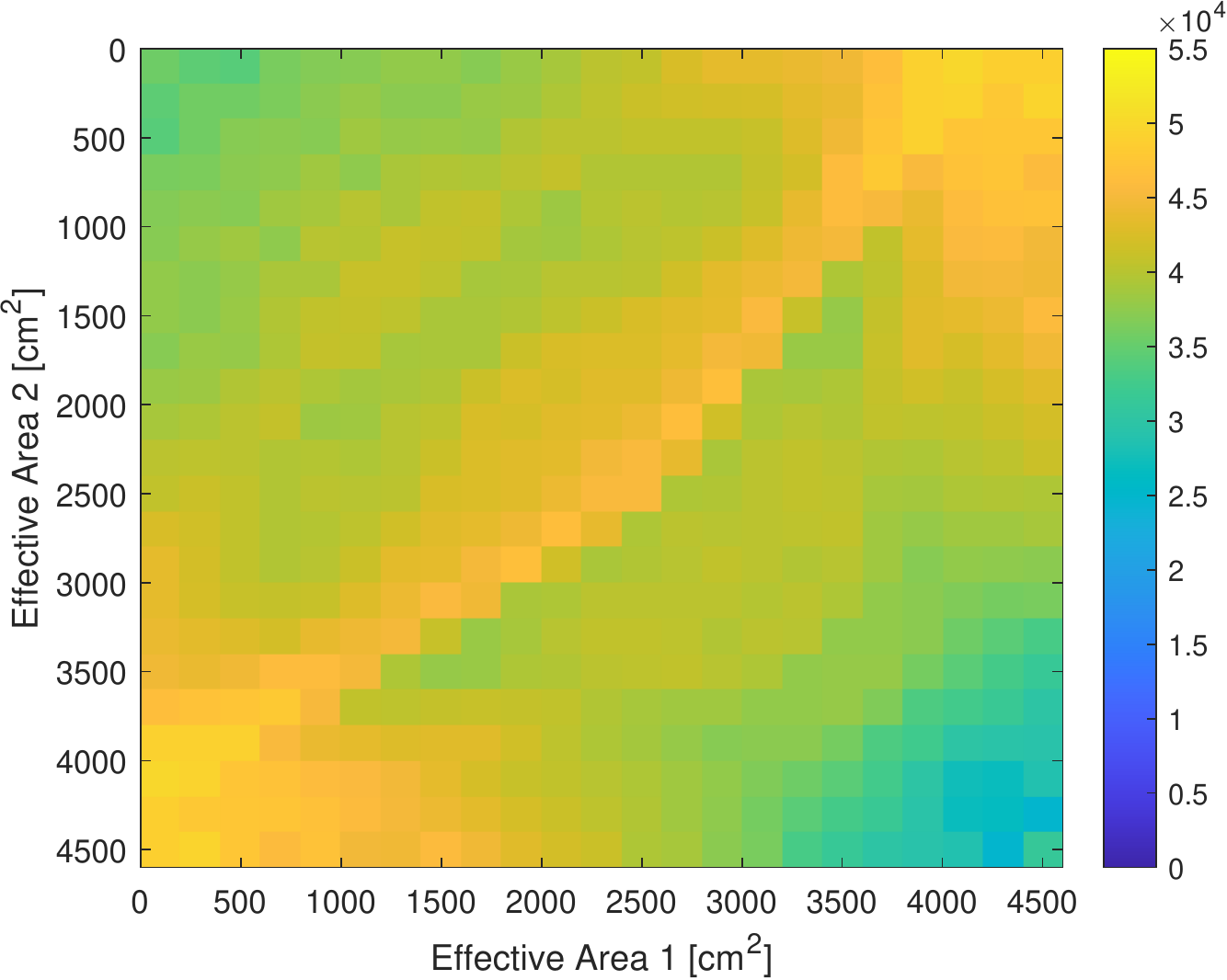}
 \includegraphics[width=0.65\columnwidth]{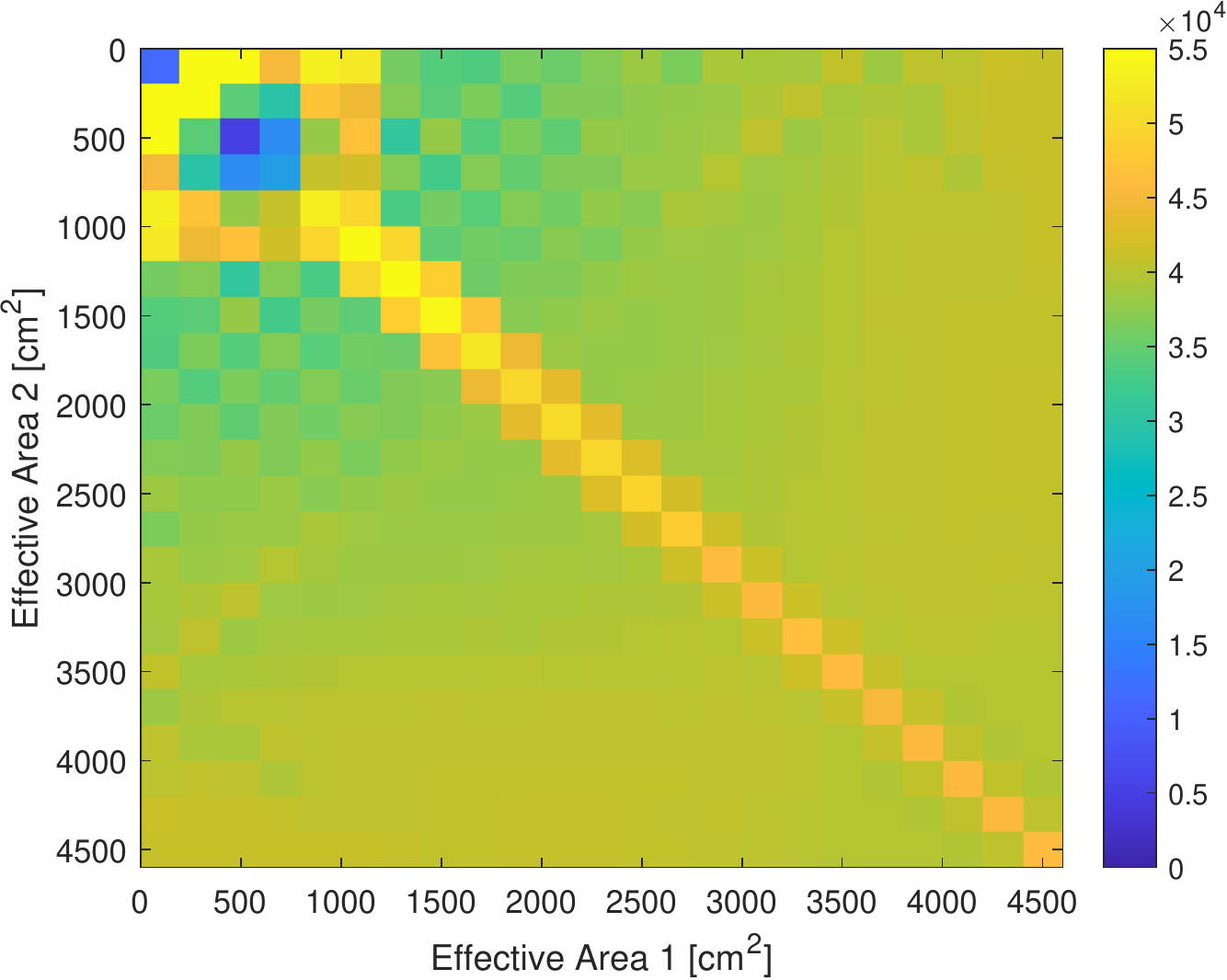}
   \caption[]{Distribution of mean baseline lengths for combinations
     of effective areas for pairs of detectors seeing
     a GRB for the geometries of detector \#01 (left) and \#02 (middle),
     and \#03 (right).
    \label{aeff_Det0102}}
\end{figure*}

Fig. \ref{aeff_Det0102} illustrates why the cube detector is so much better
in performance: 
the distribution of the mean baselines which are realized for given
pairs of detectors and their projected effective areas as
determined by their viewing direction relative to a GRB shows that
the long baselines for higher effective areas dominate clearly.
Thus, we reach sub-degree localisation accuracy for the brightest
GRB intensity bin (though for the faintest it is still of order 10\degs).

\paragraph{Detector 3}

\begin{figure*}[h]
  \includegraphics[width=0.332\textwidth, viewport=80 90 780 515, clip]{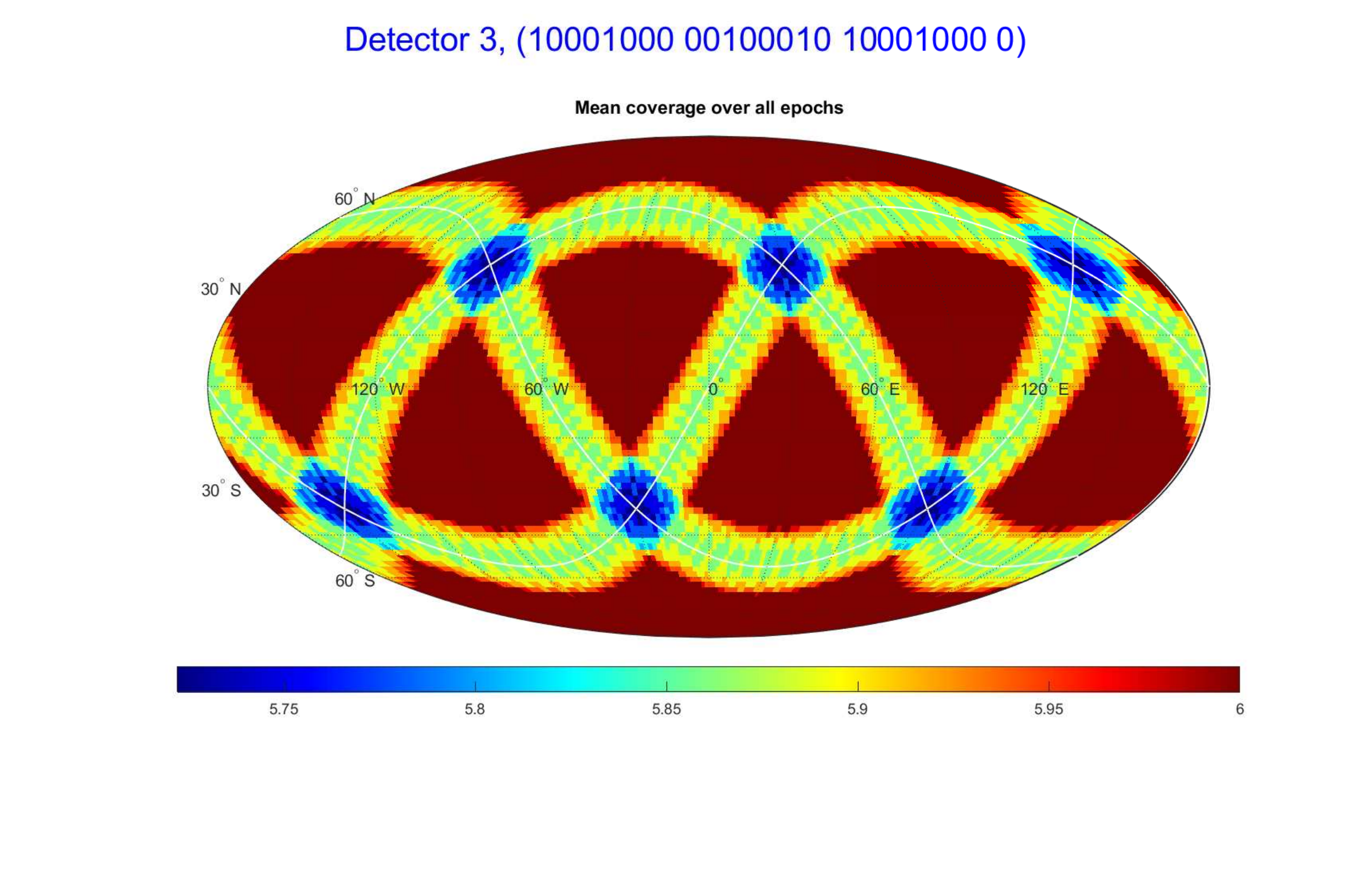}
  \includegraphics[width=0.332\textwidth, viewport=80 90 780 515, clip]{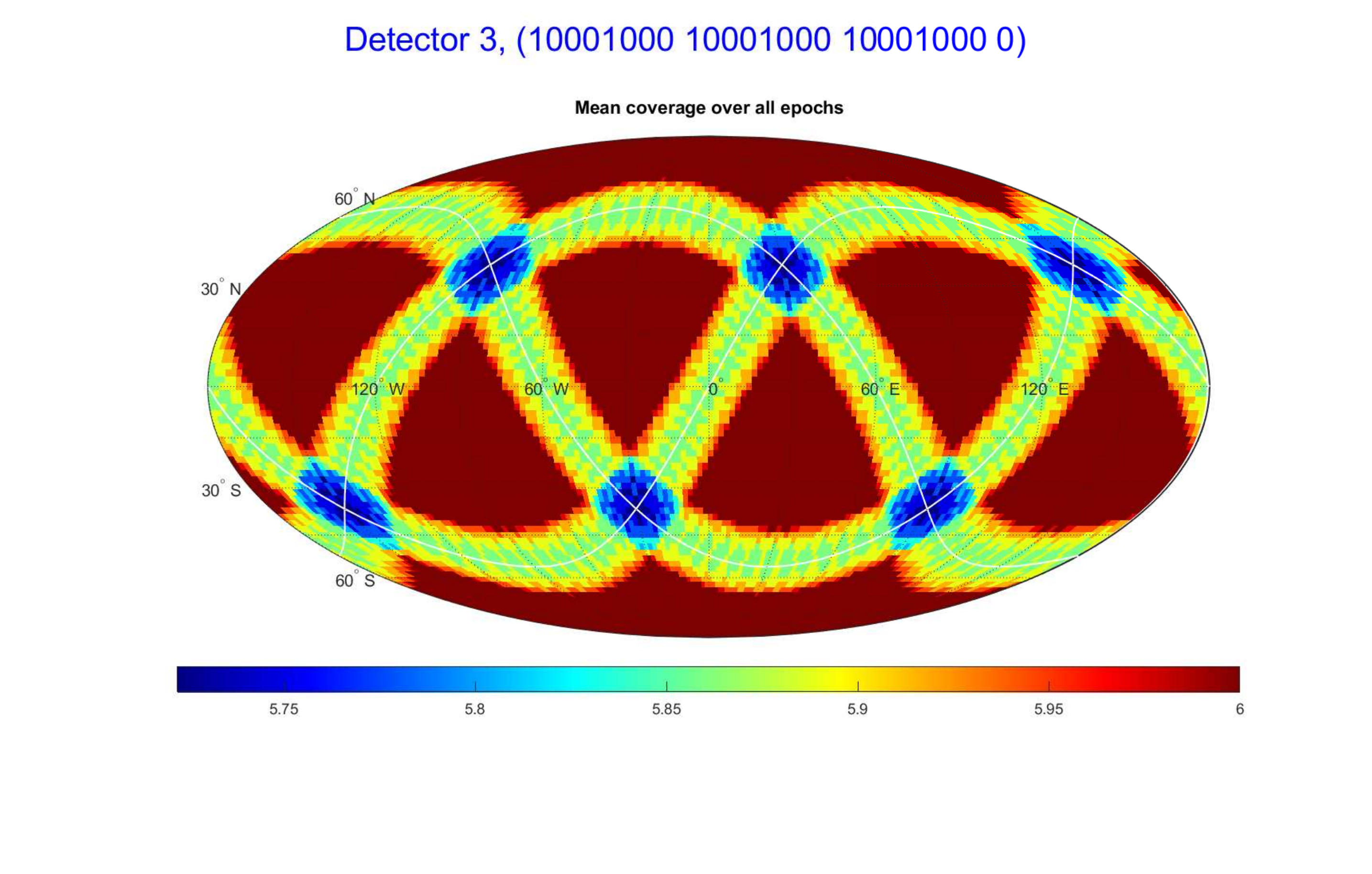}
   \includegraphics[width=0.332\textwidth, viewport=80 90 780 515, clip]{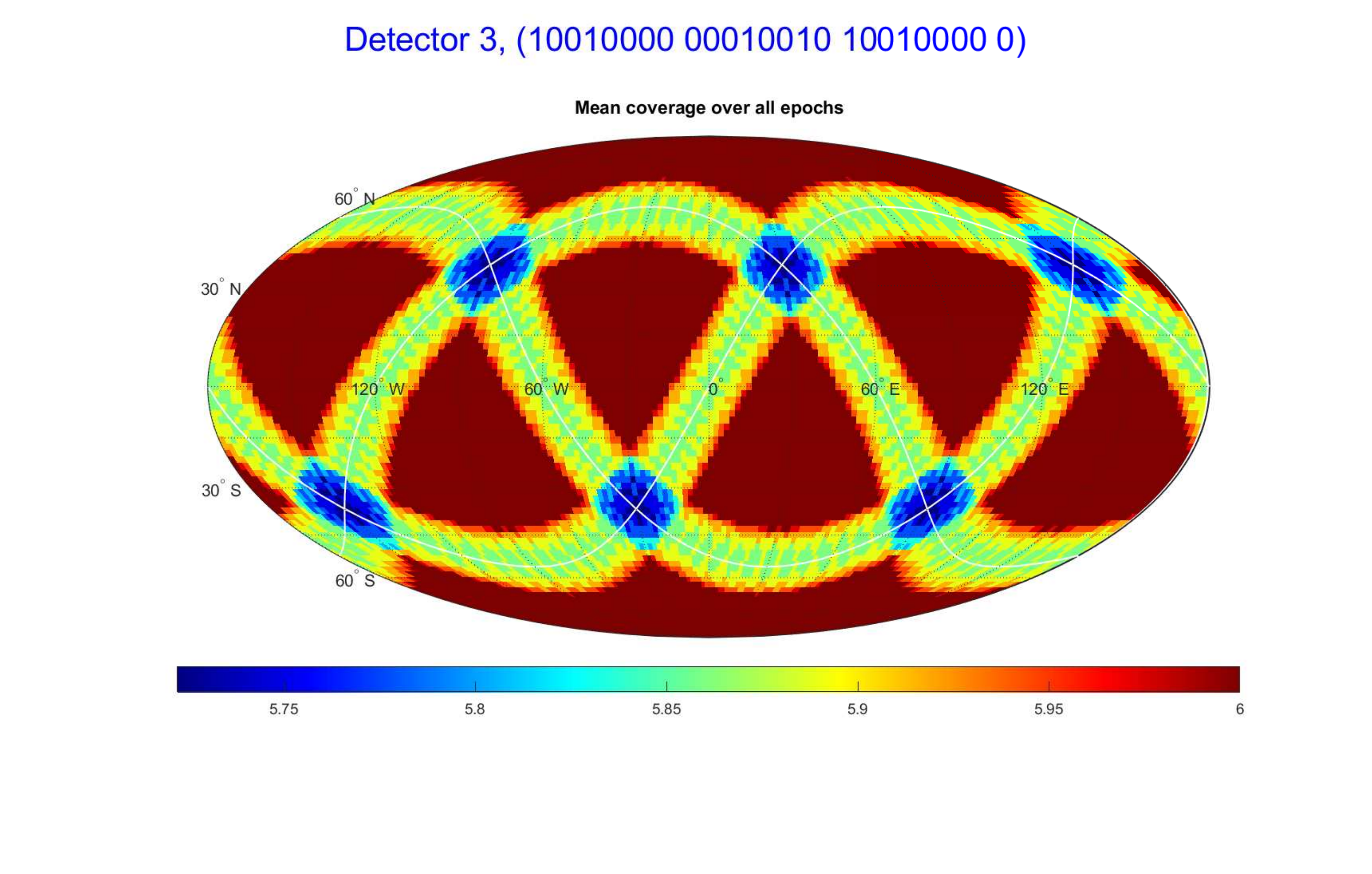}
\includegraphics[width=0.332\textwidth, viewport=80 90 780 515, clip]{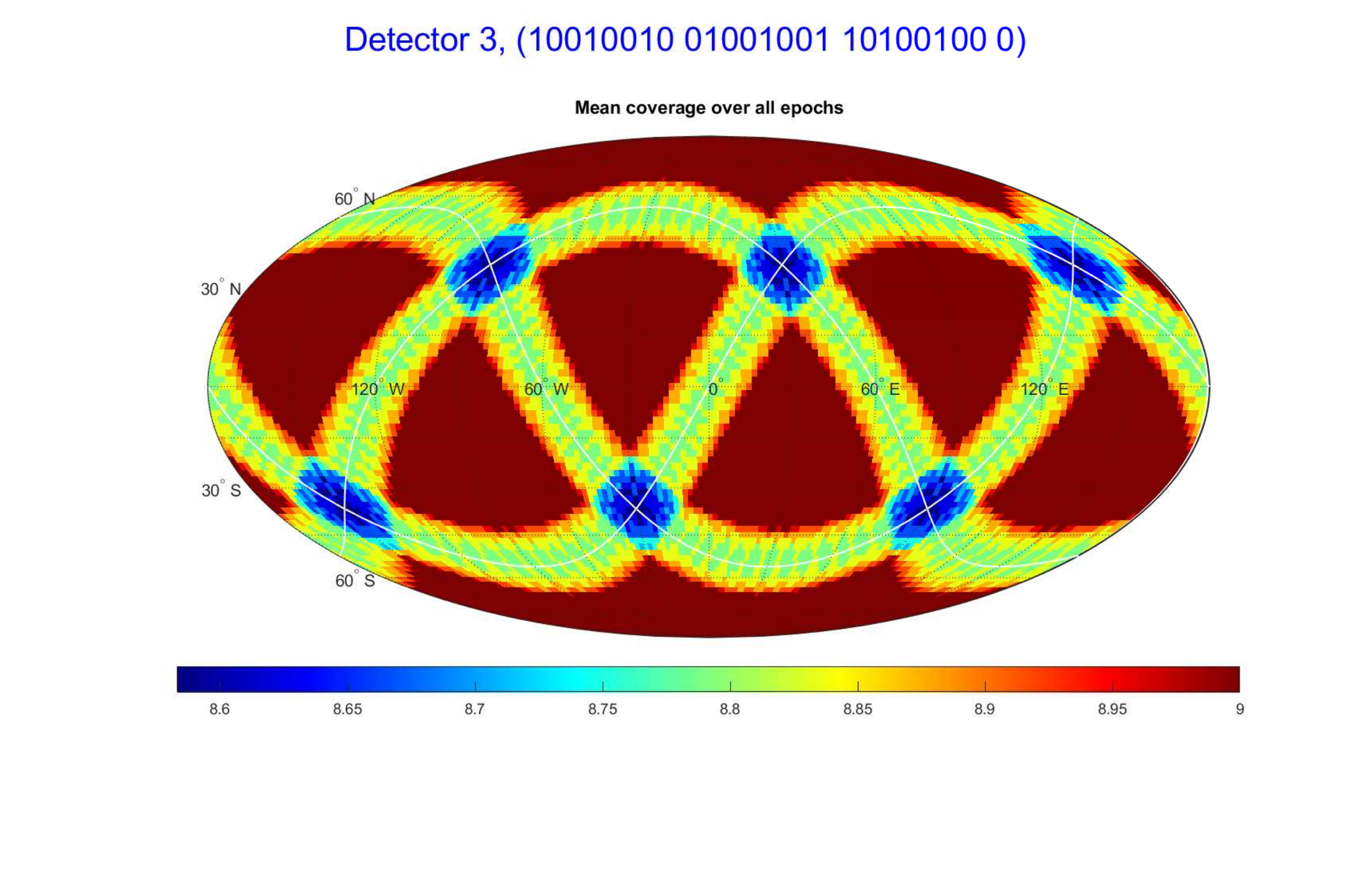}
  \includegraphics[width=0.332\textwidth, viewport=80 90 780 515, clip]{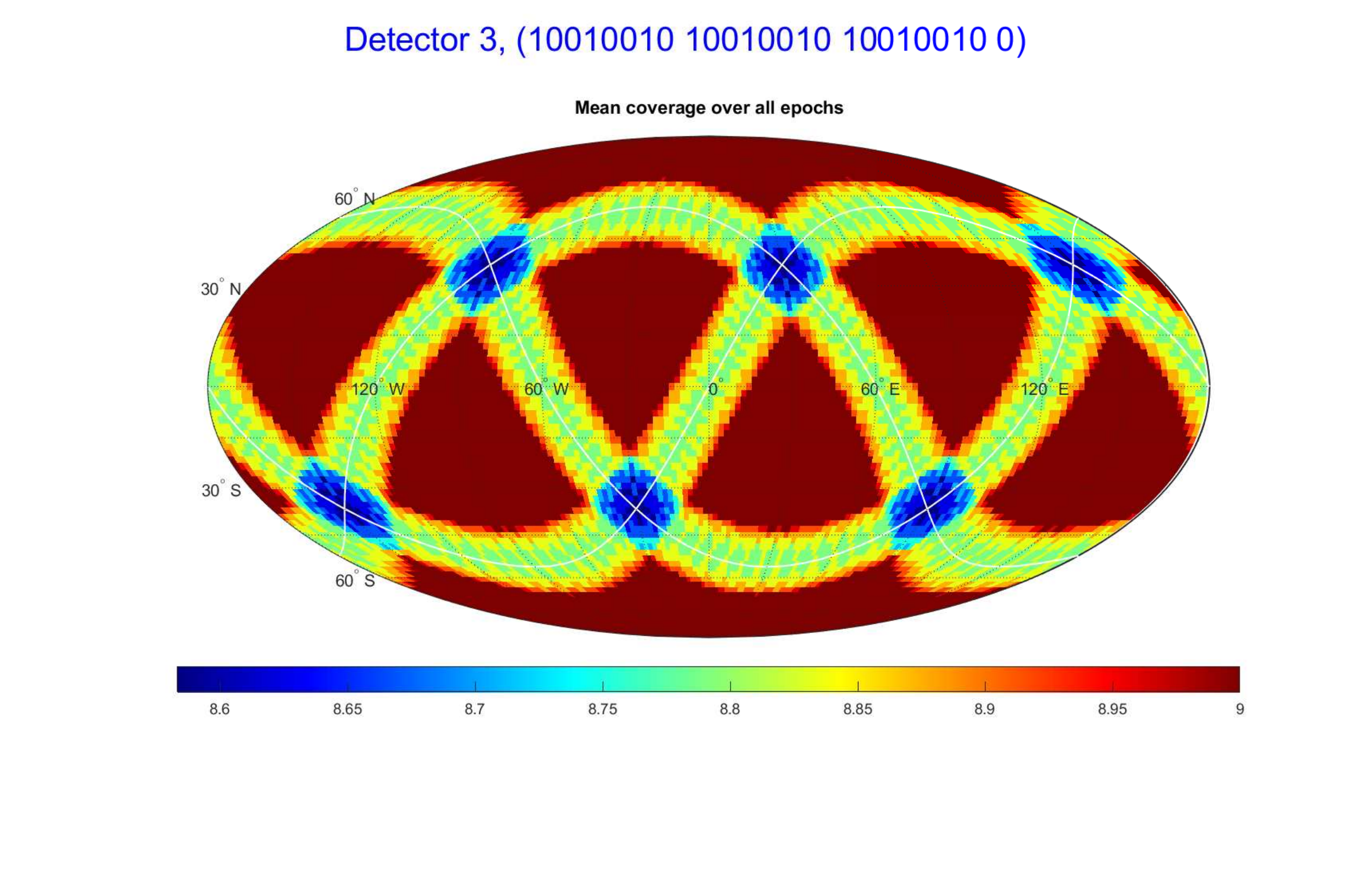}
  \includegraphics[width=0.332\textwidth, viewport=80 90 780 515, clip]{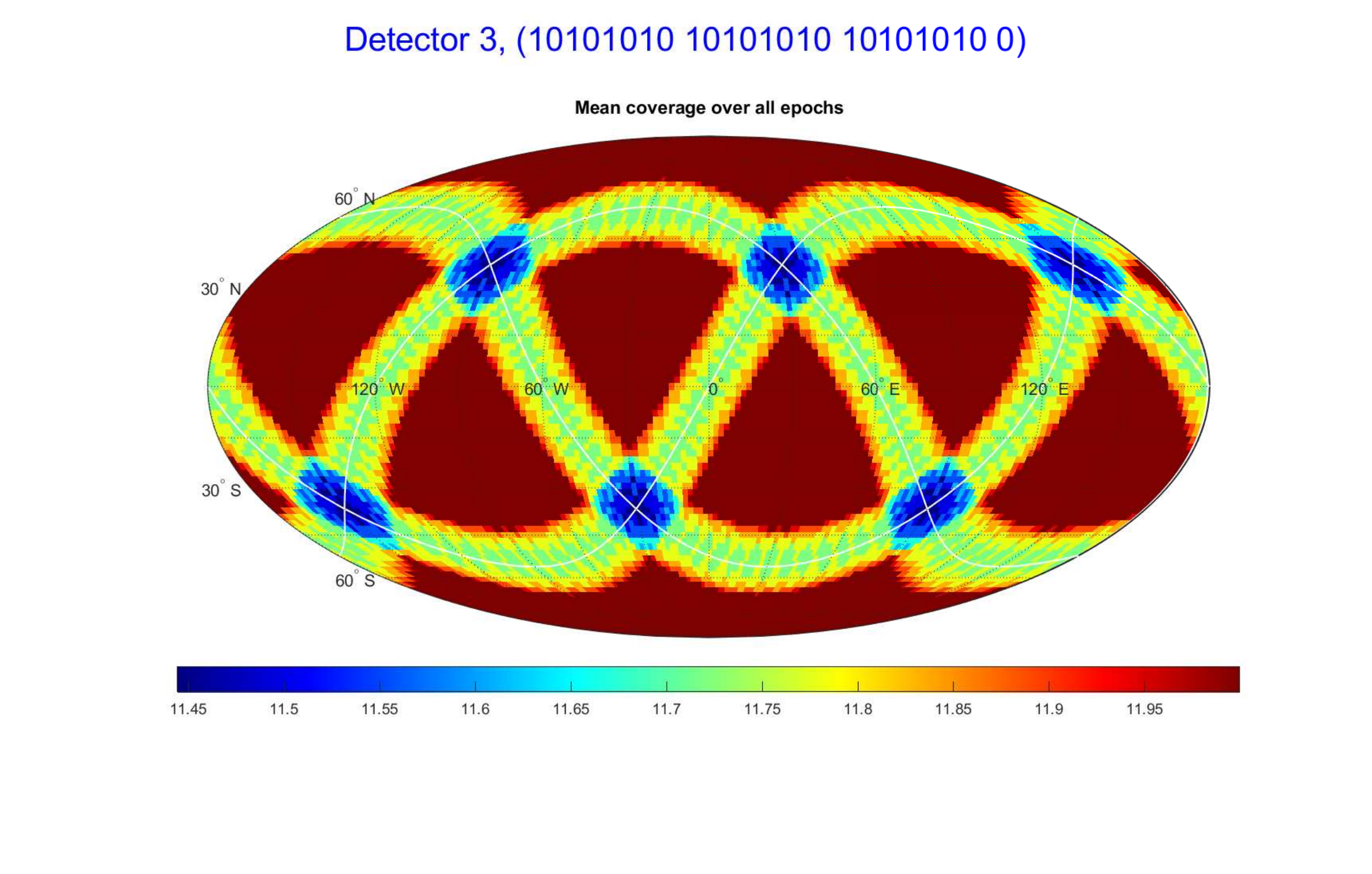}
  \caption[]{Sky coverage for 4-side detectors on 6 (top), 9 (left and middle of
   bottom row) and 12 (lower right) satellites  averaged over one orbit (right).
   Note the different distribution of satellites along each orbital plane,
   with 3 different options for the 6 satellites, and two options for 9
   satellites. The color-code provides the number of satellites which see
   a given GRB at a given time, averaged over one orbital period.
   \label{allsat_Det3_skycov}}
\end{figure*}

\begin{figure*}[h]
  \includegraphics[width=0.33\textwidth, viewport=80 90 780 515, clip]{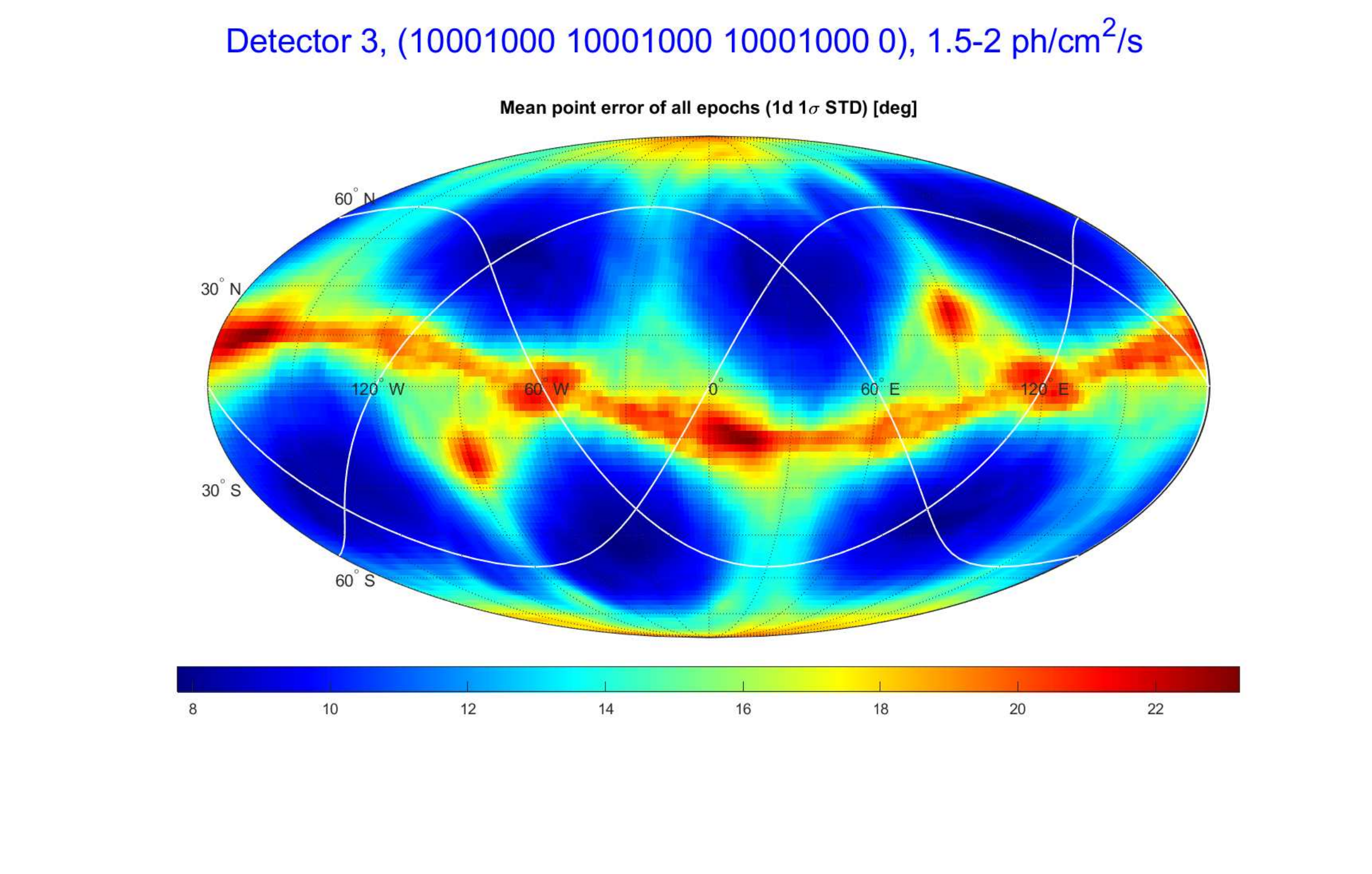}
  \includegraphics[width=0.33\textwidth, viewport=80 90 780 515, clip]{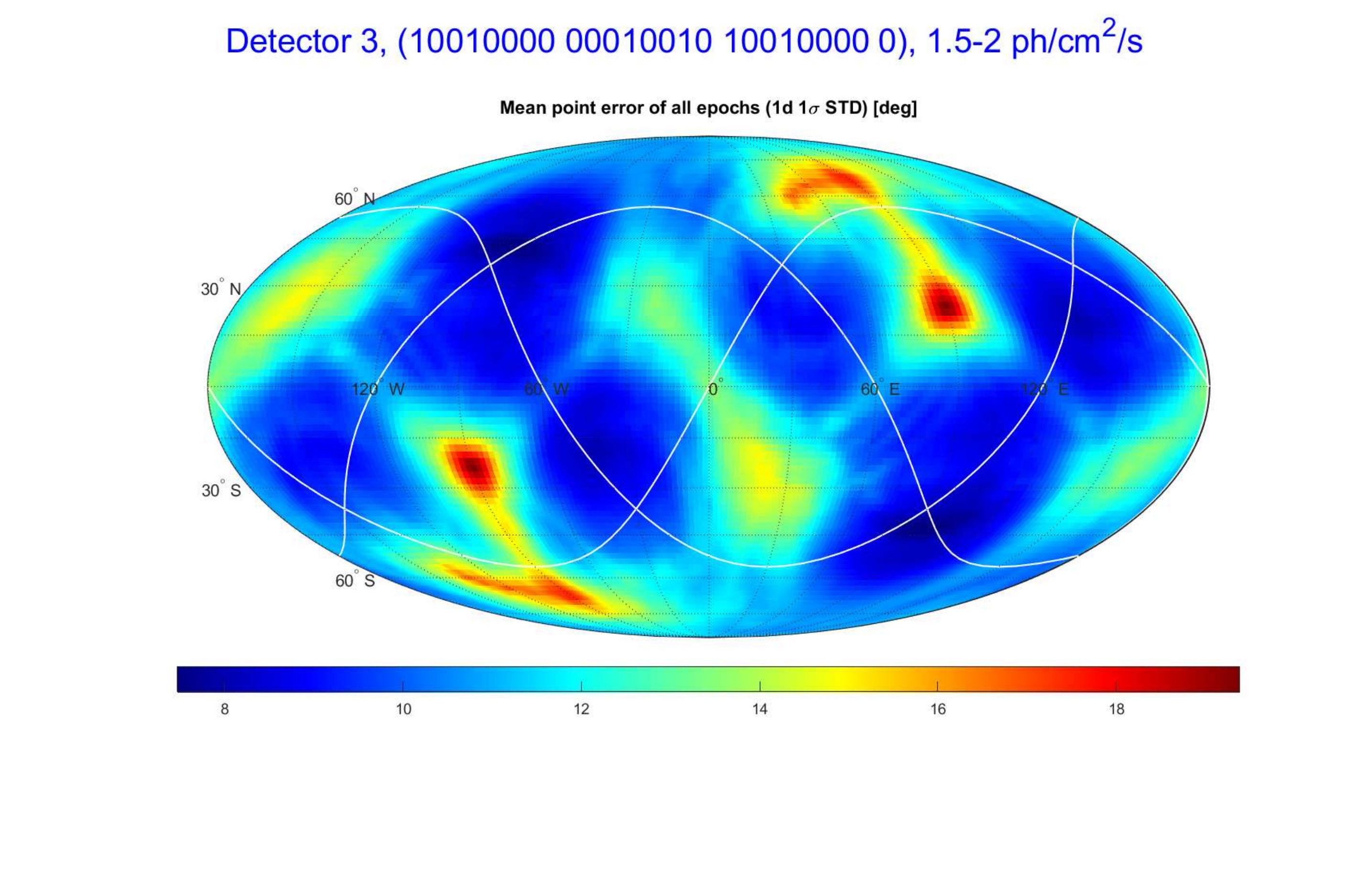}\\
  \includegraphics[width=0.33\textwidth, viewport=80 90 780 515, clip]{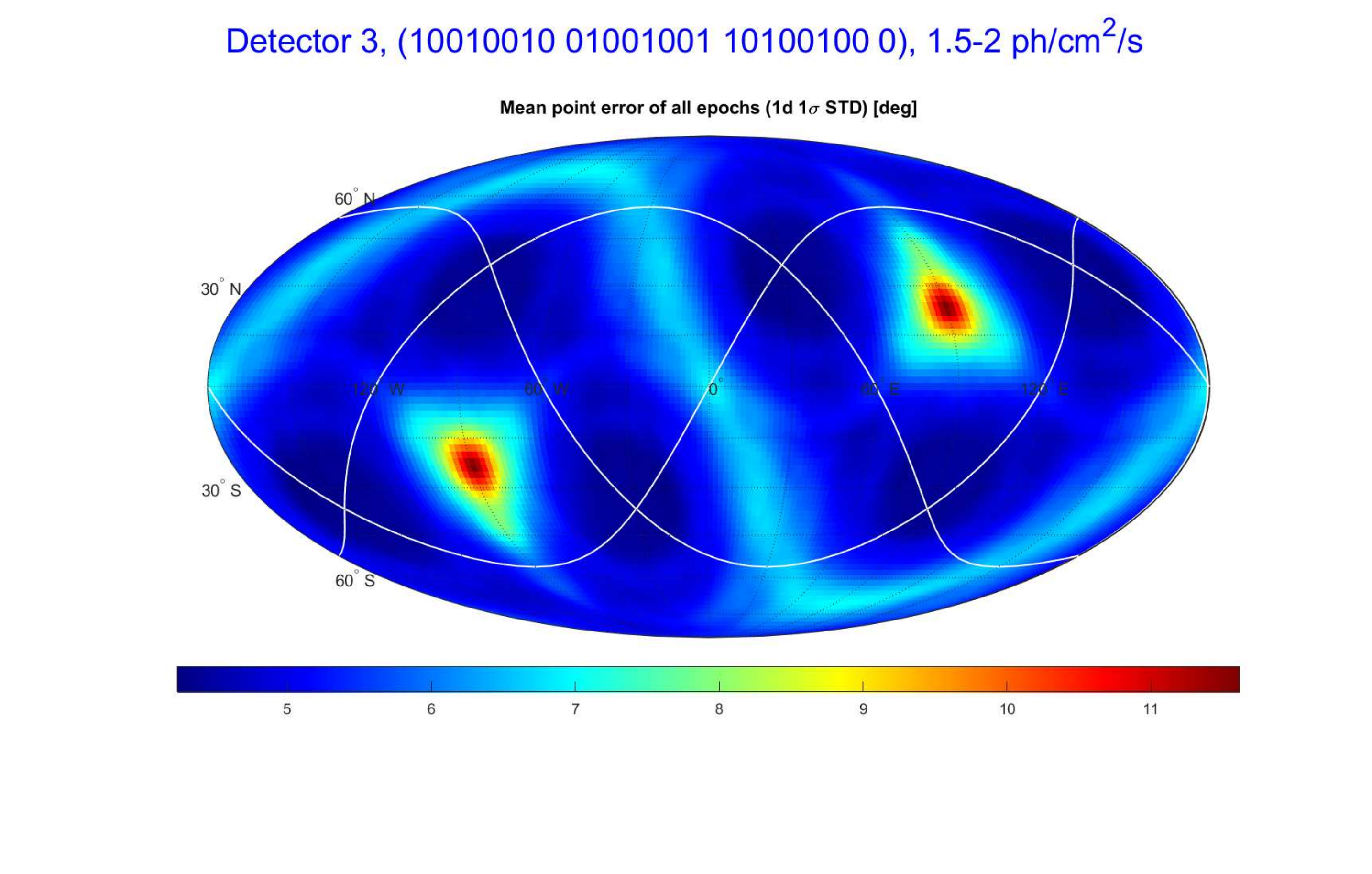}
  \includegraphics[width=0.33\textwidth, viewport=80 90 780 515, clip]{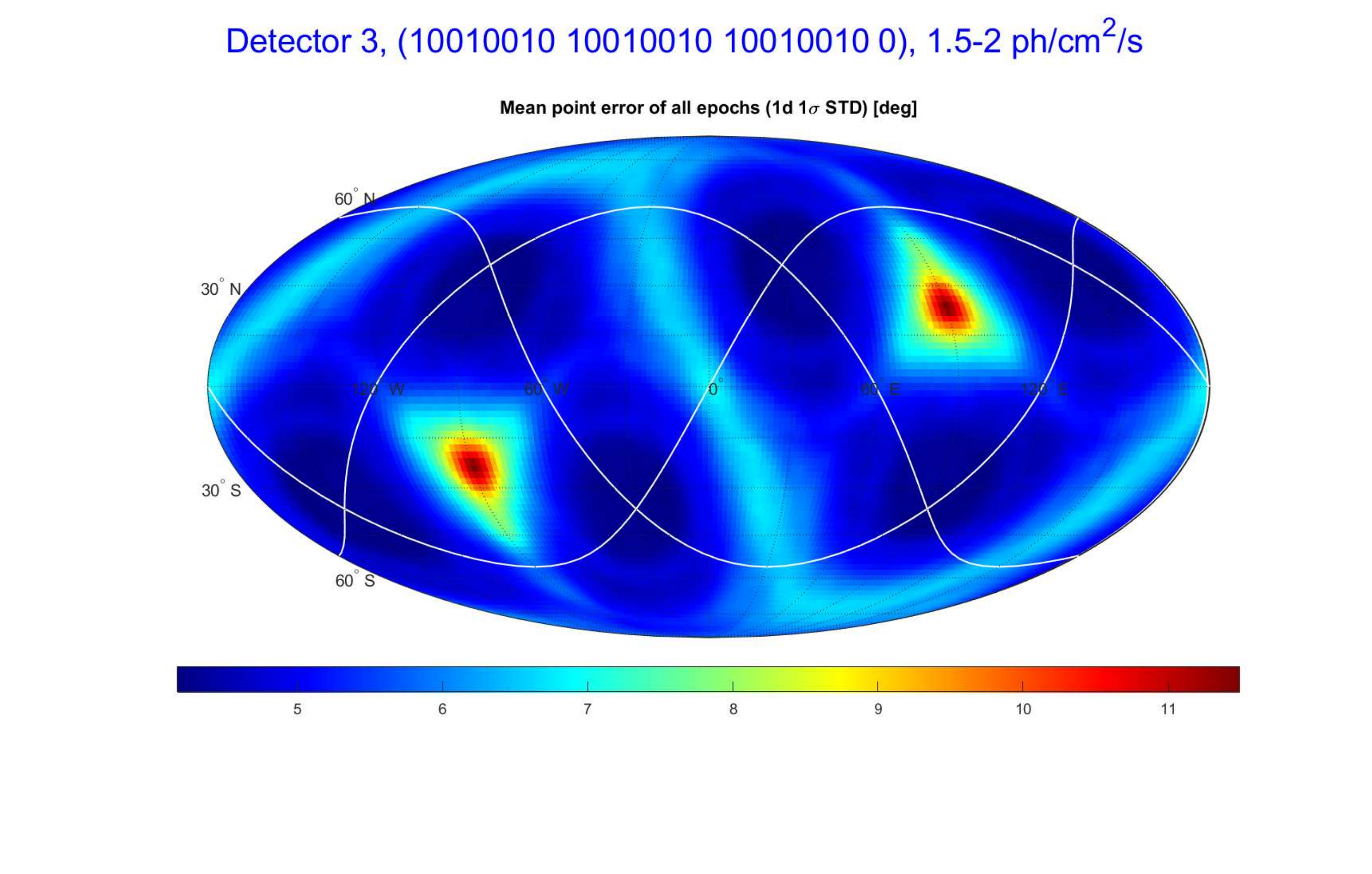}
  \includegraphics[width=0.33\textwidth, viewport=80 90 780 515, clip]{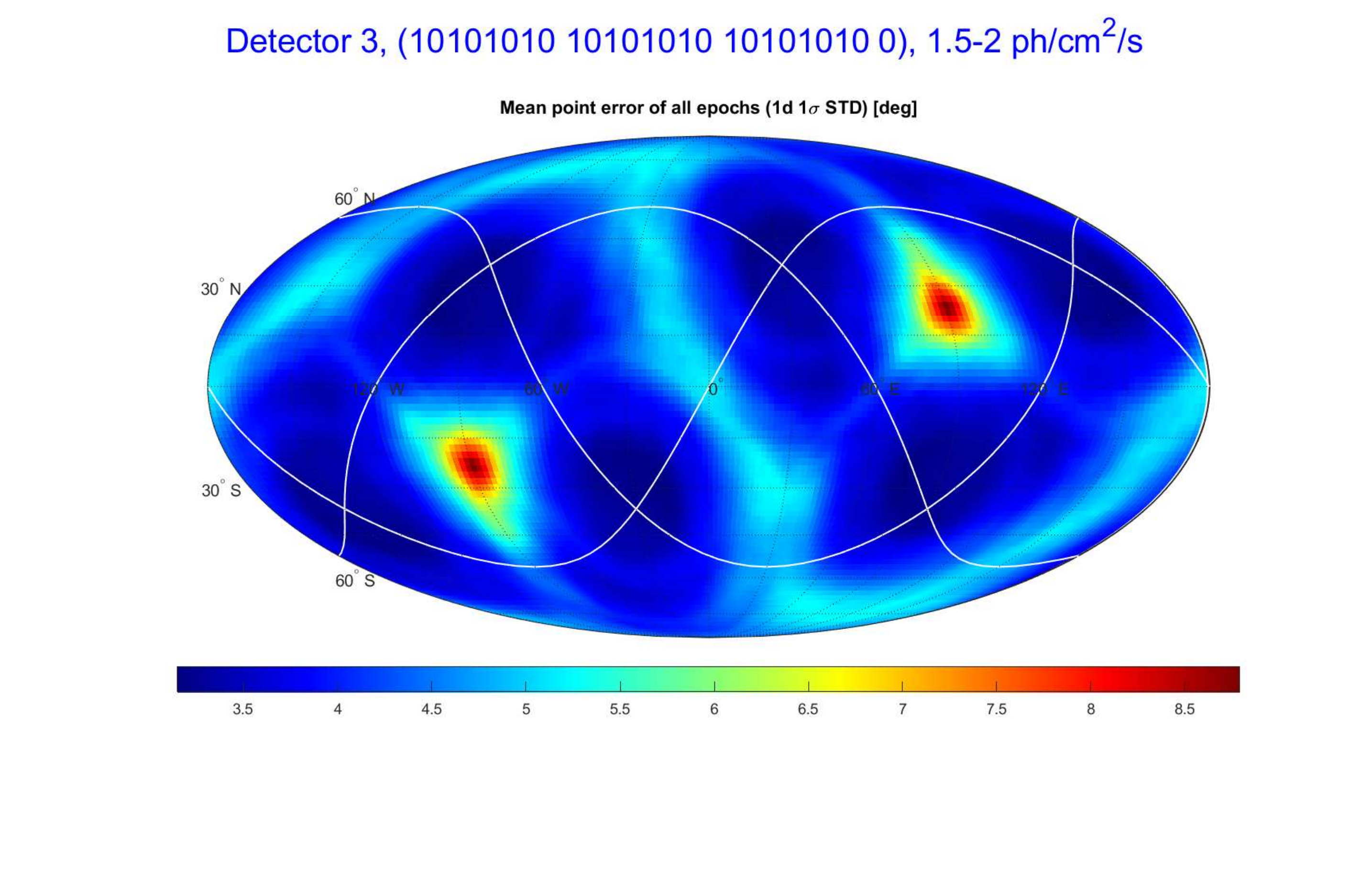}
  \caption[]{Localisation accuracy for a 4-side detector each on
    6 (top row), 9 (lower left and middle) and 12 satellites (lower right)
    for the 
    faintest intensity interval. The 6- and 9-satellite options are shown for
    two different configurations along the orbital plane.
    \label{comp_NumSat}}
  \vspace{-0.2cm}
\end{figure*}

With the zenith-looking detectors not being optimal, we now look at
an arrangement where all 4 sides of a Galileo satellite are equipped
with a 60\,cm x 60\,cm detector, with the nadir- and zenith-looking
sides without GRB detector. The localisation accuracy is very good,
see Fig. \ref{comp_NumSat}, even for this faintest GRB intensity level.
The averaged
sky coverage shows an identical sky pattern, independent of whether
we equip 6, 9 or 12 satellites with a GRB detector
(see Fig. \ref{allsat_Det3_skycov}). 
But note the different color-code normalization: obviously, when
more satellites are equipped with a detector, then there is a larger number
of satellites seeing a given GRB.

\begin{figure*}[h]
  \includegraphics[width=0.5\textwidth, viewport=80 90 780 515, clip]{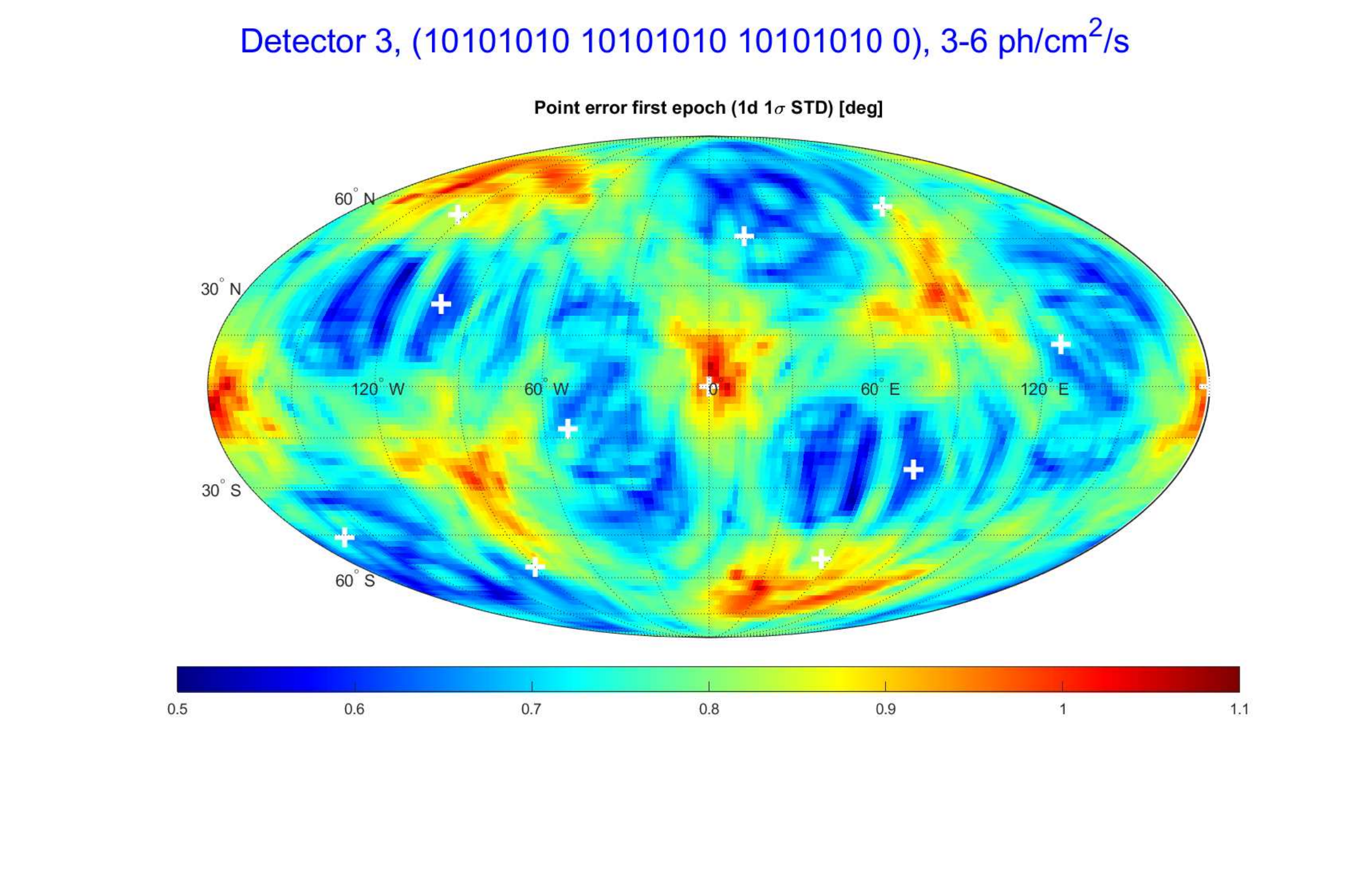}
  \includegraphics[width=0.5\textwidth, viewport=80 90 780 515, clip]{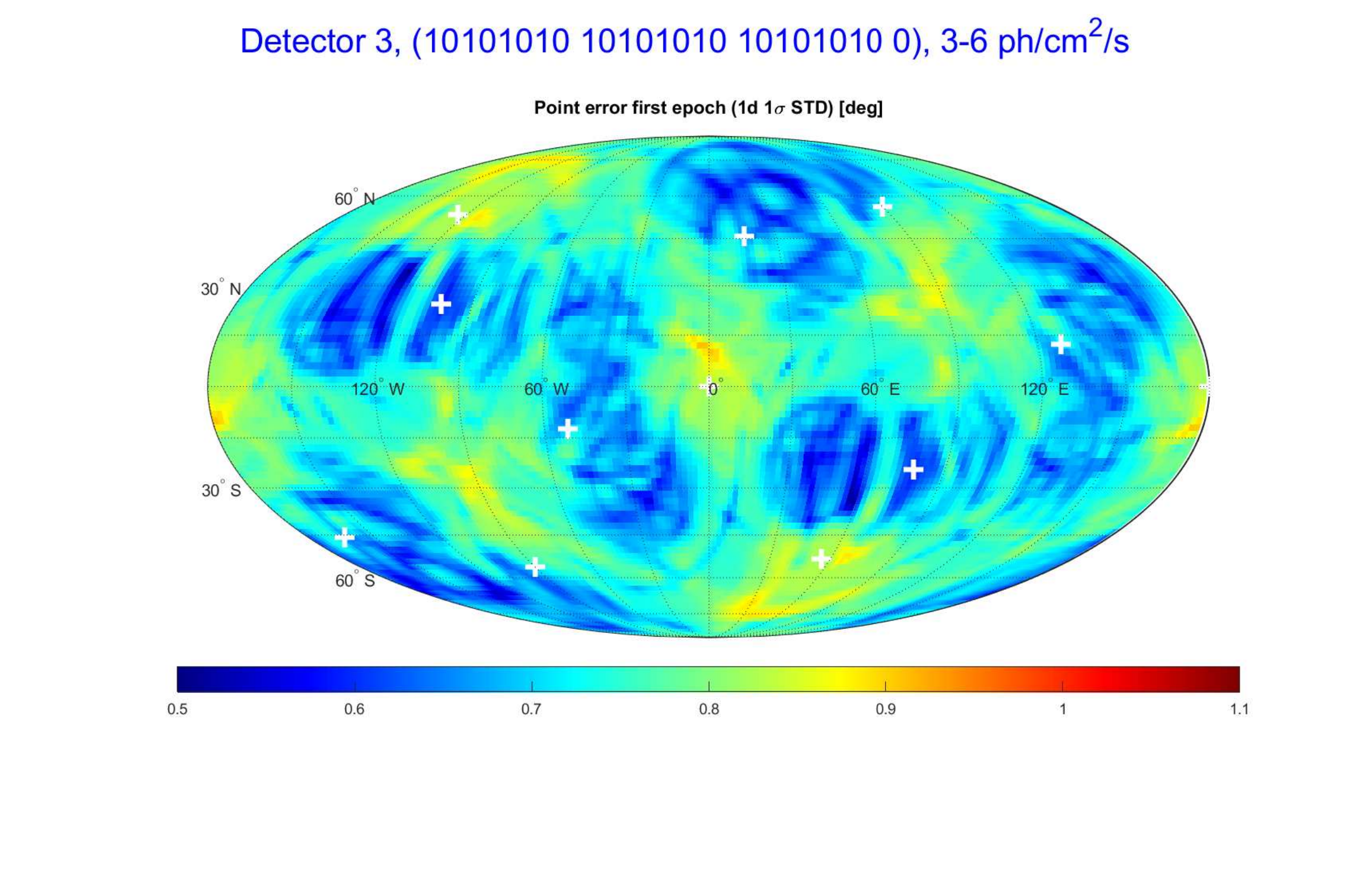}
  \caption[]{Localisation accuracy for a 4-side detector each on
    12 satellites for the first time slice, and for the second highest
    intensity interval
    for unbinned (left) and binned (right) ``accuracy matrix'', i.e.
    when the 6\,ms binned matrix performs better than the 3\,ms matrix.
    This happens for a certain intensity range where the increased S/N-ratio
    overcompensates the reduced temporal resolution of the detector.
    For yet fainter intensity intervals, the same happens for 9\,ms, and so on.
    This holds for any number of satellites equipped with a GRB detector.
    \label{comp_merged}}
\end{figure*}

\begin{figure*}[h]
   \includegraphics[width=0.5\textwidth, viewport=80 90 780 515, clip]{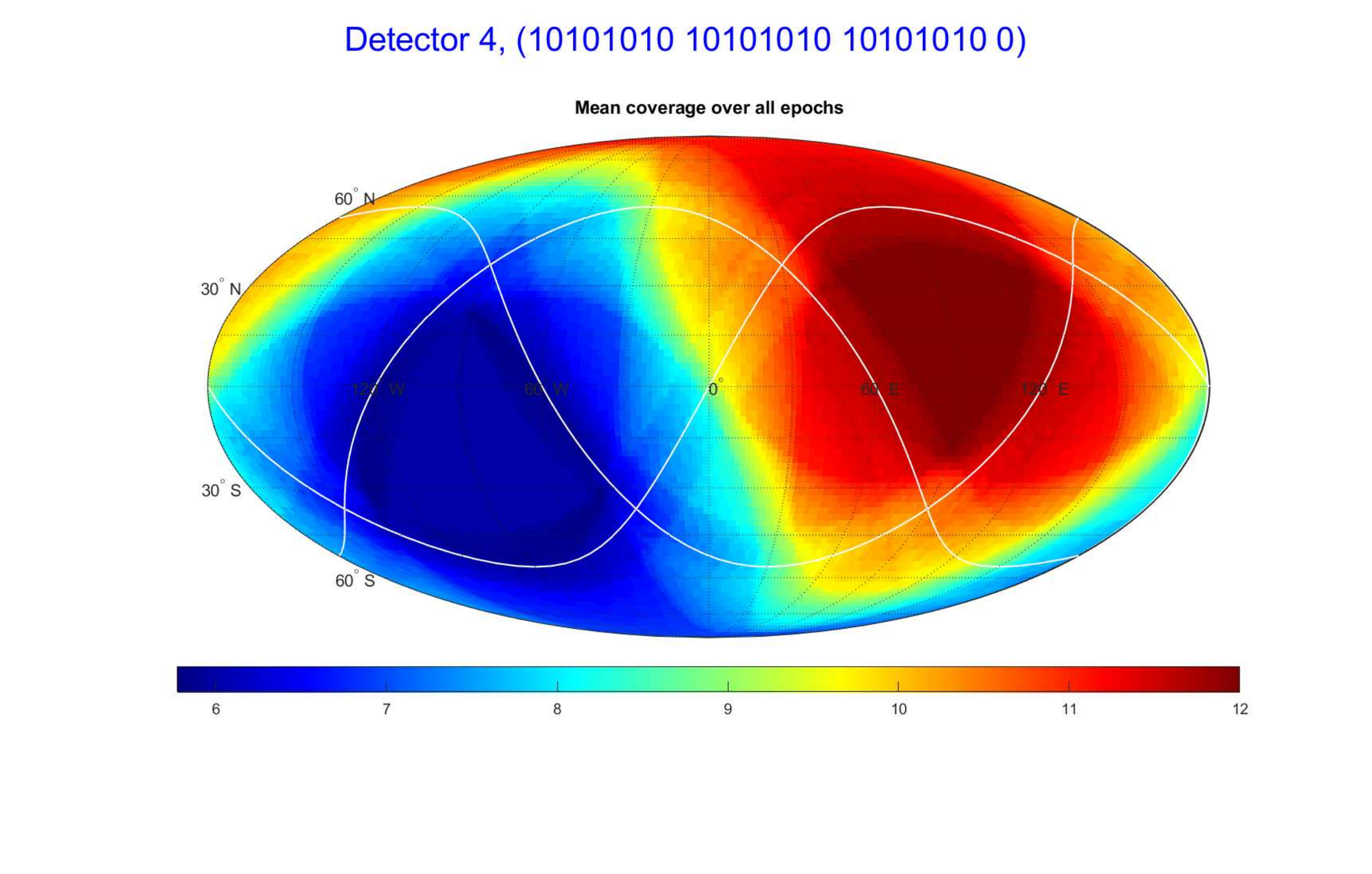}
   \includegraphics[width=0.5\textwidth, viewport=80 90 780 515, clip]{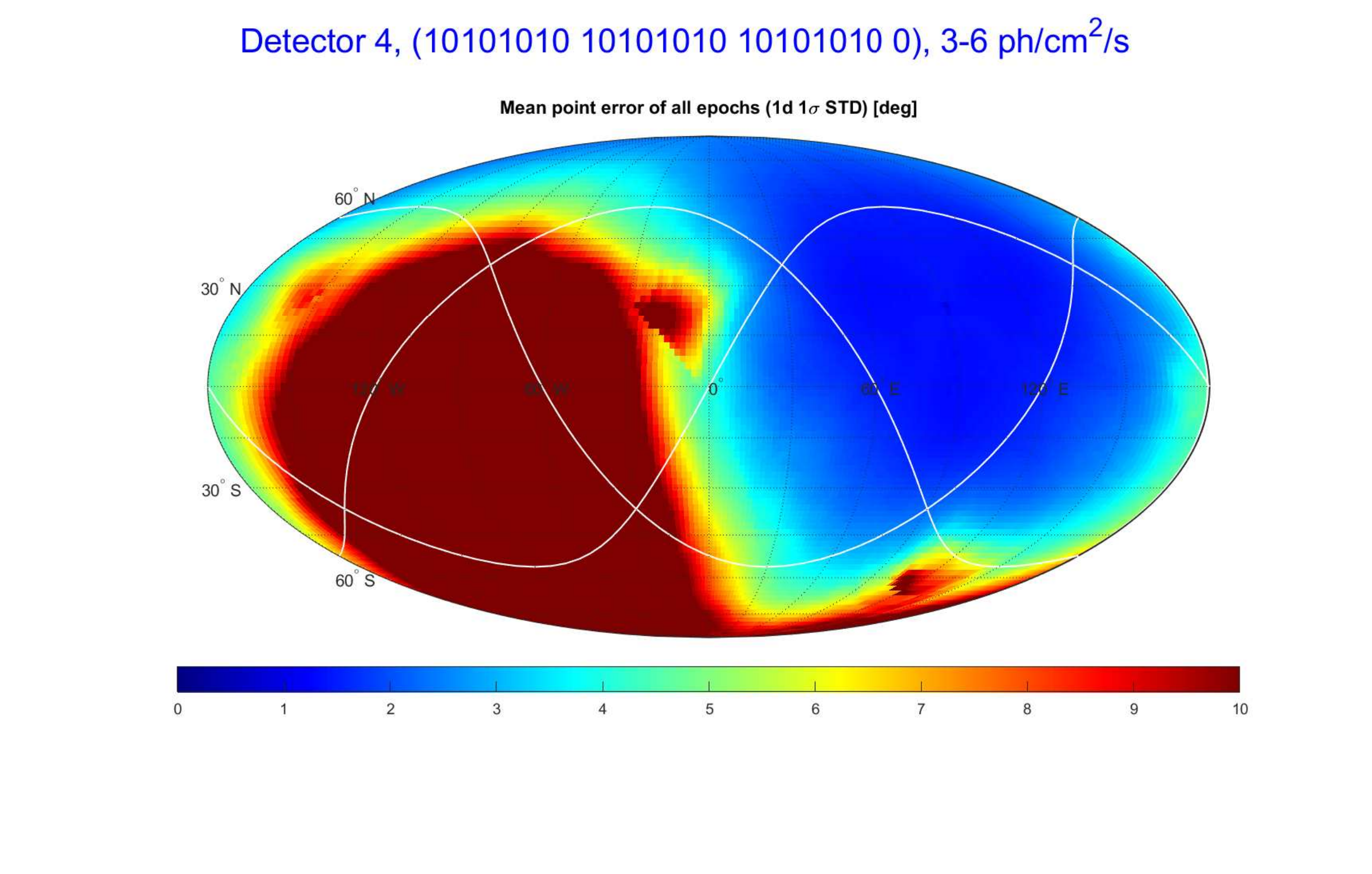}
  \caption[]{Sky coverage (left) and localisation accuracy (right; for the
    second-brightest GRB intensity interval) of
    12 satellites equipped with two detector plates on neighboring sides.
    The sky coverage  is substantially
    worse than any previous detector geometry.
    \label{Det04}}
\end{figure*}

In Fig. \ref{comp_merged} we show the effect of the so-called
"merged" configuration, i.e. the temporal re-binning to 
6\,ms whenever the 3\,ms sampling combined with the small baselines
performs worse. This is best shown with a single time slice, not
the orbit-averaged accuracy plot. The re-binning improves the
bad localisation accuracy regions (red in the left panel of
Fig. \ref{comp_merged}) by about 20\% (from 1\fdg1 to about 0\fdg9).

\paragraph{Detector 4}

With the intention to minimize the number of detector plates
on a given Galileo satellite, we included this geometry with only
two instead of 4 sides equipped with a 60\,cm x 60\,cm detector plate.
The simulations show that the sky coverage is substantially
worse (Fig. \ref{Det04}), which is a consequence of the ``eyes'' problem,
i.e. that the detector on the y-side (together with a solar
panel boom) will never look towards the Sun.
In addition, as a consequence of the yaw-steering attitude, the detector
mounted on the +X surface always looks into the hemisphere containing the Sun,
i.e., the direction towards the anti-Sun is not covered by any detector
on any satellite. The sky coverage and the localization precision thus
dramatically degrade towards this direction.

\begin{figure*}[h]
  \includegraphics[width=0.5\textwidth, viewport=80 90 780 515, clip]{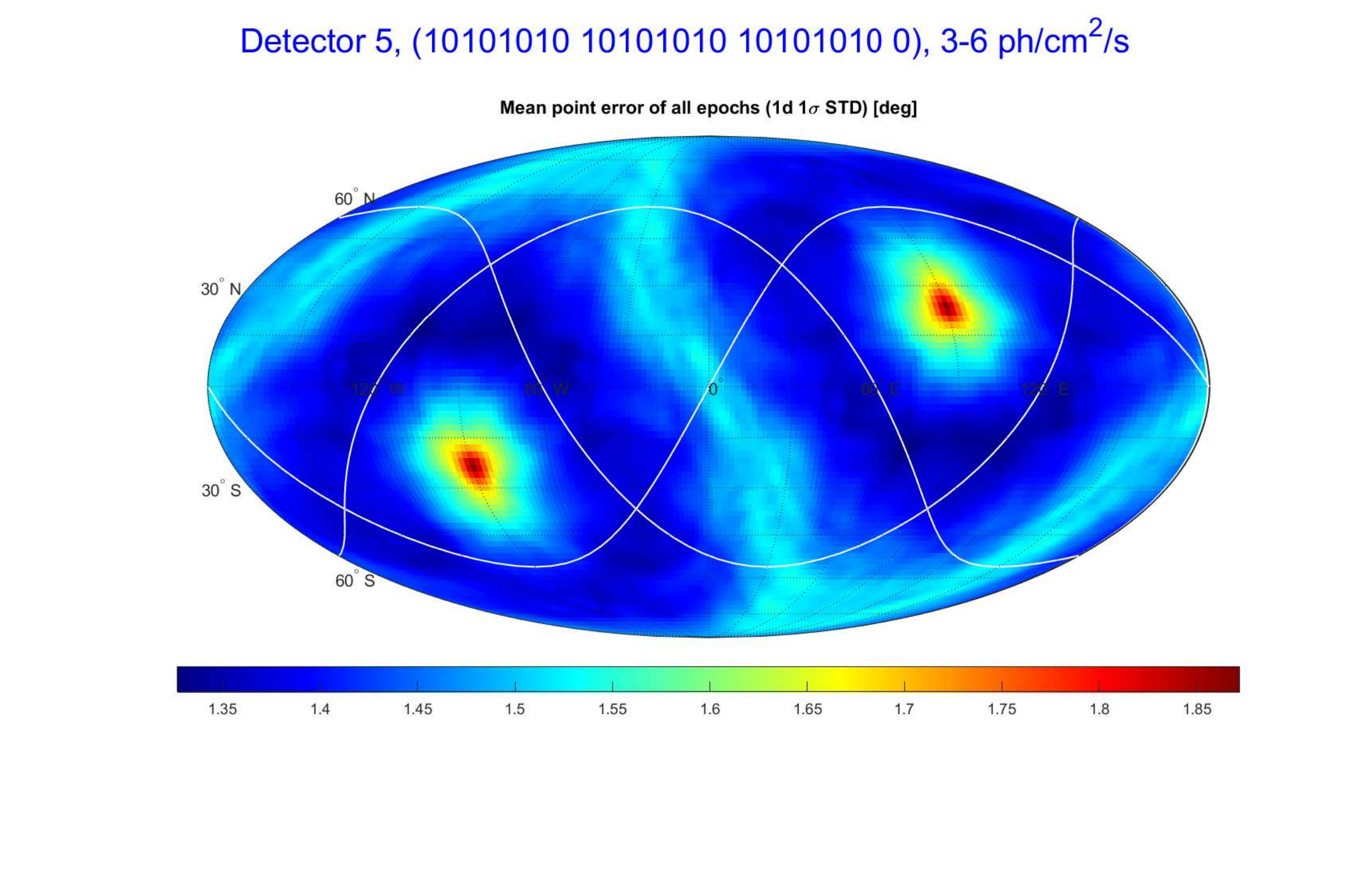}
   \includegraphics[width=0.5\textwidth, viewport=80 90 780 515, clip]{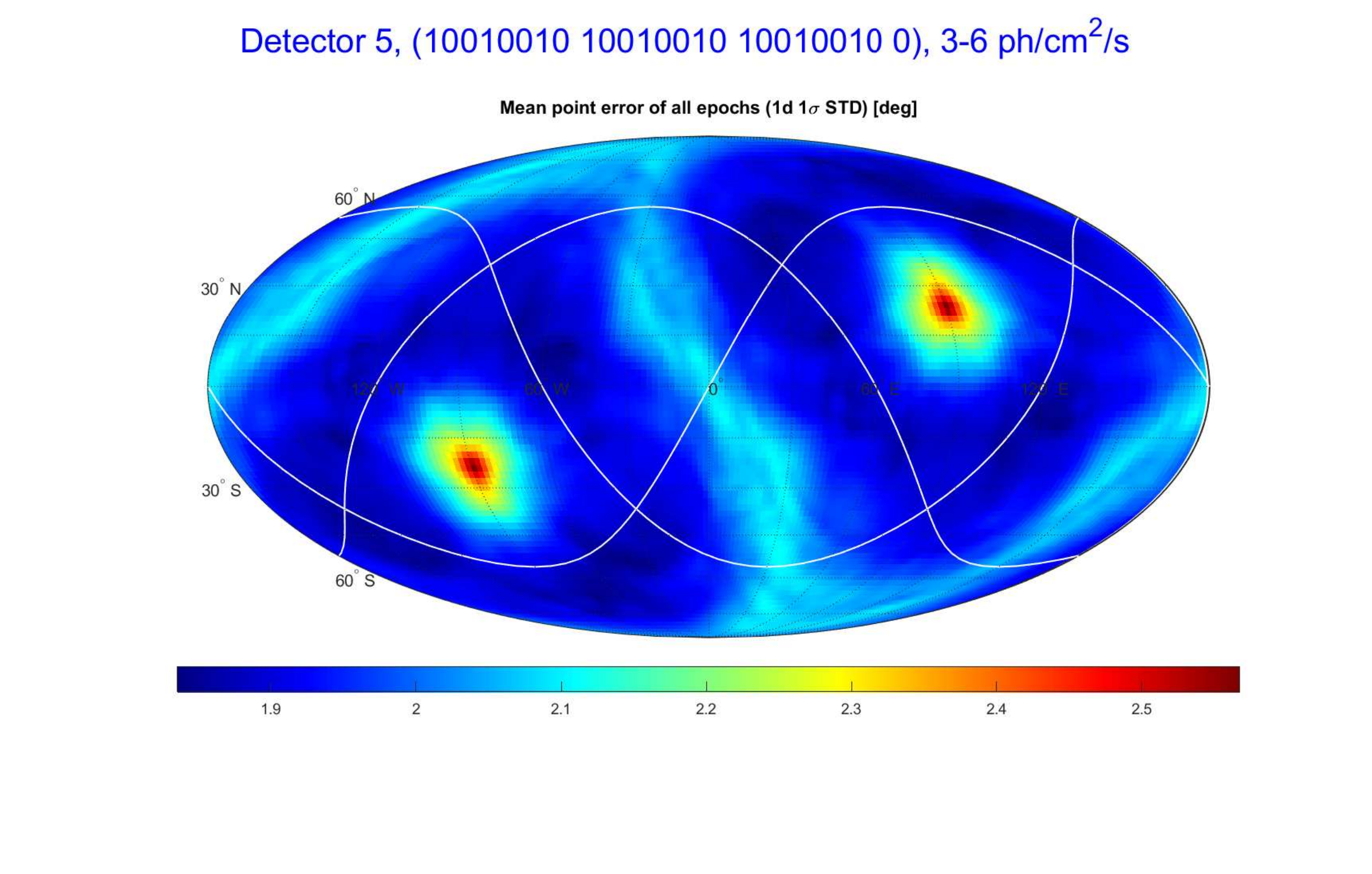}
   \caption[]{Localisation accuracy of 12 (left) and 9 (right)
    satellites equipped with 4 lateral and a zenith-looking detector, 
    for the second brightest GRB intensity interval).
    \label{Det05}}
\end{figure*}

\begin{figure*}[h]
   \includegraphics[width=0.49\textwidth, viewport=80 90 780 515, clip]{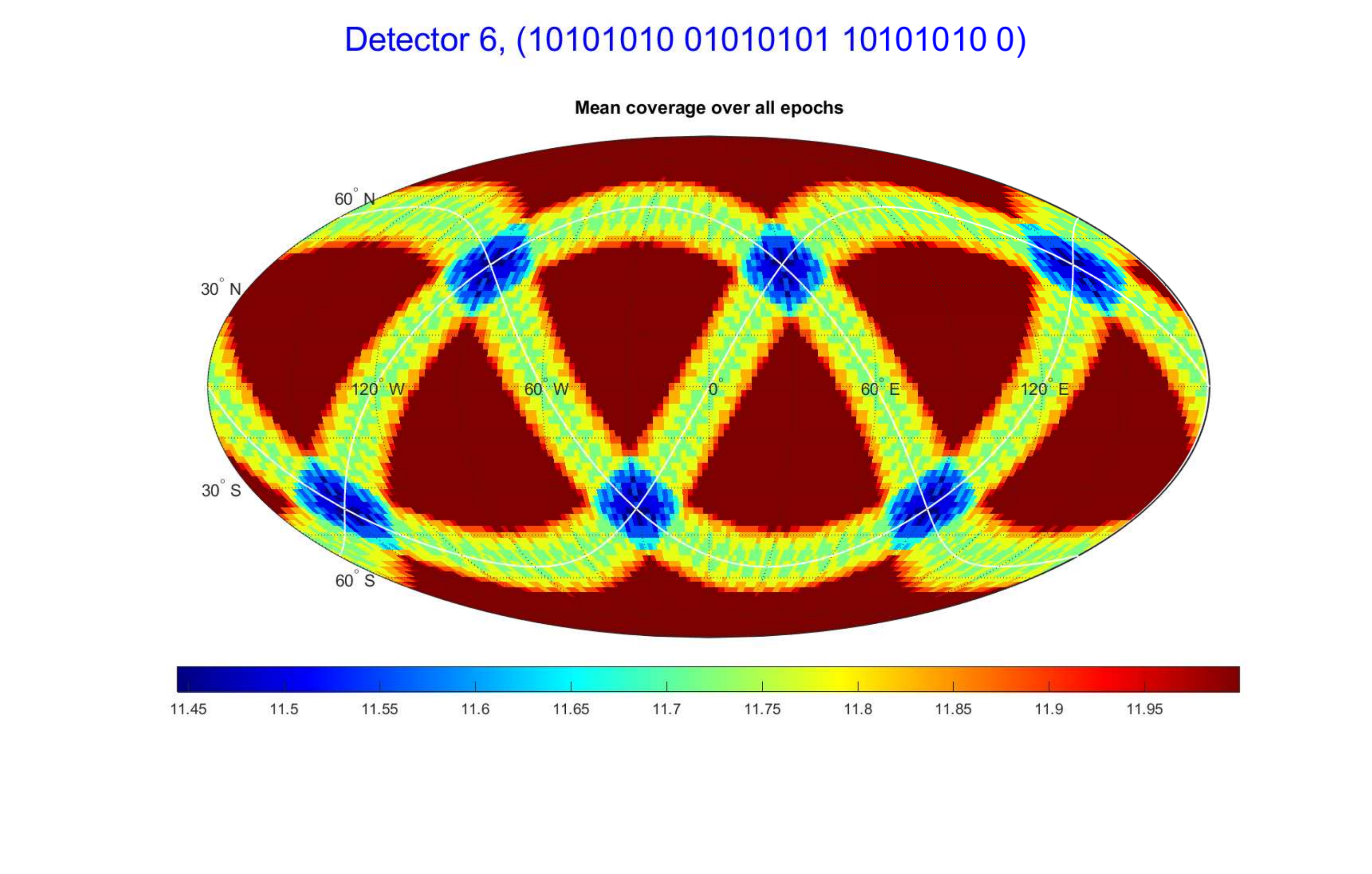}
   \includegraphics[width=0.49\textwidth, viewport=80 90 780 515, clip]{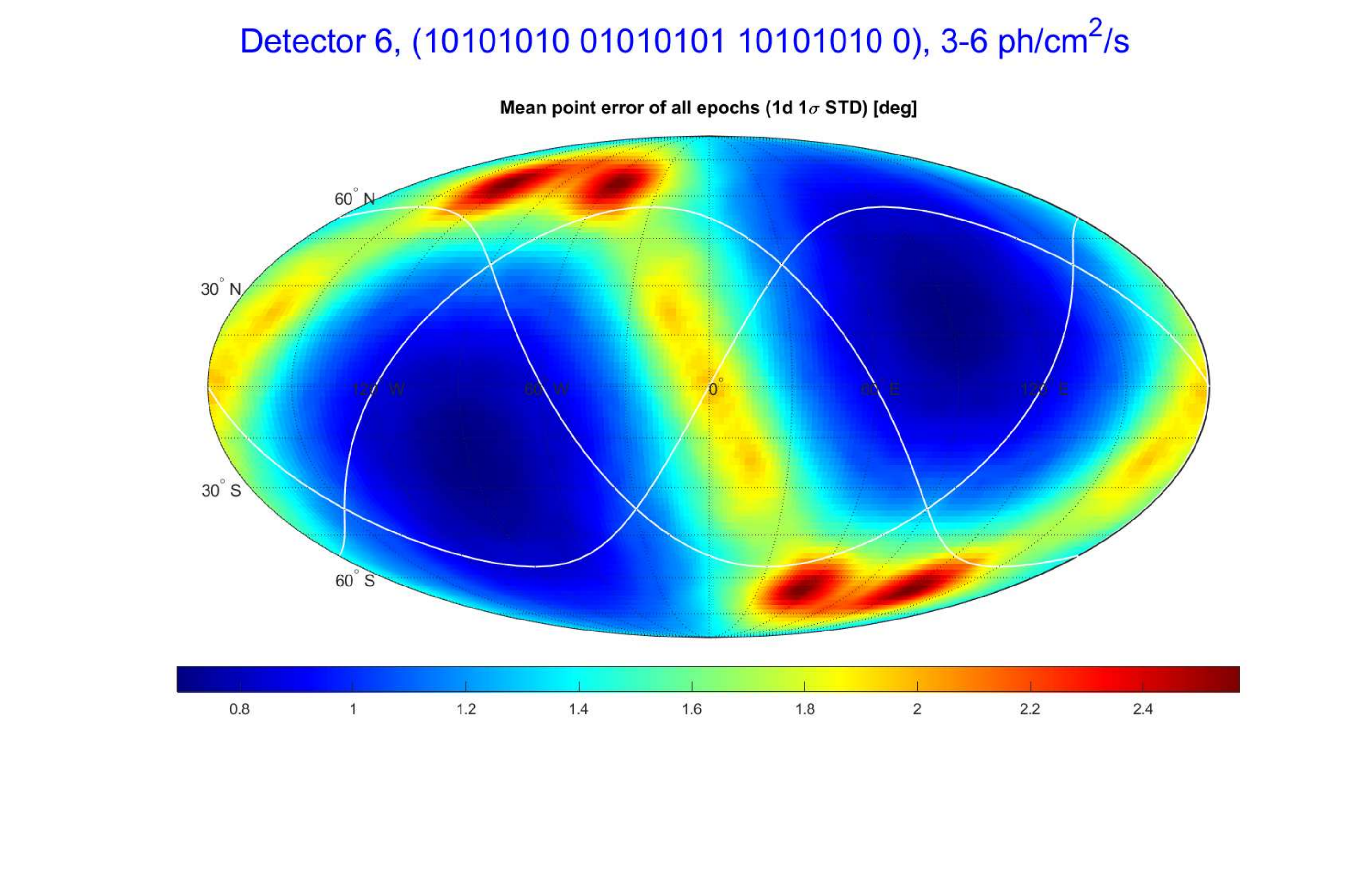}
  \caption[]{Sky coverage (left) and localisation accuracy (right; for the
    second brightest GRB intensity interval) of 12 satellites
    equipped with two opposite plus one zenith-looking detectors.
    \label{Det06}}
\end{figure*}

\paragraph{Detector 5}

\begin{figure}[h]
  \centering
  \includegraphics[width=0.9\columnwidth]{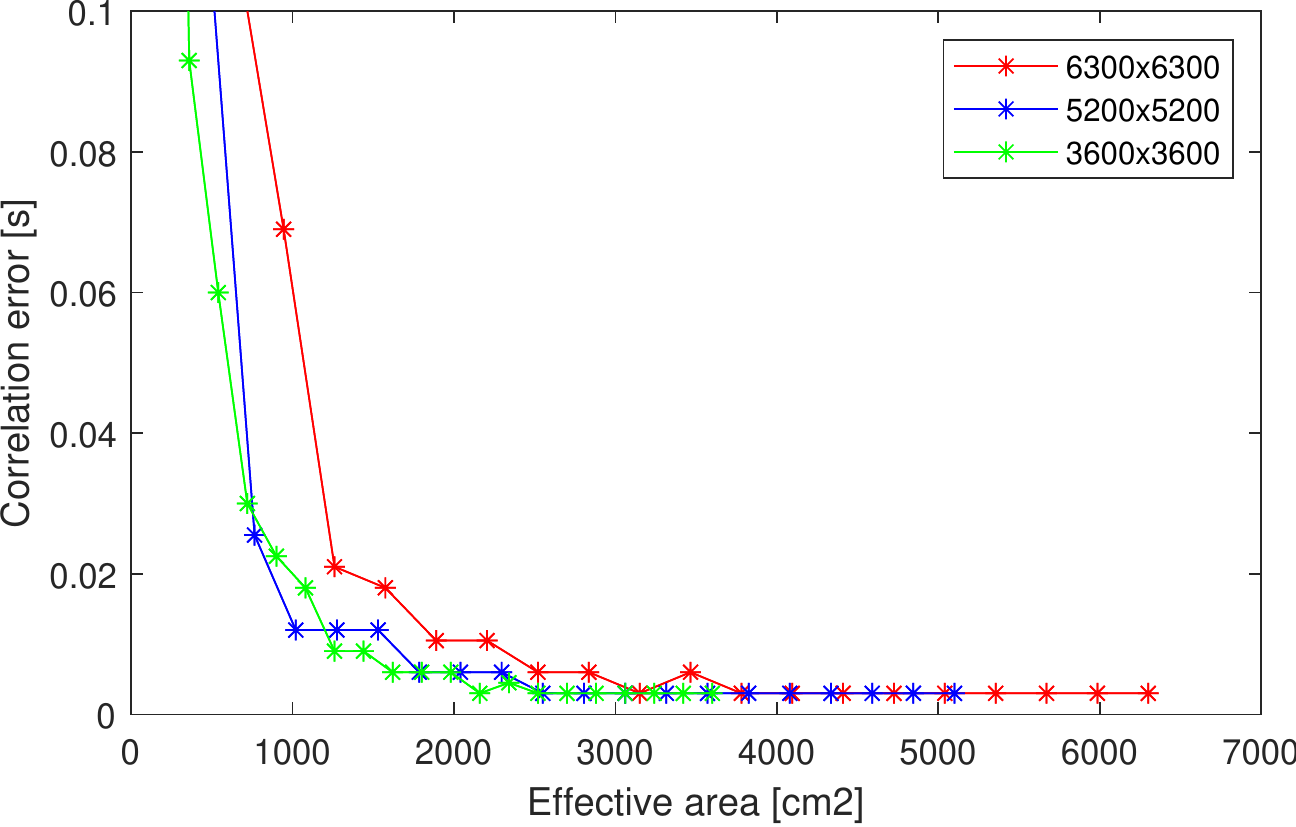}
  \caption[]{Accuracy of different detector configurations: As one
    adds perpendicular-oriented detector plates, and since the
    background radiation is isotropic and does not scale with
    the cosine of the incidence angle, the S/N-ratio depends on
    the relative inclination angles of the plates: it is best
    for a single plate (green), and gets worse for two plates
    inclined by 90 degrees (blue), and even worse for three
    plates inclined by 90 degrees (red; corresponding to the
    5-detector case for any given GRB).
        \label{comp_multi}}
\end{figure}

This is a kind of 'maximum detector' concept per satellite,
and unsurprisingly, the performance is very good (Fig. \ref{Det05}).
However, we see (Tab. \ref{tab:Det_Sat_acc}) 
that it performs slightly worse than the 4-lateral-only
detector geometry for fainter GRB intensity levels.
This is likely due to the fact that using detectors at large
inclination angles towards the GRB does not help in improving
the S/N-ratio, since co-adding the background noise of the
second (or third) plate dominates over the gain in signal.
Fig. \ref{comp_multi} shows the effect for a single plate,
and the sum of two and three perpendicular-oriented plates:
at large inclination angles, i.e. small effective area
due to the cosine effect, the S/N after combining detectors does not improve.
This calls for an optimization of the co-adding of signals from
multiple detector plates: it should not be performed on the satellite,
but on the ground, as it depends on the actual noise level
for each satellite (which we expect to vary along the orbit).
Then, cut-off angles can be applied
above which no co-addition is done.
Also for this detector we find 'eyes' in the spatial distribution of
localization precision caused by the fact that maximum 2 instead of 3
of the detector plates for any satellite can cover the Sun and anti-Sun
directions.

\paragraph{Detector 6}

This three-element option, which leaves the sides with the solar panels free,
eliminates the bad localization performance in Sun and anti-Sun directions
(``eyes'' above). However, the  Sun-equator is less well covered
(Fig. \ref{Det06}).
Otherwise, it provides a very uniform localisation capability over the sky,
at substantially improved accuracy as compared to the case of
two neighboring detectors.


\paragraph{Detectors 7--9}

Detectors \#07, \#08 and \#09 were included just for completeness
and verification purposes, and the results are given in the overview
plot of Fig. \ref{24sat_allDet}.

\paragraph{Summary of detector geometries}

Tab. \ref{tab:Det_Sat_acc} summarizes the different detector
geometries and satellite constellations considered, providing the
all-sky averaged accuracy for each of the 4 GRB intensity intervals
of Tab. \ref{tab:12Det_det1_acc}.

Thus, and quite obviously, the localisation accuracy
improves with the number of satellites equipped, since among
the detectors seeing a GRB, there is a larger likelihood of
having satellite pairs with a large distance (baseline):
only those are the ones
which improve the localisation accuracy.

The placement of detectors on satellites positioned opposite to each other
in the orbital plane causes moving patterns in the sky with reduced
localization accuracy, and thus should be avoided; this applies primarily for
low equipment rates, e.g. the 6- and 9- satellite versions discussed above. 

An interesting feature is seen in the case with 
6 satellites: when the GRB detectors are distributed isotropically,
i.e. 2 per orbit at antipodal positions, there is a pattern on the
sky at which the localisation is substantially worse (top left panel
in Fig. \ref{comp_NumSat}). This can be avoided by placing detectors
not in antipodal positions (next panel to the right in Fig. \ref{comp_NumSat}).

One special effect to comment on are the two ``eyes'' in
Fig. \ref{comp_NumSat} (lower row).
These are due to the position of the Sun in the simulation
($\alpha$ = 90\degs, $\delta$ = 23\degs) and the anti-Sun direction,
and in practice would move over the sky over the course of a year.
These are caused
by the yaw-steering motion of the satellite guaranteeing pointing of the
navigation antenna continuously to the Earth and the solar panels to the Sun.
As a consequence of this attitude mode the +Y and-Y surfaces of the satellite
where the solar panel booms are mounted never look towards the Sun.
The Sun and anti-Sun directions are thus covered only by one detector plate
per satellite with varying orientation towards the Sun. The result is a
reduced localization precision in these directions.
For the GRB/GW application this is acceptable,
since optical follow-up of the GRB or neutron star merger close to the Sun
is anyway not possible from the ground.

\begin{figure}[h]
  \vspace{-0.22cm}
  \hspace{-0.05cm}\includegraphics[width=0.52\textwidth]{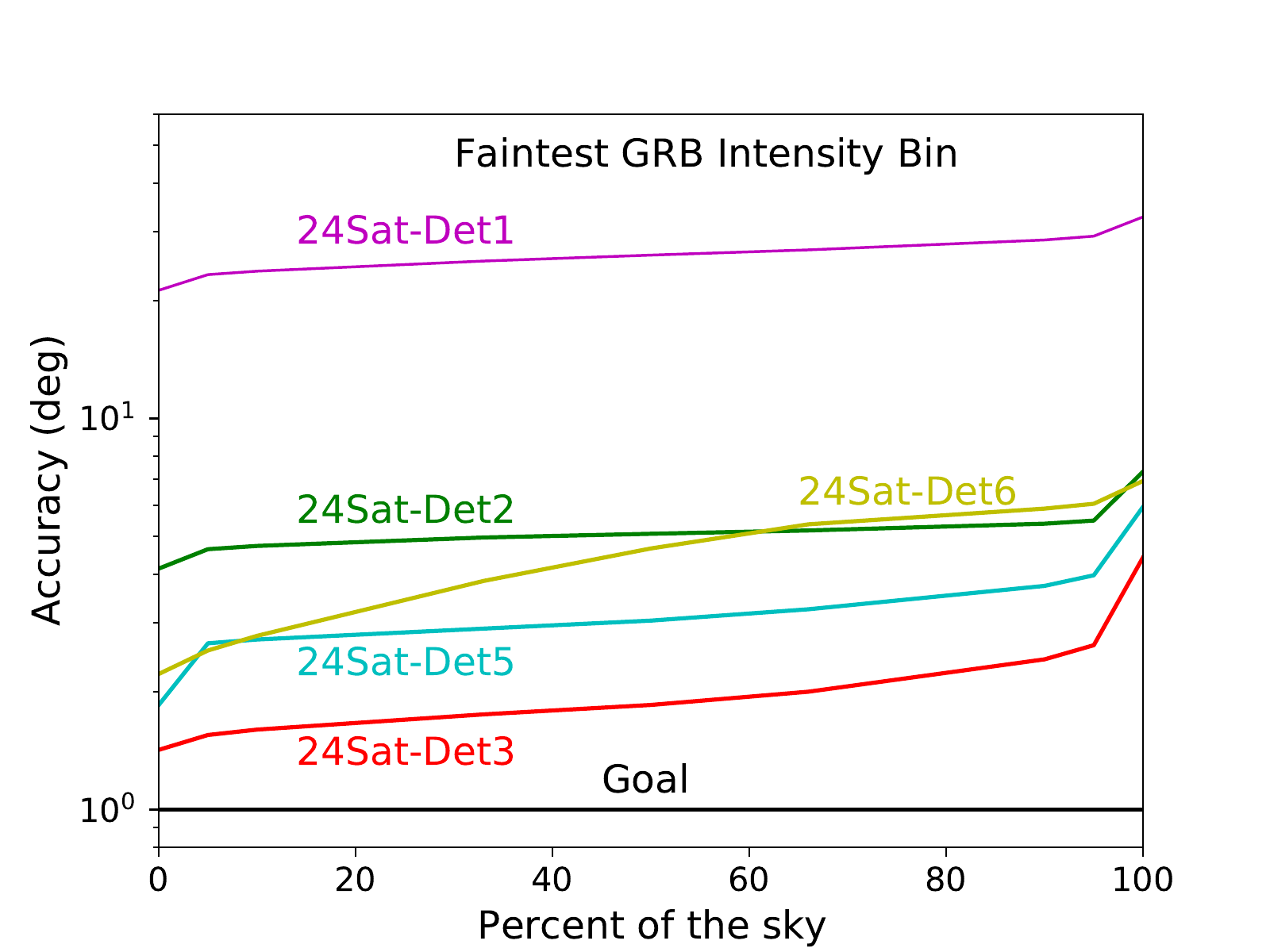}
  \hspace{-0.18cm}\includegraphics[width=0.52\textwidth]{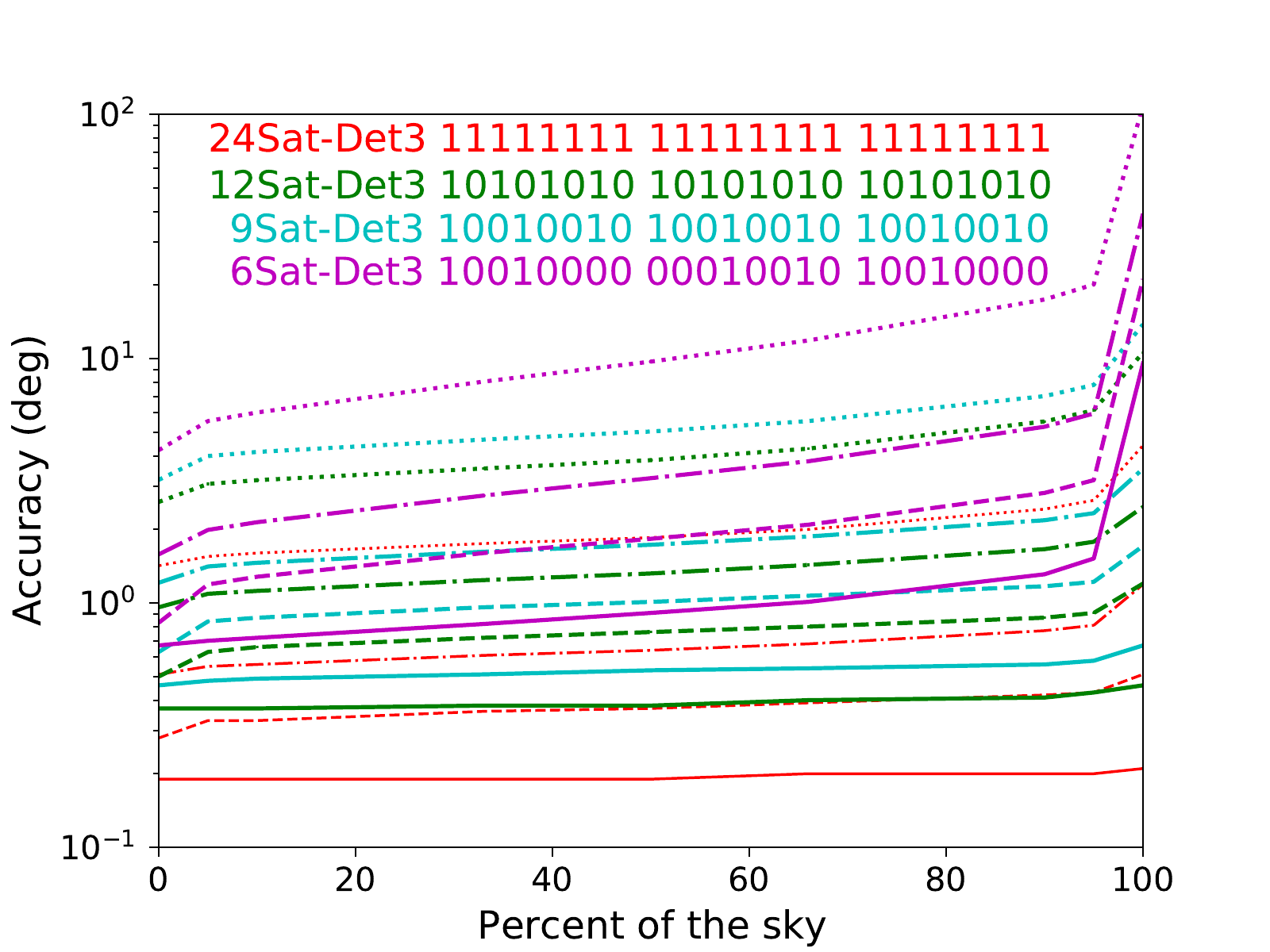}
  \vspace{-0.22cm}
  \caption[]{Accuracy of different detector and satellite configurations
    per sky fraction. Each curve shows the percentage of the sky for which the accuracy
    is better than the corresponding  y-axis in degrees.
    {\bf Top:} for the faintest intensity bin for 24 satellite equipment rate;
    {\bf Bottom: } each color represents one configuration of
    Tab. \ref{tab:acc_final} for detector \#3, with solid lines for the
    brightest GRB interval, and dotted lines for the faintest. 
        \label{acc_persky}}
\end{figure}

For obtaining the localisation accuracy in Tab. \ref{tab:Det_Sat_acc},
we averaged over the full sky.
But looking at the various figures, it becomes obvious that
there are certain small (few to 10\% of the sky) regions on the sky
which are worse than the majority of the sky.
We therefore provide a more accurate accounting of the
localisation accuracy for our best options in Tab. \ref{tab:acc_final}.
This provides the worst accuracy for the best 50\% and 90\% of the sky
(i.e. the accuracy is better than the specified value for that percentage of the sky),
respectively, as well as the best and the worst single GRB accuracy of the sky. 
For the selected best detector and satellite configurations,
we also provide a graphical representation in Fig. \ref{acc_persky}
which allows to get the accuracy for any fraction of sky coverage.

\subsection{Bayesian scheme using nazgul}

\begin{figure*}[h]
  \includegraphics[width=0.49\textwidth, viewport=90 90 750 520, clip]{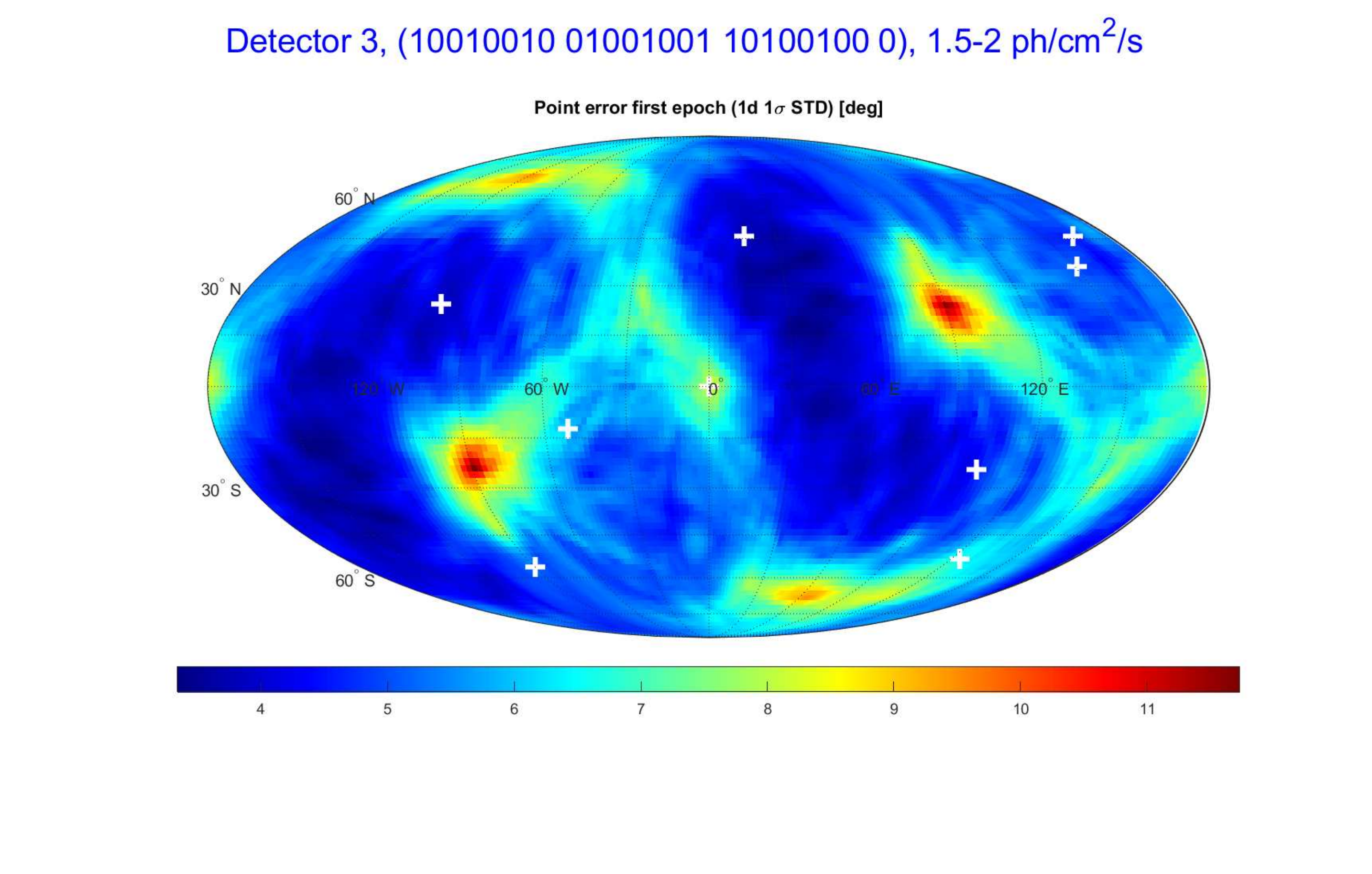}
  \includegraphics[width=0.55\textwidth]{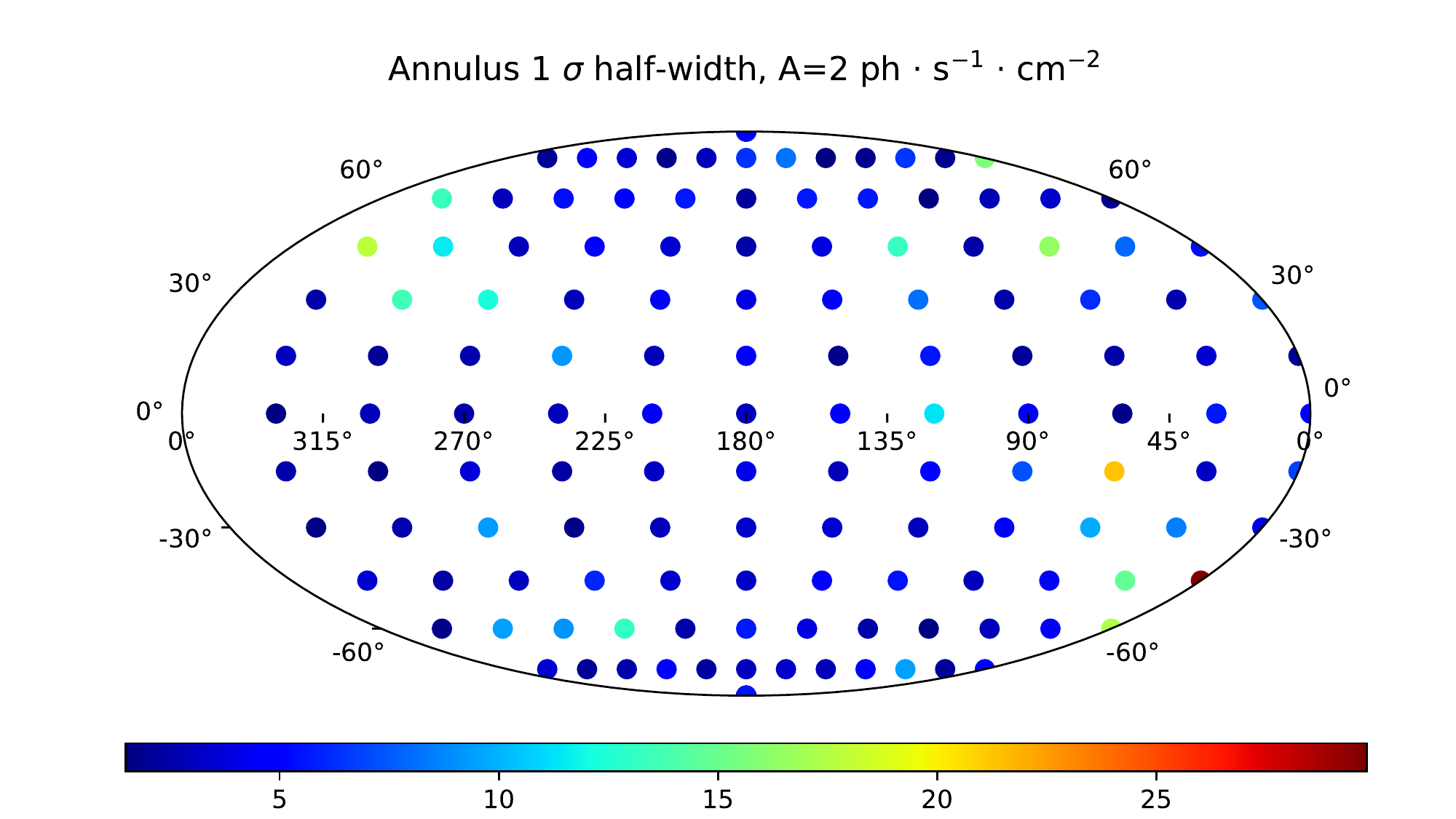}
  \caption[]{Localisation accuracy for detector \#03 (4 sides) and 9 satellites
    for the first of 72 snapshot per orbital phase and the faintest
    GRB intensity bin, computed with
    cross-correlation (left) and nazgul (right). Note that the
    Sun is not included in the nazgul
    simulation, and thus the ``eyes'' are missing.
    \label{nazgulcomp}}
\end{figure*}

Due to the massive compute-time requirements, a simulation with
nazgul was only done for one particular satellite constellation
(9 satellites, 3 in each orbital plane, equally distributed)
with one detector (\#03, looking towards 4 sides).
We use the same set-up as the one to reconstruct the time
with the cross-correlation algorithm.
Instead of 1000 different GRB light curves, we use only one light curve shape,
with 5 different flux normalization.
Also, the triangulation was only computed at 134
sky positions, instead of 10000.
From each fit we obtain a distribution of the time delay which is used
to compute both the ``best'' fit value (the median in this case)
and the 68\% probability uncertainties through the highest posterior
density interval, i.e. the shortest possible interval, necessary
to accumulate the chosen probability level.  
While the source position distribution reconstructed by nazgul is
not, in general, an annulus, for the sake of a straightforward comparison
with the classical correlation method we compute
an ``equivalent'' annulus from the fitted time delay. 
The central ring of the annulus is computed from the median of
the time delay distribution, while the width is given by the uncertainties
in an analogous way as what is done for the correlation method
\citep[see, e.g.,][]{Palshin+2013}.
This methodology, although to some extent simplistic, allows us to
compare the characteristic widths of the positional distributions
fitted by nazgul and the correlation algorithm.
The corresponding 'map' is shown in Fig. \ref{nazgulcomp}
for the faintest intensity interval, together
with the corresponding map from the cross-correlation method.
This shows, that the two methods are nicely compatible to each other.

\begin{figure}[h]
\includegraphics[width=0.5\textwidth]{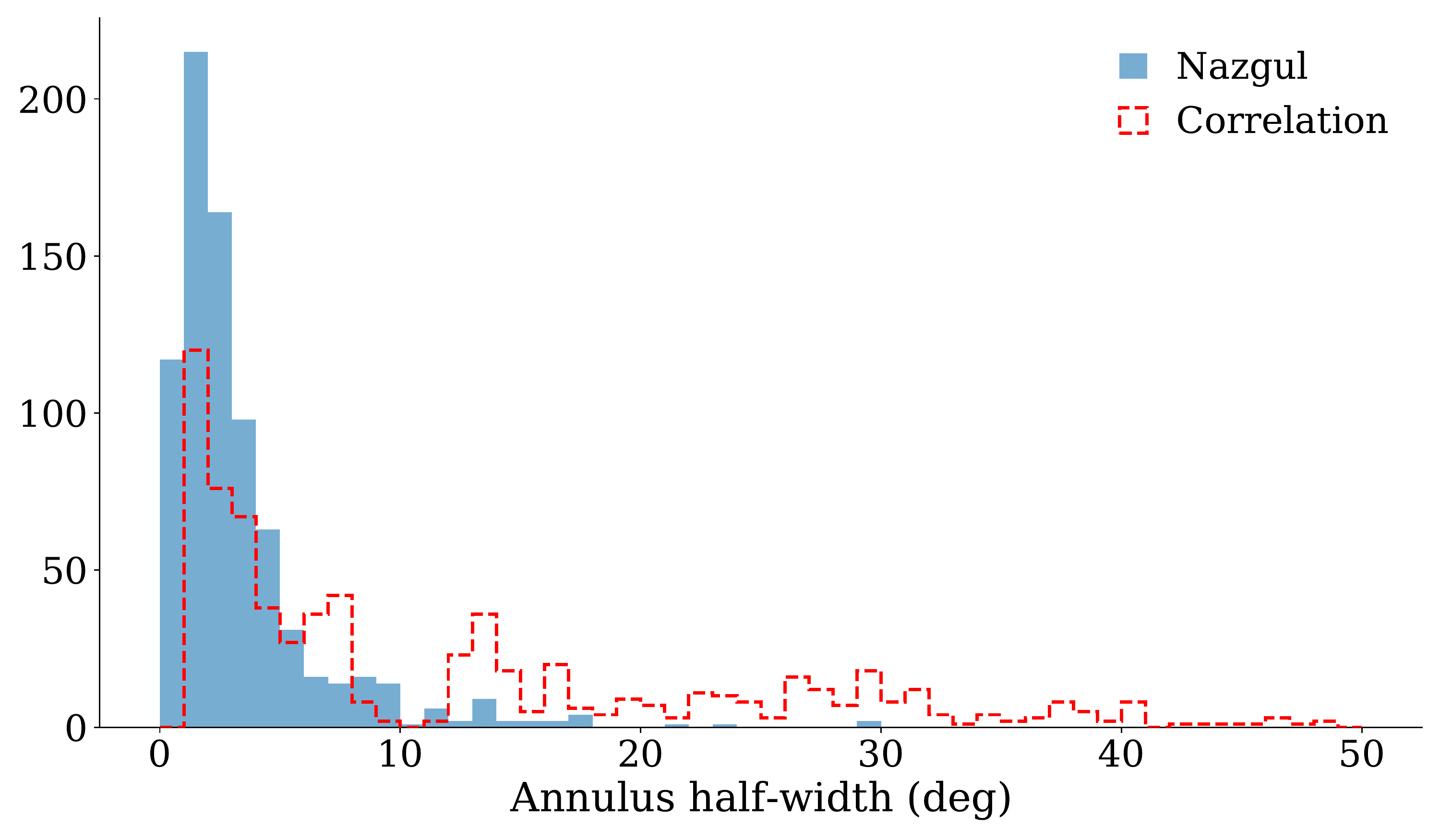}
\caption[]{Distribution of the absolute value of the difference between
    simulated and reconstructed time delays for nazgul (blue) and the
    cross-correlation method (green), again for
    detector \#03 (4 sides) and 9 satellites.
   \label{nazgulcchist}}
\end{figure}

A more quantitative comparison of the localisation accuracy is
given in Fig. \ref{nazgulcchist}, showing the histogram
of the 1$\sigma$ localization errors of nazgul vs. the cross-correlation
method. This shows,
that the nazgul distribution is a factor $\sim$2 narrower
(FWHM of about 4\degs\ vs 8\degs),
and has much less GRB reconstructions in the long tail.
Thus, the nazgul method leads to overall improvements, but is particularly
superior at the faint end of the intensity distribution.

\subsection{Comparison to previous simulations}

Recently, \cite{Hurley2020} has combined a new localisation method
with the simulation of a near-Earth network of GRB detectors.
The basic concept of this method is similar to ours, namely avoiding
cross-correlation and instead testing positions on the sky via a likelihood
method.  While this method is a substantial
improvement over the classical cross-correlation, it still suffers
from the above draw-backs (iii) and (iv) in sect. \ref{triangulation} which
is accounted for in our nazgul scheme \citep{Burgess+2021}.
In his simulations, he uses individual detectors of 100 cm$^2$ effective
area on a fleet of nine satellites, and derives localisation accuracies
for three different GRB peak intensities. His faintest and middle intensity
intervals fall in our brightest interval. In terms of sky coverage,
\cite{Hurley2020} reaches only 40\%, so a single-plate detector on each of
9 satellites is by far too little to reach all-time, all-sky coverage.
While we have not simulated such a constellation, this result is consistent
with our picture, i.e. the need to look to multiple sides at this small
satellite number. The 1$\sigma$ average localisation for his faintest
fluence GRB (16 ph/cm$^2$/s) is an ellipse with a
dimension of 4\fdg5$\times$17\fdg0, corresponding to an effective radius
(same-area circle) of 4\fdg5 (1$\sigma$). Our closest constellation
is a 1-side (zenith)
looking detector on 12 satellites, where our simulation for our brightest
intensity interval (6--100 ph/cm$^2$/s) gives 2\fdg9.
The difference between these two simulations is in effective area
(100 vs. 3600 cm$^2$), orbit radius (7000 vs. 29.000 km), and
timing accuracy (0.1 vs. 3 ms). Assuming the typical square-root
dependence on effective area, the combination of these three factors
suggests that our error should be $1 / \sqrt{36} / 29 * 7 * 30 =  1.2\times$
that of the \cite{Hurley2020} simulation, pretty close to what we obtain
(given that our intensity bin is very wide).

\subsection{Potential for improvements}

Given the comparison in the previous subsection, one could ask
whether or not reducing the time resolution in our simulations (fixed at 3\,ms)
would substantially improve our localisation accuracy? The answer is:
in theory yes, in practice likely not. The simple reason is that
the variability time scale in GRBs is at the level of a few milliseconds
\citep{MacLachlan+2013}, and not sub-milliseconds \citep{Walker+2000}.
Thus, one needs to cross-correlate the rising edge of a pulse to better
then the rise time. This is complicated even further due to the fact that the
slopes of the rise are energy dependent, i.e. detectors like the here preferred
scintillator plates will see different slopes from the same GRB as soon as
the incidence angles on two detectors are not exactly the same. This is the
reason why for instance co-adding of light curves of different GBM detectors
is of no avail.

For configurations with more than one detector plate per satellite,
we have co-added multiple single-plate detectors per
satellite. As described earlier, this does not automatically
provide better S/N-ratio, since the background radiation adds in full,
not diminished with the cosine law as the source counts. Therefore,
some optimization for adding two or three detector plates could
be implemented.
This affects mostly the faint end of the GRB intensity distribution,
and the optimization is expected to improve the localization accuracy.
On a practical note, the data from different detector plates should thus
not be combined onboard, but send down to Earth separately.
As a side effect, the onboard triggering
algorithm can make use of the separate light curve measurements to
filter out particle hits, thus dramatically reducing false triggers.

There is yet another way to improve the accuracy distribution, computed
via the classical cross-correlation analysis, beyond what we presented: namely
using systematically a re-binning procedure (in an frequentist
approach, and the corresponding $\chi^2$ analysis). In practice,
when moving from bright to fainter GRBs, there is a transition
region where re-binning from the original 3\,ms time resolution
of the detector towards, e.g., 6\,ms provides a gain over the
noise fluctuations, and leads to an improved localisation.
This has been shown above with our 'merged' map for one case
(Fig. \ref{comp_merged}).
But moving further down in intensity, the same happens for
further re-binning to 9\,ms, or 12\,ms, and so on.
The effect of this re-binning
is to optimize between noise in the light curve vs. the best
accuracy in the time delay measurement. 
Obviously, it does not improve on the
best accuracy side, but improves the bad end by of order 15\%-20\%.
This, of course, does not apply to our forward-folding  nazgul
localisation, since this is Bayesian, and the information in
``low S/N'' bins is properly accounted for.

\subsection{Inclusion of satellites beyond GNSS}

The inclusion of any satellite further out in space than the
Galileo satellites would help reducing the localisation accuracy,
as it shrinks linearly with the increase of the baseline.
Potential options are a GRB detector on
(i) the Gateway\footnote{\url{https://en.wikipedia.org/wiki/Lunar_Gateway}},
a multi-purpose space station in a highly elliptical (3000 km x 70.000 km)
seven-day near-rectilinear halo orbit  around the Moon,
presently planned to be assembled in the 2024--2028 timeframe, or
(ii) the Moon LCS (Lunar navigation and communication system),
a network of 3--4 satellites that would provide communications
and navigation services to support human and robotic exploration
on the Moon.
A GRB detector in Lunar orbit would reduce the localisation error
by a factor of $\sim$6, if the GRB detector has the same size
as discussed here for the GNSS.
Of course, this improvement would only apply in one dimension of the
error box, for GRBs coming from a direction perpendicular to the Earth-Moon
line.

\section{GNSS system requirements}

\subsection{Communication speed}

The GRB afterglow brightness fades by a
  factor of three during the first 10 min. after the burst,
  another factor of three during the next 50 min., and another
  factor of three during the next 23 hrs. The kilonova emission
  of short GRBs decays even faster. Moreover, clarifying
  the presently hottest open astrophysics questions of merging
  neutron stars such as distinguishing
  the physical source of energy input (e.g. from the central remnant
  or via radioactivity) or associated processes (e.g. internal
  shock-reheating or heating of the outer ejecta by free neutrons)
  requires ground-based optical/near-infrared spectroscopy during
  the first 12 hrs \citep{Metzger2020}. Thus, rapid communication
  at a timescale of minutes is required in order to support
  the identification of the kilonova.

GRBs occur at unpredictable time and sky position.
For the GRB position to be determined via triangulation,
we need the measured data of each of the $>$4 satellite detectors 
on one computer. In order to be scientifically useful,
the data should be downlinked within of order a few minutes.
Thus, we require that at any time every
Galileo satellite can send off its measured data, either
directly or via another satellite to a ground-station.
Since only
6 TT\&C stations around the world  are responsible for collecting
and sending the telemetry data that was generated by the
Galileo satellites,
relaying data between
different Galileo satellites to the one (or few) which
do have ground contact is a viable solution.
This should be done dynamically,
without the need of commanding, i.e. each satellite (computer)
should know at any time its acting relay satellite.

We distinguish two data transmission rates:
(1) full rate to be downlinked to Earth, within minutes: for a typical
GRB, this implies sending 0.5--1 MB over a time period of a few minutes,
e.g. 4--8 kB/s over 2 minutes per satellite.
(2) reduced rate for quick-look localisation.
As described above, this would reduce the data amount by a factor
of 100--1000.

The inter-satellite data transmission rate is likely slower than
a satellite-ground contact rate. Assuming 4 Galileo satellites
without ground-contact shall send data to one other satellite with 
ground-contact leads to a required transfer rate of 4 MB
on a time-scale of minutes. In an ideal case, this can be done in parallel.
If not, the above 4--8 kB/s are a minimum requirement.

Rapid up-link capability is not needed, since the GRB detectors should
be self-triggering.

\subsection{Ground segment}

The light curve data as measured by the multiple Galileo satellites
should be collected at one place on Earth, where the triangulation
(and thus GRB localisation) can be computed.
We suggest that the final localisation is made publicly
available immediately -- the GRB community is using the GCN
(Gamma-ray Burst Coordinate Network) for this since decades,
which would guarantee distribution to every interested user
in the world.
This would typically happen automatically, but oversight through
a, or several, (GRB) astronomer(s) is certainly not a bad idea.
This could be organized via forming a small group of interested
scientists, similar to groups which collaborate in the follow-up
observations of GRBs at optical or radio observatories.
In parallel, also the raw data should be made publicly available
at the shortest possible delay time, to allow other groups
with potential access to other long-baseline GRB data to
use those data. The high-energy mission archive at ESA would
be a logical place, but other satellite data centers
in Europe might be alternative options.

\section{Conclusions}

The GNSS provides a close-to-perfect satellite system for
the localisation of gamma-ray bursts (GRBs) via triangulation.
It provides a very promising
compromise between satellite baselines (not too long to suffer
data transmission restrictions), number of satellites, and 
required size of GRB detectors to reach sub-degree localizations.
It is the combination of detector geometry (to how many of the
six directions of
the Galileo surfaces are the detectors facing) and the number
of satellites to be equipped, which provides a scientifically
useful GRB triangulation network.

Sideways looking detectors are an extremely crucial ingredient.
We suggest to equip at least 12 satellites, four per orbital plane,
with a 4-side (excluding nadir and zenith) looking detector,
each side with 3600 cm$^2$ and 1 cm thickness.
This will provide sub-degree localisation of GRBs, in particular
faint short-duration GRBs such as GRB 170817A,
as expected from binary neutron star mergers
to be routinely measured at a rate of dozens per year in the upcoming
runs of the worldwide gravitational wave detectors.
Instead, a flat, zenith-facing detector provides only 10-20\degs\
localizations. 
Equipping only 9 Galileo satellites with such a GRB detector leads
to a 30\% loss in localisation accuracy, while the 24-satellite solution
improves it by a factor 2.

Such a configuration should be feasible to implement given the moderate
requirements
of mass ($<$20 kg) and power ($\sim$20 W) of a single detector plate
(i.e. $<$80 kg and $\sim$80 W for the 4-side detector)
as compared to the overall budget of a Galileo satellite,
though we note that this corresponds to about 10\% of
the satellite mass.
The realization of such a large-format GRB detector
plate is also technologically feasible: scintillators of the proposed type
have been flown since 40 years (TRL 9), and the Si detectors
for read-out have also seen their first space applications.

Equipping second generation Galileo satellites with GRB detectors
would turn the navigation
constellation into an observatory supporting the research on
fundamental astrophysical and cosmological problems.

\begin{acknowledgements}

JMB acknowledges support from the Alexander von Humboldt foundation.
We are grateful to Dr Javier Ventura-Traveset, Dr Erik Kuulkers, Dr Luis 
Mendes and Dr Francisco Amarillo for  their excellent scientific and 
technical support, as part of the European Space Agency supervision in the 
execution of this research activity.

The work reported in this paper has been partly funded by the EU under a 
contract of the European Space Agency in the frame of the EU Horizon 2020 
Framework Programme for Research and Innovation in Satellite Navigation. 
The view expressed herein can in no way be taken to reflect the official 
opinion of the European Union and/or the European Space Agency. Neither 
the European Union nor the European Space Agency shall be responsible for 
any use that may be made of the information it contains.

\end{acknowledgements}

\begin{appendix}

\onecolumn
  
\section{Implementation of simulated light curves \label{avalanche}}

For a realistic distribution of single vs. multi-pulse light curves,
we implement a pulse avalanche, a linear Markov
process, as proposed by \cite{SternSvensson1996}.
Here, each pulse acts as a parent pulse giving rise to a number of baby
pulses $\mu_b$, sampled from a Poisson distribution
$p_2(\mu_b) = \mu^{-1} {\rm exp}(-\mu_b/\mu)$, with the average number
being $\mu$.
A baby pulse is assumed to be delayed by a time $\Delta$t with respect to
the parent pulse. The probability distribution for the Poisson delay is
parameterized as
$p_3(\Delta t) = (\alpha \tau)^{-1} {\rm exp}(-\Delta t/\alpha \tau)$,
where $\tau$ is the time constant of the baby pulse and $\alpha$
is the delay parameter.
From observed GRBs, the time constant $\tau$ of baby pulses is of the same
order of magnitude, but shorter than the time constant $\tau_1$ of the
parent pulse. This allows the process to converge, since the pulse
avalanche eventually reaches an arbitrary short timescale, where a
natural frequency cutoff should exist.
The corresponding probability distribution is considered to be uniform in
log($\tau/\tau_1$), and parameterized as 
$p_4[{\rm log}(\tau/\tau_1)] = \abs{\delta_2 - \delta_1}^{-1}$
in the range [$\delta_1, \delta_2$] with $\delta_1 < 0$,
$\delta_2$ \gax\ 0, and $\abs{\delta_1} > \abs{\delta_2}$.
The number of spontaneous pulses $\mu_s$ is sampled from a Poisson
distribution
$p_5(\mu_s) =  \mu^{-1} {\rm exp}(-\mu_s/\mu_o)$, with $\mu_s$ the
average number of spontaneous pulses per GRB.
Lastly, the probability distribution of the time constants $\tau_0$
of spontaneous pulses is taken as $p(\tau_0) \propto 1/\tau_0$,
corresponding to a 1/f flicker noise spectrum.
Observations imply an maximum $\tau_{\rm max}$ for $\tau_0$.
We then sample ${\rm log}\tau_0$ uniformly between $\tau_{\rm min}$ and
$\tau_{\rm max}$, i.e.
$p_6({\rm log}\tau_0) = ({\rm log}\tau_{max} - {\rm log}\tau_{\rm min})^{-1}$,
where $\tau_{\rm min}$ should be smaller than the time resolution.
Varying $\tau_{\rm max}$ rescales all average avalanche properties in time.
Since more than one parent pulse is allowed per GRB, these spontaneous
primary pulses are all assumed to be delayed with different time intervals
$t$ with respect to a common invisible trigger event. We parameterize
the probability distribution for the Poisson delay $t$ of a given
spontaneous pulse as
$p_7(t) = (\alpha \tau_0)^{-1} {\rm exp}(-t/\alpha \tau_0)$, where $\alpha$
is the constant delay parameter used for all pulses and $\tau_0$
is the time constant of the spontaneous pulse.
Each spontaneous pulse gives rise to a pulse avalanche, and it is the
overlap of $\mu_s$ pulse avalanches that form a GRB.
From the analysis of about 600 CGRO/BATSE GRBs, \cite{SternSvensson1996}
suggest the following parameters: 
$\mu = 1.2$, $\alpha = 4$, $\delta_1 = -0.5$, $\delta_2 = 0$,
$\mu_0 = 1$, $\tau_{\rm max} = 26$ s. Differently than
\cite{SternSvensson1996}, we pick $\tau_{\rm min} = 0.2$ s
to allow for a better time resolution for the Galileo detector.
We do simulate short and long GRBs separately to better tune some
of the $\tau_i$ parameters, see example light curves in Fig. \ref{example-lc}.

\section{Location accuracy}

\begin{table*}[htb]
  \centering
  \vspace{-0.3cm}
  \caption{Average localisation accuracy for different detector geometries
    on a differing number of satellites.  \label{tab:Det_Sat_acc}}
  \begin{tabular}{lcccccc}
    \hline
    \noalign{\smallskip}
    Detector geometry & No. of & Orbit configuration & \multicolumn{4}{c}{Error radius} \\
                      & Satellites &  & \multicolumn{4}{c}{(1$\sigma$, deg)} \\
    \hline
    \noalign{\smallskip}
                      &               &   & 2.3--3 & 3--4.6 & 4.6--9.2 & 9.2--154 \\
                      &               &  & \multicolumn{4}{c}{ (10$^{-7}$ erg/cm$^2$/s)} \\
    \noalign{\smallskip}
    \hline
    \noalign{\smallskip}
    01 -- zenith-looking & 24 & 11111111 11111111 11111111 & 26.2 & 11.7 & 3.9 & 1.0 \\
                         & 12 & 10101010 10101010 10101010 & 63  & 31  & 11.0 & 2.9 \\
    02 -- Cube at zenith & 24 & 11111111 11111111 11111111 & 5.1 & 1.6 & 0.8 & 0.4 \\
                         & 12 & 10101010 10101010 10101010 & 10.3 & 3.2 & 1.6& 0.7 \\
    03 -- 4 lateral sides& 24 & 11111111 11111111 11111111 & 1.9 & 0.7 & 0.4 & 0.2 \\
                   & 12 & 10101010 10101010 10101010 & 4.2 & 1.4 & 0.8 & 0.4 \\
                   & 9 & 10010010 10010010 10010010 & 5.4 & 1.8 & 1.0 & 0.5 \\
                   & 9 & 10010010 01001001 10100100 & 5.4 & 1.8 & 1.0 & 0.5 \\
                   & 6 & 10001000 10001000 10001000 & 12.9 & 4.1 & 2.2 & 1.0 \\
                   & 6 & 10001000 00100010 10001000 & 24.6 & 8.2 & 4.4 & 2.2 \\
                   & 6 & 10010000 00010010 10010000 & 11.0 &3.5 & 2.0 & 1.0 \\
    04 -- 2 neighbouring sides & 24 & 11111111 11111111 11111111 & 28.2 & 16.2 & 6.5 & 1.9 \\
                   & 12 & 10101010 10101010 10101010 & 57 & 34.4 & 14.7 & 4.4 \\
    05 -- 5 sides  & 24 & 11111111 11111111 11111111 & 3.2 & 1.1 & 0.5 & 0.2 \\
                   & 12 & 10101010 10101010 10101010 & 10.5 & 4.5 & 1.4 & 0.4 \\
                   & 9 & 10010010 10010010 10010010 & 14.5 & 6.2 & 2.0 &0.6 \\
                   & 9 & 10010010 01001001 10100100 & 14.7 & 6.3 & 2.0 &0.6 \\
                   & 6 & 10001000 10001000 10001000 & 27.9& 12.2 &3.8 &1.1 \\
                   & 6 & 10001000 00100010 10001000 & 45& 20.6 &6.5 &1.8 \\
                   & 6 & 10010000 00010010 10010000 & 28.2& 12.1& 3.7&1.0 \\
    06 -- 2 opposite sides \& zenith & 24 & 11111111 11111111 11111111 & 4.5 & 1.5 & 0.8 & 0.3 \\
                   & 12 & 10101010 01010101 10101010 & 7.9 & 2.5 & 1.3 & 0.6 \\
    08 -- 1 side \& zenith & 24 & 11111111 11111111 11111111 & 15.7 & 6.5 & 2.4 & 0.7 \\
    09 -- 2 opposite sides & 24 & 11111111 11111111 11111111 & 12.0 & 5.7 & 1.8 & 0.5 \\
    \noalign{\smallskip}
    \hline
    \end{tabular}
\end{table*}    

\begin{longtable}{ccccccccc}
  \caption{Localisation accuracy for selected detector geometries
    according to fractional sky area. Brightness bins are according to
    Tab. \ref{tab:12Det_det1_acc},
    with 1 the faintest, and 4 the brightest. \label{tab:acc_final}}\\
    \hline
    \noalign{\smallskip}
    $\!\!\!$Name$\!\!\!$ & Detector geometry & $\!\!$No. of$\!\!\!\!$     & Orbit configuration & $\!\!\!$Intens.$\!\!\!$& \multicolumn{4}{c}{precision $<$p[\degs] for \% of sky} \\
          &         & Sats &      & Intvl  & best & 50\% & 90\% & worst$\!\!\!\!$ \\
    \noalign{\smallskip}
    \hline
    \noalign{\smallskip}
    \endfirsthead
    \caption{continued.}\\
    \hline
    \noalign{\smallskip}
    $\!\!\!$Name$\!\!\!$ & Detector geometry & $\!\!$No. of$\!\!\!\!$     & Orbit configuration & $\!\!\!$Intens.$\!\!\!$& \multicolumn{4}{c}{precision $<$p[\degs] for \% of sky} \\
          &         & Sats &      & Intvl  & best & 50\% & 90\% & worst$\!\!\!\!$ \\
    \noalign{\smallskip}
    \hline
    \noalign{\smallskip}
    \endhead
    \hline
    \endfoot
    01 & zenith        &24 & 11111111 11111111 11111111 & 1 & 21.3~~& 26.2~~& 28.6~~& 32.8~~ \\
     &               &24 & 11111111 11111111 11111111 & 2 & 9.1   & 11.7~~& 13.0~~& 14.4~~ \\
     &               &24 & 11111111 11111111 11111111 & 3 & 2.9   & 3.8   & 4.6   & 5.4 \\
     &               &24 & 11111111 11111111 11111111 & 4 & 0.7   & 1.0   & 1.2   & 1.3 \\
  01 & zenith        &12 & 10101010 10101010 10101010 & 1 & 34.5~~& 57.2~~& 87.8~~& 180\\
     &               &12 & 10101010 10101010 10101010 & 2 & 17.0  & 26.2~~& 47.8~~& 180\\
     &               &12 & 10101010 10101010 10101010 & 3 & 4.8   & 8.6   & 17.9~~& 180\\
     &               &12 & 10101010 10101010 10101010 & 4 & 1.0   & 2.2   & 4.4   & 180\\
  02 & cube @ zenith &24 & 11111111 11111111 11111111 & 1 & 4.1 & 5.1 & 5.4 & 7.3 \\
     &               &24 & 11111111 11111111 11111111 & 2 & 1.4 & 1.6 & 1.6 & 2.4 \\
     &               &24 & 11111111 11111111 11111111 & 3 & 0.7 & 0.8 & 0.9 & 1.1 \\
     &               &24 & 11111111 11111111 11111111 & 4 & 0.3 & 0.4 & 0.4 & 0.4 \\
  02 & cube @ zenith &12 & 10101010 10101010 10101010 & 1 & 6.9 & 10.1~~& 11.9~~& 18.2~~\\
     &               &12 & 10101010 10101010 10101010 & 2 & 2.5 & 3.1   & 3.6   & 5.3 \\
     &               &12 & 10101010 10101010 10101010 & 3 & 1.1 & 1.6   & 1.8   & 2.4\\
     &               &12 & 10101010 10101010 10101010 & 4 & 0.5 & 0.7   & 0.8   & 0.9\\
  03 & 4 lateral sides &24 & 11111111 11111111 11111111 & 1 & 1.4 & 1.9 & 2.4 & 4.4 \\
     &                 &24 & 11111111 11111111 11111111 & 2 & 0.5 & 0.6 & 0.8 & 1.2 \\
     &                 &24 & 11111111 11111111 11111111 & 3 & 0.3 & 0.4 & 0.4 & 0.5 \\
     &                 &24 & 11111111 11111111 11111111 & 4 & 0.2 & 0.2 & 0.2 & 0.2 \\
  03 & 4 lateral sides &12 & 10101010 10101010 10101010 & 1 & 2.6 & 3.8 & 5.5 & 10.6~~ \\
     &                 &12 & 10101010 10101010 10101010 & 2 & 1.0 & 1.3 & 1.7 &  2.5 \\
     &                 &12 & 10101010 10101010 10101010 & 3 & 0.5 & 0.8 & 0.9 &  1.2 \\
     &                 &12 & 10101010 10101010 10101010 & 4 & 0.4 & 0.4 & 0.4 &  0.5 \\
  03 & 4 lateral sides & 9 & 10010010 10010010 10010010 & 1 & 3.2 & 5.0 & 7.0 & 13.8~~\\
     &                 & 9 & 10010010 10010010 10010010 & 2 & 1.2 & 1.7 & 2.2 & 3.5 \\
     &                 & 9 & 10010010 10010010 10010010 & 3 & 0.6 & 1.0 & 1.2 & 1.7 \\
     &                 & 9 & 10010010 10010010 10010010 & 4 & 0.5 & 0.5 & 0.6 & 0.7 \\
  03 & 4 lateral sides & 9 & 10010010 01001001 10100100 & 1 & 3.2 & 5.2 & 7.1 & 16.0~~\\
     &                 & 9 & 10010010 01001001 10100100 & 2 & 1.1 & 1.8 & 2.2 & 3.6 \\
     &                 & 9 & 10010010 01001001 10100100 & 3 & 0.6 & 1.0 & 1.2 & 1.6 \\
     &                 & 9 & 10010010 01001001 10100100 & 4 & 0.5 & 0.5 & 0.6 & 0.7 \\
 03 & 4 lateral sides & 9 & 10010010 10010010 10010010 & 1 & 3.2 & 5.0 & 7.0 & 13.8~~ \\
    &                 & 9 & 10010010 10010010 10010010 & 2 & 1.2 & 1.7 & 2.2 &  3.5 \\
    &                 & 9 & 10010010 10010010 10010010 & 3 & 0.6 & 1.0 & 1.2 &  1.7 \\ 
    &                 & 9 & 10010010 10010010 10010010 & 4 & 0.5 & 0.5 & 0.6 &  0.7 \\
 03 & 4 lateral sides & 9 & 10010010 01001001 10100100 & 1 & 3.2 & 5.2 & 7.1 & 16.0~~ \\
    &                 & 9 & 10010010 01001001 10100100 & 2 & 1.1 & 1.8 & 2.2 &  3.6 \\
    &                 & 9 & 10010010 01001001 10100100 & 3 & 0.6 & 1.0 & 1.2 &  1.6 \\
    &                 & 9 & 10010010 01001001 10100100 & 4 & 0.5 & 0.5 & 0.6 &  0.7 \\
 03 & 4 lateral sides & 6 & 10010000 00010010 10010000 & 1 & 4.2 & 9.7 &17.5~~& 109 \\
    &                 & 6 & 10010000 00010010 10010000 & 2 & 1.6 & 3.2 & 5.3 &  39.2~ \\
    &                 & 6 & 10010000 00010010 10010000 & 3 & 0.8 & 1.8 & 2.8 &  21.1~ \\
    &                 & 6 & 10010000 00010010 10010000 & 4 & 0.7 & 0.9 & 1.3 &   9.6 \\
 04 & 2 neighbouring sides& 24& 11111111 11111111 11111111 & 1 & 3.5 &12.2~~&80.5~~& 180 \\
    &                     & 24& 11111111 11111111 11111111 & 2 & 1.1 & 4.5 & 41 & 180 \\
    &                     & 24& 11111111 11111111 11111111 & 3 & 0.5 & 1.9 & 11.8~~& 180\\
    &                     & 24& 11111111 11111111 11111111 & 4 & 0.2 & 0.6 & 3.0 & 180\\
 04 & 2 neighbouring sides& 12& 10101010 10101010 10101010& 1 & 6.4 &27.7~~&180 & 180 \\
    &                     & 12& 10101010 10101010 10101010& 2 & 2.2 &11.3~~&109 & 180 \\
    &                     & 12& 10101010 10101010 10101010& 3 & 0.9 & 4.1 & 32.3~~& 180 \\
    &                     & 12& 10101010 10101010 10101010& 4 & 0.4 & 1.1 & 8.1 & 180 \\
 05 & 4 sides + zenith &24& 11111111 11111111 11111111 & 1 & 1.9 & 3.0 & 3.7 & 5.9 \\ 
    &                  &24& 11111111 11111111 11111111 & 2 & 0.7 & 1.1 & 1.3 & 2.0 \\
    &                  &24& 11111111 11111111 11111111 & 3 & 0.4 & 0.5 & 0.6 & 0.7 \\
    &                  &24& 11111111 11111111 11111111 & 4 & 0.2 & 0.2 & 0.2 & 0.2 \\
 05 & 4 sides + zenith &12& 10101010 10101010 10101010 & 1 & 8.3 &10.3~~&12.1~~&15.7~~\\ 
    &                  &12& 10101010 10101010 10101010 & 2 & 3.7 & 4.5 & 5.1 & 7.1 \\
    &                  &12& 10101010 10101010 10101010 & 3 & 1.1 & 1.4 & 1.6 & 2.0 \\
    &                  &12& 10101010 10101010 10101010 & 4 & 0.4 & 0.4 & 0.4 & 0.5 \\
 05 & 4 sides + zenith & 9& 10010010 10010010 10010010 & 1 &10.1~~&14.3~~&16.8~~&25.0~~\\
    &                  & 9& 10010010 10010010 10010010 & 2 & 4.5 & 6.1 & 6.4 & 10.7~~\\
    &                  & 9& 10010010 10010010 10010010 & 3 & 1.5 & 1.9 & 2.0 & 2.9 \\
    &                  & 9& 10010010 10010010 10010010 & 4 & 0.5 & 0.6 & 0.6 & 0.8 \\
 05 & 4 sides + zenith & 9& 10010010 01001001 10100100 & 1 & 10.0~~&14.5~~&17.4~~&26.0~~\\
    &                  & 9& 10010010 01001001 10100100 & 2 & 4.5 & 6.2 & 7.4 & 11.5~~\\
    &                  & 9& 10010010 01001001 10100100 & 3 & 1.5 & 2.0 & 2.3 & 3.0 \\
    &                  & 9& 10010010 01001001 10100100 & 4 & 0.5 & 0.6 & 0.6 & 0.7 \\
 05 & 4 sides + zenith & 6& 10001000 10001000 10001000 & 1 &13.2~~&23.9~~&38.4~~& 180 \\
    &                  & 6& 10001000 10001000 10001000 & 2 & 6.0 & 10.2~~&16.5~~& 180 \\
    &                  & 6& 10001000 10001000 10001000 & 3 & 1.8 & 3.2 & 5.0 & 180 \\
    &                  & 6& 10001000 10001000 10001000 & 4 & 0.6 & 0.9 & 1.4 & 180 \\
 05 & 4 sides + zenith & 6& 10001000 00100010 10001000 & 1 &13.0~~&30.4~~&92.8~~& 180 \\
    &                  & 6& 10001000 00100010 10001000 & 2 &5.9 & 13.0~~&40.7~~& 180 \\
    &                  & 6& 10001000 00100010 10001000 & 3 &1.6 & 3.9 & 12.4 & 180 \\
    &                  & 6& 10001000 00100010 10001000 & 4 &0.6 & 1.1 & 3.5 & 180 \\
 05 & 4 sides + zenith & 6& 10010000 00010010 10010000 & 1 &12.0~~&26.2~~&39.5~~& 180 \\
    &                  & 6& 10010000 00010010 10010000 & 2 & 5.3 & 11.2~~&16.8~~& 110 \\
    &                  & 6& 10010000 00010010 10010000 & 3 & 1.7 & 3.9 & 5.2 & 39.8~~ \\
    &                  & 6& 10010000 00010010 10010000 & 4 & 0.7 & 1.1 & 1.4 & 11.7~~ \\
 06 & $\!\!\!\!$2 opposite sides+zenith$\!\!$&24 & 11111111 11111111 11111111 & 1 & 2.2 & 4.7 & 5.9 & 6.9 \\
    &                                        &24 & 11111111 11111111 11111111 & 2 & 0.8 & 1.5 & 1.9 & 2.4 \\
    &                                        &24 & 11111111 11111111 11111111 & 3 & 0.4 & 0.8 & 1.0 & 1.2 \\
    &                                        &24 & 11111111 11111111 11111111 & 4 & 0.2 & 0.4 & 0.4 & 0.6 \\
 06 & $\!\!\!\!$2 opposite sides+zenith$\!\!$&12 & 10101010 01010101 10101010 & 1 & 4.5 &10.5~~ &17.4~~ & 38.4~ \\
    &                                        &12 & 10101010 01010101 10101010 & 2 & 1.5 & 3.2 & 5.3 & 17.1~ \\
    &                                        &12 & 10101010 01010101 10101010 & 3 & 0.8 & 1.6 & 2.6 &  9.3 \\ 
    &                                        &12 & 10101010 01010101 10101010 & 4 & 0.5 & 0.7 & 1.1 &  6.6 \\
 07 & 1 side          & 24 & 11111111 11111111 11111111 & 1 & 4.0 &45.6~~& 180 & 180 \\ 
    &                 & 24 & 11111111 11111111 11111111 & 2 & 1.8 &23.8~~& 180 & 180 \\ 
    &                 & 24 & 11111111 11111111 11111111 & 3 & 0.7 & 7.0 &  180 & 180 \\ 
    &                 & 24 & 11111111 11111111 11111111 & 4 & 0.2 & 1.8 &  180 & 180 \\ 
 08 & 1 side + zenith & 24 & 11111111 11111111 11111111 & 1 & 2.2 & 14.0~~& 29.4~~& 39.5~~\\
    &                 & 24 & 11111111 11111111 11111111 & 2 & 0.8 & 5.3 & 13.0~~& 16.7~~\\
    &                 & 24 & 11111111 11111111 11111111 & 3 & 0.4 & 2.3 & 4.5 & 6.2 \\
    &                 & 24 & 11111111 11111111 11111111 & 4 & 0.2 & 0.7 & 1.2 & 1.6 \\
 09 & 2 opposite sides& 24 & 11111111 11111111 11111111 & 1 & 4.0 & 10.6~~&21.5~~& 24.6~~\\
    &                 & 24 & 11111111 11111111 11111111 & 2 & 2.0 & 5.7 & 10.9~~&13.3~~\\
    &                 & 24 & 11111111 11111111 11111111 & 3 & 0.7 & 1.5 & 3.3 & 4.3 \\
    &                 & 24 & 11111111 11111111 11111111 & 4 & 0.2 & 0.4 & 0.8 & 1.2 \\
 \noalign{\smallskip}
    \hline
\end{longtable}

\end{appendix}

\end{document}